\documentclass[final,3p,times]{elsarticle}

\journal{Physics Reports}

\usepackage{amsmath,amsthm,amssymb}
\usepackage{bbold}
\usepackage{mathtools}
\usepackage{slashed}
\usepackage{multicol}
\usepackage{blkarray}
\usepackage{enumerate}
\usepackage{graphicx}
\usepackage{pdfpages}
\usepackage{comment}
\usepackage{booktabs}
\usepackage{subfig}
\usepackage{nccmath} 
\usepackage{etoolbox}
\usepackage{hyperref} 
\usepackage{array}
\usepackage{url} 
\usepackage{array}
\usepackage{epsfig,multicol,bbm}
\usepackage{colortbl}
\usepackage{color}
\usepackage{tcolorbox}
\usepackage{xcolor}
\usepackage{empheq}
\newcommand*\widefbox[1]{\fbox{\hspace{2em}#1\hspace{2em}}}
\usepackage{placeins}

\newcommand{\comments}[1]{}
\newcommand{\ba}{\begin{align}}
\newcommand{\ea}{\end{align}}

\newcommand{\vo}{\mathcal{V}}

\def\flux{{\scriptscriptstyle {\rm flux}}}
\newcommand{\mc}{\mathcal}

\def\be{\begin{equation}}
\def\ee{\end{equation}}
\def\bea{\begin{eqnarray}}
\def\eea{\end{eqnarray}}

\def\ni{\noindent}
\def\bi{\begin{itemize}}
\def\ei{\end{itemize}}
\def\ben{\begin{enumerate}}
\def\een{\end{enumerate}}

\def\Mp{M_{\rm Pl}}

\def\lp{\left(}
\def\rp{\right)}
\def\lb{\left[}
\def\rb{\right]}

\def\cN{\mathcal{N}}
\def\cR{\mathcal{R}}
\def\cC{\mathcal{C}}
\def\cP{\mathcal{P}}
\def\red{\textcolor{red}}
\usepackage{colortbl}
\definecolor{green2}{cmyk}{0, 1, 0.5, 0}
\definecolor{lightgreen}{cmyk}{0.2, 0, 0.2, 0.2}
\definecolor{lightgray}{cmyk}{0.1,0.2,0,0.1}
\definecolor{lightgray2}{cmyk}{0.4,0.4,0,0.8}
\definecolor{black}{cmyk}{1.0,1.0,1.0,1.0}
\usepackage[makeroom]{cancel}
\usepackage{booktabs} 
\usepackage{float} 
\usepackage{graphicx} 
\usepackage{epstopdf} 
\usepackage{paralist} 
\usepackage[framemethod=TikZ]{mdframed}

\makeatletter
\patchcmd{\upbracefill}{\m@th}{\scriptstyle\m@th}{}{}
\patchcmd{\upbracefill}{$\braceld$}{$\scriptstyle\braceld$}{}{}
\patchcmd{\upbracefill}{\bracelu}{\bracelu\mkern-1mu}{}{}
\patchcmd{\upbracefill}{\hfill\braceru}{\hfill\mkern-1mu\braceru}{}{}
\makeatother

\begin{document}

\begin{frontmatter}

\title{{\bf String Cosmology: from the Early Universe to Today}}

\author[bol,infn]{\small Michele Cicoli}
\ead{michele.cicoli@unibo.it}

\author[ox]{\small Joseph P. Conlon }
\ead{joseph.conlon@physics.ox.ac.uk}

\author[alla]{\small Anshuman Maharana}
\ead{anshumanmaharana@hri.res.in}

\author[liv]{\small Susha Parameswaran}
\ead{Susha.Parameswaran@liverpool.ac.uk}

\author[cam]{\small Fernando Quevedo}
\ead{F.Quevedo@damtp.cam.ac.uk}

\author[swan]{\small Ivonne Zavala\corref{corriz}}
\cortext[corriz]{Corresponding author}
\ead{e.i.zavalacarrasco@swansea.ac.uk}

\address[bol]{%
        Dipartimento di Fisica e Astronomia, Universit\`a di Bologna, via Irnerio 46, 40126 Bologna, Italy.%
}

\address[infn]{INFN, Sezione di Bologna, viale Berti Pichat 6/2, 40127 Bologna, Italy.}

\address[ox]{
        Rudolf Peierls Centre for Theoretical Physics, Beecroft Building, Clarendon Laboratory, University of Oxford, Parks Road, OX1 3PU, UK
        }

\address[alla]{
        Harish Chandra Research Institute, A CI of Homi Bhabha National Institute, Chattnag Road, Jhunsi, Allahabad - 211019, India
}

\address[liv]{
        Department of Mathematical Sciences, University of Liverpool, Liverpool L69 7ZL
}

\address[cam]{
        DAMTP, University of Cambridge, Wilberforce Road, Cambridge, CB3 0WA, UK
}

\address[swan]{
        Department of Physics, Swansea University, Swansea, SA2 8PP, UK
}

\begin{abstract}
We review applications of string theory to cosmology, from primordial times to the present-day accelerated expansion. Starting with a brief overview of cosmology and string compactifications, we discuss in detail moduli stabilisation, inflation in string theory, the impact of string theory on post-inflationary dynamics (reheating, moduli domination, kination), dark energy (the cosmological constant from a string landscape and models of quintessence) and various alternative scenarios (string/brane gases, the pre big-bang scenario, rolling tachyons, ekpyrotic/cyclic cosmologies, bubbles of nothing, S-brane and holographic cosmologies). The state of the art in string constructions is described in each topic and, where relevant, connections to swampland conjectures are made. The possibilities for novel particles and excitations (axions, moduli, cosmic strings, branes, solitons, oscillons and boson stars) are emphasised. Implications for the physics of the CMB,  gravitational waves, dark matter and dark radiation  are discussed along with potential observational signatures.
\end{abstract}


\end{frontmatter}

\tableofcontents

\clearpage

\section{Introduction}

We are living in a golden age for cosmology. The exquisite precision with which the power spectrum of density perturbations of the cosmic microwave background (CMB) has been measured over the past 25 years is simply spectacular. The surprising discovery of the accelerated expansion of the universe in the present epoch, also around  25 years ago, has given rise to arguably the biggest puzzle in physics: dark energy. This, combined with ever improving observations of the large-scale structure of the universe and the compelling evidence for the existence of dark matter,  has made cosmology a precision science dominated by big data at all scales. The standard model of cosmology, $\Lambda$CDM, 
has only a handful of parameters but provides an accurate match to most observations. A successful scenario, inflation, has emerged as the standard description of early universe cosmology that addresses the main puzzles of the Big Bang model (flatness, horizon, monopole problems) and, most importantly, provides the seed for the density perturbations imprinted in the CMB. Its theoretical predictions fit remarkably well with observations.

However, unlike the Standard Model of particle physics, which is a well defined theory with concrete predictions, the standard model of cosmology, including inflation, is not based on an underlying theory. This is a fundamental  issue given  that early universe cosmology involves  temperatures and energy scales which are  higher than those probed in our  laboratories. At these scales, we do not have a complete theory. Furthermore, in contrast to particle physics, gravity cannot be neglected when addressing questions in cosmology. Therefore, in order to address the physics of the early universe, we need to have a well defined theory of gravity which also includes all other interactions. There are also puzzles at the longest wavelengths.
Formulating quantum mechanics in an accelerating spacetime (such as the present universe) is subtle as it is tied to various conceptual issues of quantum gravity.  Over more than 35 years, string theory has emerged as the most promising candidate for providing a consistent quantum 
 theory combining gravity with all other particles and interactions. Yet, string theory still lacks concrete predictions that can be tested experimentally with today's technology   and needs to be developed further so that it can be
confronted with potential observations in the not too distant future. Establishing the connections between string theory and cosmology is therefore one of primary importance for fundamental physics

Not surprisingly, there have been many attempts to extract information from string theory regarding its potential cosmological implications. This is not an easy task as our understanding of string theory is still incomplete. It is not yet possible to answer questions tied to the Big Bang singularity in the
context of string theory. Nevertheless, there are many cosmological questions that can be addressed by string theory. In particular, deriving models of cosmological inflation from string theory is a difficult but achievable task, as is the physics from the end of inflation to the present epoch which
covers a range of energies and temperatures many orders of magnitude and may, in a logarithmic scale, correspond to up to half of the expansion of the universe.  In addition, string theory can lead to various exotic phases in the (early) universe or consistency conditions that have  distinct implications for cosmology. These {\it alternatives} are the least understood but some of  the  most exciting directions to explore in string cosmology.

String cosmology is a natural meeting point for many disciplines. String theory has a large number of degrees of freedom in addition to those associated with the Standard Model and gravity. Many of these can be  light and of direct relevance for cosmology. Of particular importance are {\it moduli} -- the fields that control the shape and size of the extra dimensions, thereby setting the magnitude
 of couplings in the 4-dimensional effective field theory. Moduli are also ideal candidates for inflatons and often acquire non-trivial time dependence in post-inflationary string cosmology leading to differences from the standard cosmological timeline. In the present epoch, all moduli must be pinned to their minima or be very slowly rolling. Thus, whether it be the  early universe or the present epoch, understanding the potential energy functional for moduli fields and their dynamics is central to string cosmology. Therefore, string cosmology requires a deep  knowledge of string compactifications (dimensional reduction and  derivation of low energy effective actions that
 arise from string theory) -- this  involves  many aspects of modern  mathematics. The study of
 alternatives often requires  understanding string theory in novel regimes and has the potential to provide an  answer to the question:  What is string theory? Furthermore, addressing  questions such as the dimensionality of spacetime and, as most theorists believe, how spacetime itself may emerge from a fundamental theory, may lie in the domain of string cosmology. 
 
 Finally, from the point of view of a pragmatic cosmologist, string theory can be thought of as a black box which continually generates interesting models and scenarios. These have served as 
 a useful driver for both theory and observational targets in cosmology.  String cosmology thus brings together many areas and its study is not only central to our understanding of fundamental physics but also for advances in these areas.
 
This review aims to give a concise overview of the state of the art in  the subject. It is structured as  follows. Sec. 2 provides a brief review of
cosmology. After quickly going through Freedman-Lemaitre-Robertson-Walker (FLRW) cosmology and the history of the universe in the standard model of cosmology, we describe the physics of inflation. We discuss how inflation provides a theory for generating inhomogeneities in the universe, thereby allowing  it to connect to precision observational cosmology. We also introduce 
quintessence -- the possibility that the acceleration of the present universe is due to a slowly rolling scalar field. 

Sec. 3 deals with string compactifications and moduli fields. After providing a general overview of moduli, we discuss moduli stabilisation in various string theories. A summary of various scenarios to obtain de Sitter space (as a model of the universe in the present epoch) is provided, emphasising the general achievements and challenges. 

Sec. 4 is on inflation in string theory.  Here, we begin by describing why it is necessary to embed models of inflation
into theories of quantum gravity and the challenges for inflation in string theory. We present a list of well-established string theoretic models of inflation classified according to the form of their potential. We also give a ``report card" in the form of a table which shows how each of these models fares when confronted with observations. 

Sec. 5 is on the post-inflationary epoch between the end of inflation and the start of the Hot Big Bang. We start by discussing reheating in the context of string cosmology and identify the cosmological moduli problem, which is a generic outcome of string cosmology. We go on to describe modifications of the standard cosmological timeline (such as epochs of {\it moduli domination} and {\it kination}) which are natural in string cosmology. Opportunities and challenges in the context of dark matter and dark radiation are summarised and concrete sources of inhomogeneities and gravitational waves are identified, such as {\it oscillons}.

Sec. 6 is on dark energy (the present day constituent of the energy budget of the universe that is driving acceleration) in string theory. It is divided into two main parts -- dark energy arising from a cosmological constant term (a de Sitter solution) and dynamical dark energy (quintessence). In the
first part, after describing the cosmological constant problem we discuss how the {\it string landscape} (an enormously large number of string vacua with finely spaced values of the cosmological constant) offers a potential, if controversial, solution to the problem. In the second part, interesting avenues to construct models of quintessence in string theory are discussed and the associated challenges are outlined.

Sec. 7 deals with alternatives to the standard cosmology. We discuss string gas cosmology, the ekpyrotic/cyclic universe, rolling tachyon cosmology, pre-Big Bang cosmology, S-branes, holographic models and  models including creation or decay to nothing. This section also discusses  the {\it swampland} approach, which aims at determining consistency conditions (and their physical implications) that an effective field theory must satisfy so that it can be embedded in string theory or any theory of quantum gravity. We conclude in Sec. 8.

 \newpage

\section{Cosmology}
\label{SecCO}

The dynamics of our universe is described by Einstein equations in the presence of matter. The Friedmann-Lemaitre-Robertson-Walker (FLRW) metric describing the evolution of the Universe  is based upon the assumption of homogeneity and isotropy, which is approximately true on large scales. These assumptions  determine the metric up to an arbitrary function of time, $a(t)$, the scale factor, which measures the time evolution of the Universe, and a discrete parameter $k = -1,0,1$, which determines the 3-dimensional curvature of the Universe, namely whether it is respectively open, flat or closed.

Small deviations from homogeneity at early epochs played a very important role in the dynamical history of our universe. Small initial density perturbations grew via gravitational instability into the structures that we observe today in the universe. The temperature anisotropies observed in the Cosmic Microwave Background (CMB) are believed to have originated from quantum fluctuations generated during an inflationary stage in the early universe, which we review in Sec. \ref{sec:infla}. In this section we review the main features of the homogeneous and isotropic cosmology necessary for the subsequent sections. For dedicated accounts of the standard $\Lambda$CDM cosmology and the growth of cosmic structure, we also refer the reader to e.g. \cite{Dodelson:2003ft, Weinberg:2008zzc, Mukhanov:2005sc, Baumann:2022mni}. More technical summaries of recent progress and challenges can be found in  \cite{Chou:2022luk, Green:2022hhj, Flauger:2022hie}.
\bigskip

The FLRW metric can be written as:
\be\label{eq:FLRW}
\setlength\fboxsep{0.25cm}
\setlength\fboxrule{0.4pt}
\boxed{
ds^2 = -dt^2 + a(t) \left[dr^2 + f^2_k(r) \left(d\theta^2 +\sin^2{\theta}\,d\phi^2\right) \right] \,,
}
\ee
where
\begin{equation*}
f_k(r) =  \left\{
\begin{array}{rl}
\sin r  & \text{if }\,\, k = +1 ,\\
r & \text{if }\,\, k = 0,\\
\sinh r & \text{if } \,\, k =-1.
\end{array} \right.
\end{equation*}
The dynamics associated with the scale factor $a(t)$ is determined by Einstein's equations:
\be\label{eq:Einstein4D}
\setlength\fboxsep{0.25cm}
\setlength\fboxrule{0.4pt}
\boxed{
G_{\mu\nu} \equiv R_{\mu\nu} -\frac{1}{2} g_{\mu\nu} R = 8\pi\, G_4 \, T_{\mu\nu}\,,
}
\ee
 provided the matter content encoded in the energy-momentum tensor $T_{\mu\nu}$  is specified. Let us consider an ideal perfect fluid as the source of
the energy momentum tensor. In this case we have:
\be\label{eq:EMT4D}
 T^\mu_{\,\,\nu} = {\rm diag}\left(-\rho, p, p, p\right)\,,
 \ee
 where  $\rho$ and  $p$ are the {\em energy density} and {\em pressure} of the fluid, respectively.

Einstein's equations for the metric \eqref{eq:FLRW} and energy-momentum tensor \eqref{eq:EMT4D} give the two independent equations:
\bea
H^2 &=& \frac{8\pi G_4}{3} \rho - \frac{k}{a^2} \,, \label{eq:Fried4DH} \\
\frac{\ddot a}{a} &=& -\frac{4\pi G_4}{3} \left(\rho + 3 \,p\right) \label{eq:Fried4Da}\,,
\eea
 where $H$ is the Hubble parameter (function), defined as
 \be\label{eq:Hdef}
 \setlength\fboxsep{0.25cm}
\setlength\fboxrule{0.4pt}
\boxed{
 H \equiv \frac{\dot a}{a}\,.
 }
 \ee
 The energy momentum tensor is conserved by virtue of the Bianchi identities, $\nabla_\mu T^{\mu\nu} =0$, leading to the continuity equation
 \be\label{eq:EMconsFLRW}
 \setlength\fboxsep{0.25cm}
\setlength\fboxrule{0.4pt}
\boxed{
 \dot\rho + 3H (\rho+p) =0 \,,
 }
 \ee
 which can be derived also from  Einstein's equations above, \eqref{eq:Fried4DH}, \eqref{eq:Fried4Da}. Notice already that eq. \eqref{eq:Fried4Da} implies that in order to have {\em accelerated expansion}, that is $\ddot a>0$, the energy density and pressure must be such that
 \be
 \setlength\fboxsep{0.25cm}
\setlength\fboxrule{0.4pt}
\boxed{
\label{eq:rho3p}
 (\rho+3p) <0\,.
}
\ee

 One can write eq.~\eqref{eq:Fried4DH} in the form
 \be
\label{eq:Omega4D}
 \Omega(t) -1 = \frac{k}{(aH)^2}\,,
 \ee
where we defined the dimensionless density parameter
\be
\setlength\fboxsep{0.25cm}
\setlength\fboxrule{0.4pt}
\boxed{
\label{eq:Omegadef}
\Omega(t) \equiv \frac{\rho(t)}{\rho_c(t)} \,, \qquad \rho_c(t) \equiv \frac{3H^2(t)}{8\pi G_4}\,,}
\ee
with $\rho_c$ the critical density. From here we can see that the matter distribution determines the spatial geometry of our universe:  
  \begin{subequations}\label{eq:OmegaGeo}
\begin{empheq}[box=\widefbox]{align}
 &\Omega >1 \quad {\rm or } \quad  \rho>\rho_c \quad  \Rightarrow \,\,\, k = +1\,, \nonumber \\
 &\Omega =1 \quad {\rm or } \quad \rho=\rho_c \quad   \Rightarrow \,\,\, k = 0\,, \nonumber \\
 & \Omega <1 \quad {\rm or } \quad \rho<\rho_c \quad  \Rightarrow \,\,\, k = -1\,.
 \end{empheq}
\end{subequations}

Observations indicate that the current universe is very close to a spatially flat geometry \cite{Planck:2018vyg}. This is actually a natural result from inflation in the early universe (see below). Hence, in this section we consider a flat universe ($k =0$, $\Omega \simeq 1$). But we will keep an open mind regarding the spatial curvature when we discuss string constructions.

\subsection{Evolution of the Universe Filled with a Perfect Fluid}

Let us now consider the evolution of the universe filled with a barotropic perfect fluid with an equation of state of the form
 \be
	\label{eq:eos}
 p = \omega\, \rho\,,
 \ee
 where $\omega$ is  a constant when the perfect fluid corresponds to  matter, radiation, and vacuum domination (see Tab. \ref{tab:scalef_evolution}).

Using the equation of state we can solve Einstein's equations to obtain (for $\omega\ne-1$)
\bea\label{eq:FLRWsols}
&& H = \frac{2}{3(1+\omega)(t-t_0)} \,, \\
&& a(t) \propto (t-t_0)^{\frac{2}{3(1+\omega)}} \,, \\
&& \rho \propto a^{-3(1+\omega)} \,.
\eea
For $\omega =-1$, we see from eq.~\eqref{eq:EMconsFLRW} that the energy density is constant. In this case, the Hubble rate \eqref{eq:Fried4DH} is also constant and so the evolution for the scale factor is:
\be
\label{eq:dSa}
a \propto e^{Ht} \,,
\ee
which is a de Sitter universe. We show in Tab. \ref{tab:scalef_evolution} the behaviour of $\rho$ and $a(t)$ for typical equations of state.
Using the equation of state in eq. \eqref{eq:Fried4Da}, we see that an accelerated expansion occurs  whenever
\be
\label{eq:omega_acc}
\omega <-1/3 \,.
\ee
 In order to explain the current acceleration of the universe, we require an energy density, `dark energy', with equation of state satisfying eq.~\eqref{eq:omega_acc}.

\begin{table}[htbp!]
\begin{center}
\centering
\begin{tabular}{| l | c  | c | c | c | }
\hline
\cellcolor[gray]{0.9}  {\bf Stress Energy} &  \cellcolor[gray]{0.9} {\bf $\omega$ } &  \cellcolor[gray]{0.9} {\bf Energy Density } &  \cellcolor[gray]{0.9} {\bf Scale Factor $a(t)$}   \\
\hline \hline
 Matter   & $\omega =0$ & $\rho_m \sim a^{-3}$ & $a(t) \sim t^{2/3}$  \\
\hline
Radiation & $\omega =1/3$ & $\rho_r \sim a^{-4}$  & $a(t) \sim t^{1/2} $ \\
\hline
Kinetic energy & $\omega = 1$ & $\rho_{\rm KE} \sim a^{-6}$ & $a(t) \sim t^{1/3} $ \\
\hline
 Vacuum ($\Lambda$) & $\omega =-1$    & $\rho_\Lambda \sim \frac{\Lambda}{8\pi G_4}$  & $a(t) \sim {\rm exp}(\sqrt{\Lambda/3} \,t)$    \\
\hline
\end{tabular}
\end{center}
\caption {Scale factor and energy density behaviour for matter, radiation, kinetic energy and vacuum dominated universes for $k=0$.}
\label{tab:scalef_evolution}
\end{table}

The different equations of state satisfied by radiation, matter and dark energy (see Tab. \ref{tab:scalef_evolution}) imply that their relative abundances differed in the past universe, since their energy densities evolve very differently as the universe expands.

The current measurements of the present-day Hubble scale, $H_0$, tell us the present value of the total energy density $\rho_T=\sum_i\rho_i$, of the universe. The present value of the Hubble parameter is measured to be\footnote{The constant $h_0 \approx 0.73$ accounts for the uncertainty in $H_0$.} $H_0=100\,h_0\,{\rm km s^{-1}Mpc^{-1}} $, which gives, via \eqref{eq:Fried4Da} with $k=0$,
\be
\rho_{\rm tot} \sim \rho_{\rm c} =\frac{3}{8\pi G_4} H_0^2 \sim 10^{-27}{\rm kg/m^3}\,.
\ee
The Friedman equation \eqref{eq:Fried4Da} can then be rewritten as:
\be
\setlength\fboxsep{0.25cm}
\setlength\fboxrule{0.4pt}
\boxed{
\sum_i \Omega_i =1\,,
}
\ee
with $\Omega_i$ (see eq.~\eqref{eq:Omegadef}) the present-day fraction of energy density contributed by each fluid component and with $i$ running over all components. At present, there is good evidence for the following four components of the cosmic fluid:
\begin{enumerate}[a)]
\item  Radiation, with equation of state parameter $\omega=1/3$ and whose energy density is dominated by CMB photons. The total energy density of radiation today is a small fraction of the present total energy density with $\Omega_r \simeq10^{-4}$.

\item  Baryons, with equation of state parameter $\omega=0$, corresponding to ordinary matter (i.e.~nucleons, atoms), whose fraction is $\Omega_{\rm B} \simeq 0.04$.

\item Dark Matter, also governed by an equation of state parameter $\omega=0$, whose fraction is observationally determined to be
$\Omega_{\rm DM} \simeq 0.27$. Since both baryons and matter have the same equation of state, they can be put together to give the total matter density fraction as $\Omega_m=\Omega_{\rm B}+\Omega_{\rm DM} \simeq 0.31$.

\item Dark Energy, with equation of state parameter $\omega=-1$. Over the last two decades, the evidence for the current accelerated expansion of the universe has accumulated, giving the largest contribution to the total energy density, $\Omega_\Lambda \simeq 0.69$.

\end{enumerate}

\begin{figure}[ht]
    \centering
    \includegraphics[width = 0.7\textwidth]{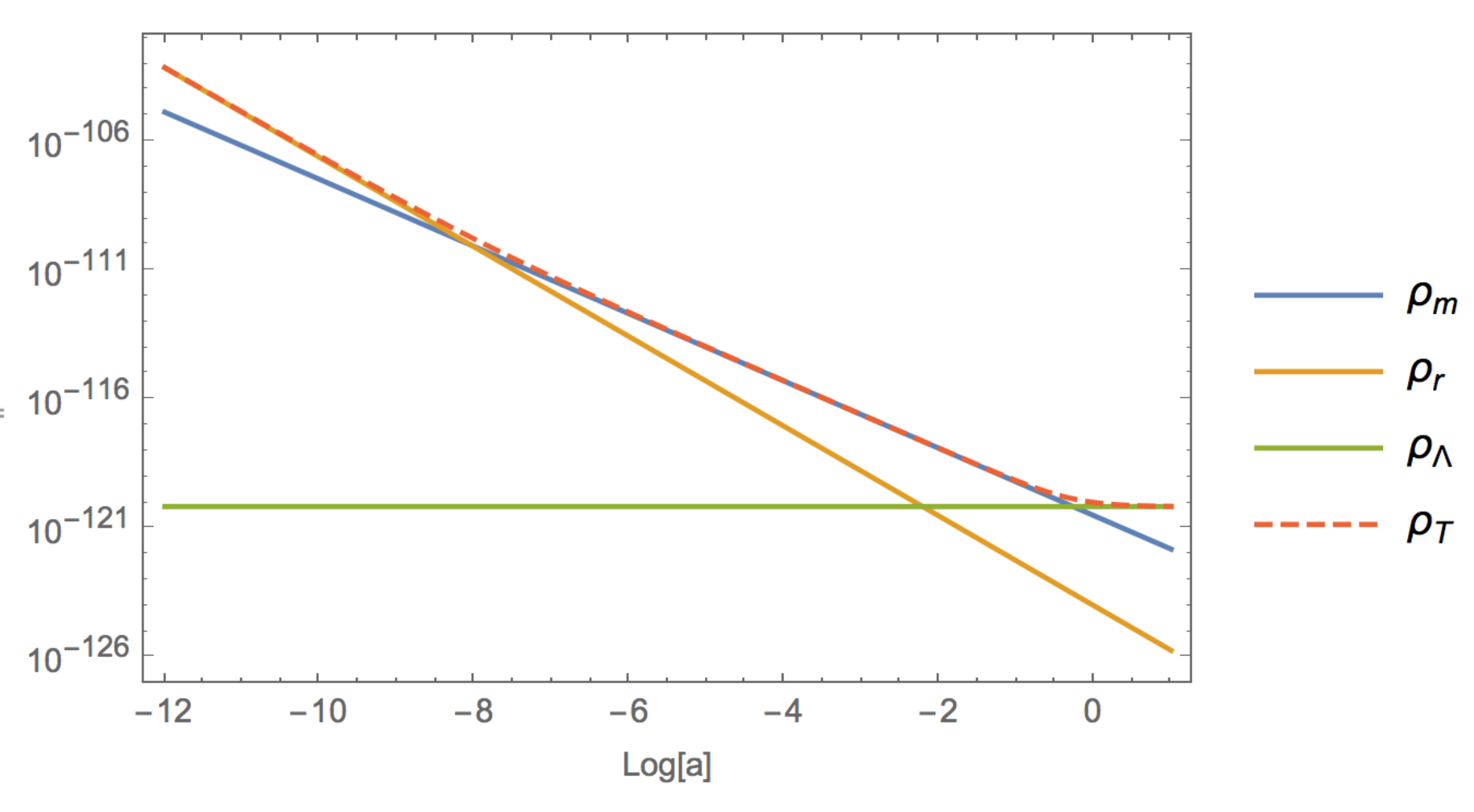}
    \caption{Energy density evolution for radiation $\rho_r$, non-relativistic matter $\rho_m$, (constant) dark energy $\rho_\Lambda$ and the total energy density, $\rho_T$ as a function of the scale factor in Planck units with $a_0=1$ today.}
    \label{fig:RhoEvol}
\end{figure}

Using the present day values, we can write the Hubble parameter more generally as:
\be
\label{eq:FriedEfolds}
\setlength\fboxsep{0.25cm}
\setlength\fboxrule{0.4pt}
\boxed{
H^2 = \frac{8\pi G}{3} \rho_T = 3H_0^2 \,\Omega_m \,e^{-3N} + 3H^2_0 \, \Omega_r \,e^{-4 N} +
3H^2_0  \,\Omega_\Lambda\,,}
\ee
where we introduced the number of {\em e-foldings}, $N\equiv \ln a$, and $a_0=1$ today.
Because each term varies so differently with time, the history of the universe can be decomposed into different epochs during which one or another term dominates the expansion and so controls the overall change in $\rho_T$, as we show in Fig.~\eqref{fig:RhoEvol}

\subsection{Major Events}
\label{subsecME}

The Hot Big Bang model for cosmology assumes that the universe was initially a hot soup of elementary particles at a very high temperature. In broad terms, the subsequent  evolution describes the cooling of this hot soup as the universe expands.  Indeed, conservation of entropy (for relativistic particles with a constant number of species) implies that temperature falls as 
\be
T(t)=T_0 \left(\frac{a_0}{a(t)}\right) \,,
\ee
and can be used as an alternative to time to parameterise the history of the universe.
There are two main consequences of such an expansion and cooling:
\begin{enumerate}
\item Reaction rates in dilute systems are generically proportional to the number of participants per unit volume, because the reactants must be able to find one another before they are able to react. Since  particle densities fall as the universal volume grows, reaction rates also fall. Thus interactions between particles freeze out when the interaction rate drops below the expansion rate. This implies that one of the main trends of cosmology is that, as the universe ages, thermal and chemical reactions fall out of equilibrium.

\item A consequence of the previous point is the appearance of bound states of particles as the universe ages. Although the reactions forming bound states can always occur, at the earliest epochs temperatures are high enough to ensure that collisions very efficiently destroy these bound states, leaving very few to survive in equilibrium conditions. As the temperature drops, the inter-particle collisions become less violent and eventually the reactions of formation can dominate to leave a population of primordial relic bound states. Moreover, in an expanding universe, broken symmetries in the laws of physics may be restored at high energies. At very early epochs, phase transitions are also expected to play an important role in the cosmic evolution, but as yet there is no direct evidence that such transitions took place.
\end{enumerate}

\begin{figure}[ht]
    \centering
    \includegraphics[width = 0.9\textwidth]{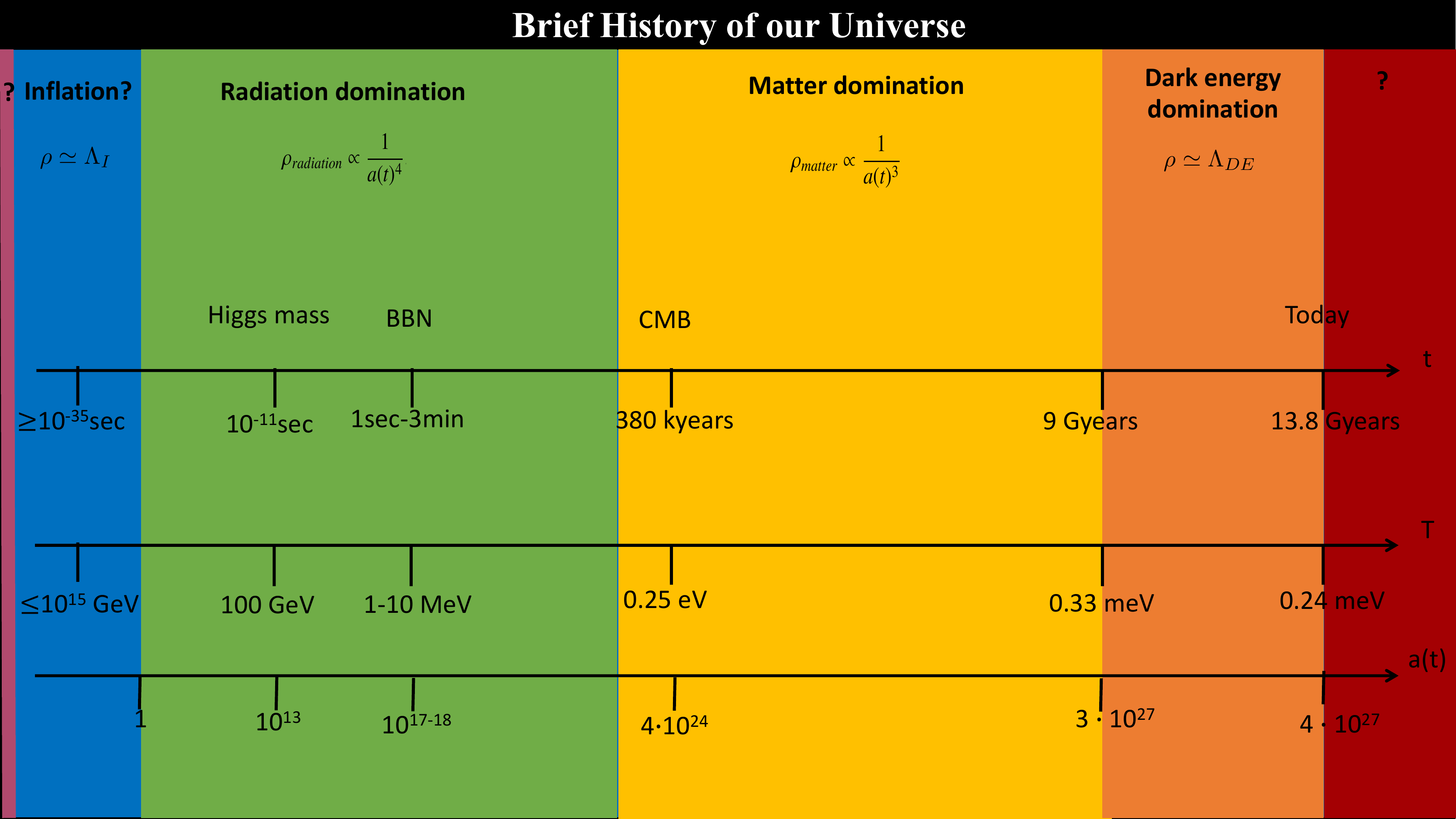}
    \caption{A schematic representation of the different epochs and their temperatures within the history of the universe in
    the standard $\Lambda$CDM cosmological model.}
    \label{fig:History}
\end{figure}

The  main events constituting the history of our universe can be summarised as follows  (see Tab. \ref{tab:universe_evol} and Fig.~\ref{fig:History}).
\begin{itemize}
\item At  $t\sim 10^{-43}$ s ($10^{19}$ GeV), we are near the Planck scale, where we expect  quantum gravity effects, such as those of string theory,
to dominate and general relativity not to be valid.  One of the fundamental issues of spacetime structure at the Planckian scale is  the question of cosmic singularities.  It is expected that these problems will be  addressed in the, as yet not definitively known, non-perturbative quantum gravity theory.

\item The period from  $t\sim 10^{-43}-10^{-14}$ s corresponds to temperatures  of around $T \sim10^{19}$ GeV - $10^4$ GeV, which are not foreseeably accessible by accelerators. In this sense, the universe can be used to test fundamental physics relevant at this scales, such as supersymmetry,  grand unification, string theory, extra dimensions, and other theories. Perhaps the most interesting phenomenon in the above energy range is the accelerated expansion of the early universe, inflation, which, as will be discussed below, likely occurred  somewhere near grand unification scales.

\item The epoch from  $t\sim 10^{-14}-10^{-10}$ s, corresponding to temperatures of $T\sim 10^4$ GeV - $100$ GeV, may be accessible by accelerators.
In particular, the standard model of the electroweak and strong interactions  is applicable here.

\item At  $t\sim 10^{-5}$ s, the corresponding temperature $T\sim200$ MeV, the QCD scale, where the  quark-gluon transition takes place.

\item Between  $t\sim0.2$ s and $200-300$ s (where $T\sim 1-2$ MeV at the start and $T\sim0.05$ MeV at the end), we have temperatures at the nuclear physics scale. Two important events happen during this period. First, the primordial neutrinos decouple from the other particles and subsequently propagate without further scatterings. Second, the process of {\em primordial nucleosynthesis}  takes place.  The initial conditions for this are set by the `freeze out' of the ratio of neutrons to protons, when the interactions that keep these particles in chemical equilibrium become inefficient; the number of the surviving neutrons subsequently determines the abundances of the primordial elements. As nuclear reactions become efficient, previously free protons and neutrons form helium and other light elements. The abundances of the light elements resulting from Big Bang Nucleosynthesis (BBN) are in very good agreement with observations, and this strongly supports our understanding of the universe's evolution back to the first second after the big bang.
 
\item $t\sim10^{11}$ s ($T\sim$ eV). This time corresponds to matter-radiation equality, which separates the radiation-dominated epoch from the matter-dominated epoch.

\item  At $t\sim 10^{12}-10^{13}$ s another two related important event happens. During so-called `recombination', nearly all free electrons and protons combine to form neutral hydrogen.  At this stage, the photons decouple and the universe becomes transparent to the background radiation. The Cosmic Microwave Background (CMB) temperature fluctuations, induced by the slightly inhomogeneous matter distribution at photon decoupling, form and survive to the present day, delivering direct information about the state of the universe at the last scattering surface.

\item Finally, at our present time $t\sim 10^{16}-10^{17}$ s,  galaxies and their clusters have formed from  small primordial inhomogeneities as a result of gravitational instability. An important question regarding this period is the nature of both dark matter and also the dark energy which is driving the present day accelerated expansion.
\end{itemize}

\begin{table}
[H]
\begin{center}
\centering
\begin{tabular}{| l |  m{8em}  | m{10em}| m{10em} | m{10em} | }
\hline
\cellcolor[gray]{0.9}  {\bf Temperature} &  \cellcolor[gray]{0.9} {\bf Time }&  \cellcolor[gray]{0.9} {\bf Particle Physics } &  \cellcolor[gray]{0.9} {\bf Cosmological Event}   \\
\hline \hline
$10^{19} {\rm GeV}$   & $10^{-43} {\rm s}$ & Quantum Gravity & Gravitons decouple?   \\
\hline
$10^{19} {\rm GeV}$ -  $10^2 {\rm GeV}$   &  $10^{-43} {\rm s}$ - $10^{-12} {\rm s}$ & Grand Unification?  Desert? String theory? Extra dimensions?  &  Inflation? Topological defects? Baryogenesis?\\
\hline
 $10^2$ GeV & $10^{-12}$ s   & Electroweak Breaking & Baryogenesis?    \\
\hline
 $0.3$ GeV & $10^{-5}$ s   & QCD scale & Quark-Hadron transition    \\
\hline
 $10-0.1$ MeV & $10^{-2}$ - $10^2$ s   & Nuclear Physics Scale & Nucleosynthesis, Neutrinos decouple     \\
\hline
$10$ eV & $10^{11}$ s   & Atomic Physics Scale & Atoms formed, CMB, Matter domination     \\
\hline
\end{tabular}
\end{center}
\caption {Brief history of our universe. Temperature units can be transformed to Kelvin using the conversion factor $1\,{\rm GeV}\, =1.16 \times 10^{13}$ K.}
\label{tab:universe_evol}
\end{table}

The standard cosmological model just discussed describes a simple and consistent picture of the relatively recent universe,  which is able to account for the many available observations of the overall structure and evolution of the universe. This picture bears up to scrutiny very well, at least for all times after the epoch of BBN. This success however, comes with some drawbacks, which  can be summarised as follows:

\begin{itemize}

\item  {\em The horizon problem}.  The CMB radiation, first discovered in 1964, is known with excellent precision and  is landmark evidence of the Big Bang origin of the universe. 
One of its most striking features is that its variations in intensity across the sky are tiny, less than 0.01\% on average.
It follows from this that the universe was extremely homogeneous at the time of recombination. Assuming the standard expansion of the universe,
we receive the same physical information from causally disconnected regions of space. It is (apparently) a puzzle why the radiation is so uniform.

\item {\em The flatness problem.}  The most recent results from the CMB are consistent with a flat universe. Namely, the  position and height of the first acoustic peak on the spectrum of the CMB provides evidence for $\Omega=1$ (see \eqref{eq:OmegaGeo}) \cite{Planck:2018vyg}.
 The flatness problem refers to the fact that for $\Omega$ to be so close to one at present, it had to be essentially one in the early universe to extraordinarily high precision, which also constitutes an apparent puzzle.

\item {\em Dark matter \& Dark Energy}. The standard cosmological model, supported by the most recent data \cite{Planck:2018vyg},  postulates the existence of two new forms of matter, namely dark matter and dark energy,  for which there is  no direct evidence from particle physics or from Earth-based experiments.

\begin{itemize}

\item Dark matter: Besides CMB evidence for dark matter, the survey and study of the behaviour of matter, such as rotation curves of galaxies, at many different scales, has given evidence that there should be a new kind of matter, not present in the standard model of particle physics. This plays  an important role in the explanation of the large scale structure formation. We still do not know what dark matter is: is it a particle, or some sort of  massive compact object present in the universe?

\item Dark energy: Recent results form the study of high redshifted supernovae, combined with CMB results provide strong evidence for the fact that the universe is  accelerating today ($\ln a\sim -0.34$, see Fig. \ref{fig:RhoEvol}). This indicates that there should be a form of `dark energy' satisfying eq.~\eqref{eq:rho3p} $(\rho+3 p)<0$ and thus causing the universe to accelerate today. Either an effective cosmological constant or a time varying scalar field,
called {\em quintessence},  are the main proposals for this dark energy.
\end{itemize}
\end{itemize}

All of these problems are strong guides as to the nature of necessary extensions beyond the Hot Big Bang, and in general to the need for physics beyond that contained in the Standard Model of particle physics.

\subsection{Cosmological Inflation}
\label{sec:infla}

Cosmic Inflation was initially motivated as a way to address the flatness and horizon problems above.  Quite compellingly, it was later found that it also provides a simple explanation for the origin of the primordial density fluctuations which seeded the observed temperature fluctuations of the CMB and the formation of galaxies through gravitational collapse.

The main idea behind inflation is that the  universe underwent a period of accelerated expansion at some point in its very distant past. If the inflationary period is long enough, it rapidly flattens the universe, solving the flatness problem. It also explains why some regions could be in causal contact with each other, solving the horizon problem. Requiring that inflation solves both the flatness and horizon problems, one can estimate that inflation should last for $N\gtrsim 60$ e-foldings.

\bigskip

An accelerated expansion implies that
\be
\ddot a >0\,.
\ee
Using \eqref{eq:Hdef} we can express this condition as
\be
\label{eq:ddota}
\frac{\ddot a}{a} = H^2\lb1-\epsilon \rb >0 \,,
\ee
where we introduce the {\em slow-roll} parameter $\epsilon$, defined as
\be
\label{eq:epsdef}
\setlength\fboxsep{0.25cm}
\setlength\fboxrule{0.4pt}
\boxed{
\epsilon\equiv -\frac{\dot H}{H^2}\,,
}
\ee
and thus the condition for an accelerated universe is encoded in the requirement that
\be
\label{eq:epscond}
\epsilon<1 \,.
\ee
Using  \eqref{eq:Fried4DH} and \eqref{eq:EMconsFLRW} in \eqref{eq:epsdef}, we can write $\epsilon$ as
\be
\setlength\fboxsep{0.25cm}
\setlength\fboxrule{0.4pt}
\boxed{
\epsilon \equiv \frac{3}{2}(1+\omega)\,,
}
\ee
and thus $\epsilon <1$ implies the condition  \eqref{eq:omega_acc} for an accelerated expansion, as seen previously.

This  is equivalent to the statement that  the {\em comoving Hubble radius} $(aH)^{-1}$ shrinks in accelerated expansion,
rather than the growing behaviour of radiation and matter dominated phases. That is,
\be
\frac{d}{d t}(aH)^{-1} = -\frac{1}{a} \lb 1-\epsilon \rb <0.
\ee
In a universe dominated by a fluid with equation of state $p=\omega \rho$, the comoving Hubble radius behaves as
\be
\frac{1}{aH} \sim t^{\frac{\epsilon-1}{\epsilon}}\,,
\ee
and thus again we see that $\epsilon<1$ implies that the comoving Hubble radius decreases, while for $\epsilon>1$, it increases.
For example, during matter domination $\omega=0$ and $\epsilon = 3/2$, while during radiation domination $\omega=1/3$ and $\epsilon=2$.
Note that as soon as the condition $\epsilon<1$ fails, inflation ends and thus we can define the end of inflation as $\epsilon\sim 1$.

\begin{figure}[ht]
    \centering
    \includegraphics[width = 0.7\textwidth]{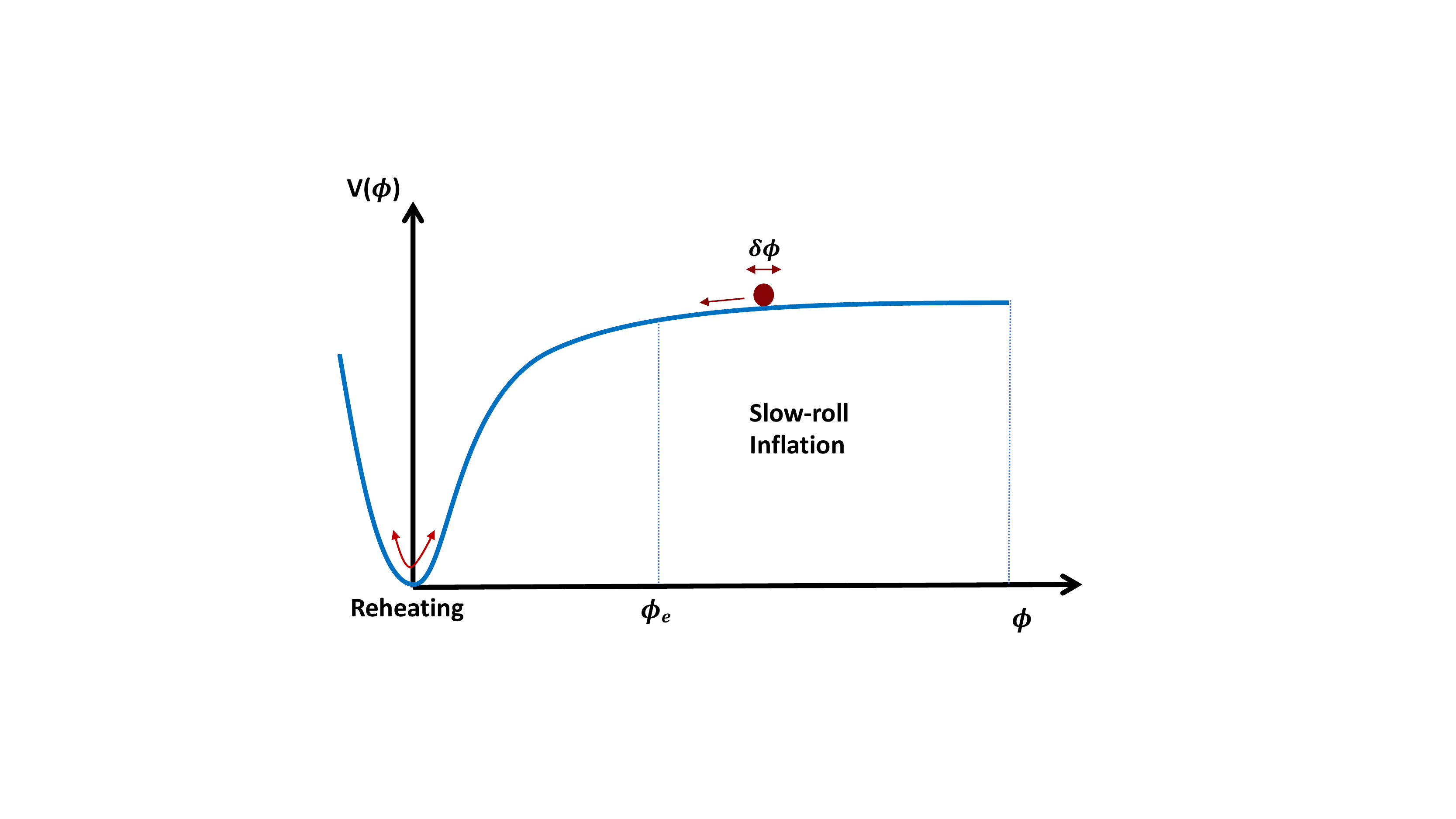}
    \caption{An illustration of the standard picture of slow-roll inflation ending in fast roll of the inflation to a minimum and subsequent reheating of the universe.}
    \label{fig:Inflation}
\end{figure}

In the de Sitter limit, $\epsilon\to0$, the space grows exponentially as in \eqref{eq:dSa}. More generally, an inflationary expansion requires a somewhat unconventional matter content. Indeed, from \eqref{eq:Fried4Da} we see that, for a universe supported by a perfect fluid, the energy density and pressure should satisfy
\be
\rho+3p <0 \,,
\ee
that is, the overall pressure of the universe should be negative $p <-\rho/3$, which corresponds 
to a violation of the {\em strong energy condition} (SEC)\footnote{The SEC for a perfect fluid states that $\rho+p\geq 0$ \cite{Hawking:1973uf}.}.This occurs in neither radiation nor matter dominated phases (for which $p=\rho/3, p=0$ respectively).
However, one simple energy source that can drive inflation is the  positive potential energy  of a single (canonically normalised) scalar field with negligible kinetic energy (see Fig.~{fig:Inflation} for an illustrative example). As we will encounter later, other alternatives are also possible.

\subsubsection{Slow-roll conditions}

Let us consider a single (canonically normalised) scalar field,  {\em the  inflaton}, with potential energy $V$, coupled to gravity.  Its 
action reads
\be
\label{eq:scalarS}
S= \int{d^4x \sqrt{- g} \left[\frac{1}{8\pi\,G_4} \,\frac{R_4}{2}  - \frac{1}{2} \partial_\mu\varphi\, \partial^\mu\varphi - V(\varphi)\right]} \,.
\ee
Although the inflaton can in principle depend on both time and space, inflation rapidly smooths out spatial variations, and thus for the background evolution, it suffices to study\footnote{The spatial dependence will be relevant later for the quantum fluctuations of the inflaton.} $\varphi=\varphi(t)$.
In a spatially flat FLRW spacetime \eqref{eq:FLRW} the variation of the action \eqref{eq:scalarS} with respect to $\varphi$ gives
\be
\label{eq:phieq}
 \ddot \varphi + 3H\dot\varphi+  V_{,\varphi} =0  \,.
\ee
The energy momentum tensor of the field derived from \eqref{eq:scalarS} gives
\be
T_{\mu\nu} = \partial_\mu\varphi\partial_\nu\varphi -g_{\mu\nu}\lb \frac12 (\partial\varphi)^2 +V(\varphi)\rb\,.
\ee
In the flat FLRW background, the energy density and pressure of the scalar are found to be
\begin{subequations}
 \begin{align}
 \label{eq:rhoscalar}
\rho_\varphi& = \frac{1}{2} \dot \varphi^2 +V(\varphi)\,, \\
p_\varphi &= \frac{1}{2} \dot \varphi^2 -V(\varphi)\,.\label{eq:Pscalar}
\end{align}
\end{subequations}
With this, eqs. \eqref{eq:Fried4DH} and \eqref{eq:Fried4Da} yield
\bea
&&H^2 = \frac{8\pi\,G_4}{3} \left(\frac{\dot \varphi^2}{2}  + V(\varphi)\right) \,, \label{eq:Hscalar} \\
&& \frac{\ddot a}{a} = -\frac{8\pi\,G_4}{3} \lp \dot\varphi^2 -V(\varphi) \rp  \,.
\label{eq:Rayscalar}
\eea
We now introduce the slow-roll conditions. A nearly exponential expansion can be ensured by the requirement that the fractional change of the Hubble
parameter per e-fold $N$ is small, that is $\epsilon \ll1 $ (see eq.~\eqref{eq:epsdef}). In terms of the inflaton,
$\varphi$, this can be written as (from now on we use $\Mp$ rather than $G_4$)
\be
\label{eq:epsfi}
\epsilon = \frac{\dot\varphi^2}{2 \Mp^2 H^2} \ll 1\,.
\ee
Requiring that inflation lasts for a sufficiently long time that the horizon problem is solved is equivalent to requiring that $\epsilon$ remain small for a sufficient number of Hubble times, which is measured by the {\em second slow-roll} parameter, $\eta$, defined as
\be
\label{eq:etafi}
\setlength\fboxsep{0.25cm}
\setlength\fboxrule{0.4pt}
\boxed{
\eta\equiv \frac{\dot \epsilon}{H\epsilon} = \frac{\ddot H}{H\dot H} + 2\epsilon = 2\frac{\ddot \varphi}{H\varphi} + 2\epsilon\,.
}
\ee
This then implies that $\delta_\varphi\ll1$, where we defined 
\be
\label{eq:deltafi}
\setlength\fboxsep{0.25cm}
\setlength\fboxrule{0.4pt}
\boxed{
\delta_\varphi \equiv \frac{\ddot \varphi}{H\dot\varphi}  \,.
}
\ee
 Using the Friedman equation \eqref{eq:Hscalar}, we  see that the first slow-roll condition \eqref{eq:epsfi}, implies
that $\dot\varphi^2 \ll V$ and therefore we can write \eqref{eq:Hscalar} as
\be
\label{eq:Hslowr}
H^2 \simeq \frac{V(\varphi)}{3 \Mp^2} \,.
\ee
Moreover, using \eqref{eq:deltafi}, we can write \eqref{eq:phieq} as
\be
\label{eq:fislowr}
3H\dot \varphi + V_{,\varphi} \simeq0\,.
 \ee

In the present case of a single scalar field, we can write the slow-roll conditions \eqref{eq:epsfi} and \eqref{eq:deltafi} (equivalently \eqref{eq:etafi}) solely in terms of the scalar potential and its derivatives as follows.  From the condition \eqref{eq:epsfi}, using \eqref{eq:fislowr} and \eqref{eq:Hslowr}, we arrive at
\be\label{eq:epsV}
\setlength\fboxsep{0.25cm}
\setlength\fboxrule{0.4pt}
\boxed{
  \epsilon_V\equiv \frac{\Mp^2}{2} \lp\frac{V_{,\varphi}}{V}\rp^2 \simeq \epsilon \,,
  }
\ee
which is the first {\em potential slow-roll parameter}.
Next, using the conditions \eqref{eq:fislowr} and  \eqref{eq:Hslowr} in \eqref{eq:deltafi}, we obtain
\be
\Mp^2 \frac{V_{,\varphi\varphi}}{V} + \epsilon \ll 1\,,
\ee
and so therefore introduce the second potential slow-roll parameter,  $\eta_V$,
\be\label{eq:etaV}
\setlength\fboxsep{0.25cm}
\setlength\fboxrule{0.4pt}
\boxed{
\eta_V \equiv \Mp^2 \,\Bigg|\frac{V_{,\varphi\varphi}}{V} \Bigg|\,.
}
\ee
Thus, {\em in single field inflation}, the slow-roll parameters can be written in terms of the scalar potential and its derivatives, which need to be  small during  inflation:
\be
\epsilon_V \ll1 \,,\qquad \eta_V\ll1 \,.
\ee
Note that in this case, the smallness of the $\eta_V$-parameter (which in the present single field case is equivalent to $\eta$ and $\delta_\varphi$),  implies that the mass of the inflation, $|m_{\rm inf}^2| \sim |V_{,\varphi\varphi}| \ll H^2$ 
(as we will see, this conclusion no longer holds when more scalar fields are present \cite{Chakraborty:2019dfh,Aragam:2021scu}). The required smallness of the slow-roll parameters, and in particular the mass of the inflaton, is vulnerable to quantum corrections, as we will discuss in detail when we consider UV complete models in Sec. \ref{sec:infla}.

\subsubsection{Primordial fluctuations}\label{sec:PrimF}

As we have seen, the early universe is supposed to have been rendered very nearly uniform by a primordial inflationary epoch. According to our current understanding, structures in the universe originated from tiny `seed' perturbations, which grew to form  all the structures we observe today.
Observations of the CMB support this view, indicating that at the time of decoupling the universe was very nearly homogeneous with small inhomogeneities at the $10^{-5}$ level.
The best candidate for the origin of these perturbations is quantum fluctuations produced during  inflation  in the early universe.
 These perturbations extend from extremely short scales to cosmological scales by the stretching of space during inflation.
\begin{center}
\begin{figure}[H]
\includegraphics[width=14cm]{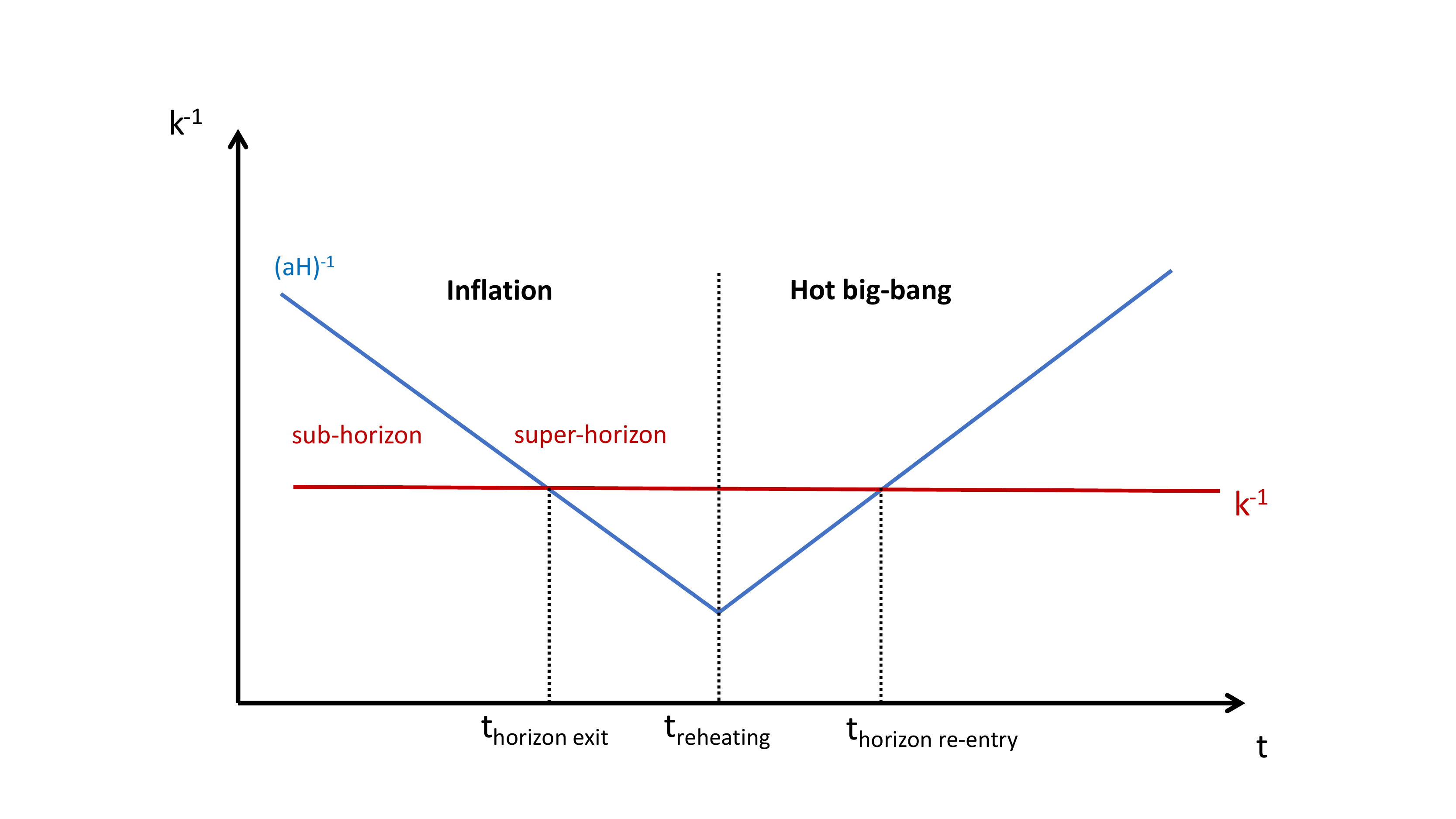}
\caption{Horizon exit and re-entry of a density perturbation with wave number $k$.}
\label{Fig:perturbations}
\end{figure}
\end{center}
The shrinking of the comoving Hubble radius (Hubble horizon) during inflation implies that fluctuations leave the horizon at some point (see Fig.~\ref{Fig:perturbations}).  Once inflation  ends,  the Hubble radius increases and  the fluctuations eventually reenter it during the radiation -- or matter -- dominated epochs. Fluctuations that exit the horizon around 60 e-foldings or so before the end of inflation, reenter with physical wavelengths in the range accessible to cosmological observations, with the CMB probing around 7-10 e-folds (note that the number 60 here depends on the post-inflationary evolution which, as discussed in Sec. \ref{reheating}, can be quite different in stringy scenarios compared to the vanilla picture of immediate reheating; see \cite{Bhattacharya:2017ysa, Bhattacharya:2017pws} for discussions in a stringy context). The spectra generated for density perturbations and gravitational waves during inflation provide a distinctive signature and can be measured by analysing the microwave background radiation anisotropies.

During inflation,  the inflaton field dominates the energy density of the universe, and thus any perturbation on it  implies  a perturbation of the  energy-momentum tensor
\be
\delta\varphi \quad  \Longleftrightarrow  \quad \delta T^{\mu\nu}\,.
\ee
A perturbation in the energy-momentum tensor then implies, via Einstein's equations of motion, a perturbation of the metric
\be
\delta G_{\mu\nu} =  \lp \delta(R_{\mu\nu}) -\frac12\delta(g_{\mu\nu}\,R)\rp = 8\pi G \delta T^{\mu\nu} \, ,
\ee
and so we have
\be
\delta\varphi \quad  \Longleftrightarrow \quad \delta g_{\mu\nu}\,.
\ee
The metric perturbations can be decomposed according to their spin into {\em scalar}, {\em vector} and {\em tensor} perturbations with respect to rotations of spatial coordinates on hypersurfaces of constant time. At  linear order, the scalar, vector and tensor perturbations evolve independently (decouple) and it is thus possible to analyse them separately. Vector perturbations do not get excited during inflation because there are no rotational velocity fields. In what follows, we summarise the analysis of scalar and tensor perturbations in inflation. For more details see e.g.~\cite{Mukhanov:1990me,Riotto:2002yw}.

\subsubsection*{Gauge choice}

An important  subtlety in the study of cosmological perturbations is that  the split into background and perturbations  is not unique, but depends on the choice of coordinates or the gauge choice.  It is important to note that there is no preferred gauge.  
To eliminate this ambiguity, one has two choices:
either identify  {\em gauge invariant} quantities or choose a given gauge and perform the calculations in that gauge.
Both options have advantages and drawbacks. By selecting a certain gauge, the calculations might be made technically simpler, but there is a risk that doing so introduces gauge artifacts or unphysical perturbations. On the other hand, a gauge-invariant computation may be technically more involved, but has the advantage of dealing only with physical quantities.

\subsubsection*{Gauge-invariant variables}

As we discussed above, it is helpful to provide gauge-invariant combinations of metric and matter perturbations in order to avoid the problem of spurious gauge modes. There are three gauge invariant quantities that are usually  defined in calculations of inflation: \\

\begin{tcolorbox}[title={\bf Gauge invariant variables}]

\begin{enumerate}[i.]
\item The {\em comoving curvature perturbation}. This is given by
\be\label{eq:curvatureR}
{\cal R} =\Psi + H\, \frac{\delta\varphi}{\dot\varphi} \,,
\ee
where $\Psi$ is the spatial {\em curvature perturbation}. In geometrical terms, $\cal R$  measures the spatial curvature of comoving hypersurfaces.

\item The {\em curvature perturbation on slices of uniform energy density}. This is given by
\be
\zeta = \Psi -\frac{\delta\rho}{3(\rho+p)}\,.
\ee
Geometrically, $\zeta$ measures the spatial curvature of constant-density hypersurfaces.
 For a scalar field, $(\rho+p) =\dot\varphi^2$. Moreover, during inflation   $\delta\rho \simeq-3H\,\dot\varphi\,\delta \varphi$.  Thus
 $\zeta$ and $\cal R$ are  equal during slow-roll inflation. As we will see they are also equal on {\em super-horizon scales} and therefore the correlation functions of $\zeta$ and $\cal R$ are  the same at horizon crossing. Moreover,  both  are conserved on super-horizon scales during slow-roll inflation.

\item Scalar field perturbations in  {\em spatially flat gauge}. The {\em spatially flat gauge} is defined as the slicing where there is no curvature $\Psi=0$. It gives a  gauge-invariant measure of inflaton perturbations and is given by
\be
Q = \delta\varphi +\frac{\dot\varphi}{H}\,\Psi \,. 
\ee

\end{enumerate}

\end{tcolorbox}

\bigskip
One can compute the curvature perturbation generated during inflation on super-Hubble scales,  $\zeta$ or ${\cal R}$, either using a particular gauge and computing the gauge-invariant curvature in that gauge, or by doing a fully gauge-invariant calculation. The results are equivalent.

The gauge-invariant curvature perturbation $\cal R$ defined above   is conserved outside of the horizon. Thus, we can compute it   at {\em horizon exit} and remain ignorant about the sub-horizon physics during and after {\em reheating} until horizon re-entry of a given $\cal R$-mode, $k$.

The equation of motion for the curvature perturbation $\cal R$, takes a simple harmonic oscillator form  and thus it can be quantised by promoting the classical field $\cal R$ to a quantum operator and then quantising it. One can then compute the power spectrum of curvature fluctuations at horizon crossing.

We summarise the results and refer the reader to the bibliography for the details on the computations \cite{Riotto:2002yw}.

\subsubsection*{Scalar perturbations}

The mode equation of motion for the Fourier components of $\cal R$ is given by
\be
\label{eq:R}
\setlength\fboxsep{0.25cm}
\setlength\fboxrule{0.4pt}
\boxed{
\cR''_k + 2\frac{z'}{z}\,\cR_k'+k^2\,\cR_k=0\,,
}
\ee
where here a prime denotes derivative with respect to  conformal time
$\eta$, $d\eta = dt/a(t)$; $k$ is the wavenumber and $z\equiv a\,\dot\varphi/H$, sometimes referred to as the {\em pump field}, which satisfies
\be
\label{eq:pumpf}
\setlength\fboxsep{0.25cm}
\setlength\fboxrule{0.4pt}
\boxed{
\frac{z'}{z} = aH\lp 1+\epsilon-\delta\rp\,,
}
\ee
where we have defined\footnote{Note from \eqref{eq:etafi} that $\eta=-2\delta+2\epsilon$. Note also the difference between $\delta$, determined by the full energy density and $\delta_\varphi$, which is associated only to the dynamics of a scalar fluid(s) component.}
\be
\setlength\fboxsep{0.25cm}
\setlength\fboxrule{0.4pt}
\boxed{
\delta\equiv -\frac{\ddot H}{2H\dot H}\,.
}
\ee
Let us note that fluctuations are created on all length scales, $\lambda$.
Relating the length scale  with its wavenumber $k$, as $\lambda = 2\pi a/k$
this means that the fluctuations are created with a spectrum of wavenumbers, $k$. Fluctuations that are cosmologically relevant start their lives inside the horizon (i.e.~Hubble radius), that is $k/aH\gg 1$.
However, while the comoving wavenumber is constant the comoving Hubble radius shrinks during  inflation. Scales for 
which  $k/aH\ll 1$ are outside the Hubble radius; eventually, all fluctuations exit the horizon. Thus we refer to the scales as follows (see Fig.~\ref{Fig:perturbations}):

\begin{subequations}\label{eq:scales}
\begin{empheq}[box=\widefbox]{align}
 &\frac{k}{aH}\gg 1 \qquad \Rightarrow \qquad \text{sub-horizon scales} \nonumber\\
&\frac{k}{aH}\ll 1 \qquad \Rightarrow \qquad \text{super-horizon scales} \nonumber
\end{empheq}
\end{subequations}

\ni For scales well outside the horizon, the solutions to \eqref{eq:R} are given by
\be
\label{eq:Rsuperhor}
\cR_k(\eta) = \cC_1 + \cC_2 \int\frac{d\eta}{z^2}\,,
\ee
where $\cC_1$ and $\cC_2$ are integration constants. From \eqref{eq:pumpf} we have
\be\label{eq:zsol}
z(a) = z_0 \,{\rm exp}\lb \int{ (1+\epsilon-\delta) \,d\ln a}\rb\,,
\ee
and therefore we see that during slow-roll,  when $\epsilon, \delta\ll 1$, $z\sim a$. Since  in this case $a\sim -1/(H\eta)$ we see that the term proportional to $\cC_2$ in \eqref{eq:Rsuperhor} decays rapidly as $a^{-3}$  outside the horizon, and is thus called the {\em decaying mode}. The  curvature perturbation is conserved at super-horizon scales and controlled by the {\em constant mode} $\cC_1$.
We thus see that the constancy of ${\cal R}_k$ depends on $\epsilon$ and $\delta$ doing nothing dramatic even after horizon crossing.
However, a more dramatic situation can  arise from a failure of slow-roll. If at any time after horizon crossing the friction term in \eqref{eq:R} changes sign,  becoming a negative driving term, the decaying mode can become a growing mode with interesting cosmological implications \cite{Leach:2000yw,Leach:2001zf,Ozsoy:2018flq}. This change of sign can occur whenever $z$ reaches a local maximum, that is,  whenever $1+\epsilon-\delta=0$. Since $\epsilon$ is always positive, $\delta$ must be at least one for this to happen. This can occur during a transient period of fast-roll, ultra slow-roll  or non slow-roll period. We review below briefly this possibility.

\begin{figure}[t]
\begin{center}
\includegraphics[width=145mm,height=105mm]{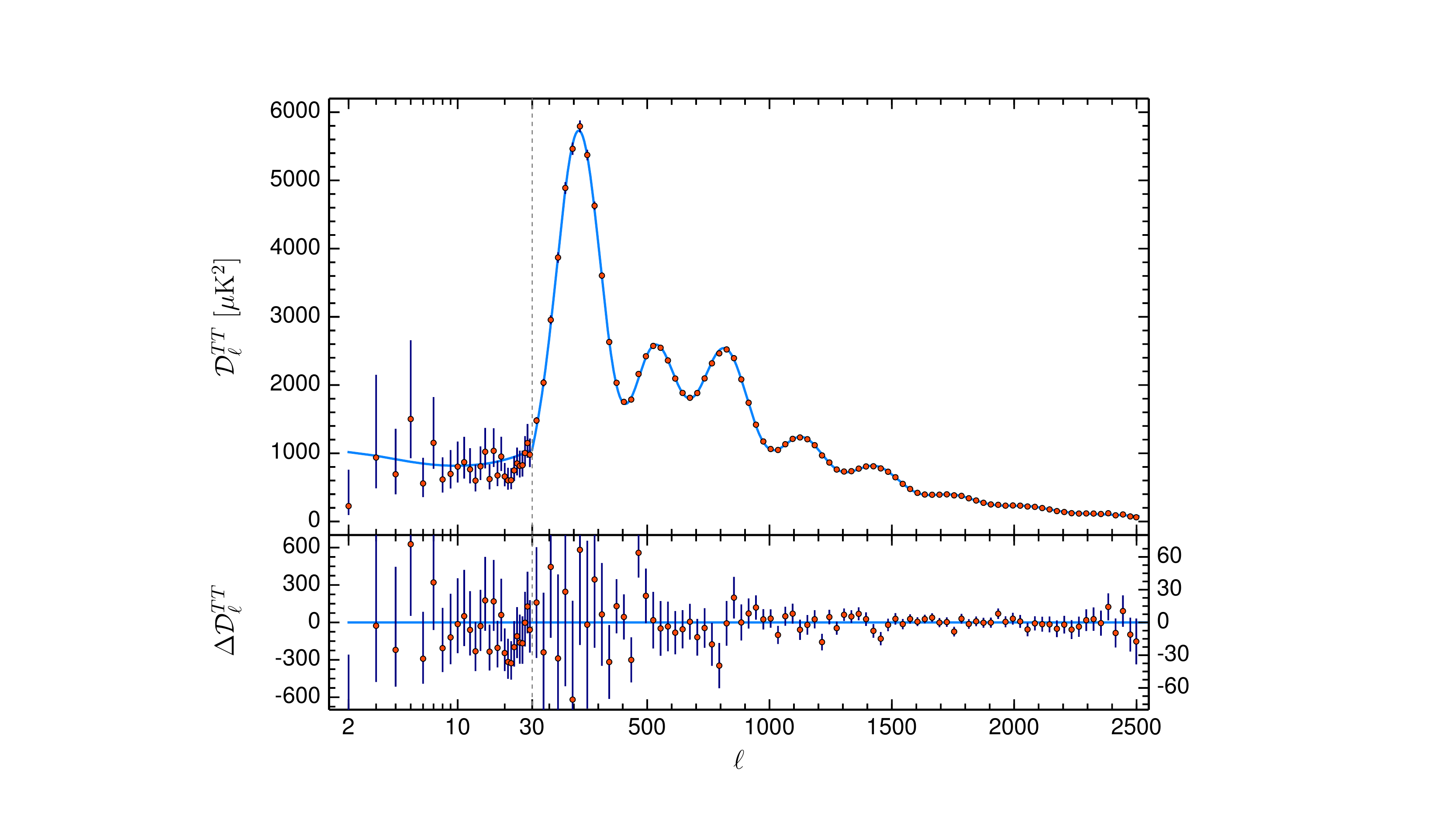} 
\caption{The power spectrum of the scalar fluctuations of the cosmic microwave background including error bars for the $\Lambda$CDM model. This is one of most impressive examples of the precision level ($\simeq 10^{-5}$)  reached in cosmology so far, in which only a handful of parameters is enough to explain such a large number of data points (taken from \cite{Planck:2018vyg}). } \label{Fig:Spectrum} 
\end{center}
\end{figure}

The  amplitude of the scalar power spectrum at leading order in slow-roll can be obtained by matching the super-horizon solution with the Bunch-Davies vacuum at sub-horizon scales, to obtain\footnote{This is sometimes denoted as $\cP_\cR$ or $\Delta_s$.}:
\be
\setlength\fboxsep{0.25cm}
\setlength\fboxrule{0.4pt}
\boxed{
\cP_\cR = \frac{ H^4}{(2\pi)^2\,\dot\varphi^2}\,\Bigg|_{k=aH} = \frac{H^2}{8\pi^2\Mp^2 \,\epsilon }\,\Bigg|_{k=aH}\,,}
\ee
where all quantities are evaluated at {\em horizon crossing}, $k=aH$ and we have used \eqref{eq:epsfi} in the last equality. The power spectrum of the  cosmic microwave background scalar fluctuations is shown in Figure \ref{Fig:Spectrum}. 

\subsubsection*{Primordial tensor perturbations}

Quantum fluctuations in the gravitational field are generated in a similar fashion as the scalar perturbations discussed so far.
In general, the linear tensor perturbations may be written as
\be
ds^2 = a^2(\eta)\lb -d\eta^2 +(\delta_{ij}+h_{ij}) dx^idx^j\rb\,,
\ee
with $h_{ij}\ll1$. If the energy momentum tensor is diagonal, as is the case in the simplest inflationary model we have discussed so far,  the tensor modes do not have any source and their action is that of two (not yet canonically normalised) independent massless scalar fields\footnote{The tensor $h_{ij}$ has six degrees of freedom, but  tensor perturbations are traceless,  $h^i_i=0$, and transverse $\partial^ih_{ij}=0,  (i = 1,2,3)$. These are four constraints, that leave only  two physical degrees of freedom, or polarisations.}.

The corresponding canonically normalised field (dropping $ij$ indices),
\be
v_k \equiv  \frac{a}{2}\Mp \, h_k\,,
\ee
satisfies the equation of motion
\be
\label{eq:tensoreq}
v_k'' +\lp k^2 -\frac{a''}{a}\rp v_k =0\,,
\ee
which is the equation of motion of a massless scalar field  in a quasi-de Sitter epoch. It is interesting to note that, contrary to the scalar case, no  interesting effects arising from transient violations of slow-roll can occur for gravitational waves in standard GR.\footnote{However, in general scalar-tensor theories there can be non-trivial effects as discussed in \cite{Mylova:2018yap}.} This can be most easily seen as follows. Defining the field
\be
\psi_k = \frac{v_k}{a}\,,
\ee
Eq.~\eqref{eq:tensoreq} becomes
\be\label{eq:tensoreq2}
\psi_k'' + 2 aH \psi_k' + k^2 \psi_k =0\,.
\ee
As the `pump field' $a$ increases for all time, the constancy of the gravitational wave amplitude after horizon crossing is guaranteed until horizon re-entry \cite{Leach:2000yw}.

The  amplitude of the tensor power spectrum is  found to be
\be\label{eq:PT}
\setlength\fboxsep{0.25cm}
\setlength\fboxrule{0.4pt}
\boxed{
\cP_T = \frac{2}{\pi^2}\frac{H^2}{\Mp^2}\Bigg|_{k=aH}\,.}
\ee
Note that  this differs from the scalar power spectrum by depending only on the value of $H$ and not additionally on the slow-roll parameter $\epsilon$. Consequently, a comparison of both scalar and tensor modes  amplitudes provides a direct measure of the slow-roll parameter $\epsilon$. A more precise statement of this comparison is usually phrased in terms of the parameter $r$, defined as {\em tensor-to-scalar ratio} of the power spectra
\be\label{eq:rdef}
\setlength\fboxsep{0.25cm}
\setlength\fboxrule{0.4pt}
\boxed{
r\equiv \frac{\cP_T}{\cP_\cR} = 16 \,\epsilon\,.
}
\ee

\subsubsection{Scale dependence}

The scale dependence of the power spectra  is given by the spectral tilt indices and follows from the time-dependence of the Hubble parameter. The  scalar and tensor spectral indices are given, respectively, by
\be
\setlength\fboxsep{0.25cm}
\setlength\fboxrule{0.4pt}
\boxed{
n_s -1 \equiv \frac{d\ln \cP_\cR}{d\ln k} \,, \qquad \qquad n_t \equiv \frac{d\ln \cP_T }{d\ln k}\,.
}
\ee
Using that $d\ln k = H dt + d (\ln H)$ one finds, to first order in the Hubble slow-roll parameters
\begin{subequations}
\label{eq:nsnT}
\begin{empheq}[box=\widefbox]{align}
 n_s-1 &= -2\epsilon -\eta  \,,\\
n_T &= -2\epsilon\,,
\end{empheq}
\end{subequations}
where $\epsilon, \eta$ are defined in eqs.~\eqref{eq:epsdef} and \eqref{eq:etafi} respectively and these quantities are defined at horizon crossing.

We see that single-field slow-roll models satisfy a {\em consistency condition} between the tensor-to-scalar ratio $r$ and the tensor tilt $n_T$:
\be\label{eq:ccinfla}
\setlength\fboxsep{0.25cm}
\setlength\fboxrule{0.4pt}
\boxed{
r=-8\, n_T \,.
}
\ee
 If this relation were to be falsified by future observations of the CMB anisotropies, it would indicate that inflation was not driven by a single field.

 \subsubsection{Lyth bound}

Note that from eqs.~\eqref{eq:rdef} and \eqref{eq:epsfi}, we see that  the tensor-to-scalar ratio relates directly to the evolution of the inflaton as a function of the number of e-foldings $N=\int H dt$:
 \be
 r = \frac{8}{\Mp^2} \lp\frac{d\varphi}{dN}\rp^2\,.
 \ee
 Therefore,  the total field evolution, between the time when CMB fluctuations left the horizon at $N_{\rm hc}$ and the end of inflation at $N_{\rm end}$, is given by
 \be
 \frac{\Delta\varphi}{\Mp} = \int^{N_{\rm hc}}_{N_{\rm end}}{ dN \,\sqrt{\frac{r}{8}} } \,.
 \label{LythBound}
 \ee
 Making the conservative assumption that $r$ remains approximately constant during the inflationary period probed by the CMB, the inflaton must satisfy the so-called {\em Lyth bound} \footnote{Taking into account that the fact that $r$ does not
remain constant gives a much stronger bound \cite{Garcia-Bellido:2014wfa}.} \cite{Lyth:1996im,Boubekeur:2005zm}:
\be\label{eq:Lythbd}
\setlength\fboxsep{0.25cm}
\setlength\fboxrule{0.4pt}
\boxed{
\frac{\Delta\varphi}{\Mp} \gtrsim  \,2\times \lp\frac{r}{0.01} \rp^{1/2}\,.
}
\ee
This relation indicates that `large' values of the tensor-to-scalar ratio, $r \sim 0.01$, correlate with $\Delta\varphi \sim \Mp$, or {\em large-field inflation}. The vulnerability of large-field inflation to quantum corrections will be discussed in Sec. \ref{sec:infla}.

Using Eqs. \eqref{eq:PT} and \eqref{eq:rdef}, one can also immediately relate the Hubble parameter during inflation to the tensor-to-scalar ratio or slow-roll parameter $\epsilon$:
\be
H_{\rm inf} = \sqrt{8 \pi^2 \cP_\cR\,\epsilon}\, \Mp \,.
\ee
The observational constraints that we are about to summarise might then make a high GUT-scale, large-field inflation seem more likely in the context of a single inflaton field.

\subsubsection{Current inflationary constraints}

In this section we summarise the most recent CMB experimental results that have tested the physics of inflation \cite{Planck:2018jri} (see Fig.~\ref{Fig:SpectralIndex}). Let us start by providing the current best-fit value for the power spectrum amplitude, defined through
 \be
 \cP_\cR = A_s \lp\frac{k}{k_*}\rp^{n_s-1}\,,
 \ee
 where $k_*$ is a pivot scale taken at $k_*=0.05 {\rm Mpc}^{-1}$ in the Planck analysis \cite{Planck:2018vyg}, and found to be
 \be
A_s = (2.100 \pm 0.030) \times 10^{-9}  \qquad \text{(68 \%, Planck TT,TE,EE+lowE+lensing)}\,.
\ee
The spectral tilt \cite{Planck:2018jri} index and latest bound on the tensor-to-scalar ratio \cite{BICEP:2021xfz} given by
\bea
n_s &=& 0.9649 \pm 0.0042 \qquad\quad  (68 \%, \text{Planck TT,TE,EE+lowE+lensing}), \\
 \alpha_s &=&  -0.0045 \pm 0.0067 \qquad\,\, (68 \%, \text{Planck TT,TE,EE+lowE+lensing}) \\
r_{0.05} &<&0.036 \qquad \qquad \qquad\quad \,\,(\text{at 95\% confidence})\,,
\eea
where $\alpha_s$ constrains the scale dependence of the scalar spectral index and is defined by
\be
\alpha_s \equiv \frac{d n_s}{d\ln k}\,.
\ee

\subsubsection{Inflationary models, a selection}\label{subsIM}

\begin{figure}[t]
\begin{center}
\includegraphics[width=140mm,height=85mm]{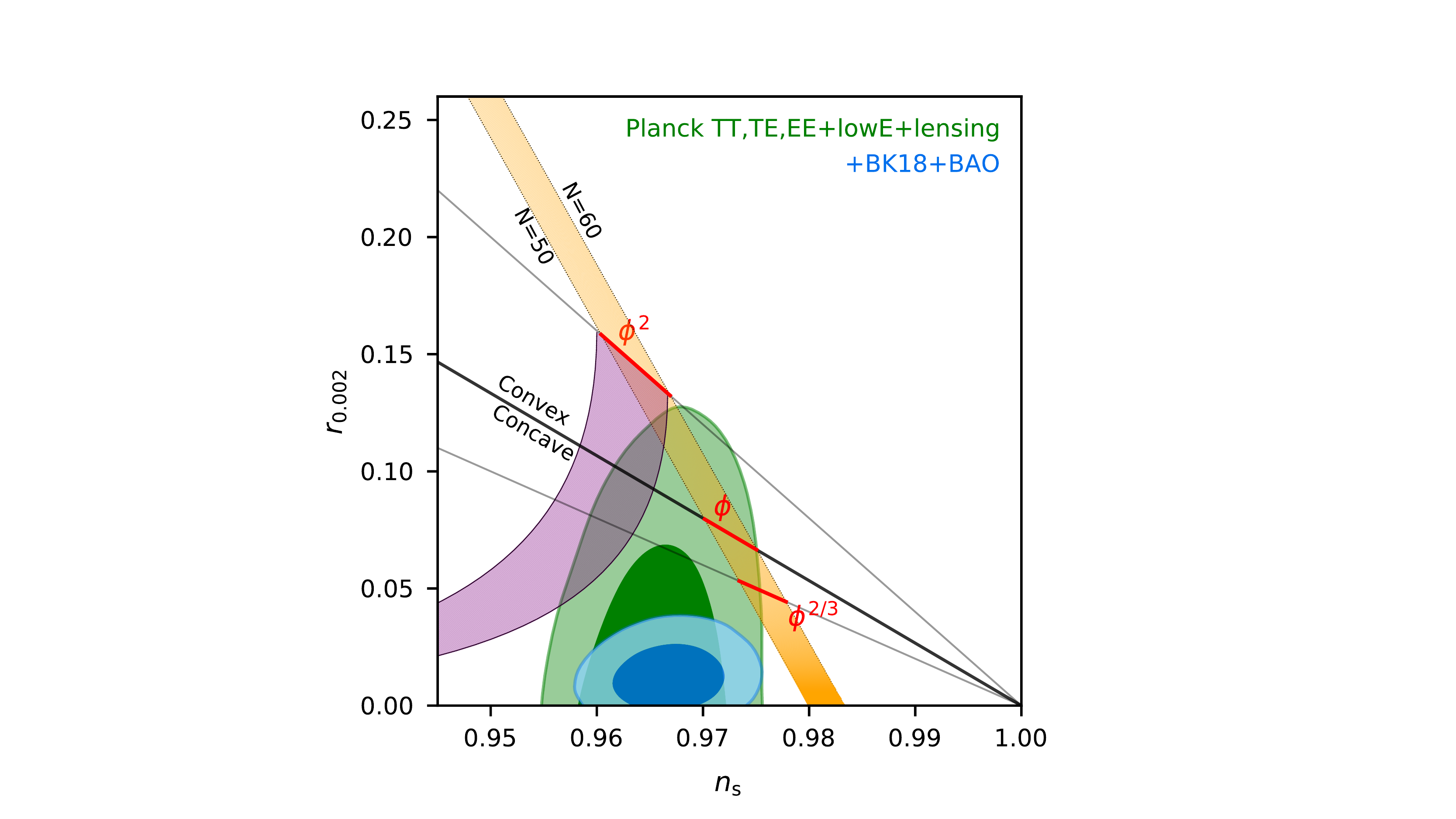} 
\caption{The most recent constraints on inflationary models in the tensor-to-scalar ratio $r$ and spectral index $n_s$ plane, taken from \cite{BICEP:2021xfz}. A substantial amount of inflationary models are already in tension with observations. } \label{Fig:SpectralIndex} 
\end{center}
\end{figure}

In the box below we illustrate three prototypical vanilla single field inflationary models together with their predictions for $n_s, r, \Delta\varphi$. All these examples have monotonically increasing slow-roll parameters and can be considered as large field inflation.  Notice, indeed, that whereas super-Planckian field ranges correspond to around $r\gtrsim 10^{-2}$ in the conservative Lyth bound \eqref{eq:Lythbd}, once the spectral tilt is taken into account, super-Planckian field ranges are obtained already around when $r\gtrsim 10^{-5}$ \cite{Garcia-Bellido:2014wfa}.  We also comment that, whilst the squared monomial and natural inflation models are in tension with the latest cosmological data, the Starobinsky model is well within the current data (see Fig.~\ref{Fig:SpectralIndex}). 

\bigskip

\begin{tcolorbox}[title={\bf Selected inflationary models}]
\begin{enumerate}
    \item  Monomial inflation\footnote{All observables are computed at $N_*=60$.}: $V= V_0 \,\frac{\phi^2}{2}$
    \be\label{eq:monoinf}
    n_s = 0.9666 \,, \quad r = 0.133\,, \quad  \Delta\varphi = 14.411\,\Mp
    \ee

    \item Starobinsky inflation: $V = V_0\lp1-e^{-\sqrt{2/3} \varphi} \rp^2$
    \be\label{eq:staroinf}
    n_s = 0.9674 \,, \quad r = 0.003\,, \quad  \Delta\varphi = 4.809\,\Mp
    \ee

\item Natural inflation: $V = V_0\,\lp1-\cos (\varphi/f )\rp$
 \be\label{eq:natuinf}
    n_s = 0.9626 \,, \quad r = 0.069\,, \quad  \Delta\varphi = 12.903\,\Mp
    \ee
($f=7\,\Mp$).

    \end{enumerate}
\end{tcolorbox}

\subsection{Multi-field Inflation}
\label{sec:multinf}

So far, we have discussed the simplest (vanilla) inflationary scenario, where a single canonically normalised scalar field drives inflation. However, as discussed at more length in Sec. \ref{ModuliSection}, in string theory there are usually many scalar fields (as well as other fields of various spin) which may either drive inflation or act as spectator fields with interesting cosmological implications. In what follows, we briefly review these possibilities from a pure field theory perspective, which will be important for our discussion on string inflation in Sec. \ref{sec:infla}.

Let us consider the  Lagrangian for several scalar fields, minimally coupled to gravity\footnote{For a recent review on multi-field inflation in field theory see \cite{Gong:2016qmq}.}:
\be\label{4Daction}
S= \int{d^4x \sqrt{-\tt g} \left[\Mp^2 \frac{R}{2}  - \frac12\,g_{ab}(\phi^c) \partial_\mu\phi^a \partial^\mu\phi^b - V(\phi^a)\right]} \,,
\ee
where $g_{ab}$ is the {\em field space metric}.  The equations of motion derived from this action are given by
\bea
&&H^2 = \frac{1}{3\Mp^2} \left(\frac{\dot \varphi^2}{2}  + V(\phi^a)\right) \,, \label{H} \\
&& \ddot \phi^a + 3H\dot\phi^a + \Gamma^a_{bc} \dot\phi^b\dot\phi^c + g^{ab} V_{,b} =0  \,,
\label{phis}
\eea
where
\be\label{varphi}
\dot\varphi^2 \equiv g_{ab} \dot \phi^a\dot\phi^b\,
\ee
and the Christoffel symbols in \eqref{phis} are computed using the scalar manifold metric $g_{ab}$, while $V_{,a}$ denotes derivatives with respect to the scalar field $\phi^a$.

The slow-roll conditions and rapid turning in multi-field inflation  can be understood neatly  using a  kinematic basis to decompose the inflationary trajectory into {\em tangent} and {\em normal} directions  (i.e.~adiabatic and entropic). Focusing on the two field case for concreteness, we introduce   unit tangent (adiabatic) and normal (entropic) vectors  $T^a$ and $N^a$, as follows:
\be
T^a = \frac{\dot\phi^a}{\dot\varphi}\,, \qquad T^aT_a =1\,,\qquad
N^aT_a=0, \qquad N^aN_a=1\,.
\ee
The  equations of motion \eqref{phis} for the scalars $\phi^a$ projected along these two directions become:
\bea
 \ddot\varphi + 3H\dot\varphi + V_T & = &0 \,, \label{varphiT}\\
D_t T^a  + \Omega N^a &=&0\,,\label{varphiN}
\eea
 where $V_T = V_{,a}T^a$, $V_N = V_{,a}N^a$ and the {\em turning rate} parameter, $\Omega$, is defined as
\be\label{Omega}
\setlength\fboxsep{0.25cm}
\setlength\fboxrule{0.4pt}
\boxed{
\Omega \equiv \frac{V_N}{\dot\varphi}\,.}
\ee
The field-space covariant time derivative is  defined as:
 \be\label{Dt}
 D_tT^a \equiv \dot T^a + \Gamma^{a}_{bc} T^b \dot \phi^c \,.
 \ee

Using the equations of motion, we can  write  the projections of the Hessian elements along the tangent vector as  \cite{Achucarro:2010da,Hetz:2016ics,Christodoulidis:2018qdw,Chakraborty:2019dfh}:
\be\label{VTT1}
\frac{V_{TT}}{3H^2} =  \frac{\Omega^2}{3H^2} + 2\,\epsilon -\frac{\eta}{2} -
\frac{\xi_\varphi}{3}   \,,
\ee
as well as the projection along $T $ and $N$ as \cite{Aragam:2021scu}:
\be
\label{VTN1}
\frac{V_{TN}}{3H^2} = w\left( 1-\epsilon+\frac\eta3+\frac\nu3 \right).
\ee
In these equations  we  introduced  the slow-roll parameter
\be
\xi_\varphi\equiv \frac{\dddot\varphi}{H^2 \dot\varphi}\,,
\ee
as well as the {\em dimensionless turning rate}, $w$, which measures the {\em non-geodesicity} of the trajectory:
\be\label{eq:w}
\setlength\fboxsep{0.25cm}
\setlength\fboxrule{0.4pt}
\boxed{
w \equiv \frac{\Omega}{H}\,,}
\ee
and a new {\em slow-roll} parameter, which arises only in the multi-field case, $\nu$:
\be\label{eq:nu}
\setlength\fboxsep{0.25cm}
\setlength\fboxrule{0.4pt}
\boxed{
\nu\equiv \frac{\dot w}{H\,w}\,.}
\ee

\smallskip

\ni Note  that the expressions \eqref{VTT1}, \eqref{VTN1}  {\em are exact}, as we have not made use of any slow-roll approximations.
 On the other hand, $V_{NN}$ depends on the inflationary trajectory in a model-dependent fashion.

\subsubsection{Slow-roll in multi-field inflation}

Let us revisit the slow-roll conditions in the case that more than one field is present. These  require the  slow-roll parameters $\epsilon, \eta, \delta_\varphi $ above, to be much smaller than one in order to guarantee long-lasting, slow-roll inflation, that is,  $\epsilon, \eta, \delta_\varphi, \xi_\varphi \ll 1$. It is easy to check that these conditions imply
\bea
&&H^2 \simeq \frac{V}{3M_{Pl}} \,, \\
&& 3 H\dot \varphi + V_T \simeq 0 \,, 
\eea
and thus that the tangent projection of the derivative of the potential needs to be  small:
\be\label{epT}
\epsilon_T \equiv \frac{M_{Pl}^2}{2} \left(\frac{V_T}{V}\right)^2 \ll1  \,.
\ee
On the other hand, the normal projection $V_N$ can be large,  and it is related to the turning rate defined in eq.~\eqref{Omega}.
Moreover, from \eqref{VTT1} we see that during slow-roll, the equations of motion imply
\be\label{VTTsr}
\frac{V_{TT }}{3H^2} \sim \frac{\Omega^2}{3H^2}\,,
\ee
while from \eqref{VTN1}, requiring $\eta\ll1$ (equivalently $\delta_\varphi \ll 1$) implies that (barring possible cancellations)
\be\label{VTN_slowroll}
   \frac{V_{TN}}{3H^2}\sim \frac{ \Omega}{H} \,, \qquad {\rm and} \qquad
     \nu \ll 1\,.
\ee
Hence, we see that $\nu$ behaves as a new slow-roll parameter in multi-field inflation: the turning rate is guaranteed to be slowly varying during slow-roll \cite{Aragam:2020uqi,Aragam:2021scu}.

We see then that in the multi-field case, slow-roll inflation does not require small eigenvalues of the Hessian \cite{Chakraborty:2019dfh,Aragam:2021scu}, as usually believed. Namely, defining the multi-field generalisation of the $\eta_V$ parameter \eqref{eq:etaV} as
\be\label{etaVmulti}
\setlength\fboxsep{0.25cm}
\setlength\fboxrule{0.4pt}
\boxed{
\eta_V^{ m} \equiv \Mp^2 \left|{\rm min\,\,\, eigenvalue}  \left(\frac{ \nabla^a \nabla_b V}{V}\right)\right|, }
\ee
it is clear that $\eta_V^m$ does not need to be small and indeed can  be much  larger than one in multi-field inflation \cite{Chakraborty:2019dfh,Aragam:2021scu}. This implies that in multi-field inflation, all inflatons can be heavier than the Hubble scale \cite{Chakraborty:2019dfh}.
Thus the $\eta$-problem in multi-field inflation may manifest itself, if present, in a different form. For example, the curvature of the field space metric, $R_{\rm fs}$, may in general be non-zero and in particular could be large (in Planck units). Therefore, this introduces a new scale, and the flatness of the potential may be constrained over this new scale (see e.g.~\cite{Renaux-Petel:2021yxh}). However, in general, we expect $R_{\rm fs}$ to be of order one in Planck units.

\subsubsection*{ Multi-field inflation and swampland constraints}

Let us finally make some comments  between multi-field  inflation and the recently proposed dS conjectures \cite{Obied:2018sgi,Garg:2018reu,Ooguri:2018wrx}, which require that (see Section \ref{Sec:Swamp})
\bea\label{eq:swamp1}
 \frac{\nabla V}{V} \geq \frac{c}{\Mp}  \qquad {\text{or}}  \qquad 
  \frac{{\rm min} (\nabla^a \nabla_b V)}{V} \leq -\frac{c'}{\Mp^2},
\eea
where $c, c'$ are  ${\cal O}(1)$ constants.
From our discussion on  slow-roll inflation in the multi-field case, it can be easily seen that  the first condition in \eqref{eq:swamp1} can be satisfied, so long as the  turning rate $\Omega/H$  is sufficiently large \cite{Achucarro:2018vey}.
Indeed, the potential slow-roll parameter \eqref{eq:epsV} in  the multi-field case is given by
\be\label{epsimulti}
\epsilon_V^m \equiv \frac{\Mp^2}{2} \frac{V^aV_a}{V^2} = \epsilon_T + \frac{\Omega^2}{9H^2}\epsilon \,.
\ee
For $\epsilon_T\simeq \epsilon$,  one arrives at the relation  \cite{Achucarro:2018vey,Hetz:2016ics}:
\be\label{epsV2}
\epsilon_V^m \simeq \epsilon\left(1  + \frac{\Omega^2}{9H^2}\right)\,,
\ee
and therefore we see that in a multi-field inflationary model, where $\Omega\ne 0$, for sufficiently large turning rate  $\Omega/H$  $\epsilon_V^m$ can be comparable to or larger than one.
On the other hand, the second condition in \eqref{eq:swamp1} is precisely the requirement that $\eta_V^m$  be large (with negative eigenvalue). As we discussed above, this can happen in mulltifield slow-roll inflation without disrupting it\footnote{A field theory example with large values of $c'$ is given in \cite{Christodoulidis:2018qdw}. Supergravity examples with large $c'$ are given in \cite{Aragam:2021scu}.}.
In summary, multi-field inflation allows for  new inflationary attractors, which do not need to satisfy the single field potential flatness conditions \eqref{eq:epsV}, \eqref{eq:etaV}.

\subsubsection{Cosmological perturbations, the multi-field case}

The presence of more than one field changes the kinds of primordial fluctuations which are possible, because with several fields there can be perturbations,  for which the total energy density, $\delta\rho=0$, remains unchanged. Such fluctuations are called  {\em isocurvature} fluctuations, in contrast to the {\em adiabatic} fluctuations involving nonzero $\delta\rho$ considered in the single field case.

There are strong observational constraints on the existence of isocurvature fluctuations and current observations are consistent with purely adiabatic oscillations at horizon re-entry  \cite{Planck:2018vyg,Planck:2018jri}. Primordial isocurvature modes need not be a problem for an inflationary model even if they are generated at horizon exit, provided they are subsequently erased before horizon re-entry. Moreover, the presence of more than one field can give rise to large non-Gaussianties in the power spectrum, which are also constrained by observations \cite{Planck:2019kim}. At the same time, additional scalars -- as well as other higher spin fields -- may give rise to interesting phenomenology and are thus of great interest for future 
experiments\footnote{See \cite{Arkani-Hamed:2015bza,Lee:2016vti,Alexander:2019vtb} for studies  of the imprints of new (higher spin) particles on the non-gaussianities of the cosmological fluctuations and \cite{Alexander:2020gmv,Criado:2020jkp,Jenks:2022wtj,Gondolo:2021fqo} for proposals of dark matter as higher spin fields, and their phenomenology.
}.

\subsubsection{Spectator fields during inflation}

Scalar fields acting as spectators during inflation are well motivated not only from a purely field theory perspective, but also from a phenomenological point of view. Any additional fields may give rise to interesting features that could produce observable effects which would then be of great importance. The same can be said for spin one fields such as $U(1)$ vector fields, producing for example   anisotropies, as well as $SU(2)$ gauge fields, potentially sourcing  tensor perturbations, and evading the Lyth bound discussed above (see e.g.~\cite{Maleknejad:2012fw} for a review).
We will not review here the vast literature on the subject, but will mention some interesting possibilities in the context of string cosmology in Sec. \ref{sec:infla}.

\subsection{Quintessence}\label{sec:quint}

As we mentioned before (see Sec.~\ref{subsecME}),  current observations provide strong evidence for the current acceleration of the universe.
In the $\Lambda$CDM standard model of cosmology, this is due to a constant vacuum energy, $\Lambda$.
However, the constant vacuum energy appears to be  much smaller than  would be expected from estimates based on quantum field theory. This has led to the  widespread speculation that the vacuum energy may not be constant, but it may now be small because the universe is old. Such a  time-varying vacuum energy is called {\em quintessence} \cite{Wetterich:1987fm,Peebles:1987ek,Ratra:1987rm}.\footnote{The name quintessence was coined in \cite{Caldwell:1997ii}. For reviews and several references see e.g.~\cite{Copeland:2006wr,Linder:2007wa,Tsujikawa:2013fta}.}

The natural way to introduce a time-varying vacuum energy is to assume the existence of one or more scalar fields, on which the vacuum energy depends, and whose cosmic expectation values change with time. We have seen that scalar fields of this type  play a crucial role in cosmological inflation and thus the discussion of quintessence uses several of the concepts which already appear in inflation.
Considering a single scalar field, the idea is that its dynamics drive the present epoch of accelerated expansion. Dark energy started to dominate relatively recently, namely less than a single e-fold ago (see Fig.~\ref{fig:RhoEvol}). It may therefore seem easy to have a sufficiently flat scalar potential, which starts dominating the energy density less than an e-folding ago.
The original and simplest example is provided by the following potential \cite{Wetterich:1987fm,Peebles:1987ek,Ratra:1987rm}
\be
\label{eq:QV}
V(\phi) = M^{4+\alpha} \phi^{-\alpha},
\ee
where $\alpha$ is positive but otherwise arbitrary, and $M$ is a constant with units of mass. We can call this type of potential  {\em runaway} like. We give a brief discussion here, but explore this class of runaway potentials in string theory more in Sec. \ref{reheating} and Sec. \ref{sec:DE}, in the context of both transient post-inflationary runaways and also quintessence dynamics in the late-time universe. 
The model \eqref{eq:QV} can be solved in some detail as a concrete example of quintessence (see e.g.~\cite{Weinberg:2008zzc} and further discussion in Sec. \ref{sec:DE}).

Any successful quintessence must have the property that at early times, the energy density of the quintessence field is subdominant over radiation to avoid conflict with BBN. With potential \eqref{eq:QV}, one can show that the field has a solution at early times during radiation domination such that
\be
\rho_\phi\propto t^{-2\alpha/(\alpha+2)}\,,
\ee
and thus at early times ($t\ll 1$) $\rho_\phi$ is indeed less than $\rho_r$ which goes as $t^{-2}$. This solution turns out 
to be an attractor, known as a {\em tracker solution}. After radiation domination, the universe undergoes an epoch of matter domination, but the tracker solution of $\phi$ continues to have energy density falling as $\rho_\phi\sim t^{-2\alpha/(\alpha+2)}$, and since $\rho_r$ and $\rho_m$ decrease faster ($t^{-8/3}, t^{-2}$ respectively in matter domination) eventually, both will fall below $\rho_\phi$. 

At late times, 
one finds $\rho_\phi\propto t^{-\alpha/(2+\alpha/2)}$, or $\ln a \propto t^{2/(2+\alpha/2)}$ and the expansion is  dominated by the quintessence field.  The point when $\rho_m=\rho_\phi$ is given by
$t_c\approx M^{-(4+\alpha)/2}G_4^{-(2+\alpha)/4}$, which gives $\phi(t_c)\approx G_4^{-1/2}$. 
Finally, to achieve agreement with observations, setting the critical time $t_c$ at which $\rho_m\approx \rho_\phi$ to be close to the present moment $t_0 \approx 1/H_0$,  requires the constant factor in $V(\phi)$ to take the value
\be
M^{4+\alpha} \approx G_4^{-1-\alpha/2}H_0^2\,.
\ee
There is, however, no fundamental explanation as to why this should be the case.

Although quintessence has been proposed as a potential candidate to explain the current acceleration, it still faces several challenges. We will discuss later on  in more detail how these manifest  in the context of string theory (see Sec.~\ref{sec:ChallengesQ}), but let us briefly mention here some of the main challenges for any model of quintessence.
\bi
 \item {\em Fine-tunings}. As can be evident from the above discussion, quintessence models need to explain why the field has to be exactly at the point where $V(\phi_0) \simeq \rho_0 \simeq (0.003 \,{\rm eV})^4$ today. Moreover, it is not hard to check that a successful quintessence model requires the mass of the field to be extremely light, $m_\phi\simeq H_0\simeq 10^{-33}\,{\rm eV}$ (at least in the case of single field), which must be protected against quantum corrections.

\item {\em Phenomenological constraints}. Phenomenological problems to realise quintessence arise as the quintessence field must be extremely weakly coupled to ordinary matter;  otherwise, its exchange would generate observable long-range forces, which are severely constrained by experiments.
The quintessence field can be a scalar as we saw above, or a pseudoscalar, such as an axion.\footnote{Other fields have also been proposed as quintessence, however the same problems arise as for the scalar case.} The advantage of axions is that they can avoid  fifth-force constraints, but a typical axion potential requires a trans-Planckian decay constant to drive any successful period of accelerated expansion.
\ei

\subsection{Possible Tensions with $\Lambda$CDM?}

We end this overview section on cosmology with mention of various possible hints towards tensions between Planck observations of the CMB and other cosmological probes.  Although the statistical significance of these tensions is not definitive, and even still under debate, if any are confirmed and not due to systematics, this would be exciting evidence of new physics beyond the $\Lambda$CDM model.  

The most famous of these tensions is the Hubble tension.  Planck constraints on today's Hubble parameter, which is obtained by assuming the standard six parameter $\Lambda$CDM model, yields $H_0 = 67.44 \pm 0.58 {\rm km}\,{\rm s}^{-1}\,{\rm Mpc}^{-1}$ \cite{Planck:2018vyg}.  This is to be contrasted with recent direct local distance ladder measurements of $H_0$ from the SH0ES collaboration, which gives instead $H_0 = 74.03 \pm 1.42 {\rm km}\,{\rm s}^{-1}\,{\rm Mpc}^{-1}$ \cite{Riess:2019cxk}. This amounts to a $4.3\sigma$ discrepancy, with other direct measurements going in the same direction.  For a recent review on the $H_0$ discrepancies and phenomenological solutions see \cite{DiValentino:2021izs}; see \cite{Efstathiou:2020wxn} for a critical perspective. Proposals to resolve the tension include increasing the number of effective neutrino species $N_{\rm eff}$, modifying the dark energy equation of state, and the presence of some early dark energy; for a systematic comparison of several proposals and their relative success see \cite{Schoneberg:2021qvd}.  A review of all the current discordances, covering $H_0$, the $\sigma_8$--$S_8$ tension, and other less statistically significant anomalies, together with an experimental outlook, is given in \cite{Abdalla:2022yfr}.

\newpage

\section{Moduli}
\label{ModuliSection}

\subsection{String Compactifications}

Even if a full non-perturbative understanding of string and M-theory is still lacking, it has long been understood that at low energies there are 5
different limits of string theory in 10-dimensional flat space, which are related to each other by duality transformations. 
The M-theory picture also leads to a sixth limit, namely 11-dimensional supergravity, often referred to as the low-energy limit of M-theory, the still-not-fully-defined theory that encompasses all the string theories as different limits. 

What these limits have in common, and arguably the \emph{single most important physical implication of string theories}, is the existence of extra dimensions. The process of starting from a higher-dimensional theory and then obtaining a 4-dimensional effective theory is known as compactification, and over the past 35 years string compactifications have been studied in much detail. Starting from a 10-dimensional theory, the different fields have to be decomposed into their components in the 4 non-compact dimensions and also their ones in the extra compact dimensions. For instance, the 10-dimensional graviton $ g_{_{MN}} $ splits into the 4-dimensional graviton $g_{\mu\nu}$, a set of scalar fields $g_{mn}$ that correspond to moduli fields and potentially also vector fields $g_{\mu n}$. Notice that from the 4-dimensional perspective the indices $m,n$ are just internal indices, as in compactification the extra dimensions are regarded as no longer directly visible from the 4-dimensional perspective:
\begin{equation}
g_{_{MN}}=\begin{pmatrix} g_{\mu\nu} & g_{\mu n} \\
g_{n\nu} & g_{mn}
\end{pmatrix}\qquad \mu,\nu =1, \cdots, 4\ ; \qquad m,n=1,\cdots, 6
\end{equation}
A similar decomposition is performed with the higher-form antisymmetric tensors $B_{MN}$, $C_{MNP}$, etc present in each of the 6 theories, with the form content of each theory shown in Tab. \ref{tab:boson_spectrum}.

\begin{table}[H]
\begin{center}
\centering
\begin{tabular}{ | c | c | c | c | }
\hline
\cellcolor[gray]{0.9}  {\bf Theory} &  \cellcolor[gray]{0.9} {\bf Dimension } &  \cellcolor[gray]{0.9} {\bf Supercharges} &  \cellcolor[gray]{0.9} {\bf Massless Bosons}   \\
\hline \hline 
{Heterotic}   & 10  & 16  & $g_{_{MN}}, B_{_{MN}}, \phi $  \\
$E_8 \times E_8$ & & &$A_{_M}^{ij}$ \\
\hline
{Heterotic}   & 10  & 16  & $g_{_{MN}}, B_{_{MN}}, \phi $  \\
$SO(32)$ & & &$A_{_M}^{ij}$ \\
\hline
{Type I}   & 10  & 16  & $g_{_{MN}},  \phi, A_{_M}^{ij} $  \\
$SO(32)$ & & &$C_{_{MN}}$ \\
\hline
{Type IIA}   & 10  & 32  & $g_{_{MN}}, B_{_{MN}}, \phi $  \\
& & &$C_{_M}, C_{_{MNP}}$ \\
\hline
{Type IIB}   & 10  & 32  & $g_{_{MN}}, B_{_{MN}},\phi $  \\
& & &$C_{_0}, C_{_{MN}}, C_{_{MNPQ}}$ \\
\hline
{M-Theory}   & 11  & 32  & $g_{_{MN}}, C_{_{MNP}}$ \\
\hline
\end{tabular}
\end{center} 
\caption {\footnotesize{The massless bosonic spectrum of the five string theories and of $11$-dimensional supergravity. The corresponding massless fermionic spectrum is determined by supersymmetry. Moduli fields all originate from these simple spectra in 10d, reduced on the internal manifold. There are also matter states, which in IIA and IIB string theories come from D-brane intersections and in heterotic string theory come from solutions of the Dirac equation with non-trivial gauge configuration. Further moduli, such as open string moduli from separation between D-branes or closed string bundle moduli, can also be present.} }
\label{tab:boson_spectrum}
\end{table}

The most studied compactifications are those that preserve $\cN=1$ supersymmetry. These offer a greater degree of control over the effective action compared to non-supersymmetric theories, while also allowing for the presence of chiral fermions and sufficient dynamics to 
allow for hierarchies and a non-supersymmetric vacuum state.

These correspond in the case of the heterotic or type I theories to the internal space being a Calabi-Yau (CY) manifold. These are manifolds of $SU(3)$ holonomy (or vanishing first Chern class). CY manifolds are complex K\"ahler manifolds, meaning that the metric can be written as a second derivative of a K\"ahler potential $K(z_i, \bar z_{\bar\jmath})$: $g_{i\bar\jmath}=\partial_i\partial_{\bar\jmath} K$. 
However, since they do not have isometries, except for a few numerical examples, there are no known analytic metrics 
for compact CY manifolds of complex dimension greater than one.  Instead, we rely mostly on their topological structure 
(and indeed, the full details of the internal metric are not needed for most parts of the 4-dimensional effective Lagrangian). 
The most relevant topological quantities are the non-trivial homological cycles. 
Their number are given by the corresponding Hodge numbers $h^{p,q}$.  

\begin{figure}[t]
\begin{center}
\includegraphics[width=120mm,height=80mm]{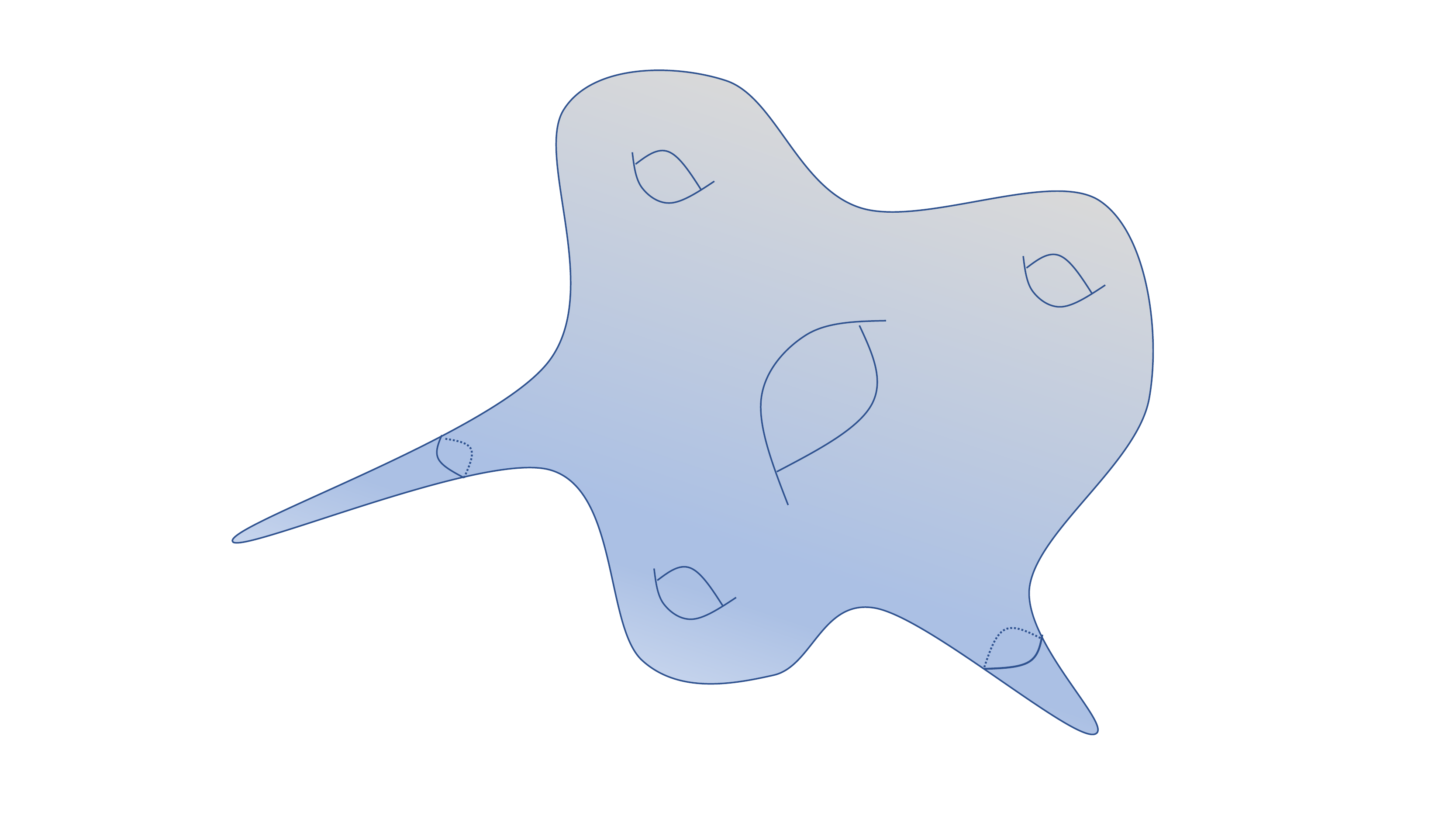} 
\caption{A schematic representation of the complicated geometry and topology of higher-dimensional Calabi-Yaus.} \label{Fig:CY2} 
\end{center}
\end{figure}

The simplest CY manifold is the one complex dimensional case corresponding to the torus. This has only two non-trivial homological cycles ($h^{0,1}=h^{1,0}=1$) and its homological structure is summarised by the Hodge diamond:
\bea
\setlength\fboxsep{0.25cm}
\setlength\fboxrule{0.4pt}
\boxed{
\begin{array}{  c  c c}
& h^{1,1}&\\
h^{1,0}& & h^{0,1}\\
& h^{0,0}&
\end{array}
=
\begin{array}{  c  c c}
& 1&\\
1& & 1\\
&1&
\end{array}
}
\nonumber
\eea
Compactifying a string theory on a 2-torus gives rise to two geometric moduli. These are the \emph{K\"ahler modulus} $T$ and the \emph{complex structure modulus} $U$ corresponding to 
\be
\setlength\fboxsep{0.25cm}
\setlength\fboxrule{0.4pt}
\boxed{
T=\sqrt{g}+i B_{12}, \qquad U=\frac{\sqrt{g}}{g_{22}}+i\, \frac{g_{12}}{g_{22}},
}
\ee
where $g_{ij}, B_{ij}$ are the components of the metric and the antisymmetric tensors, with $g ={\rm det} \, g_{ij}$. Roughly speaking, Re $T$ determines the size of the torus and Re $U$ the shape. These simple properties (K\"ahler and complex structure moduli, which respectively correspond to size and shape of the compact space) generalise to the more complicated CY 3-folds and 4-folds which are relevant for string and F-theories. A schematic representation of a Calabi-Yau is figure \ref{Fig:CY2}.

The corresponding Hodge diamond for a 3 complex dimensions CY manifold is:
\bea
\setlength\fboxsep{0.25cm}
\setlength\fboxrule{0.4pt}
\boxed{
\begin{array}{  c  c  c  c  c c c}
& & & h^{3,3} & & &\\
& &  h^{3,2}& & h^{2,3} & &\\
& h^{3,1} & & h^{1,1} & &h^{1,3} &\\
h^{3,0}& & h^{2,1} & & h^{1,2} & & h^{0,3}\\
& h^{2,0} & & h^{2,2} & & h^{0,2} &\\
& & h^{1,0} & & h^{0,1} & &\\
& & & h^{0,0} & & &
\end{array}
=
\begin{array}{  c  c  c  c  c c c}
& & & 1 & & &\\
& & 0 & & 0 & &\\
& 0 & & h^{1,1} & &0 &\\
1& & h^{2,1} & & h^{2,1} & & 1\\
& 0 & & h^{1,1} & &0 &\\
& & 0 & & 0 & &\\
& & & 1 & & &
\end{array}
}
\nonumber
\eea

The relevant numbers here are $h^{1,1}$, counting the number of K\"ahler moduli $T_i, i=1, \cdots, h^{1,1}$ (volumes of 4-cycles or their dual 2-cycles), and $h^{1,2}$, counting the number of complex structure moduli $U_\alpha, \alpha=1,\cdots, h^{1,2}$ (number of 3-cycles). There exist databases of millions of Calabi-Yau manifolds with different values of $h^{1,1}$ and $h^{1,2}$. Typically, these numbers can be as high as hundreds or thousands, see e.g.  \cite{Kreuzer:2000xy}. A recent package, CYTools \cite{Demirtas:2022hqf}, provides tools to compute various topological properties of CY manifolds efficiently.

Another `universal' modulus of great importance is the dilaton $\phi$ (see the table) whose vacuum expectation value $\langle \phi \rangle $ determines the  string coupling, $g_s$. This reflects the fact that string theory has no free parameters and so the strength of string interactions, $g_s$, is itself the expectation value of a field.

In addition to these `universal' moduli, there are also normally other moduli present in the effective theory. These include open string moduli associated to the motion and deformation of branes and corresponding gauge moduli, such as bundle moduli, associated to deformations of vector bundles present in the compactification.

\subsection{General Properties of Moduli}
   
We have said above that the existence of extra dimensions is the most important physical implication of string theory. As moduli are the way these extra dimensions manifest themselves in the 4-dimensional effective field theory, 
moduli are arguably the most important type of particle arising in string compactifications. The 
moduli are scalar degrees of freedom in the effective action of the 4-dimensional observer and describe low energy
excitations in the extra dimensions (such as shape and size of the extra dimensions). They are 
gauge singlet scalars, typically with gravitational strength interactions. In the simplest supersymmetric compactifications with extended supersymmetry, the potential remains flat and the moduli are massless. Besides other issues such as the 
absence of chiral matter, such models are automatically ruled out since these massless moduli would mediate unobserved long-range scalar gravitational-strength interactions (fifth forces).

 Luckily, for models with $\cN= 1$ or $0$ supersymmetry (which, in any case, are the ones of phenomenological interest), there
exist `moduli stabilisation mechanisms'. These lift the flat potentials, give them a mass and allow for the construction of phenomenologically
viable models. Even though moduli are gauge singlets and hard to detect experimentally, their role in string cosmology cannot be over-emphasised. 

Why? Moduli are, in a stringy context, the most natural candidates to be inflaton fields or to drive any alternative early universe cosmology. This is already very important. But what is perhaps even more important, and highly relevant for the later cosmological evolution of the universe, is that in this context the inclusion of moduli into the spectrum has a unique ability to undercut and render invalid the pre-existing cosmology. As we discuss in detail
in chapters \ref{sec:infla}, \ref{reheating} and \ref{sec:DE}, moduli fields can potentially help address many important unanswered questions such as the nature of dark energy, dark matter and dark radiation. 

 Furthermore,  vacuum expectation values of moduli also determine the low energy effective action of a model. As mentioned above, string theory  has no free dimensionless parameters: couplings and ratios of scales in the low energy effective action are set by the 
 values taken on by the moduli. Thus, the task of computing moduli potentials and finding their minima lies at the heart of \emph{string} phenomenology.

 At some levels, moduli are simply examples of scalar fields. The discovery of the existence of an
apparently fundamental scalar (the Higgs) confirms the existence of scalar fields in nature and gives
further motivation for studying their properties. However in many ways, the properties of moduli are crucially different from more familiar scalars such
as the Higgs, and intuitions carried over from the Standard Model electroweak theory or the 
Minimal Supersymmetric Standard Model (MSSM) are misleading when applied to moduli.

Let us list some of these differences:

\begin{itemize}

\item{Moduli are uncharged under Standard Model gauge fields}

It is a basic feature of string moduli that they are neutral under the Standard Model, and also normally under
any additional hidden sector gauge groups that may be introduced. Neutrality under gauge interactions is key to some of the most interesting
features of moduli, as it implies they have no `quick' decay modes.\footnote{In some cases a modulus may be non-linearly charged under an anomalous $U(1)$. This only applies for moduli with shift-symmetries (see below). In this case, the real part of the modulus enters the Fayet-Iliopoulos D-term, and the
axionic part of the modulus is eaten by the massive $U(1)$. The D-term condition then fixes a combination of the real part of the modulus
and the matter fields that are charged under the anomalous $U(1)$, with the orthogonal combination remaining massless.}

\item{The couplings and interactions of moduli come with $\Mp^{-1}$ factors}

Even neutral fields can decay rapidly if they have renormalisable couplings to Standard Model (SM) matter, for example $\Phi h h h$, where $\Phi$ is some modulus field, and $h$ some scalar SM field (such as the Higgs).
However string moduli often descend from higher-dimensional modes of the graviton. This means that all their couplings -- both self-couplings and
couplings the Standard Model sector -- are `really' non-renormalisable.
This includes apparently renormalisable couplings such as
\be
\lambda \,\Phi^4 \in \mc{L}\,.
\ee
In such a coupling, $\lambda$ is dimensionless. However, for moduli $\lambda$ is given by $\frac{m_{\Phi}}{\Mp}$ or $\frac{m_{3/2}}{\Mp}$, where $m_{3/2}$ is the gravitino mass. In this case, $\lambda$ carries hidden factors of $\Mp^{-1}$ and so is numerically extremely small.

The fundamental scale in string theory is the string scale, $M_s$, and not the 4-dimensional Planck scale $\Mp$. In cases
where the string scale and Planck scale are widely separated -- for example with a large compactification volume -- the difference can be significant. Moduli that control local properties of the extra dimensions, such as the size of blow-up cycles, have interactions suppressed by 
the string scale, whereas moduli that control global properties, such as the overall volume, have interactions suppressed by the 4d Planck scale.
Cosmologically it is the latter that are most relevant, as they have the weakest interactions and so survive for the longest time period.

Combined with the neutrality of moduli, the consequence of the $\Mp$ suppression is that moduli always interact weakly. They are hard to produce -- but once produced, they are also hard to get rid of, as they do not thermalise and so live for a long time.

\item{There is generally no concept of `zero VEV' for moduli}

Many scalar fields have a well-defined notion of zero VEV, which acts as a preferred locus in field space. The zero VEV location often
corresponds to the restoration of a broken symmetry. For example, for gauge charged scalars, the gauge symmetry is spontaneously broken for non-zero VEV and restored at zero VEV. An example of this behaviour is the Higgs field, for which the VEV signals electroweak symmetry breaking.

This is not true of moduli. The VEVs of string moduli should instead be interpreted as the values of compactification parameters. For example, the vev of the volume modulus corresponds to the compactification volume (with the canonically normalised field $\Phi$ being defined as $\Phi = - \Mp \ln \mc{V}$) and the VEV of the dilaton corresponds to the string coupling. There is no preferred value for these fields and thus there is no notion of zero VEV for the moduli.\footnote{One can of course define the 'zero vev locus' as the position of the modulus at the full minimum of the potential. However, this is semantics as this locus does not have any \emph{a prior} significance prior to the determination of the potential.} In the low energy theory, the values of these VEVs set the coupling `constants', such as the gauge couplings and the Yukawa couplings.

Note that one exception to this idea are the blow-up moduli (note the universal moduli such as the volume or dilaton do \emph{not} fall into these categories). These moduli control the blow-up of singularities in the geometry, from zero size to a finite radius. These are commonly associated to orbifold points (where the blow-up moduli are also called twisted sector moduli). In this case, the notion of zero VEV does make sense -- zero VEV corresponds to the singular limit in which the cycle is blown down to zero size, whereas finite VEV corresponds to the resolution of the singular geometry into a smooth space.

\item{The `infinite VEV' limit represents a decompactification limit}

There are generally certain directions in field space -- specifically for the volume and dilaton modulus -- where the moduli space is unbounded.
Along such directions, the VEV of the volume or dilaton modulus can be increased arbitrarily, while remaining within the low energy 4-dimensional effective field theory. Even if towers of states descend exponentially in mass as per the swampland distance conjecture \cite{Ooguri:2006in, Ooguri:2018wrx}, as long as there remains a hierarchical separation between moduli and KK masses, so that $m_{\Phi}/m_{KK} \ll 1$, the 4-dimensional effective field theory remains a good description of the low-energy physics. 

In the simplest 1-modulus example, the metric on moduli space is set by $K = - 3 \ln (T + \overline{T})$ (for the overall volume modulus) or $- \ln (S + \overline{S})$ (for the dilaton), giving the kinetic terms
\begin{subequations}
\label{eq:nsnT1}
\begin{empheq}[box=\widefbox]{align}
3 \frac{\partial_{\mu} \tau_R \partial^{\mu} \tau_R}{4 \tau_R^2} + 3 \frac{\partial_{\mu} \tau_I \partial^{\mu} \tau_I}{4 \tau_R^2} & \hbox{\,\, (volume modulus), } \\
\frac{\partial_{\mu} S_R \partial^{\mu} S_R}{4 S_R^2} + \frac{\partial_{\mu} S_I \partial^{\mu} S_I}{4 S_R^2} & \hbox{ \,\,(dilaton) },
\end{empheq}
\end{subequations}
where $T= \tau_R+i\,\tau_I$, \,\, $S=S_R+i\,S_I$.
The canonically normalised fields are $\Phi_T = \sqrt{\frac{3}{2}} \ln \tau_R$, $\Phi_S=\sqrt{\frac{1}{2}} \ln S_R $, and from this
it follows that the limits $\tau_R \to 0$, $\tau_R \to \infty$, $S_R \to 0$ or $S_R \to \infty$ are all at infinite distance in field space
from any finite value of $\tau_R$ or $S_R$. The infinite limits correspond to a decompactification limit, in which the string scale $M_s = \frac{g_s^{3/4}}{\sqrt{4\pi}} \frac{\Mp}{\sqrt{\mathcal{V}_s}}$ (where $\mathcal{V}_s$ is the compactification volume in string frame related to the volume in Einstein frame $\mathcal{V}$ as $\mathcal{V}_s= g_s^{3/2} \mathcal{V}$ after Weyl rescaling to 10-dimensional Einstein frame\footnote{ The relation between the metrics in string and Einstein frames in 10 dimensions is convention dependent and in general is given by  $G_{MN}^S = e^{\frac{\phi-\phi_0}{2}}G_{MN}^E$, the above choice of conventions corresponds to $\phi_0=0$. See \cite{ValeixoBento:2023afn} for a guide into frames conventions used in string compactifications.} for a ) is infinitely smaller than the 4-dimensional Planck scale $\Mp = 2.4 \times 10^{18} \hbox{GeV}$: $M_s/\Mp \to 0$.

\item{Moduli often carry shift symmetries}

When the low-energy effective field theory preserves $\cN=1$ supersymmetry, it is common for the moduli representing the scalar part of the chiral multiplet to carry shift symmetries. For example, in type IIB D3/D7 compactifications, the dilaton and K\"ahler moduli carry shift symmetries
while the complex structure moduli do not (in type IIA compactifications, it is the complex structure moduli which have shift symmetries). Moduli with shift symmetries generally enter the gauge kinetic functions, where their imaginary parts
are the \emph{axions} of the corresponding gauge groups.

In such cases, the origin of the shift symmetry is normally that  the real part of the modulus corresponds to the size of a cycle, $\Sigma_i$, whereas the imaginary part of the modulus corresponds to the reduction of an RR form field on this cycle, $\int_{\Sigma_i} C_i$. The shift symmetry of the modulus arises from the fact that the RR fields have no perturbative couplings to the string modes, and only couple to branes.

As chiral multiplets, the statement of the shift symmetry for a modulus $T$ is that the perturbative action is invariant under $T \to T + i \,c$,  where $c$ is a constant, and is thus a function only of $T + \overline{T}$. The imaginary part of these moduli, $\hbox{Im}(T)$, are axions, and are massless in perturbation theory.

In terms of moduli stabilisation, the significance of a shift symmetry is that a potential for the modulus cannot be generated perturbatively, and 
can only appear via non-perturbative effects such as brane instantons or gaugino condensation \cite{Dine:1986vd}. In a weakly coupled theory, where such non-perturbative effects are automatically small, this implies that such moduli are light.

\end{itemize}

With this, we end our general discussion of the properties of moduli fields and now turn to moduli stabilisation. Our discussion aims to
provide an overview of the subject, without getting into the technicalities. We refer the reader to the review articles  and lecture notes \cite{ Silverstein:2004id,  Grana:2005jc, Douglas:2006es, Denef:2007pq, Denef:2008wq, Weigand:2018rez, Hebecker:2020aqr} for more technical discussions. The book \cite{Ibanez:2012zz} provides comprehensive introduction to string phenomenology (along with a discussion on moduli stabilisation). We defer related discussions around conjectured swampland constraints on low energy effective field theories to Sec. \ref{Sec:Swamp}.

\subsection{Moduli Stabilisation}
\label{sec:MS}

The low energy effective action of string theories in ten dimensions can be organised in a double expansion: the $\alpha'$ and $g_s$ expansions. The former captures the effect of integrating out heavy string modes (i.e the massive string states) whereas the latter describes string loops. At leading order, the effective low-energy actions are the 10-dimensional supergravity theories (and 11-dimensional supergravity in the case of M-theory). 

The simplest compactified vacuum configurations are those in which the internal flux fields vanish and the scalar fields in the 10-dimensional
actions are constant. As a result, the 10-dimensional matter stress-tensor vanishes, leading to a Minkowski compactification with a Ricci flat internal manifold. A requirement that some supersymmetry is preserved then implies that the internal manifold is a Calabi-Yau. Upon dimensional reduction, this
leads to massless (complex) scalars whose wavefunctions in the extra dimensions are given by harmonic forms on the Calabi-Yau (K\"ahler deformations and axionic fields that arise from the dimensional reduction of form fields pair up as complex scalars, whereas complex structure deformations are intrinsically complex). 

As mentioned earlier, these massless scalars are disastrous for phenomenology and so construction of phenomenologically viable models requires incorporating
effects that stabilise the moduli. This requires going beyond the simplest solutions and incorporating various additional effects into
the  effective action.  The analysis depends on the type of string theory. Before getting into the details for each case, we give
a qualitative description of the key ingredients. As the appearance of moduli within simple compactifications is due to the presence of flat directions in the low energy effective field theories' scalar potential, to lift them we need to include effects that lead to a non-trivial energy profile along these directions.

\begin{itemize}

\item {\it Fluxes}: A $p$-form flux can thread a $p$-cycle, $\Sigma_p$, in the internal manifold. The threading is characterised by integers, as the Dirac quantisation condition forbids continuous deformations.
The presence of  background flux can  lead to a non-trivial energy profile along various directions in field space. For instance, for the 
overall radius of the compactification $(R)$, a $p$-form flux contributes to the potential (see \cite{Giddings:2003zw,Silverstein:2004id} 
for derivations of these different scalings) as 
$$
   V(R) \propto R^{-6 -2p},
$$
lifting the flatness of the radial direction. These fluxes are crucial to all flux compactifications, and also appear in e.g. the maximally supersymmetric $AdS_5 \times S^5$ solution used in the AdS/CFT correspondence \cite{Maldacena:1997re}.

\item {\it Localised objects}: Space filling D-branes and orientifold planes are consistent with maximal symmetry in four dimensions and contribute
to the moduli potential. For a $p$-dimensional localised object, the contribution to the potential for the radial mode scales as 
$$
  V(R) \propto T_p R^{p -15},
$$
where $T_p$ is the tension of the object. We note that this tension is negative for O-planes.

\item {\it Extra dimensional curvature}: Backgrounds with non-trivial matter stress-tensors have non-vanishing curvature in the
extra dimension, which also contributes to the effective potential. For the radion,
$$
  V(R) \propto {1 \over R^{8} },
$$
with a positively curved internal space making a negative contribution to the potential (e.g.~in the $S^5$ in the AdS/CFT $AdS_5 \times S^5$ solution).

\item{ \it $\alpha'$ and loop corrections:} The effective potential receives contributions order by order in the $\alpha'$ and $g_s$ expansions. These
can lift directions which are flat in the leading order approximation. For instance, the leading $\alpha'$ correction in type IIB  \cite{Becker:2002nn} makes a contribution to the radion potential which behaves as
$$
  V(R) \propto {1 \over R^{18}}.
$$
Such $\alpha'$ corrections are crucial in e.g. the Large Volume Scenario.

\item {\it Non-Perturbative effects:} Non-Perturbative effects such as gaugino condensation or wrapped Euclidean branes play a key role
in stabilising flat directions associated with axionic shift symmetries. These symmetries are broken non-perturbatively and  potentials are induced which are exponential functions of the moduli.

\item  {\it Supersymmetry breaking:} Breaking of supersymmetry leads to (low energy) loop corrections to the potential which will themselves depend on the moduli VEVs.

\end{itemize}

 Understanding how these effects can combine to yield vacua where all moduli are stabilised is highly non-trivial.\footnote{Given this, one interesting alternative avenue is to look for string vacua which from the onset have either a small number of moduli or no moduli at all. Asymmetric orbifolds (see e.g. \cite{Narain:1986qm, Blumenhagen:2000fp, Silverstein:2001xn}) are one approach in this direction. However, to the best of our knowledge, there is no yet a single 4-dimensional construction without moduli.} Furthermore, there are various problems and no-go theorems, which clarify the challenges involved, while at the same time providing guidance on the necessary ingredients for any successful stabilisation mechanism.

\begin{figure}[t]
\begin{center}
\includegraphics[width=120mm,height=80mm]{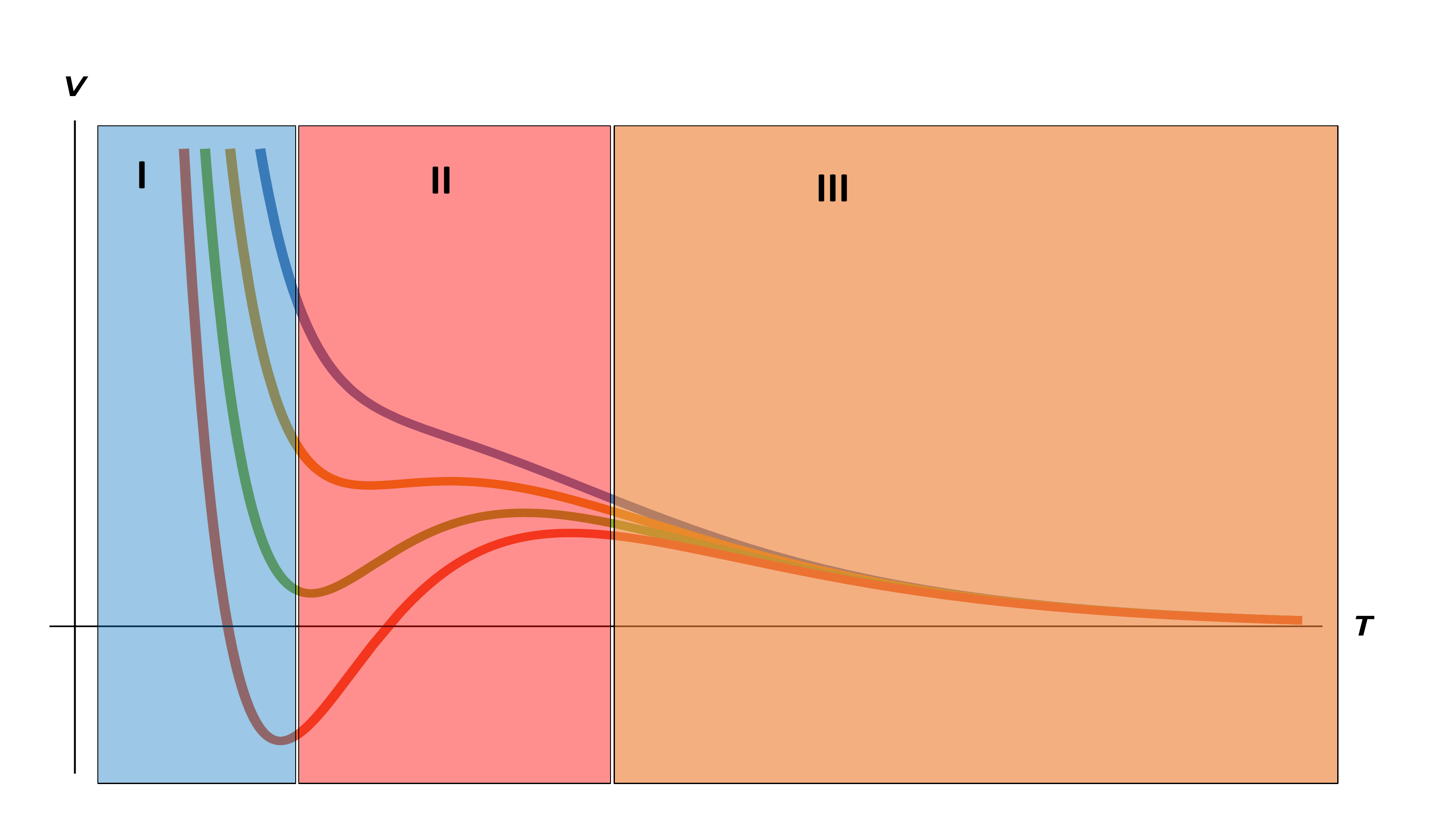} 
\caption{Dine-Seiberg problem. The scalar potential as a function of the volume or dilaton modulus vanishes asymptotically. Since these fields are the $\alpha'$ and loop expansion parameters respectively, the only region in which these expansions are under full control is the runaway region III. If there is a non trivial minimum it would naturally fall in the small volume/strong coupling region I. In order to obtain reliable minima in the desired region II in which hierarchies  and weak couplings exist (as seen in nature),  compactification parameters, such as integer fluxes or ranks of gauge groups, need to be used.} \label{Fig:CY} 
\end{center}
\end{figure}

\smallskip
\ni {\it {The Dine-Seiberg problem \cite{Dine:1985he}:}}  The Dine-Seiberg problem (illustrated in figure \ref{Fig:CY}) stems from the fact that the 
parameters for the loop and $\alpha'$ expansions are themselves moduli VEVs (the dilaton and the volume of the compactification). Leading order flatness in these directions implies that stabilisation must necessarily involve competition between subleading terms. The Dine-Seiberg problem is the claim that if two terms in the $g_s^{-1}$ or $R^{-1}$ expansions are competing, then all terms should compete and thus any resulting vacuum, except for the runaway-one corresponding to the 10-dimensional free theory,  is at strong coupling and so not trustable. In epigraph form: `If the expansion can be trusted, the modulus is not stabilised; if the modulus is stabilised, the expansion cannot be trusted.'

Key to this formulation is the idea that stabilisation of either dilaton or K\"ahler moduli involves comparing terms from a single expansion in which there is no extra structure to suppress the strength of each term. Later, we will discuss how this problem is alleviated in the context of type IIB flux compactifications (either through additional moduli or the use of the no-scale structure present in the compactification).\\

\noindent {\it{Constraints from positivity conditions of the stress tensor:}}  Positivity conditions obeyed by stress 
tensors of low energy effective actions lead to interesting constraints on the allowed solution space. As the 10 and 11-dimensional
supergravity theories that arise in string theory obey the strong energy condition, this leads to an important no-go theorem \cite{Gibbons:1984kp, deWit:1986mwo, Maldacena:2000mw}.

The general 10-dimensional metric consistent with maximal symmetry in 4 dimensions is of the form
$$ 
  ds^{2} = g_{MN} dx^{M} dx^{N} = e^{2A(y)} \tilde{g}_{\mu \nu} dx^{\mu} dx^{\nu} + g_{mn}(y) dy^{m} dy^{n},
$$
where $\tilde{g}_{\mu \nu}$ is the metric of a maximally symmetric space in 4 dimensions. The trace reversed Einstein equation in the non-compact directions reads
\begin{equation}
 R_{\mu \nu} = \tilde{R}_{\mu \nu} - \tilde{g}_{\mu \nu} \left( \nabla^{2} A + 2 ( \nabla A)^{2} \right) = T_{\mu \nu} - {1 \over 8} e^{2A} \tilde{g}_{\mu \nu} T^{L}_{\phantom{L} L} \ .
\end{equation}
Contracting this with $g^{\mu \nu}$ one finds
\begin{equation}
\label{gauss}
  \tilde{R} + e^{2A} \hat{T} = 2 e^{-2A} \nabla^{2}  e^{2A},
\end{equation}
where we have defined
\begin{equation}
\label{hatT}
  \hat{T} =  - T^{\mu}_{\phantom{\mu} \mu} + {1 \over 2} T^{L}_{\phantom{L} L}
   = {1 \over 2} \left( - T^{\mu}_{\phantom{\mu} \mu} + {1 \over 2} T^{m}_{\phantom{m} m} \right) \ .
\end{equation}
It is easy to check that $\hat{T}$ is positive semi-definite for all $p$-form flux configurations which are consistent with maximal symmetry. In particular, it is positive semi-definite for $p \geq 1$ and vanishes for $p=1$. Now, multiplying (\ref{gauss}) by $e^{2A}$ and integrating over the compact manifold one concludes that dS compactifications are impossible while Minkowski compactifications allow only 1-form flux. This `no-go theorem' can be evaded by making use of local terms in the supergravity action. For example, orientifold planes in IIB string theory carry negative tension and provide  (localised) negative contributions to $\hat{T}$. Furthermore higher derivative corrections, loop corrections and non-perturbative effects will in general modify (\ref{gauss}). 

The above result has been generalised to various settings in  \cite{Hertzberg:2007wc, Green:2011cn, Gautason:2012tb, Quigley:2015jia, Kutasov:2015eba, Basile:2020mpt, Andriot:2022xjh}.
In particular,  \cite{Kutasov:2015eba} obtained a no-go theorem for dS solutions (of any dimensionality) to all orders in the $\alpha'$ expansion
in heterotic strings  (recently this was generalised to one loop in a specific setting involving string theory without spacetime supersymmetry \cite{Baykara:2022cwj}). Implications for other string theories follow from dualities. 

Attempts to construct dS space in string theory always involve the inclusion of corrections to the effective action which allow for the evasion of the no-go theorems. That said, we note that there also exist arguments that the obstacles to dS are deeper and more fundamental than simply the absence of particular objects in the low-energy supergravity theory. For examples of these principled obstructions to dS, see 
\cite{ Obied:2018sgi, Banks:2012hx, Banks:2019oiz, Brennan:2017rbf, Danielsson:2018ztv}  (the Swampland dS conjectures will be discussed in
Sec. 7) and also \cite{Sethi:2017phn} for an alternative perspective.

We next discuss the generation of moduli potentials -- moduli stabilisation -- in various string theories. Our focus will be on the form of the effective potential and Minkowski/AdS/dS solutions in four dimensions. Time-dependent cosmological solutions will be discussed in the later sections. We start with arguably the most developed constructions, those of type IIB string theory.

\subsection{Moduli Stabilisation: IIB}

\label{sec:dSstrings}

Moduli stabilisation in the context of semi-realistic vacua (i.e.~incorporating hierarchies and supersymmetry breaking) is best understood in the context of type IIB models and we start by discussing these models. In type IIB, for a special (albeit large) choice of the fluxes and localised sources,
one has  the knowledge of 10-dimensional solutions which incorporate the backreaction of fluxes. These fluxes also stabilise the dilaton and complex structure moduli.  This class is often referred to as pseudo-BPS. We discuss these  following \cite{Giddings:2001yu, Dasgupta:1999ss, Gukov:1999ya} (see  e.g.  \cite{hep-th/0012213, Taylor:1999ii, Michelson:1996pn} for earlier work).  The construction of these solutions starts by considering an orientifolded Calabi-Yau (all field configurations and fluctuations are required to be consistent with the orientifolding, for details see \cite{Grimm:2004uq}). The
3-form fluxes of the IIB theory, $F_3$ and $H_3$ are turned on and satisfy Dirac quantisation conditions
\begin{equation}
\setlength\fboxsep{0.25cm}
\setlength\fboxrule{0.4pt}
\boxed{
\label{fluxquanta}
{1 \over 2 \pi \alpha'}  \int_{\Sigma_{3}} F_3 \in 2 \pi \mathbb{Z},  \phantom{abc} {1 \over 2 \pi \alpha'}  \int_{\Sigma_{3}} H_3 \in 2 \pi \mathbb{Z}.
}
\end{equation}
These thread the 3-cycles of the Calabi-Yau. The 10-dimensional Einstein frame metric takes the form
\begin{equation}
\label{gkpxx}
  ds^2 = e^{-2A(y)} \eta_{\mu \nu} dx^{\mu} dx^{\nu} + e^{2A(y)} \tilde{g}_{mn}(y) dy^{m} dy^{n},
\end{equation}
where $\tilde{g}_{mn}(y)$ is the metric of the underlying Calabi-Yau  and $e^{-2A(y)}$ is the `warp factor'. Thus, one is naturally
led to warped compactifications with an internal manifold which is conformal to a (orientifolded) Calabi-Yau -- an appealing aspect since it preserves
much of the structure and intuition of pure CY compactifications.\footnote{The metric does not take the form (\ref{gkpxx}) for general solutions in IIB; this form applies only for those in the pseudo-BPS class. We will describe below the conditions that this implies on the localised sources and fluxes.}

The warp factor is sourced by 3-form flux and localised objects\footnote{These have to be space-filling to maintain Poincare invariance.
} which carry D3-charge and is determined by the equation:
\begin{equation} 
\label{warpf}
\setlength\fboxsep{0.25cm}
\setlength\fboxrule{0.4pt}
\boxed{
 -  \tilde{ \nabla}^{2} e^{-4A} =   { G_{mnp} \tilde{G}^{mnp}  \over {12 {\rm{Re }} S} } + 2 \kappa_{10}^2 T_{3} \tilde{\rho}^{\rm loc}_3
 }
\end{equation}
where $S =  g_s^{-1} - i\,C_0$ is the axio-dilaton, $G_{3} = F_3 - iS H_3$ is the complexified 3-form flux, $\tilde{\rho}^{\rm loc}_{3}$
is the localised $D3$ brane charge density and $T_3$ the tension of D3-branes in the 10-dimensional Einstein frame. The superscript
tilde is used to indicate the use of the metric $\tilde{g}_{mn}$.

The localised sources which contribute to the D3-charge are D3-branes and O3-planes which are point-like in the extra dimensions, as well as D7-branes and O7-planes which wrap four cycles of the Calabi-Yau (these are the only localised sources allowed for these pseudo-BPS solutions).   Note that the contribution of the fluxes to the right hand side of (\ref{warpf}) is positive definite. Thus, cancellation of the $D3$ tadpole condition requires the presence of carriers of negative $D3$ charge. This is provided by the O3-planes\footnote{As mentioned above, the orientifold planes are central to the very existence of these compactifications, as they have a negative tension and hence can provide a negative contribution to $\hat{T}$ (as defined in (\ref{hatT})). This evades the no-go theorem and enables the existence of warped Minkowski compactifications.} or wrapped D7-branes. Thus, the choice of flux quanta is limited to those whose total D3-charge is less in magnitude than the upper bound set by the contributions from the negative charge carriers.\footnote{The solutions can be generalised to F-theory, where, the induced D3-charge from the D7-branes D3 is given by the Euler number of the associated four-fold X, $Q_{\rm D3} = { \chi(X) \over 24}$.} While finite (see for instance \cite{Bakker:2021uqw,Grimm:2021vpn}), the number
of consistent flux configurations for a given Calabi-Yau can be enormous, leading to the idea of the string landscape
\cite{Bousso:2000xa, Feng:2000if,Susskind:2003kw, Denef:2007pq}. We will discuss this in more detail in Sec. \ref{ssec:landscape}.

The equations of motion also require that the complexified 3-form flux is imaginary self-dual, i.e.
\be
   G_{3} = i *_6 G_3 \ .
 \label{isd}
 \ee
In terms of Hodge decomposition, this implies $G_3\in(2,1)\oplus(0,3)$ within the Hodge structure of the Calabi-Yau.  
 
In general, these solutions break supersymmetry. Supersymmetry is preserved if $G_3$ is purely $(2,1)$ (it is the presence of a $(0,3)$ component that breaks supersymmetry). Even if supersymmetry is broken at this level, we note that higher order corrections in the 4-dimensional effective action can then restore supersymmetry, and we will see explicit examples of this later in the section.  
   
For a given choice of flux quanta in  (\ref{fluxquanta}), the imaginary self dual condition can be regarded as an equation for the metric and the dilaton, which fixes the values of the complex structure moduli and the dilaton.\footnote{For early work
 on complex structure moduli stabilisation in specific settings  see \cite{Dasgupta:1999ss, hep-th/0201028, Frey:2002hf, hep-th/0312104, Giryavets:2004zr, Conlon:2004ds}.  More recently, explicit studies of the flux induced potential have been carried out in \cite{Cicoli:2022vny, Coudarchet:2022fcl}.  
 An interesting recent development  is the `tadpole conjecture' (see \cite{Bena:2020xrh, Marchesano:2021gyv, Plauschinn:2021hkp, Grana:2022dfw, Lust:2021xds, Lust:2022mhk, Becker:2022hse}). This suggests a tension between stabilising complex structure moduli at generic points in moduli space and keeping the flux-induced contribution to the D3-tadpole within the bounds allowed by the orientifold.}
As a result of the presence of fluxes, complex structure moduli acquire a mass
$$
m_{{\rm cs}} \sim { \Mp \over {\mathcal{V}}}\,, 
$$
where $\mathcal{V}$ is the Einstein frame volume of the internal manifold measured in string units (our frame conventions throughout this section will be those described below equation (\ref{eq:nsnT1})). In general, there is no reason that a particular flux configuration should lead to the dilaton being 
stabilised at weak coupling. To have control over the  effective field theory, only those flux choices that lead to $g_{s} \ll 1$ are relevant. The K\"ahler moduli remain unfixed at this level.  As we will see, these K\"ahler directions can be stabilised at  large volumes by the inclusion of perturbative (in both $\alpha'$ and $g_s$) and non-perturbative corrections to the effective action.  This leads to isolated vacua at  $g_{s} \ll 1$ and large volume,
 where the effective field theory is under control. For this reason, models combining moduli stabilisation, hierarchies and supersymmetry breaking are best understood in type IIB and several proposals for the construction of dS vacua have been made in the setting. 
 
A further phenomenologically appealing feature of the solutions is the presence of regions in the internal manifold with large warping.  The compactifications therefore provide a realisation of the ideas of Randall and Sundrum \cite{Randall:1999ee, Randall:1999vf} in an ultraviolet complete setting. Large warping arises when fluxes thread a shrinking 3-cycle. For instance, if $M$ units of flux thread the shrinking `A cycle' of a conifold
singularity and $K$ units thread its dual `B cycle', then the conifold singularity is resolved and the warp factor on the minimal volume $S^{3}$ associated with the resolution is
$$
  e^{A_{\rm 0}} =  {\rm{exp}} \left( -2\pi K \big{/} 3 M g_s \right).
$$
Note that this factor is exponentially small in the integer flux quanta.  Locally, the geometry is well described by the Klebanov Strassler solution \cite{Klebanov:2000hb} and such regions of large warping are often referred to as warped throats. Warped throats have a wide variety of phenomenological applications and will appear repeatedly in our discussion of cosmological models.
    
We next turn to the 4-dimensional effective action describing the low energy fluctuations about these backgrounds (obtained once the Kaluza-Klein modes are integrated out). This is crucial for developing K\"ahler moduli stabilisation and also provides a complementary 4-dimensional perspective for looking at the complex structure moduli stabilisation. 
   
 We first describe  basic features of the effective action (see  \cite{Giddings:2001yu,  Grimm:2004uq} for  further details).
The relevant closed string fields are the   complex structure moduli $U_a$, $a=1,\cdots, h_{-}^{1,2}$ (the number of harmonic $(2,1)$-forms of the Calabi-Yau that are odd under the orientifold involution), the  axio-dilaton $S$, and the K\"ahler moduli $T_i$, $i=1,\cdots, h^{1,1}$. For simplicity, here we discuss the case where $h^{1,1}_{-} = 0$ (for a detailed analysis of the effective action for compactifications with $h^{1,1}_{-} \neq 0$ see \cite{Grimm:2004uq}, and e.g.~\cite{Cicoli:2021tzt} for a recent discussion). The K\"ahler moduli are then complexified by the definitions:
$$
\setlength\fboxsep{0.25cm}
\setlength\fboxrule{0.4pt}
\boxed{
    T_{i} = {1 \over 2} \int_{\Sigma_{i}} J \wedge J + i \int_{\Sigma_{i}} C_4 \equiv \tau_i + i \theta_i \ ,
}
$$
where $J$ is the CY K\"ahler form (in the Einstein frame, in units of the string length $\ell_s = 2 \pi \sqrt{\alpha'}$) and $C_4$ the 4-form potential. The integrals are performed over 4-cycles of the orientifolded Calabi-Yau. 

 In the language of $\mathcal{N}=1$ supersymmetry, the fields lie in chiral multiplets and the low energy effective action is specified by the K\"ahler potential and superpotential. The tree-level K\"ahler potential (in the limit of large volume) is given by 
\begin{subequations}
\label{eq:nsnT3}
\begin{empheq}[box=\widefbox]{align}
K&= K_{\rm kah} + K_{\rm dil} + K_{\rm cs} \cr
&= -2\ln\vo-\ln\left(S+\overline{S}\right)-\ln \left(-i\, \int_X \Omega\wedge \overline{\Omega}\right) ,
 \label{Ktree}
 \end{empheq}
\end{subequations}
where $\vo = \ell_s^{-6} \int_X \sqrt{g_{(6)}}\, d^6 y$ is the volume of the internal manifold (in the Einstein frame) in units of the string length $\ell_s$. The internal volume $\vo$ is a homogeneous function of degree $3/2$ of the real parts of the K\"ahler moduli $\tau_i$, the volumes of the four cycles. $\Omega$ is  the holomorphic $(3,0)$-form of the manifold. The effect of fluxes is captured 
by the Gukov-Vafa-Witten superpotential \cite{Gukov:1999ya}:
\be
\setlength\fboxsep{0.25cm}
\setlength\fboxrule{0.4pt}
\boxed{
W_\flux=\int_X G_3\wedge \Omega .
\label{Wflux}
}
\ee

Recall that the F-term supergravity scalar potential for a  superpotential $W(\Phi_I)$ and K\"ahler potential $K(\Phi_I, \overline{\Phi}_{\bar{I}})$ is  given by (in units where $\Mp = 1$):
\be
\setlength\fboxsep{0.25cm}
\setlength\fboxrule{0.4pt}
\boxed{
V_F= e^K\left(K^{I\overline{J}}\, D_{I} W {D}_{\overline{J}} \overline{W} - 3|W|^2\right) \,,
\label{VF}
}
\ee
where $K^{I \overline{J}}$ is the inverse of the K\"ahler metric and $D_{I}$ are  K\"ahler covariant derivatives:
 $D_{I} \equiv  \partial_{I} + (\partial_{I} K) W$. For the K\"ahler potential (\ref{Ktree}) and superpotential (\ref{Wflux}), the 
 potential (\ref{VF}) only depends on the K\"ahler moduli through the overall prefactor $e^K$. This comes from a combination of the following facts:

\begin{itemize}

\item The superpotential (\ref{Wflux}) is independent of the K\"ahler moduli. This is obvious from its expression, as the holomorphic 3-form only depends on the complex structure data. But, there is also a symmetry reason behind this perturbative absence of the K\"ahler moduli from the superpotential. The imaginary part of the 
K\"ahler moduli are axionic fields which enjoy shift symmetries $T_i\rightarrow T_i + i\, c_i$ with constant $c_i$'s. This, together 
with the fact that the superpotential is holomorphic, implies that the superpotential cannot depend on $T_{i}$ (within perturbation theory). This shift symmetry is preserved to all orders in perturbation theory \cite{Burgess:2005jx, Witten:1985bz, Burgess:1985zz, Dine:1986vd}, and hence the superpotential is independent of the K\"ahler moduli to all orders in perturbation theory.

\item The K\"ahler potential is of the no-scale form \cite{Cremmer:1983bf, Ellis:1983sf} i.e $K^{T_i\overline{T}_{\bar{\jmath}}} K_{T_i} K_{\overline{T}_{\bar{\jmath}}}=3$. This is a consequence of the fact that the volume $\mathcal{V}$ is a degree $3/2$ homogeneous polynomial in $\tau_i$.

\end{itemize}
With this, the potential is solely a function of the complex structure moduli and the dilaton:
$$
  V^{\rm no-scale}_{F}  = e^K K^{\alpha\overline{\beta}}\, D_{\alpha} W \overline{D}_{\overline{\beta}} \overline{W},
$$
where the indices $\alpha, \beta$ now run only over the complex structure moduli and the dilaton. The potential is minimised
by solving 
$$D_{\alpha} W = 0.$$
This can be shown to be equivalent to the imaginary self-dual condition on the fluxes (\ref{isd}).
At the minimum, the potential vanishes (consistent with the fact that these are Minkowski compactifications). Since $D_{\alpha} W$ are proportional to the corresponding F-terms, this means that the complex structure and dilaton moduli do not break supersymmetry, however the F-terms
of the K\"ahler moduli ($F_i = D_{T_{i}}W$) are non-vanishing and supersymmetry is broken unless $W=0$ (which is equivalent to $G_{3}$ being $(2,1)$ ).
These reflect the standard properties of no-scale vacua: a vanishing vacuum energy $V=0$ together with broken supersymmetry.

Let us next discuss the stabilisation of the K\"ahler moduli. As discussed above, the shift symmetries of the axionic parts of the K\"ahler moduli forbid their appearance in the superpotential to all orders in perturbation theory. However, as these moduli represent the gauge couplings for matter fields on D7-branes, non-perturbative effects like gaugino condensation on D7-branes or Euclidean D3-instantons can generate a superpotential for them (see  \cite{Blumenhagen:2009qh} for a review). The full superpotential for closed string moduli takes the form:
\be
W=W_{\rm flux}(S,U)+W_{\rm np}(S,U,T)\,.
\label{Wtotal}
\ee
The K\"ahler potential  for the K\"ahler  moduli receives various perturbative corrections. In general,
the no-scale condition $K^{T_i\overline{T}_{\bar{\jmath}}} K_{T_i} K_{\overline{T}_{\bar{\jmath}}}=3$, satisfied by the tree level K\"ahler potential, will be 
broken by these corrections. We denote the corrections by $K_{\rm p}$, i.e. 
\be
\label{kcor}
  K_{\rm kah} =  -2\ln\vo + K_{\rm p}.
 \ee
The correction terms $W_{\rm np}$ and $K_{\rm p}$ in (\ref{Wtotal}) and (\ref{kcor}) lead to a potential for the K\"aher moduli. This potential generates a minimum for the moduli,
stabilising them, and is also crucial for understanding the moduli dynamics in a cosmological context.

There are two major scenarios for fixing the K\"ahler moduli. These are the  KKLT construction \cite{Kachru:2003aw} and the Large Volume Scenario (LVS) \cite{Balasubramanian:2005zx}, which we now describe these in detail. Other proposals for K\"ahler moduli stabilisation within the ambit of IIB flux compactifications include \cite{vonGersdorff:2005bf, Berg:2005yu, Westphal:2006tn, Cicoli:2012fh,Gallego:2017dvd, Antoniadis:2018hqy, AbdusSalam:2020ywo}. The most pertinent details of these two main constructions are   
\begin{itemize}
    
\item The KKLT construction  makes use of the fact that the vacuum expectation value of the flux superpotential can be tuned to small values. 
   This serves as a small parameter, allowing for various contributions to the potential arising from $W_{\rm np}$ to compete. The result is 
   an AdS minimum which is supersymmetric.
   
\item The LVS construction makes use of the perturbative no-scale breaking effect in  (\ref{kcor}) driven by  a $\vo$-dependent $\alpha'$ correction. This competes with a non-perturbative correction on a small (blow-up) 4-cycle, resulting in a non-supersymmetric AdS minimum. At the minimum, the volume $\vo\sim e^{1/g_s}\gg 1$ is exponentially large in string units. Supersymmetry continues to be broken by the F-terms associated with the K\"ahler moduli.
\end{itemize}

\subsubsection{The KKLT construction}

As we have seen, the effect of turning on fluxes in type IIB is to generate a potential for the dilaton and complex structure moduli, while leaving the K\"ahler moduli flat. The first step in the KKLT construction involves integrating out the dilaton and complex structure moduli, and then considering the low energy effective action for the K\"ahler moduli alone.  Although a realistic model will have multiple K\"ahler moduli, we will work with a single modulus to illustrate the basic features of the model. The non-perturbative contribution to the superpotential can arise as a result of Euclidean D3-branes or gaugino condensation on wrapped D7-branes.
In both cases, the  superpotential takes the form\footnote{For a superpotential to be generated, the divisor associated with the Kahler modulus has to be
rigid or be rigidified by the presence of fluxes  see e.g. \cite{Witten:1996bn, Kallosh:2005yu, Kallosh:2005gs, Bianchi:2011qh, Blumenhagen:2009qh, Kim:2022uni,  Alexandrov:2022mmy, Kim:2022jvv, Kim:2023cbh} for technical discussions. Recently, there have been conjectures on challenges in generating such a superpotential term \cite{Lust:2022lfc}. }
$$ 
  W_{\rm np} = A(U, S) \,e^{ -a T},
$$
where the pre-factor $A$ is a function of complex structure moduli $U$ and the dilaton $S$, but has no dependence on the K\"ahler moduli. 
The constant `$a$' is equal to $2 \pi$ when the effect is generated by Euclidean D3-branes, while in the case of gaugino condensation on wrapped D7-branes  $a = 2 \pi \big{/} N$, where $N$ is the rank of the condensing gauge group. With the dilaton and complex structure moduli integrated out, the flux superpotential makes a constant contribution to the superpotential $(W_0)$. The full superpotential is then
$$
W = W_0 + A\,e^{-aT}.
$$
A key requirement of KKLT is that the fluxes are tuned so that $|W_0| \ll 1$. Working with the tree-level K\"ahler potential $K = - 3 \log \left( T + \overline{T} \right)$, the resulting potential is 
$$
 V = { |a A|^{2} \over 6 \tau}  e^{-2a \tau} + {a |A|^2 \over 2 \tau^{2} } e^{-2a\tau} + {{a \,{\rm{Re}}(A W_0^{*} e^{-i \theta})} \over 2 \tau^{2} } e^{-a\tau}.
 $$
After adjusting the phase of the axion to its minimum, one obtains a supersymmetric minimum $D_TW=0$ with 
$$
   \tau \sim { 1\over  a}   \ln  { |W_0|^{-1}  } > 1\,.
$$
From the logarithmic dependence of $\tau$ on $|W_0|$, we understand why a small $|W_0|$ is a key requirement of the construction. The existence of a large number of choices of flux quanta which lead to small values of $|W_0|$ follows from statistical considerations \cite{Denef:2004ze}. Recently, explicit constructions with $|W_0|$ as low as $10^{-120}$ have been carried out  \cite{Demirtas:2019sip,Demirtas:2021nlu} (for earlier work on obtaining 
low values of $|W_0|$ see \cite{Cole:2019enn,Denef:2004dm,hep-th/0312104, Conlon:2004ds}). Sufficiently large $\tau$ is necessary to ensure that any perturbative corrections to the K\"ahler potential yield sub-leading contributions to the
potential that we can safely ignore, and thereby justifying the use of the tree level K\"ahler potential. Furthermore, 
the masses of the K\"ahler moduli are \cite{Choi:2004sx, Choi:2005ge}
$$
  m_{T} \sim m_{3/2} \ln \left( \frac{M_P}{m_{3/2}} \right), 
$$
where the gravitino mass $m_{3/2} = e^{K/2} \vert W \vert$. Although logarithmically enhanced compared to $m_{3/2}$, this implies that the masses of the K\"ahler moduli are parametrically lighter than those of the complex structure moduli, as required for the consistency of
the 2-step procedure. For work on realising  the KKLT construction in explicit settings, see e.g. \cite{Denef:2004dm, Denef:2005mm, Demirtas:2021nlu}.

Note that the no-scale structure present in the pure flux GKP compactifications is absent from KKLT; the process of stabilisation generates a supersymmetric vacuum and returns the mass of the K\"ahler modulus to above that of the gravitino mass. This is an important qualitative difference between KKLT and LVS, which we now discuss, where the no-scale structure survives in the leading approximation.

\subsubsection{The Large Volume Scenario}
 
The starting point for the LVS construction \cite{Balasubramanian:2005zx} is the same as for KKLT, namely the low energy effective field theory after the complex structure and the axio-dilaton have been integrated out. LVS requires at least two moduli, with the Calabi-Yau having some form of `Swiss-cheese' structure, i.e described by an overall volume with subsequent moduli representing geometric `holes' corresponding to blow-up moduli. The simplest realisation is for the case of two K\"ahler moduli. Here the expression for the volume of the Calabi-Yau takes the form:
$$
  \vo=\tau_b^{3/2}-\tau_s^{3/2},
$$
where $\tau_b$ is volume of a big 4-cycle (in Einstein frame) and $\tau_s$ measures the volume of a hole in it (more precisely $\tau_s$ controls the volume of an exceptional del Pezzo divisor resolving a point-like singularity). The leading $\alpha'$ correction to the K\"ahler potential \cite{Becker:2002nn} is included,
 $$
     K = -2\ln\left(\vo+\frac{\xi}{2}\left(\frac{S+\overline{S}}{2}\right)^{3/2}\right),
 $$
with $\xi\equiv -\frac{\chi(X) \zeta(3)}{2(2\pi)^3}$ where $\chi(X)$ is the Euler number of the Calabi-Yau and $\zeta$ is the Riemann zeta function. LVS also requires a non-perturbative effect supported on the small cycle,
\bea
W &=& W_0+A_s\,e^{-a_sT_s}\,.
\eea
As discussed above, here $a_s= 2\pi \big{/} N$ in the case that the effect arises as a result of gaugino condensation and $a_s=2\pi$ in the case of wrapped
Euclidean branes. Working in the limit $\tau_b \gg \tau_s$, after  fixing the axionic partner of $\tau_s$ at its minimum, the scalar potential (\ref{VF}) takes the form:
\be
V =\frac43\frac{a_s^2A_s^2\sqrt{\tau_s}e^{-2a_s\tau_s}}{s\mathcal{V}}-\frac{2a_sA_s|W_0|\tau_s \, e^{-a_s\tau_s}}{s\mathcal{V}^2}+\frac{3\sqrt{s}\,\xi \, |W_0|^2}{8\vo^3}\,.
\label{VLVS}
\ee
Minimising the potential\footnote{Here, s is to be thought of as parameter, it is fixed along with the complex structure moduli as result of the presence of fluxes.}, one finds a minimum at
\be
\langle\vo\rangle \simeq \frac{3\sqrt{\langle\tau_s\rangle}\,|W_0|}{4a_s A_s}\,e^{a_s \langle\tau_s\rangle}\qquad\text{and}\qquad \langle\tau_s\rangle \simeq \frac{1}{g_s}\left(\frac{\xi}{2}\right)^{2/3}\,.
\label{LVSmin}
\ee
Let us stress some important aspects of the scenario:
\ben

\item The minimum arises as a balance between an $\alpha'$ correction (giving a term in the potential scaling as $\vo^{-3}$) and non-perturbative effects on the small 4-cycle (which are definitely generated for a rigid del Pezzo divisor like $\tau_s$). As a result, the overall volume is exponentially large in the size of the small 4-cycle.

\item The construction generalises to the situation with multiple blow-up moduli. In the general situation with blow up and fibre moduli, the fibre moduli can be stabilised by loop effects \cite{Cicoli:2008va} or higher derivative corrections \cite{Cicoli:2016chb}. Moduli stabilisation in the large volume scenario has been extensively studied, see \cite{Conlon:2005ki, Berg:2005yu, Berg:2007wt, Blumenhagen:2007sm, Conlon:2010ji, Cicoli:2011qg, Cicoli:2012vw,  Cicoli:2013mpa,  Cicoli:2013cha, Reece:2015qbf, Cicoli:2016xae, Cicoli:2017shd, Gallego:2017dvd, Cicoli:2017axo} and  \cite{AbdusSalam:2020ywo, Cicoli:2021dhg, Gao:2022fdi, Junghans:2022exo, Leontaris:2022rzj, Junghans:2022kxg} for recent studies.

\item In LVS models, a small value of the dilaton helps guarantee that the effective field theory is under control. For  $g_s\lesssim 0.1$, (\ref{LVSmin}) implies
that $\tau_b$ and $\tau_s$ are much larger than the string scale, and hence the use of the supergravity approximation is justified. 

\item The models can be constructed  for natural values of $|W_0|$; $W_0 \sim \mathcal{O}(1-10)$. See \cite{Cicoli:2013swa} for a discussion and  \cite{Louis:2012nb} for a concrete example. 

\item The LVS vacuum is AdS with the value of the potential at the minimum $V_{\rm LVS}\sim - m_{3/2}^3 \Mp$. It is  non-supersymmetric with the largest F-term given by $F^{T_b} \sim \tau_b \,m_{3/2}$ (inherited from the no-scale structure). Hence the Goldstino is the fermionic partner of $T_b$ in the corresponding $N=1$ chiral multiplet. This is eaten up by the gravitino which develops a non-zero mass.

\een

\subsubsection{Attractive features of IIB models}

Much of our discussion of cosmological models will be set in IIB flux vacua. There are two major reasons for this: first, the low energy effective action is well understood, and second, this low energy effective action has many attractive features from the point of view of phenomenology. Let us list some of these:

\ben

\item \emph{Controlled flux backreaction}:  The  backreaction of the fluxes on the internal geometry is well understood and leads to internal
manifolds which are conformally Calabi-Yau. The understanding of the resulting underlying moduli space is better than in other settings, with the effect of the warp (conformal) factor on the low energy effective action being negligible at large volume. For progress in computing the  form of the K\"ahler potential including the effects of warping see \cite{DeWolfe:2002nn, deAlwis:2003sn, Giddings:2005ff, Frey:2006wv, Burgess:2006mn, Douglas:2007tu, Shiu:2008ry, Douglas:2008jx,Frey:2008xw, Martucci:2014ska}.

\item \emph{Suppressed scalar potential scale}: IIB offers vacua at large volume, where there exists a clean separation between the string, Kaluza-Klein and moduli mass scales (recall that the Kaluza Klein scale $M_{\rm KK} \propto  \Mp \big{/} \vo^{2/3}$ and  $M_{\rm s} \propto  \Mp \big{/} \vo^{1/2}$). On the other hand, the mass of the complex structure moduli behave as $M_{\rm cs} \propto  \Mp \big{/} \vo$. Thus, at large volume
$$
M_{\rm moduli} \ll M_{\rm KK} \ll M_{\rm s}.
$$
The moduli effective action therefore provides a good description of the low energy dynamics.

\item \emph{Hierarchically suppressed SUSY breaking scale}: Supersymmetry is broken at tree-level by the F-terms of the K\"ahler moduli which are proportional
to the $(0,3)$ component of $G_3$. The gravitino mass is given by  $m_{3/2}= e^{K/2} W_0 \sim {W_0} \big{/} {\vo}$. This is
hierarchically smaller than the Kaluza-Klein scale for both KKLT (where $W_0\ll 1$) and  LVS models (where $\vo \gg 1$). See  \cite{Cicoli:2013swa}  for a detailed discussion.
  
\item \emph{Progress in computing higher order corrections to the effective action}: Since the K\"ahler moduli are flat at tree level, 
 higher order corrections to the effective action become crucial for their stabilisation and cosmological dynamics. There has been a lot of progress in understanding these.
 
 The first computation in this direction was the $\cN=2$ $\mc{O}(\alpha'^3)$ corrections to the K\"ahler potential $K$ \cite{Becker:2002nn}. Additional $\cN=2$ $\mc{O}(g_s^2 \alpha'^2)$ and $\mc{O}(g_s^2 \alpha'^4)$ contributions to $K$ have been obtained  in \cite{Berg:2005ja} and further advanced  in \cite{Berg:2007wt}. In this context, Ref.~\cite{Cicoli:2007xp} showed the existence of an ``extended no-scale structure" which implies that $\mc{O}(g_s^2 \alpha'^2)$ contributions to the scalar potential vanish. Moreover, Ref. \cite{Bonetti:2016dqh} reconsidered $\cN=2$ $\mc{O}(\alpha'^3)$ contributions to the K\"ahler potential incorporating the backreaction of these terms on the  geometry and found that they lead to moduli redefinitions.  Higher derivative $\cN=2$ $\mc{O}(\alpha'^3)$ terms have been computed in \cite{Ciupke:2015msa, Grimm:2017okk}. 

There has also been substantial progress in understanding  $\cN=1$ perturbative effects. Ref.~\cite{Berg:2014ama, Haack:2015pbv,Haack:2018ufg}  obtained  $\cN=1$ string loop corrections to the Einstein-Hilbert term showing that they generate $g_s^2$ corrections to a term involving the CY Euler number (see also \cite{Antoniadis:2018hqy}).  In \cite{Conlon:2009xf,Conlon:2009kt} it was shown that worldsheet 1-loop corrections can also led to field redefinitions of the K\"ahler moduli at 1-loop level. Ref. \cite{Grimm:2013gma} showed that $\cN=1$ $\mc{O}(\alpha'^2)$ corrections to the effective action  lead to moduli redefinitions, while ref. \cite{Minasian:2015bxa} found that $\cN=1$ $\mc{O}(\alpha'^3)$ effects are captured by a shift of the CY Euler number term.   Recently, a systematic treatment of corrections in the F-theory setting has been carried out in \cite{Cicoli:2021rub}.

It is often not necessary to obtain the full functional dependence of these corrections on all moduli. Rather, it is sufficient to determine their dependence on the K\"ahler moduli (which are the light ones).  The functional dependence of string loop corrections to $K$ on the K\"ahler moduli can often  be determined from generalisations of toroidal computations and low-energy arguments \cite{Cicoli:2007xp, Burgess:2010sy}. Another powerful tool is imposing the positivity of the K\"ahler metric \cite{Conlon:2006gv}. Moreover, ref. \cite{Burgess:2020qsc} used the $2$ scaling symmetries of the 10-dimensional action at tree-level to infer the dependence on the overall volume mode and the dilaton of any arbitrary perturbative correction to the effective action. Recently, the one loop correction to the K\"ahler potential has
been related to a supersymmetric index \cite{Kim:2023sfs}.

 \item \emph{Expansion parameters}:  As mentioned in our discussion of the Dine-Seiberg problem, moduli stabilisation corresponds to fixing the values of fields whose VEVs themselves give the expansion parameters of the theories. Furthermore, the asymptotic weakly-coupled 
 regions in field space correspond to runaway potentials. Thus, an interesting question is: how can small expansion parameters arise when moduli are stabilised?
To answer this, let us start by taking the superpotential and K\"ahler  potential to be of the general form in (\ref{Wtotal}) and (\ref{kcor}). We write the F-term potential as
\be
V=V_0 +\delta V\,,
\ee
where $V_0$ is the tree-level potential and $\delta V$ is the correction term in the potential arising as a result of the correction terms in (\ref{Wtotal}) and (\ref{kcor}). Since $V_0$ is independent of the  K\"ahler moduli, the minimum of the potential in the K\"ahler moduli space is determined by $\delta V$. The leading contribution to $\delta V$  takes the form \cite{Conlon:2005ki}:
\be
\delta V\propto e^K\left( W_0^2 \,K_{\rm p} + W_0\, W_{\rm np}\right)\,.
\label{eq:deltaV}
\ee 
The most obvious regime to consider is  where both the tree level contribution to the superpotential dominates over the non-perturbative part ($W_0\gg W_{\rm np}$) and also the perturbative corrections to the K\"ahler potential are much larger than the non-perturbative corrections to the superpotential
($K_{\rm p}\gg W_{\rm np}$). In this regime,  the first term in (\ref{eq:deltaV}) is  the leading order term. It would lift the potential and give  a runaway behaviour (unless terms of different order in the perturbative expansion compete,  but such a competition will always lead to problems with the
perturbative expansion if there is only one expansion parameter \footnote{Recently an exception to this rule was pointed out in \cite{Burgess:2022nbx} in the sense that there exist cases in QFT for which the perturbative expansion can be resummed as in the renormalisation group case, this may allow for a reliable perturbative minimum for $\mathcal V$ as long as the dilaton has been fixed at weak coupling.}). This is the Dine-Seiberg problem \cite{Dine:1985he} for the setting. 

Type IIB flux compactifications  provide at least  two concrete ways to alleviate the problem:

\begin{itemize}[*]

\item[*] In the KKLT construction, the large discrete degeneracy of flux vacua is used to tune $W_0$ to be exponentially small  so that $W_0\sim W_{\rm np}$. As a result,  terms which are of order $W_{\rm np}^2$  are also required to be included in (\ref{eq:deltaV}). The K\"ahler moduli are stabilised due to competition between terms which are of the order  $W_0\, W_{\rm np}$ and $W_{\rm np}^2$. Note that in this regime corrections to the K\"ahler potential can be consistently neglected since their contribution to the scalar potential is sub-leading.

\item[*] LVS models exploit another possibility. Here, the key idea is that  string compactifications can feature more than one expansion parameter (with each corresponding
to the vacuum expectation value of one of the many fields in string theory). The two terms in (\ref{eq:deltaV}) can consistently compete with each other to generate a minimum so long as each arises from a different expansion. In LVS, the first term in (\ref{eq:deltaV}) is part of an expansion in terms of inverse powers of the overall volume of the compactification $(1 \big{/} \vo)$ while the second term is part of an expansion in the size of non-perturbative effects on a small 4-cycle $(e^{-a_s\tau_s})$. These compete to yield a minimum at large volume.

\end{itemize}

\end{enumerate}

\subsubsection{De Sitter in IIB}

The vacua we have discussed so far are AdS. It is possible to obtain dS vacua either by incorporating additional effects which are part of the low energy
 effective action or taking a more general approach to finding minima of the effective action. Below, we describe various proposals for
  constructions of dS vacua\footnote{For recent summaries of the state of the art in dS constructions   and the challenges involved see for example  \cite{Kachru:2018aqn, Cicoli:2018kdo}.} in the IIB setting, illustrating such general constructions in figures \ref{Fig:CY3} and \ref{Fig:dS1}.

\begin{itemize}

\item {\it Anti-branes}: This was proposed as part of the original KKLT construction  \cite{Kachru:2003aw}.  Anti-D3-branes experience
a potential in the imaginary self dual backgrounds of \cite{Giddings:2001yu}. This drives them  to the bottoms of warped throats within the compactification.
An anti-D3-brane at the bottom of a warped throat makes a positive definite contribution to the potential. This is given by
$$
   V_{\overline{\rm D3}} \sim { e^{4A_0} \over {(T+ \overline{T})^2} },
$$
where $e^{A_0}$ is the value of the warp factor at the bottom of the throat. Such a contribution uplifts the KKLT AdS vacuum to a dS one. 
The introduction of an anti-brane takes the configuration away from the pseudo-BPS class, and
various aspects of the effective field theory remain to be understood (see  e.g. \cite{Bena:2018fqc, Dudas:2019pls, Bena:2009xk, Polchinski:2015bea, Kachru:2019dvo, Bento:2021nbb, Moritz:2018ani, Gao:2020xqh, Gautason:2019jwq, Bena:2022ive} and references therein).  Embedding of the system in a supersymmetric effective field theory by making use of the nilpotent field formalism is discussed in \cite{Kallosh:2015nia} and references therein. A recent construction
\cite{Bena:2022cwb}, provides a way to make dS constructions with anti-D3-branes minimalistic (in addition to keeping the effective field theory under control).

\begin{figure}[t]
\begin{center}
\includegraphics[width=65mm,height=80mm]{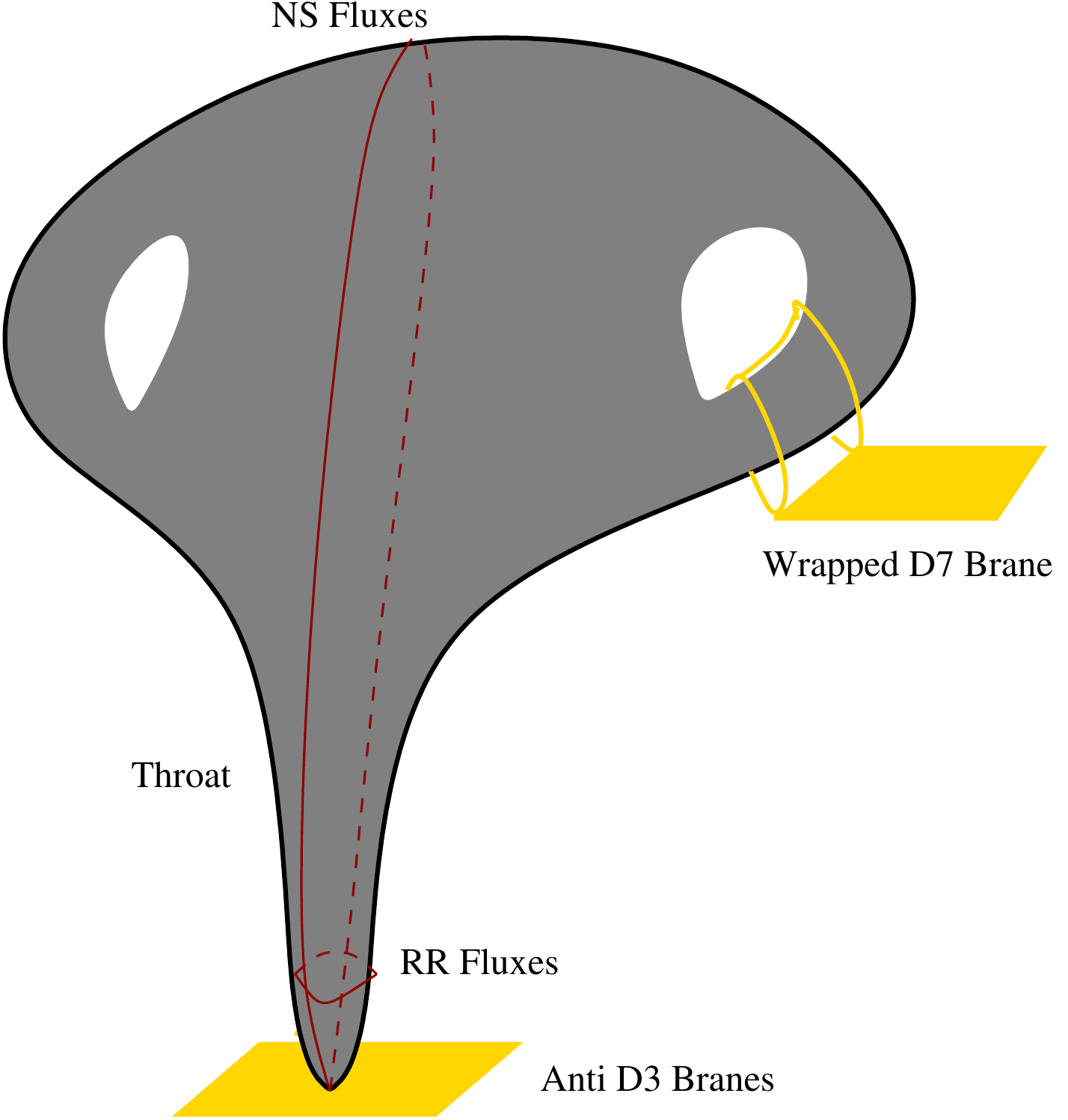} 
\caption{A cartoon representation of a typical Calabi-Yau configuration as used in KKLT and LVS scenarios. The D7-branes wrapped 4-cycles and may host the gauge theory that provides the corresponding non-perturbative effects in the superpotential. The non-trivial fluxes typically lead to the 3-cycles corresponding to long throats that give rise to warped factors in the metric and may host anti-D3-branes at their tip to provide the dS uplift.} \label{Fig:CY3} 
\end{center}
\end{figure}

\begin{figure}[t]
\begin{center}
\includegraphics[width=120mm,height=60mm]{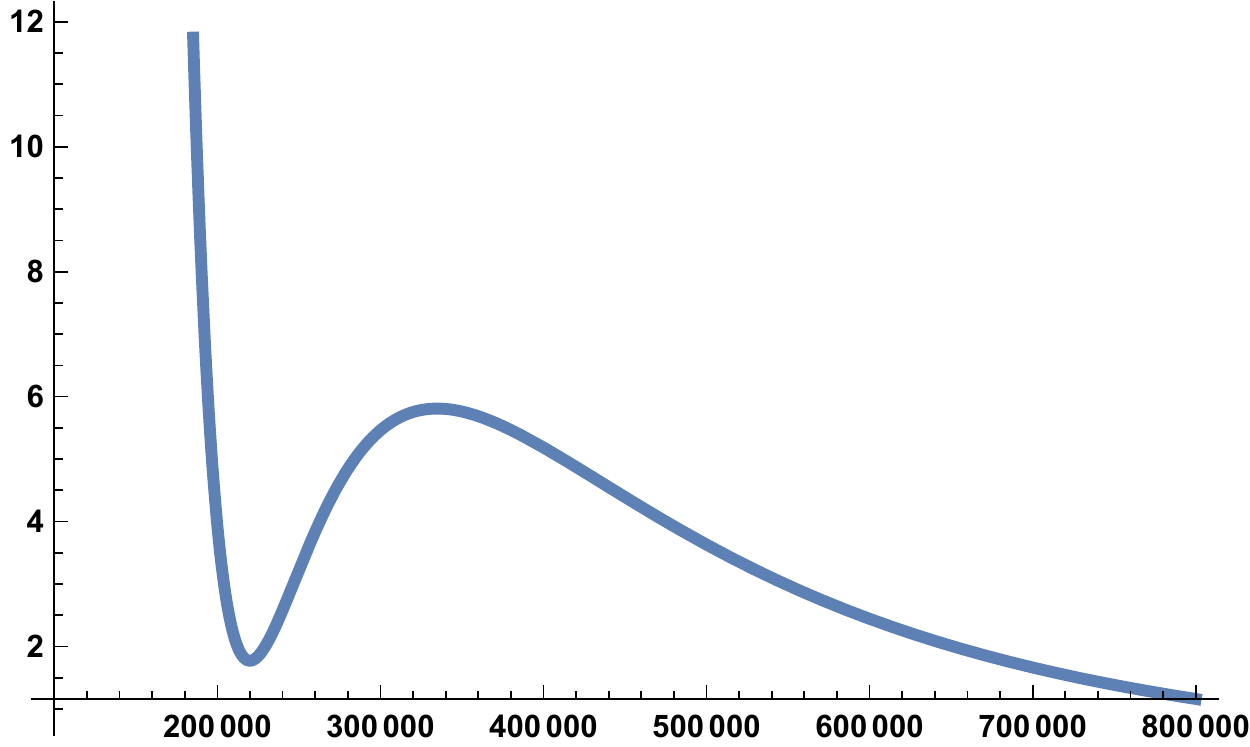} 
\caption{A typical potential giving rise to dS as illustrated by the minimum at positive value of the scalar potential. The x-axis units represent a cycle volume while the y-axis gives the scalar potential in arbitrary units.} \label{Fig:dS1} 
\end{center}
\end{figure}

\item {\it Magnetised branes} \cite{Burgess:2003ic}: Here, one considers $U(1)$ fluxes localised in a warped throat on D7-branes wrapping the $T$ (volume) modulus. If the vacuum expectation values of the matter fields charged under the $U(1)$ are zero, a term is generated in the effective potential which is exactly of the same form as that arises in the presence of an anti-D3-brane . This term corresponds to a D-term contribution in the 4-dimensional supersymmetric effective action.

\item {\it K\"ahler uplift} \cite{Westphal:2006tn, deAlwis:2011dp, Rummel:2011cd, Louis:2012nb}: The $\alpha'$ corrections to the K\"ahler potential in the KKLT construction can be made to compete with the non-perturbative effects and the flux contributions to produce solutions with positive vacuum energy.  The dS minima so obtained are in regions which correspond to the edge of the validity of the effective field theory. An explicit construction with all geometric moduli stabilised has been carried out in \cite{Louis:2012nb}. 

\item{\it T-branes} \cite{Cicoli:2015ylx}: In the presence of  supersymmetry breaking imaginary self-dual (ISD) 3-form and gauge field fluxes on D7-branes, as in a generic string compactification with orientifolds and wrapped branes, one is led to T-brane configurations. That is, the D7-brane adjoint scalars $\Phi$ are in a configuration for which $\left[ \Phi, \Phi^{\dagger} \right] \neq 0 $. Such configurations provide a positive definite contribution to the 4-dimensional potential which can uplift the AdS minima to dS. Explicit Calabi-Yau orientifold examples with T-brane uplifting in the LVS framework have been derived in \cite{Cicoli:2012vw,Cicoli:2013mpa,Cicoli:2013cha,Cicoli:2017shd,Cicoli:2021dhg}.

\item {\it Dilaton dependent non-perturbative effects} \cite{Cicoli:2012fh}: Here, dilaton-dependent non-perturbative effects coming from E($-1$)-instantons or strong dynamics on  hidden sector of D3-branes at singularities were considered. The non-perturbative term yields a positive definite contribution to the scalar potential similar to that which arises from anti-D3-branes. 

 \item{\it dS vacua from logarithmic/power law loop corrections \cite{Antoniadis:2018hqy, Antoniadis:2019rkh, Berg:2005yu}}: In the presence of 
 intersecting D7-branes,  there are loop corrections
 to the K\"ahler potential whose  contributions to the potential  are logarithmic in the volume of the compatification \cite{Antoniadis:2018hqy, Antoniadis:2019rkh}.  These arise from graviton kinetic terms related to the emission of closed strings on non-vanishing local tadpoles. The logarithmic dependence arises from infrared divergences due to  effective propagation in the two transverse directions to the D7-branes.  
 Combining  the logarithmic terms with  the (usual) power law $\alpha'$ and $g_{s}$ corrections to the potential, one obtains a non-supersymmetric AdS minimum.  dS vacua are obtained by incorporating the  effects of D-term contributions from $U(1)$ magnetic fluxes along the world-volume directions of the D7-branes.  
 
 Closely related are the constructions of \cite{Berg:2005yu}, where it was found that K\"ahler moduli can be stabilised
 by perturbative power law corrections, or those of  \cite{Balasubramanian:2004uy} which used the perturbative $\alpha^{'}$ corrections to generate a dS minimum.
 
 Again, various additional effects can lead to dS minima in the setting. These constructions can provide an avenue to obtain dS vacua without making use of  the non-perturbative part of the superpotential, whose computation involves  various subtleties \cite{Sethi:2017phn}.  

\item {\it Complex structure F-terms} \cite{Gallego:2017dvd}: The potential for the complex structure and dilaton generated by 
fluxes  has  supersymmetric minima where $D_UW_{\rm flux}=0, D_SW_{\rm flux}=0$. There can also be other minima, ones where the F-terms
associated with these fields are non-vanishing. These minima lead to dS vacua once the K\"ahler moduli are stabilised without the need of further ingredients. A concrete example with $\vo\simeq 10^4$ was constructed in \cite{Gallego:2017dvd}.

\item {\it Non-perturbative dS vacua} \cite{Blaback:2013qza}:  Here, dS minima arise from stabilising all the geometric moduli in  one step via the inclusion of  background fluxes and non-perturbative contributions to the superpotential. The key challenge is to develop a good understanding of the $S$ and $U$-moduli dependence of the prefactors of the non-perturbative effects. Duality covariance was used to constrain their form in the analysis. Finding the minima requires sophisticated numerical methods such as genetic algorithms.

\item {\it  Compactifications on Riemann surfaces }\cite{Saltman:2004jh}: This construction makes use of the fact that negative curvature in the 
extra dimensions makes a positive contribution to the 4-dimensional effective potential. In the simplest version, the compactification manifold is a product of 3 Riemann surfaces with one of them having genus greater or equal than 2.
Fluxes are introduced to stabilise the complex structure moduli. At this level, there are  tadpoles for the dilaton and the volumes of each of the surfaces. These are stabilised by the introduction of D7-branes and anti-D7-branes. Supersymmetry is broken at a high scale by the compactification manifold. A related recent development is M-theory on hyperbolic manifolds \cite{DeLuca:2021pej}

\end{itemize}

\subsubsection{Open directions in IIB}

Before moving on to moduli stabilisation in other string theories, we list some interesting open directions in type IIB

\begin{itemize}
\item \textit{Better understanding of higher derivative and $g_s$ corrections}:  Higher derivative
and $g_s$ corrections play a central role in various scenarios for moduli stabilisation. Furthermore, they are important
for checking the validity of the effective field theory used for the construction of the vacua. As we have described above, much progress has
been made in computing these corrections in type IIB. Yet, we still do not have an understanding of all the corrections
in the most general configuration with branes and fluxes. Developing such an understanding is of much 
importance. To give a specific example, it is important to
understand the precise effect  of the logarithmic corrections of  \cite{Antoniadis:2018hqy, Antoniadis:2019rkh} (when they are present; in general, such corrections are absent) in the LVS context, see e.g. \cite{Conlon:2010ji, Junghans:2022exo, Gao:2022fdi}.

\item  \textit{Non-perturbative effects}:  A full understanding of the conditions under which non-perturbative effects in the superpotential are generated  is of importance (see e.g. the discussion in \cite{Sethi:2017phn, Kachru:2018aqn}). Recent progress in this direction includes generalisation of the fermion zero mode conditions  for effective divisors in CY threefolds with singularities along rational curves \cite{Gendler:2022qof}.

\item \textit{The uplift sector}: The introduction of the uplift sector is crucial to obtain dS vacua. This ties the uplift sector
to the  issues  which arise while defining a quantum theory on dS space (see e.g. \cite{Witten:2001kn, Banks:2012hx, Maltz:2016iaw}) and the dS swampland conjecture (see Sec. \ref{Sec:Swamp} for a discussion). Thus, progress in better understanding the physics of the uplift sector  can potentially shed light on various conceptual issues in quantum gravity.

\item  \textit{F-theory moduli stabilisation}: Type IIB models can be thought of as  the weak coupling limit of  F-theory constructions. Thus study of moduli stabilisation in F-theory can potentially lead to novel scenarios and also provide better understanding of existing models. The central issue is  to address moduli stabilisation directly within the F-theory framework. For a review of the phenomenology of F-theory vacua see e.g. \cite{Maharana:2012tu}.

\item \textit{Local curvatures}: Our present understanding   allows for the computation of the volumes of 4-cycles and 2-cycles after moduli stabilisation. This gives information only about the average size of the curvatures; in principle a 2-cycle with volume which is large in string units can be anisotropic and have regions where the curvature is large. While this is a challenge, it is not expected to be a generic issue.
Progress on this front will require explicit knowledge of metrics on Calabi-Yaus, see e.g. \cite{Headrick:2005ch, Ashmore:2021ohf, Gerdes:2022nzr} for work in this direction.

\item \textit{Model scans}: Detailed understanding of models and their phenomenology requires scans over large number of models. To give a specific example, 
the construction of dS vacua requires uplift potentials of very specific magnitude (as an order one increase in the magnitude leads to a runaway). Given this,
isolating concrete models will certainly need an extensive scan over models varying the flux quanta. The large multitude of IIB vacua implies that optimisation strategies are important. Modern computational tools are promising  in this direction (see \cite{He:2017aed, Ruehle:2020jrk, Cole:2021nnt, Abel:2022nje, Abel:2021ddu}).

\item \textit{Open string sector}: Realistic phenomenology requires the introduction of  open string sector(s) which provide the Standard Model degrees of freedom.\footnote{For a recent review on construction of the Standard Model sector in IIB/F-theory see \cite{Marchesano:2022qbx}.} 
Combining stabilisation of the closed and open string modes in a controlled fashion in Calabi-Yaus  is a very demanding task. This is  because of the difficulty to solve the minimisation equations in the presence of a large number of complex structure and open string moduli. Much progress has been made in this direction \cite{Conlon:2008wa,Krippendorf:2010hj, Cicoli:2011qg, Cicoli:2012vw, Cicoli:2013mpa, Cicoli:2013cha, Cicoli:2016xae, Cicoli:2017shd, Cicoli:2017axo} but a globally consistent model with full moduli stabilisation in a controlled dS vacuum is yet be achieved.

\item \textit{The nature of the expansion}: The weak coupling expansion in quantum field theory is an asymptotic expansion. In flux compactifications, the expansion parameters are small but the tadpole constraint implies that one cannot take the limit of arbitrarily small coupling (as is true of asymptotic expansions). It would be interesting to develop the mathematical theory of such expansions and understand them better.
\end{itemize}

\subsection{Moduli Stabilisation: The full diversity of scenarios}

Although type IIB constructions are the most studied, they represent only one corner of string theory. It is therefore necessary to review moduli stabilisation also in other corners of string theory.

\subsubsection{Moduli stabilisation in type IIA}

In type IIB, there are only 3-form fluxes. These only couple to the complex structure moduli, explaining why fluxes in IIB stabilise the complex structure moduli alone.
In contrast, type IIA theory has both the NSNS 3-form and even RR forms in its spectrum. This allows for a rich structure in the
flux superpotential: the 3-form can couple to the holomorphic 3-form  of the compactification manifold (as in the IIB theory), while even RR forms
can couple to the K\"ahler form. Thus one can hope to stabilise all the geometric moduli solely through the presence of fluxes.  Here, 
key challenges are instead the lack of a full understanding of the 10-dimensional solutions incorporating the back reaction of the fluxes and the questions of how to generate hierarchies and supersymmetry breaking. In fact, it is known that the internal manifold can be neither Calabi-Yau nor conformal to a Calabi-Yau once fluxes are turned on. For supersymmetric solutions, the internal manifold has to be half flat with an $SU(3)$ structure \cite{Behrndt:2004mj, Behrndt:2004km,  House:2005yc, Lust:2004ig}.

In view of the difficulty in obtaining 10-dimensional solutions, the approach taken has been to consider configurations which satisfy the tadpole cancellation conditions and compute the energy functional for these. Critical points of this energy functional can be expected to correspond
to solutions to the 10-dimensional equations of motion in the limit of large volume. We will discuss two constructions in more detail. The first will be set in orientifolds of Calabi-Yaus \cite{DeWolfe:2005uu} (see  e.g. \cite{Kachru:2004jr, Derendinger:2004jn,  Aldazabal:2006up, Saueressig:2005es, Villadoro:2005cu, Ihl:2007ah} for related work). The second will involve compactifications on Nil manifolds \cite{Silverstein:2007ac}.

The simplest $\cN = 1$ supersymmetric IIA orientifolds involve an anti-holomorphic involution acting on a Calabi-Yau. The fixed locus of the involution is wrapped by O6-planes. The tadpole cancellation condition for D6-charge takes the form:
$$
    N_{\rm D6}  + \int_{\Sigma} F_0 \wedge H_3 = 2 N_{\rm O6}, 
$$
where $F_0$ is the mass parameter of the IIA theory and $\Sigma$ is a 3-cycle that the 3-form threads. Thus, by a suitable choice of $H_3$ the tadpole can be cancelled without the introduction of D6-branes. The other fluxes can be arbitrary. The effective action for moduli of the orientifold\footnote{
Type IIA orientifolds have $h^{2,1} + h^{1,1}_{-} +1$  scalar moduli, which are part of $\cN=1$ chiral multiplets.} 
in such configurations was computed in \cite{DeWolfe:2005uu} by making use of the formalism of \cite{Grimm:2004ua}. The following interesting features were found:
\begin{itemize}

\item The classical flux potential stabilised all the geometric moduli. The axionic partners of the complex structure moduli remain unfixed. In principle, these can be lifted by Euclidean D2-instantons but, in practice, such massless gravitationally coupled axions are phenomenologically harmless.

\item As opposed to the IIB case, the number of vacua obtained can be infinite.

\item A scaling argument showed that the solutions can be brought to the regime of arbitrarily small values of $g_s$ and arbitrarily 
large volume.\footnote{For an AdS$_3$ solution (for type IIA on $G_2$ orientifolds) with similar features see \cite{Farakos:2020phe}.}

\end{itemize}

 Various open questions remain  in connecting these DGKT configurations to 10-dimensional solutions. The analysis of  \cite{Acharya:2006ne} revealed that the solutions could be lifted to ten dimensions after the O6-plane was smeared. Another issue is the quantum interpretation of the massive IIA theory (see \cite{McOrist:2012yc} for a detailed discussion). Interestingly, no stable dS vacua have been found \cite{Caviezel:2008tf, Flauger:2008ad, Danielsson:2009ff, Caviezel:2009tu, Danielsson:2010bc, Danielsson:2011au, Andriot:2018ept} in the setting. Recently, the AdS vacua have been explored from the holographic perspective  \cite{Conlon:2021cjk, Apers:2022tfm}, revealing an interesting structure including integer conformal dimensions for fields dual to the moduli.
 
The second example we discuss is that of \cite{Silverstein:2007ac}, which is a construction of metastable dS vacua. As in \cite{Saltman:2004jh}, this exploits the fact that negative scalar curvature in the internal dimensions makes a positive contribution to the 4-dimensional scalar potential. The compactification manifold used is a product of 2 Nil 3-manifolds (these have negative scalar curvature). Orientifolds, branes, fractional Chern-Simons forms  and fluxes consistent with tadpole cancellation conditions are considered. With suitable choice of the discrete parameters, the curvature, field strengths, inverse volume, and 4-dimensional string coupling can be made parametrically small, and the dS Hubble scale can be tuned to be parametrically smaller than the moduli masses. However,  various aspects of the effective field theory remain to be explored (see \cite{Haque:2008jz} for a detailed discussion). As the compactification manifold breaks supersymmetry, high scale supersymmetry is a generic prediction of the models. 

\subsubsection{Moduli stabilisation in heterotic strings}

Heterotic string theory was the original set up for model building in string phenomenology, and as such work towards moduli stabilisation also spans across four decades.\footnote{Some early work on cosmological aspects of heterotic string theory can be found in \cite{Kogan:1986tg, Kogan:1986fx, Khlopov:2002gg}.} Nevertheless, although there are by now well-developed landscapes of heterotic Standard Models containing the exact charged spectrum of the MSSM \cite{Lebedev:2006kn, Lebedev:2008un, Anderson:2011ns, Anderson:2012yf, Anderson:2013xka}, it is fair to say that the programme of moduli stabilisation is much less developed.  Moduli stabilisation in heterotic string theories share many ingredients with type II scenarios, but there are also a number of aspects that make the stabilisation of moduli in heterotic string theory more challenging:
\begin{itemize}
\item The matter sector resides in the bulk, rather than localised on D-branes.  This means that the stabilisation of geometric moduli and the dilaton cannot be decoupled from the problem of building a realistic particle physics.  Because the visible sector gauge couplings are set by the 10D dilaton and volume modulus, there are preferred values for the moduli in order to obtain gauge coupling unification at the GUT scale: $g_s^2/\mathcal{V}=\alpha_{GUT}\approx 1/25$ or, equivalently, $\textrm{Re} S \equiv \sqrt{g} e^{-2\phi}=2$, which are at the bounds of validity of the string coupling and weak curvature expansions.  Moreover, the GUT scale is identified with $M_{\rm KK}$, which however is pushed beyond the phenomenological value by the requirement $\mathcal{V} \lesssim 25$ that follows from the previous expressions after imposing $g_s \lesssim 1$.  Indeed, for an isotropic compactification, $M_{\rm KK}\sim M_s/\mathcal{V}^\frac16 \gtrsim 8 \times 10^{16}\,$GeV, where we used $M_s^2 \sim \Mp^2/4\pi \alpha_{\rm GUT}^{-1} \approx (10^{17} \textrm{GeV})^2$, compared with $M_{\rm GUT} \sim 2 \times 10^{16}$GeV.  This can be circumvented by considering anisotropic compactifications e.g. with two large extra dimensions \cite{Hebecker:2004ce, Dundee:2008ts, Buchmuller:2007qf, Loaiza-Brito:2005bto}.

\item The supersymmetry conditions for a Minkowski-CY background, $H = -\frac12 *dJ$, force all the $H$-flux to vanish for a CY (where $dJ=0$) \cite{Strominger:1986uh}, in contrast to type IIB where a supersymmetric background can be obtained for primitive (2,1)-type 3-form flux backgrounds.  Therefore, at leading order, if fluxes are used to stabilise the complex structure moduli in heterotic string theory, one has to pay either by breaking supersymmetry or deforming away from a CY.

\item The only flux available is that from the single NS-NS 3-form $H$.  The flux superpotential is therefore independent of the dilaton; this implies ($i$) flux backgrounds do not have the no-scale cancellation seen in type IIB ($ii$) the dilaton cannot be stabilised by flux alone and non-perturbative effects such as gaugino condensation have to be relied upon.  As will be discussed further below, there is in fact an interesting interplay between $H$-flux and gaugino condensation \cite{Dine:1985rz, Derendinger:1985kk}, due to the perfect square structure in the heterotic effective supergravity theory, which is given in the Einstein frame by:
\begin{equation}
S_{\rm het} \supset \frac{1}{2\alpha'^4}\int d^{10}x\sqrt{-g}\left(-\frac{1}{12}e^{-\phi}\left(H_{MNP}-\frac{\alpha'}{16}e^{\phi/2}\,\overline{\lambda_{10}}\,\Gamma_{MNP}\lambda_{10} \right)^2\right)\,. 
\label{E:perfectsquare}
\end{equation}

\item The $H$-flux is real.  Having only half as many fluxes per modulus compared with type IIB means that one cannot use the discretuum of fluxes to tune the VEV of the flux superpotential to small values \cite{Curio:2005ew} (the existence of a metastable solution to the $2h_{1,2}$ real coupled, partial non-linear differential equations to stabilise the $2h_{1,2}$ real moduli at weak coupling, with $2h_{1,2}$ real flux parameters, leaves little room to further tune $W$ \cite{Cicoli:2014sva}; see also \cite{Anderson:2011cza} for a proof of this in the large complex structure limit); this is dangerous as a large flux superpotential induces large runaways in the K\"ahler moduli and dilaton directions, which is difficult to counterbalance with exponentially suppressed non-perturbative contributions as done in KKLT.  
\item As will be discussed below, small flux superpotentials can be obtained by considering fractional Chern-Simons contributions to $H$.  Indeed, in the heterotic string, perturbative anomalies in spacetime or on the worldsheet imply that the gauge-invariant $H$ takes the form \cite{Rohm:1985jv}:
\begin{equation}
H = dB -\frac{\alpha'}{4}\Omega_3(A)-\frac{\alpha'}{4}\Omega_3(\omega)\,, \label{E:HwithCS}
\end{equation}
with the Chern-Simons invariant $\Omega_3(A)= \textrm{tr}(A\wedge dA)+\frac23 A \wedge A \wedge A$ and similarly for $\Omega_3(\omega)$.  Note that the Kalb-Ramond and Yang-Mills Chern-Simons contributions are leading order in the derivative expansion, whereas the Lorentz Chern-Simons term is higher order.  Whilst the flat bundles from Wilson lines contribute $H$-flux at leading order, anomaly cancellation and the Bianchi identity force the Chern-Simons $H$-flux from a non-standard embedding of the gauge connection into the spin connection to appear at higher order; the latter's contribution to the superpotential in the low energy effective field theory for massless modes must thus be vanishing due to the non-renormalisation theorems \cite{Dine:1986vd}, although -- as we will see -- it can still contribute to the stabilisation of would-be moduli once these are integrated in.
\item In analogy to the open string moduli in type II theories, heterotic string theories have, on top of the geometric moduli and the dilaton, a number of vector bundle moduli.  The moduli space of complex structure and vector bundle moduli does not take the form of a direct product, rather there is some non-trivial mixing imposed by the Hermitian Yang-Mills equations \cite{Witten:1985bz, Anderson:2010mh, Anderson:2011cza, Anderson:2011ty}.  Although in a supersymmetric Minkowski-CY background the complex structure moduli cannot be stabilised by $H$-flux, they may be stabilised by the gauge bundle moduli. This will be discussed further below.
\end{itemize}

We  now review the main moduli stabilisation scenarios for heterotic string compactifications.

\begin{itemize}
 \item \emph{Heterotic orbifold compactifications.}  The MSSM has been successfully constructed in heterotic toroidal orbifold compactifications \cite{Buchmuller:2006ik, Lebedev:2006kn,Lebedev:2008un}.  The 4-dimensional $N=1$ effective supergravity theory describing these models can be derived using a combination of dimensional reduction, conformal field theory \cite{Hamidi:1986vh, Dixon:1986qv, Dixon:1989fj} and target space modular invariance that descends from the underlying torus (the latter is typically broken to some subgroup by Wilson lines) \cite{Ferrara:1989bc, Lauer:1989ax, Ibanez:1992hc}.  Although the dilaton, K\"ahler and complex structure moduli are flat directions at tree-level, non-perturbative effects tend to lift them all.  As the dilaton describes the tree level gauge couplings, it is lifted by gaugino condensation in a non-Abelian hidden gauge sector \cite{Nilles:1982ik, Ferrara:1982qs, Dine:1985rz}.  Moreover, the gauge couplings also receive threshold corrections from massive string states, which depend on several of the K\"aher moduli and all of the complex structure moduli; these are then also lifted by gaugino condensation \cite{deCarlos:1991gq, deCarlos:1992kox} (interestingly, in some cases, target space modular invariance implies that the resulting scalar potential goes to plus infinity in both the small and large modulus limits, ensuring the existence of a metastable minimum somewhere in-between along this direction and hence forcing compactification \cite{Font:1990nt, Cvetic:1991qm, Bailin:1993fm}).  Finally, worldsheet instantons described by metastable untwisted strings formed by twisted strings distributed at distant fixed points, lift all the remaining K\"ahler moduli, which describe the fixed points' separation \cite{Lust:1991yi}.  
 
 These ingredients were combined in \cite{Parameswaran:2010ec}, which studied the moduli stabilisation of all bulk moduli in top-down explicit MSSM models, with 10 real moduli.  With moduli stabilisation being induced entirely non-perturbatively, it may seem reasonable to expect that a small cosmological constant could be achieved; on the other hand, with 10 real moduli directions, a random search for a metastable minimum would only find 1 in every $2^{10}$ solutions.  In fact, although many consistent dS critical points were found (and no AdS ones), they all had large vacuum energies and tachyonic instabilities.  Aside from the search of metastable dS vacua -- or an understanding as to why they do not exist (see \cite{Olguin-Trejo:2018zun, Gonzalo:2018guu, Leedom:2022zdm} for some discussion on implications of modular invariance for the dS swampland proposal) -- there are a number of details which require further attention.  Not all the moduli dependent terms in the $\cN=1$ Lagrangian are computable with current knowledge.  The same goes for the input parameters in the 4-dimensional scalar potential, which are certain (non-Abelian) singlet matter VEVs, $A_\alpha$, that ($i$) `turn on' the worldsheet instanton contributions as $W_{yuk} \sim h_{\alpha\beta\gamma}(T_i) A_\alpha A_\beta A_\gamma$ and ($ii$) give masses via higher order Yukawa couplings to charged hidden matter, which would otherwise prevent gaugini from condensing.  Another important issue is the stabilisation of twisted moduli or `blow-up' modes.

 \item \emph{Smooth Calabi-Yau compactifications without fluxes.} Heterotic Standard Models have also been successfully built using smooth CY compactifications and algebraic-geometric techniques \cite{Braun:2005ux, Bouchard:2005ag, Braun:2005nv, Anderson:2011ns, Anderson:2012yf, Anderson:2013xka}.  Key ingredients in these constructions are the non-standard embedding of the gauge bundle into the tangent bundle, together with Wilson lines.  It may be possible that these very same ingredients help with the moduli stabilisation, without the need for the fluxes essential in type II.  Indeed, as mentioned above, the number of gauge bundle moduli depends on the values of the complex structure and, viceversa, the number of complex structure moduli depends on the values of the gauge bundle moduli \cite{Witten:1985bz, Anderson:2010mh, Anderson:2011cza, Anderson:2011ty}. The choice of gauge bundle may then be such that all the complex structure moduli and a large number of other moduli are stabilised \emph{at tree level}; in the most restrictive case only one combination of the K\"ahler moduli and dilaton is left as a flat direction in an ${\cal N}=1$ Minkowski vacuum \cite{Anderson:2011cza}.  

 In a bit more detail \cite{Anderson:2010mh, Anderson:2011cza, Anderson:2011ty}, the $\cN=1$ supersymmetry conditions impose that the gauge fields must obey the Hermitian Yang-Mills equations of zero slope: $F_{ij} = 0 = F_{\bar{\imath}\bar{\jmath}}$ and $g^{i\bar{\jmath}}F_{i\bar{\jmath}}=0$.  These conditions become, respectively, F-term and D-term conditions associated with a 4D $\cN=1$ effective scalar potential, which descends from the term in the 10D action: $-\frac{1}{2\kappa_{10}^2}\alpha'\int d^{10}x \sqrt{-g} \left(-\frac12 \textrm{tr}(g^{i\bar{\jmath}}F_{i\bar{\jmath}})^2+\textrm{tr}(g^{i\bar{\imath}} g^{j\bar{\jmath}} F_{ij} F_{\bar{\imath}\bar{\jmath}})\right)$.  For a given choice of vector bundle moduli, the F-term condition that the vector bundle be holomorphic with respect to a given complex structure fixes some -- possibly all -- complex structure moduli. The D-term condition corresponds to the vector bundle being poly-stable with vanishing slope, and it evidently depends on the K\"ahler moduli via the metric components $g^{i\bar{\jmath}}$.  In fact, it also carries a dilaton dependence via one-loop corrections in the weakly coupled string.  Thus the D-term condition fixes some -- possibly all -- K\"ahler moduli and dilaton, with the exception of the overall volume modulus.  In fact, if the mass scale of these stabilised moduli is of the same order as the KK scale, they should not be considered as moduli at all; so the number of moduli in such a compactification is less than that derived from the CY Hodge numbers.  The vector bundle moduli themselves could be stabilised by the non-perturbative effects, where the Pfaffian prefactors in the superpotential are polynomials in the bundle moduli \cite{Buchbinder:2002ic, Buchbinder:2002pr}.

 Ref. \cite{Anderson:2011cza} proved a no-go theorem that a single surviving flat direction could be stabilised by non-perturbative effects, due to the constraints imposed by the U(1) symmetries associated with the D-term stabilisation.  However, if two flat directions survive, they could be stabilised non-perturbatively by gaugino condensation and/or instantons; as this starts from a genuine $\cN=1$ Minkowski vacuum, these non-perturbative effects do not suffer the usual subtleties associated with rolling solutions \cite{Sethi:2017phn}.  See also \cite{Becker:2004gw, Braun:2006th} for related work in the context of heterotic M-theory.  It remains an open question whether or not explicit realisations of this attractive scenario can be realised.

 \item \emph{Heterotic flux compactifications on non-K\"ahler manifolds.}  As noted above, the introduction of $H$-flux in leading order heterotic compactifications deforms away from $\cN=1$ supersymmetric Calabi-Yaus, either by breaking supersymmetry or by leading to a non-K\"ahler internal space \cite{Strominger:1986uh}.  Indeed, the supersymmetry condition $H=-\frac12*\textrm{d}J$, implies that -- for supersymmetric backgrounds -- flux induces non-K\"ahlerity and can only be (2,1) and (1,2).  Going beyond CY manifolds to manifolds with $SU(3)$ structure (a manifold with $SU(3)$ structure is one that admits a globally defined spinor, and thus preserves some supersymmetry) leads one to consider `geometric fluxes' or `torsion', which have even degree and thus the potential to stabilise the K\"ahler moduli at leading order \cite{Strominger:1986uh, LopesCardoso:2002vpf, Becker:2003sh, Becker:2003yv, Gauntlett:2003cy}.  Manifolds with $SU(3)$ structure are characterised in terms of 5 so-called torsion classes $\mathcal{W}_i$ ($i=1,\dots,5$), with $dJ \in \mathcal{W}_1 \oplus \mathcal{W}_3 \oplus \mathcal{W}_4$ and  $d\Omega \in \mathcal{W}_1 \oplus \mathcal{W}_2 \oplus \mathcal{W}_5$  (see \cite{Grana:2005jc} for a review).  The moduli space of general $SU(3)$ structure manifolds is not so well understood.  The simplest non-CY $SU(3)$ structure manifolds are nearly K\"ahler manifolds, which have only the first torsion class $\mathcal{W}_1$ non-vanishing.
 
  Half-flat manifolds arise as the type II mirrors of Calabi-Yaus with NSNS flux \cite{Gurrieri:2004dt}; they have vanishing $\mathcal{W}_1^-$, $\mathcal{W}_2^-$ (imaginary parts of $\mathcal{W}_1$ and $\mathcal{W}_2$), $\mathcal{W}_4$ and $\mathcal{W}_5$ and their moduli space of metrics is identical to that of the mirror Calabi-Yau.  In \cite{Gurrieri:2004dt, Gurrieri:2007jg} it was shown that dimensional reduction of the heterotic string theory on half-flat mirror manifolds, in the large complex structure limit, yields a 4-dimensional $\cN=1$ supergravity whose K\"ahler potential is identical to the mirror Calabi-Yau and whose superpotential is a generalisation  of the well-known Gukov-Vafa-Witten expression, $W \sim \int \Omega \wedge (H+idJ) \sim e_i T^i + \epsilon_a Z^a + \mu^a \mathcal{G}_a$, with, respectively, geometric, NSNS electric and NSNS magnetic flux parameters $e_i, \epsilon_a$ and $\mu^a$; thus we see that all the geometric moduli can be stabilised in these backgrounds.  This result is expected to extend to all $SU(3)$ structure manifolds (see also \cite{LopesCardoso:2003dvb, Becker:2003sh}).  Progress in constructing gauge bundles for $SU(3)$ structure manifolds has been made for the case of nearly K\"ahler manifolds given as group or group coset manifolds \cite{Chatzistavrakidis:2009mh, Lechtenfeld:2010dr, Klaput:2011mz}; backreaction of the gauge fields induced by the Bianchi identity at order $\alpha'$ can further help with the moduli stabilisation \cite{Klaput:2012vv}.  Due to the runaway dilaton, compactifications on nearly K\"ahler or half-flat manifolds yield domain wall solutions at leading order \cite{Lukas:2010mf, Klaput:2011mz}; a maximally symmetric vacuum could be found by introducing other effects, such as gaugino condensation \cite{Klaput:2012vv}.  However, as mentioned, it would be difficult to balance the tree-level $H$-flux against non-perturbative effects that are exponentially suppressed at weak coupling. 
 
 \item \emph{Heterotic flux compactifications with Chern-Simons terms.}
$H$-flux appears in the 10-dimensional supergravity action in a perfect square together with the gaugino fermion bilinear (\ref{E:perfectsquare}).  The supersymmetry conditions on a CY background require this bilinear to vanish \cite{Dine:1986vd, LopesCardoso:2003sp, Frey:2005zz} but by breaking supersymmetry spontaneously, a background $H$-flux and gaugino condensate are compatible with a Minkowski $\times$ CY compactification, with complex structure and dilaton stabilised \cite{Rohm:1985jv}.   Working out how these effects, understood within the 4D effective field theory, are captured by the 10D supergravity theory, has not yet been fully achieved \cite{LopesCardoso:2003sp, Frey:2005zz}.  Note that satisfying the Minkowski condition would require balancing the background $H$-flux against non-perturbative effects that are exponentially suppressed at weak coupling; as discussed above, this cannot be obtained with integer quantised $H$-flux.  Ref. \cite{Gukov:2003cy} therefore considered the $H$-flux sourced by background discrete Wilson lines, via their contributions to the Chern-Simons term in $H$ (\ref{E:HwithCS}); this $H$-flux can be fractionally quantised instead of integer valued (see Ref. \cite{Apruzzi:2014dza} for the computation of Chern-Simons invariants from Wilson line backgrounds, and Ref. \cite{Anderson:2020ebu} for the same from non-standard embeddings).  Vector bundle and K\"ahler moduli could also be stabilised into a supersymmetric AdS solution once threshold corrections are taken into account.  These latter solutions involve some subtleties as they pass through a strong coupling singularity \cite{Gukov:2003cy}, which can be avoided by considering instead worldsheet instantons \cite{Curio:2005ew}.  
 
 \item \emph{Heterotic Large Volume Scenario.} Ref.~\cite{Cicoli:2013rwa} put several of the ingredients previously discussed -- fractional Chern-Simons fluxes, the requirement of a holomorphic slope-stable gauge bundle, tree-level and non-perturbative (including one-loop threshold corrections) superpotentials -- together with $\alpha'$ corrections to the K\"ahler potential, to compose a heterotic version of the Large Volume Scenario.  In this scenario, the complex structure and gauge bundle moduli are stabilised supersymmetrically at leading order, the dilaton non-perturbatively by gaugino condensation and/or worldsheet instantons, and the K\"ahler moduli by subleading perturbative corrections, ultimately breaking supersymmetry spontaneously in a Minkowski or dS vacuum.  If the complex structure moduli are stabilised by fractional flux, then $|W| \sim \mathcal{O}(0.1-0.01)$ and the supersymmetry breaking is high-scale with $m_{3/2} \sim M_{\rm GUT}$; if they are instead stabilised by the gauge bundle, then $|W|$ can be exponentially suppressed and the supersymmetry breaking scale can be low.  It would be important to work out if these scenarios can be realised in a controlled way with explicit top-down constructions.

\end{itemize}
 
\subsubsection{Moduli stabilisation in type I}

Moduli stabilisation in type I has been developed in \cite{Antoniadis:2004pp, Antoniadis:2005nu, Antoniadis:2006eu, Kumar:2006er, Antoniadis:2007jq, Antoniadis:2009bg}. So far, the studies have been mostly in a toroidal setting ($T^{6} \big{/} Z_2$ backgrounds).
 Magnetic fluxes  are introduced  on D9-branes that wrap the internal manifold. The boundary term in the open string action 
plays a  crucial role. This modifies the open string Hamiltonian and its spectrum, and leads to constraints on the closed string background fields due to their couplings to the open string action.  The supersymmetry conditions, together with conditions which define a meaningful world-volume theory, put restrictions on the values of the moduli and fix them.  The presence of branes and magnetic fluxes reduces  the $\cN=4$ theory to  an $\cN = 1$ supersymmetric one.

 The simplest version of the model involves an O9-plane and several stacks of D9-branes. Some of the key features are:
 \begin{itemize}
 
 \item   The introduction of oblique (non-commuting) magnetic fields. These are necessary to fix the off-diagonal components of the metric.
 
 \item  The non-linear part of the Dirac-Born-Infeld (DBI) action is used to fix the overall volume.
 
 \item The large radius limit can be attained by appropriate choice of the magnetic flux quanta.
  
 \item The magnetised  branes  lead to D5-brane charge, for which the associated tadpole has to be cancelled (this can be done without
 the introduction of D5- and O5-planes).
  \end{itemize}

  At this level, the axio-dilaton remains unfixed.   It can be fixed by turning on NSNS and RR fluxes. The choice of fluxes can be made so that the background continues to preserve $\cN=1$ supersymmetry and remain in the weakly coupled regime $(g_s \ll 1)$. 

Interesting open directions for type I strings are:
\begin{itemize}

\item Study of the analogous non-supersymmetric backgrounds.

\item Developing a complete understanding of open string moduli stabilisation in the setting.

\item To see if dS vacua can be obtained. For this, understanding various higher order corrections to the effective action are important. One advantage is that a full string theory description is available and this can be used to compute the relevant corrections. The methods developed in  \cite{Bianchi:2007fx} can be useful to compute non-perturbative effects.

\item Implementing the analogous mechanisms in CY compactifications.

\item Counting the number of such vacua and developing an understanding of their statistical properties.
\end{itemize}

\subsubsection {Moduli stabilisation in M-theory} 

M-theory compactifications on manifolds of G2 holonomy lead to 4-dimensional theories with $N=1$ supersymmetry. Phenomenologically viable compactifications with non-Abelian gauge symmetries and chiral fermions arise if the  G2 manifolds have specific types of singularities. Non-Abelian gauge fields are localized on 3-dimensional submanifolds inside the internal space \cite{Acharya:1998pm}. Chiral fermions are supported at points in the internal manifold where there are conical singularities of a particular type \cite{Acharya:2001gy}.  Although many examples of smooth G2 manifolds have been constructed \cite{Kovalev:2001zr, Joyce:2002eb}, an explicit construction of compact G2 manifolds with the singularities required  has not been carried out so far. But such singular manifolds are believed to exist based on dualities between M-theory and  heterotic and type IIA compactifications. The phenomenological studies of these involve assuming the existence of the singularities and exploring the implications.

Upon reduction to 4 dimensions, geometric moduli corresponding to volumes of 3-cycles, $s_i$, pair up  with axionic fields that arise from the dimensional reduction of the $C_3$ potential to yield complex scalars  which are part of $\cN=1$ chiral multiplets. Let us summarise some of the key features of the 4-dimensional effective action that describes moduli dynamics (see e.g. \cite{Friedmann:2002ty} for further details). The K\"ahler potential is
$$
    K =  -3 \ln V_7,
$$
where $V_7$ is the volume of the 7-manifold (in units of the eleven-dimensional Planck length). The volume is known to be a degree $7/3$ homogeneous polynomial in $s_i$, the 3-cycle volumes. We will focus on scenarios in which the moduli are stabilised as a result of condensation in gauge theories which are localised on three-dimensional submanifolds.\footnote{For constructions involving non-trivial background $C_3$ flux see e.g. \cite{Acharya:2002kv,  deCarlos:2004ses, House:2004pm}} For an $SU(N)$ gauge theory, this leads to a superpotential 
\begin{equation}
\label{Mhidden}
   W =  A\, {\rm{exp}} \left( {-2 \pi f_{h} \over N} \right),
\end{equation}
where $f_{h}$ is the hidden sector gauge kinetic function and $A$ is related to the cutoff of the supersymmetric gauge theory. The gauge kinetic functions are related to the moduli fields by relations of the form $f_{h} = N^{i} \phi_i$; where $\phi_i$ are the moduli (in conventions in which the imaginary parts correspond to the volumes of 3-cycles and the real parts are axionic) and $N^{i}$ are integers. The sum runs over all moduli. In general, there are multiple hidden sectors; the superpotential is the sum of contributions of the type (\ref{Mhidden}) arising from each one of them. This leads to a moduli
potential which stabilises all of them. In fact, all the moduli can be stabilised by considering superpotentials that arise from two hidden
sector contributions \cite{Acharya:2006ia, Acharya:2007rc, Acharya:2011kz}. The vacua obtained are AdS. The value of the cosmological constant can be tuned by varying the ranks of the condensing gauge groups. As there are no fluxes, the landscape of vacua is scanned  by scanning through
these ranks. 

Candidate dS vacua have been constructed in  \cite{Acharya:2007rc, Acharya:2008hi}. This was done by including matter fields charged under the hidden sector gauge symmetries.   The dominant contribution to the vacuum energy then arises  from  F -term contributions of  hidden matter fields and there are solutions with a positive value for the cosmological constant. See \cite{Acharya:2012tw} for  a detailed review of these models. 
 
    The most important open direction is to develop an understanding of the  nature of the singularities that can arise in compact  G2 manifolds. With this,
  one can carry out a systematic derivation of the matter content and the associated effective field theory.

 \subsection{Holographic Approaches to Moduli Stabilisation}
 
The standard approach to moduli stabilisation, as just described in this section, operates through the language and techniques of
effective field theory. It starts with a 10-dimensional quantum string theory. Using dimensional reduction and effective field theory, it aims to describe the dynamics of this theory in a 4-dimensional language in which the string and Kaluza-Klein modes have been integrated out. Vacua of
 this 4-dimensional theory are found and then claimed to be also vacua of the full 10-dimensional theory. This claim is justified
 by computing corrections to the 4-dimensional theory, in both the $\alpha'$ and $g_s$ expansion, and showing that they are small. The smaller we these corrections are computed to be, the better we think we control the vacuum.

Recently, a complementary approach to thinking about moduli-stabilised vacua has appeared. Most scenarios of moduli stabilisation give, at least initially, AdS vacua and additional ingredients are necessary to uplift these vacua to dS solution. We know from the AdS/CFT correspondence that AdS vacua often have an interpretation in terms of dual conformal field theories. Even without an explicit construction of a dual CFT, the symmetries of AdS map onto the symmetries of CFTs \cite{Heemskerk:2009pn}. It is therefore interesting to re-express the properties 
of phenomenologically appealing moduli-stabilised AdS vacua in terms of CFT quantities.

States in the AdS theory correspond to operators in the CFT. Single-particle excitations in the AdS theory correspond to single-trace primary operators in the CFT, while interactions map onto higher-point correlators in the CFT. Although the whole AdS theory can be mapped into the CFT, the simplest relation concerns masses and conformal dimensions,
\be
\setlength\fboxsep{0.25cm}
\setlength\fboxrule{0.4pt}
\boxed{
\Delta_{\Phi} (\Delta_{\Phi} - d) = m_{\Phi}^2 R_{AdS}^2,
}
\ee
where $\Phi$ denotes both a field in $AdS_{d+1}$ and its corresponding dual operator in $CFT_{d}$, $\Delta$ is the conformal dimension and $R_{AdS}$ is the AdS radius.

The aims of this program are
\begin{itemize}
\item
By rewriting the vacua of moduli-stabilisation scenarios in an alternative language, search for hidden structures that may be opaque when written in the more conventional language of supergravity and dimensional reduction. From the perspective, AdS moduli-stabilised vacua at many different scales appear very similar from a CFT perspective (e.g. \cite{Conlon:2018vov, Conlon:2020wmc} ). Another example of this is the discovery that the mass spectrum of moduli in type IIA DGKT vacua all correspond to integer conformal dimensions in any dual CFT. This is a striking feature whose origin is not yet understood \cite{Conlon:2021cjk, Apers:2022tfm, Apers:2022zjx, Quirant:2022fpn, QuirantPellin:2022vyp, Plauschinn:2022ztd, Apers:2022vfp}.

\item
In the study of CFTs, the bootstrap program has been able to develop some sharp bounds on allowed consistent regimes for CFTs for quantities such as operator dimensions, both ruling out certain possible ranges and isolating some allowed islands which contain consistent theories (see e.g \cite{Simmons-Duffin:2016gjk}). By writing moduli-stabilising scenarios in CFT language, this provides a chance of ruling out particular scenarios (in the case that they could be proved inconsistent through CFT techniques) or giving strong evidence for their consistency (if they were found to exist in an allowed island).

\item Certain swampland criteria applicable to AdS theories may be given a sharper understanding when viewed holographically from the perspective of a dual CFT. Examples include the weak gravity conjecture \cite{Montero:2018fns, Furuuchi:2017upe, Harlow:2018tng, Andriolo:2020lul, Aharony:2021mpc, Palti:2022unw}, and the Swampland Distance Conjecture \cite{Baume:2020dqd, Perlmutter:2020buo}.

\item
The use of CFT techniques may eventually allow an end-run around the need to use perturbative effective field theory arguments to argue for the consistency of such vacua. This would be appealing as, in practice, it is unlikely that all $\alpha'$ or $g_s$ corrections could ever be computed for realistic string vacua with $N=0$ supersymmetry.

\item
Moduli-stabilised vacua exhibiting hierarchies have some unusual features from a CFT perspective. The presence of a small vacuum energy implies the central charge of the CFT is parametrically large. As the string and Kaluza-Klein modes are parametrically heavier than the moduli fields, the resulting CFT is scale-separated: it contains a parametrically large gap between the conformal dimensions of the single-trace operators corresponding to the moduli and the conformal dimensions of the heavy fields corresponding to the string and KK modes. This is a striking property as no such CFTs are currently known.

There are two lessons one could draw from this. The pessimistic one is that this may lead to the exclusion of standard moduli stabilisation scenarios such as DGKT, KKLT and LVS via holographic methods. The optimistic one is that the distinctive properties of these moduli stabilisation scenarios may act as guides enabling a direct construction of scale-separated CFTs, which would represent a major breakthrough in conformal field theory.

\end{itemize}

Although this program is still relatively new, it offers considerable promise as a way of uncovering deeper structure in compactifications and also as a potential gateway to the discovery of scale-separated CFTs.

\subsection{Other Approaches to dS}

\begin{itemize}

\item {\it Non-critical strings} \cite{Maloney:2002rr}: This construction is based on asymmetric orbifolds of super-critical strings
(it builds upon \cite{Silverstein:2001xn}).
 Taking  the  number of dimensions to be large and turning on fluxes\footnote{In $D$ dimensions, the number of RR fields
 is $2^{D}$. They dominate the spectrum.}, dilaton potentials are generated leading to nontrivial minima at arbitrarily small cosmological constant and string coupling. These are  separated by a barrier from a flat-space linear dilaton region. 
Estimates of the width of the vacua for decay to the flat space region via  instanton  processes yield a time scale which is larger than the Poincare recurrence time.\footnote{The Poincare recurrence time for dS space, $t_{rec} \sim H^{-1} e^{S_{dS}}$ with$S_{dS} \sim \frac{\Mp^2}{H^2}$ the entropy of the cosmological horizon, is the time-scale in which the finite entropy system violates the second law and enters briefly a very low entropy state.  It is expected on general grounds that any dS space must decay in a time shorter than $t_{rec}$ \cite{Goheer:2002vf, Kachru:2003sx}.} Much remains to be understood about the effective field theory of these models.

\item {\it Non-geometric vacua:}  T-duality transformations on flux vacua leads to non-geometric vacua i.e. configurations
where the metric is  defined on the internal manifold modulo  duality transformations,  see e.g \cite{Kachru:2002sk, Shelton:2005cf, Grana:2006kf} and \cite{Plauschinn:2018wbo} for an overview. Non-geometric constructions can yield dS vacua without tachyons \cite{deCarlos:2009fq, Danielsson:2012by, Blaback:2013ht, Damian:2013dq, Damian:2013dwa}. However the exact form of the moduli space is unknown, and this is required to check the 
 validity of the approximations. Moreover, the  tree-level stabilisation procedure typically leads to a 4-dimensional potential of order the string scale with the volume of the internal manifold also close to the string scale, putting under pressure the validity of the KK truncation used to define the 4-dimensional effective field theory. There are also issues associated
 with tadpole cancellation. For a detailed discussion of the challenges in obtaining non-geometric dS vacua, see \cite{Plauschinn:2020ram}.
 
 \item{\it dS vacua from RG running:} The 4-dimensional low energy effective actions obtained from string theory are Wilsonian effective actions that arise after integrating out the string and Kaluza-Klein modes. The Wilsonian scale associated with these actions is the Kaluza-Klein scale,
 which is much larger than the scales of cosmological observations of the energy density of the universe. To obtain the effective
 action relevant at the cosmological scales, all particles with masses between the Kaluza-Klein and cosmological scales have to be integrated 
 out. As a result of this, an AdS minimum of the high scale action can correspond to a solution with positive energy density at the cosmological scales.
 This possibility has been analysed in \cite{deAlwis:2021zab, DeAlwis:2021gou}.  Conditions on the spectrum of LVS models
 necessary to achieve this have been  studied in detail. In this picture, a high density of AdS vacua of the high scale effective action 
 translates to  finely spaced dS vacua at low scales. This can be used to tune the observed value of the cosmological constant.

 \item{\it dS from a decaying AdS:} This scenario is a variation of the Randall-Sundrum construction. In the 
 Randall-Sundrum construction, two identical AdS$_5$ vacua are separated by a 3-brane. The five-dimensional graviton has a zero mode confined on the brane and this leads to the observers on the 3-brane effectively experiencing gravity in 4 dimensions (despite the brane being embedded in a five-dimensional space). Localisation of the graviton wavefunction is also responsible for solving the hierarchy problem between the Planck and weak scales. To obtain an effective dS space for 4-dimensional observers, \cite{Banerjee:2018qey, Banerjee:2019fzz} considered the setting with a metastable AdS$_5$ vacuum decaying to a supersymmetric one via bubble nucleation.  A spherical brane separates the two phases and observers on the brane experience  an effective positive cosmological constant coupled to matter and radiation. Various aspects of the proposal have been studied in \cite{Banerjee:2019fzz}, and see \cite{Berglund:2021xlm} for a related construction.

\item{\it dS as a coherent state:} It has been proposed that 4-dimensional dS space is a coherent state (in particular a Glauber-Sudarshan state) over a supersymmetric Minkowski vacuum (see e.g. \cite{Brahma:2020htg, Bernardo:2021rul}).  The claim is that this is realizable in full string theory, but only with the inclusion of with temporally varying degrees of freedom and quantum corrections arising from them.  Fluctuations over the dS space is governed by a generalized graviton (and flux)-added coherent state. The realisation of dS space as a state, and not as a vacuum, potentially resolves many conceptual issues associated it. 
\ei

\subsection{Vacuum Transitions}

One of the most robust and striking outcomes of moduli stabilisation is the fact that our universe is clearly unstable. Either it corresponds to one of the many flux vacua with a positive cosmological constant which indicates an asymptotically dS space in four dimensions or it corresponds to a time varying configuration in which one or many fields runaway towards the 10-dimensional vacuum with vanishing string coupling. In both cases the universe is unstable, either by the runaway in the latter case or by tunneling from the dS minimum towards other vacua including the lower energy one corresponding to flat 10 dimensions. This is known as a vacuum transition. For a general review see \cite{Weinberg:2012pjx,Rubakov:2002fi}.

Therefore the term moduli stabilisation is relative since at some point the 4-dimensional vacuum is not stable but at best metastable. It is then important to understand the possible final outcomes of each of these vacua. Over the past 40 years vacuum transitions have been studied in field theory and gravity following the seminal work of Sidney Coleman and collaborators. It is a subject in which quantum aspects of gravity play a crucial role and together with the study of black holes it is the main arena to explore in which to learn quantum gravitational physics. At some point the full machinery of string theory should play a role in understanding properly this phenomenon. In the meantime, following the effective field theory approach of this chapter we may explore the
decay rates for the string vacua discussed in this section, such as type IIA and KKLT and LVS in type IIB flux compactifications.

Given the apparent immense nature of the flux landscape vacuum transitions can play a crucial role in string cosmology. In this scenario the beginning of our universe would most probably be a vacuum transition from another universe and a potential end of (at least a portion of) our universe could also be through a vacuum transition to yet another universe.

\subsubsection*{Transitions without gravity}

 We all know the well understood tunneling effect in quantum mechanics for which a particle can cross a potential barrier with a non-vanishing  probability. The transition rate can be explicitly computed using the semi-classical WKB approximation $\psi\simeq e^{iS/\hbar}$ with $S=S_0+\hbar S_1+ \cdots $. The transition rate is exponentially small due mainly to the exponential decay of the wave function in the region under the barrier.
\be
\Gamma = \frac{|\psi_T|^2}{|\psi_I|^2}\simeq e^{-B/\hbar}; \qquad B=\int_{x_1}^{x_2} dx \sqrt{2m(V(x)-E)}
\ee
Where $\psi_{I,T}$ are the incident and transmitted wave functions, $V(x)$ the potential energy and $E$ the energy. Notice that since, under the barrier, the argument of the square root is positive, it hints at  a classical negative kinetic energy and so imaginary speed, which in turn would indicate an imaginary time that has lead to a Euclidean rather than Lorentzian approach when this is extended to field theory.

 For quantum field theory Coleman and collaborators \cite{Coleman:1977py, Callan:1977pt} found that this WKB approach in quantum mechanics can be generalised considering a scalar potential with at least two different minima, the false ($V_{FV}$) and true ($V_{TV}$) vacua depending on the value  of the potential at the minima, separated by a barrier of height $\Delta V=V_{FV}-V_{TV}$ (illustrated in figure \ref{Fig:BN}).  Since in field theory the energy is an integral over the full spacetime volume of the energy density determined by the scalar potential and the scalar field gradients $ U(\phi)=\int d^3x\left(\frac{1}{2}(\nabla \phi)^2+V(\phi)\right)$ the transition amplitude vanishes (since it corresponds to an infinite potential barrier). Instead, the transition happens locally by bubble nucleation, in the sense that a bubble of the true vacuum can materialise in a background of the  false vacuum and then expand. Since there are many paths that the field can follow to interpolate between the two vacua, Coleman and collaborators developed a formalism to estimate the transition rate per unit volume $\Gamma$ using instanton techniques in Euclidean field theory. The path connecting the two vacua that minimises the action corresponds to an instanton that, due to the time reversal invariance of the equations, start at one point and finishes at the same point, and hence is known as a {\it bounce}.

 The bounce can be found by solving the Euclidean action field equations:
 \be
 \frac{d^2\varphi}{d\tau^2}+\nabla^2\varphi-\frac{dV}{d\varphi}=0
 \ee
 The transition rate can be estimated numerically and can be explicitly computed in the thin-wall approximation $\epsilon =V_{FV}-V_{TV}$ much smaller than the height of the barrier. The transition probability per unit volume takes the form
\be
\Gamma\simeq e^{-B/\hbar}; \qquad B=S(\varphi_b)- S(\varphi_{FV})
\ee
where $S(\varphi_b)$ corresponds to the action evaluated at the bounce configuration and $S(\varphi_{FV})$ the action at the background false vacuum.

\begin{figure}[t]
\begin{center}
\includegraphics[width=120mm,height=72mm]{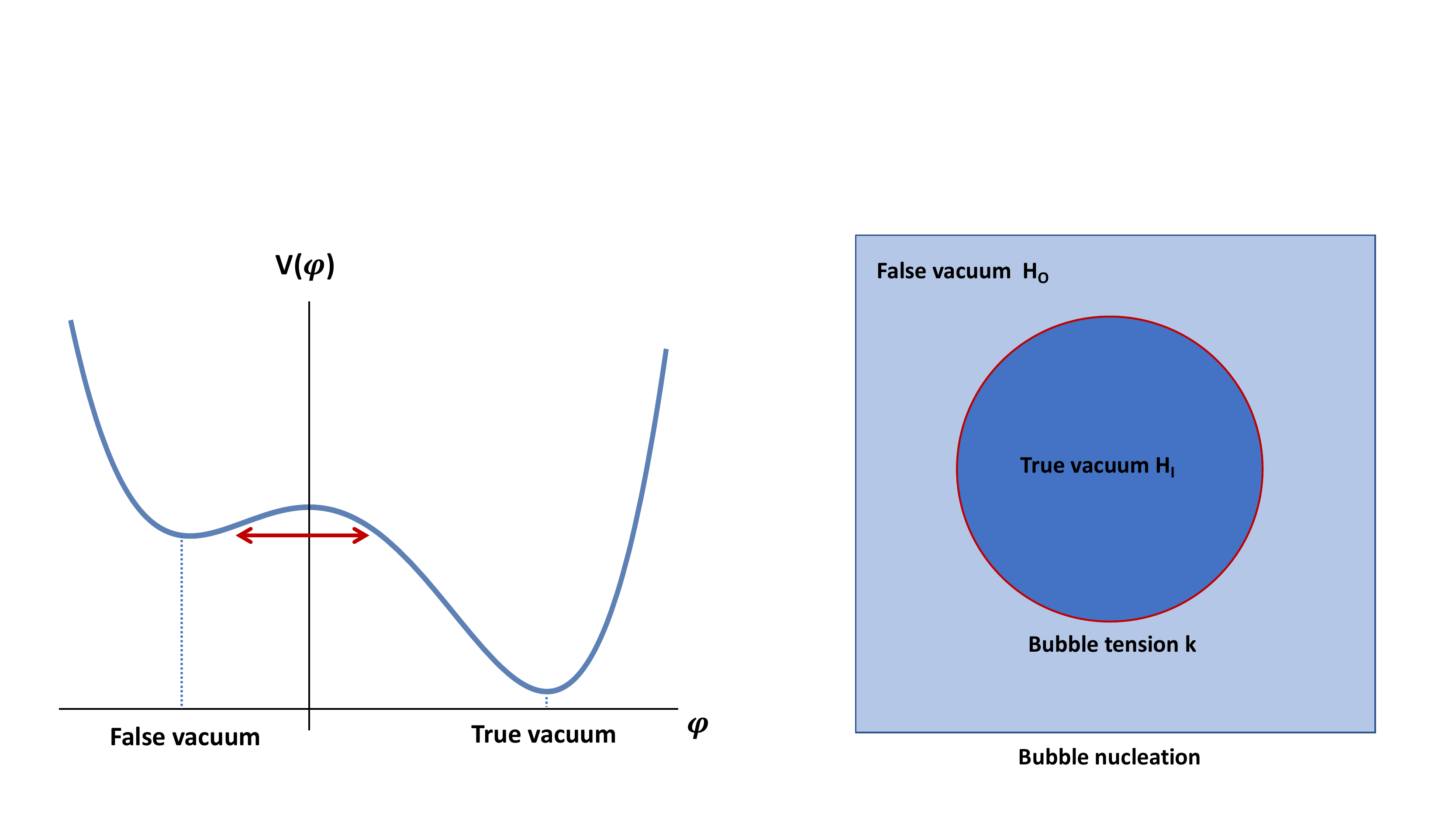} 
\caption{Vacuum transitions in field theory and gravity.} \label{Fig:BN} 
\end{center}
\end{figure}

An important property of the bounce is that the field equations (and boundary conditions) are not only spherically symmetric, corresponding to the symmetry of the bubble, but have an enhanced $O(4)$ symmetry since they can be formulated in terms of the $O(4)$ invariant parameter $\xi^2= \tau^2+r^2$. Coleman and collaborators used an $O(4)$ ansatz $\varphi=\varphi(\xi)$ to find the bounce and later on proved that it is actually the minimum configuration \cite{Coleman:1977th}. However, in order to study the further evolution of the bubble after materialisation, a Wick rotation needs to be performed $\tau\rightarrow it$ and the $O(4)$ symmetry becomes $O(3,1)$. The region outside the light-cone of the origin can be properly treated in terms of the coordinates $t,r$. But the region inside the light-cone does not exist in the Euclidean coordinates and in Lorentzian coordinates are better described by the change of coordinates $(\hat t, \hat r)=(\sqrt{t^2-r^2}, r/\sqrt{t^2-r^2})$. It can be easily seen  that in terms of these coordinates the Minkowski metric takes the form
 \be
 ds^2=d\hat t^2- \hat t^2\left[\frac{d\hat r^2}{1+\hat r^2}+\hat r^2\left(d\theta^2+\sin^2\theta d\phi^2\right)\right]=d\hat t^2- \hat t^2 d\Omega^2_{(k=-1)}
 \ee
corresponding to a an open ($k=-1$) FLRW universe.

\subsubsection*{Transitions including gravity}

Let us then briefly review the subject of vacuum transitions in gravitational theories (also illustrated in figure \ref{Fig:BN}). Following the Euclidean approach, in a seminal contribution,  Coleman and de Luccia (CDL) \cite{Coleman:1980aw} further generalised the vacuum transition formalism  to include the effects of gravity. Contrary to the case without  gravity, in which the formation and evolution of the bubble corresponds to opposite contributions from internal pressure and the tension of the wall, once gravity is included it also contributes to the stability of the bubble. In particular it allows up-tunneling transitions from the true vacuum to the false vacuum in which pressure and tension both tend to compress the bubble but gravity, through the presence of the cosmological constant, contributes towards the expansion of the bubble. actually, Lee and Weinberg \cite{Lee:1987qc} estimated the transition rate from true to false vacuum.

A further important observation on the gravity case is that there is no proof that  the optimal instanton configuration is the $O(4)$ symmetric bounce. Nevertheless CDL assumed this symmetry and managed to estimate the transition rates for both dS to Minkowski and Minkowski to AdS. Further generalisations were done later on.

A related but different approach to vacuum transitions was performed by Brown and Teitelboim (BT) \cite{Brown:1988kg} in which instead of a scalar potential with different minima, they studied bubble nucleation in terms of a theory with fluxes, similar to the ones discussed in the previous sections. They concentrated on the case of 3-index antisymmetric tensors coupled to gravity in which even though they do not correspond to propagating degrees of freedom, they can give rise to the nucleation of membranes in 4 dimensions. A chain of transitions may be induced reducing the fluxes and then the value of the cosmological constant. Their estimate of the transition rates were performed in the Euclidean approach and agree with the CDL calculations in the thin wall approximation. In particular for a transition from dS space with cosmological constant $\lambda_{\rm O}=3H_{\rm O}^2$ to another vacuum with $\Lambda_{\rm I}=3H_{\rm I}^2$ nucleating a bubble of tension $\kappa$ is $\Gamma=e^{-B/\hbar}$ with 
\be
B  =-\frac{\pi}{G}\left[ \frac{\left[ (H_{{\rm O}}^{2}-H_{{\rm I}}^{2})^{2}+\kappa^{2}(H_{{\rm O}}^{2}+H_{{\rm I}}^{2})\right] R_{{0}}}{4\kappa H_{{\rm O}}^{2}H_{{\rm I}}^{2}}-\frac{1}{2}\left(H_{{\rm I}}^{-2}-H_{{\rm O}}^{-2}\right)\right] \label{eq:SdS}
\ee
where
\be
R_{0}^{-2}  =\frac{1}{4\kappa^{2}}\left[(H_{{\rm O}}^{2}-H_{{\rm I}}^{2})^{2}+2\kappa^{2}(H_{{\rm O}}^{2}+H_{{\rm I}}^{2})+\kappa^{4}\right]\label{eq:R0}\\
\ee
Here the indices I and O refer to inside and outside the bubble. This includes both down and up-tunneling between the 2 dS spacetimes. An interesting property of these transitions is that the ratio of up to down-tunneling can be easily computed and give $\Gamma_{\rm up}/\Gamma_{\rm down}=e^{-(S_{TV}-S_{FV})}$ with $S_{TV}, S_{FV}$ corresponding to the entropies of each of the vacua, true and false respectively. Recall that the entropy of dS space is given by $S=\pi/(GH^2)$. Therefore this is a statement of detailed balance in which in equilibrium the relative transition rates  are weighted by the corresponding entropy. Similar to the case without gravity, the evolution of the bubble after materialisation requires an analytic continuation from a Euclidean to a Lorentzian metric that describes an open universe. Note, however, that this result fully relied on the underlying $O(4)$ symmetry of the bounce, something which is far from being shown in the case with gravity.

Similar results for the transition rates hold for AdS down tunneling. However here a condition on the relative value of the cosmological constants and the wall tension has to be satisfied: 
$\kappa<\sqrt{|H_{\rm I}^2}-\sqrt{|H_{\rm O}^2|}$. Furthermore up-tunneling from AdS  is not allowed. Nor from Minkowski to dS. However a proposal by Farhi, Guth and Guven (FGG) \cite{Farhi:1989yr} for {\it creation of universes in the laboratory} meaning nucleating a dS bubble from Minkowski, has been considered. Following the standard Euclidean approach it was found that a singular instanton is needed in order to achieve the bubble nucleation. 

This prompted a Hamiltonian approach by Fischler, Morgan and Polchinski \cite{Fischler:1990pk} in which the Minkowski to dS transition was reconsidered and concluded that it is actually possible to realise the FGG proposal in Lorentzian spacetimes. A key ingredient of this Hamiltonian formalism is to assume only the standard $O(3)$ spherical symmetry to describe the bubble and its evolution. This symmetry automatically include Schwarzschild black holes solutions with a given mass parameter $M$. As long as $M\neq 0$ the up-tunneling from Minkowski is allowed. Recent generalisations of this result have been developed \cite{Bachlechner:2016mtp,DeAlwis:2019rxg}. The Hamiltonian approach is so far very much restricted to the extreme thin-wall approximation and further developments need to be performed to include scalar potentials with different minima. The calculations are also performed in the global dS slicing that when studying the further evolution of the bubble wall fits with a closed rather than open universe. This put at least into question the general claim from the Euclidean approach that the universe within the bubble has to correspond to an open universe. This is an important point for the string landscape since if the open universe claim holds in general, it may be conceivable to rule out the full string landscape if in the future is found that the universe is not open \cite{Dyson:2002pf, Freivogel:2005vv, Cespedes:2020xpn}. This subject needs further studies before arriving to a concrete conclusion. 

\subsubsection*{Transitions in string scenarios}

\begin{figure}[t]
\begin{center}
\includegraphics[width=120mm,height=72mm]{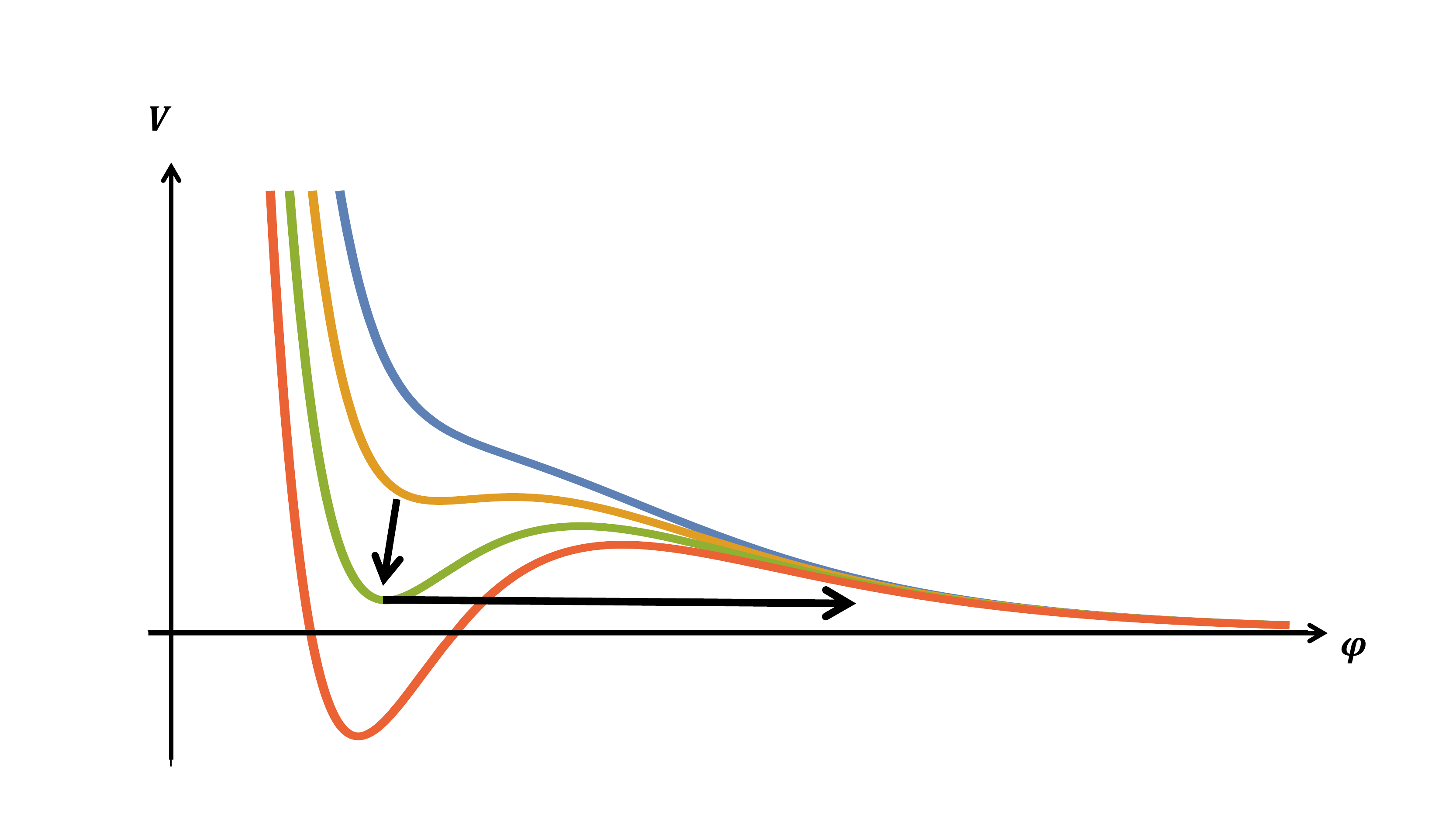} 
\caption{Vacuum transitions in string theory. First bubble nucleation from flux/D3 brane charge transitions illustrated by the vertical arrow. Then a CDL-like transition crossing the potential barrier illustrated by the horizontal line. In string theory this transition may correspond towards decompactification.} \label{Fig:BN2} 
\end{center}
\end{figure}
In string theory (illustrated in figure \ref{Fig:BN2}) we may consider the scenarios discussed in the previous sections. First we should point out that there can be at least two different types of transitions:
\begin{enumerate}
\item
Transitions of the Brown-Teitelboim type corresponding to transitions among vacua corresponding to the different quantised fluxes. Concretely, given two vacua with values of three-form fluxes 
$H_{mnp}$, $F_{mnp}$ in terms of integers $(K,M)$. For a transition from a $(K,M)$ vacuum to a $(K',M')$ vacuum, a brane carrying D3-brane charges $(K-K',M-M')$ can be nucleated to mediate between the two vacua. These are precisely D5 (or NS5 ) branes that can wrap the 3-cycles carrying the fluxes and with the remaining 2 spatial dimensions corresponding to an $S^2$ wall separating the two different 4-dimensional vacua.  

This provides an elegant string theory implementation of the 4-dimensional vacuum nucleation picture. It fits nicely in the sense that D3 and D7 branes being BPS and codimension 4 have been used to host the Standard Model and/or hidden sectors, whereas D5-branes do not play a direct role as long as supersymmetry is preserved (since supersymmetry requires codimension 4 branes). But D5-branes naturally couple to 3-forms and furthermore the extra dimensions nicely fit with the dimension of a wall separating different 4-dimensional vacua. From the 4-dimensional EFT, there is not a scalar potential which connects the two vacua so this cannot be explicitly described in terms of the CDL bounce solution within the 4-dimensional EFT. But it fits nicely in the BT formalism.

\item Transitions {\it a la}  CDL are also implemented since any potential dS vacuum should coexist with the runaway vacuum corresponding to infinite volume and vanishing string coupling. The potential for the volume modulus connects both vacua and the transition may be estimated.
\end{enumerate}

Transition rates have been estimated for type IIA \cite{Narayan:2010em} as well as type IIB flux compactifications  (for both KKLT and LVS) vacua \cite{Kachru:2003aw, Westphal:2007xd, deAlwis:2013gka}. The probability amplitude $\Gamma\sim e^{-24\pi^2/\Lambda_0}$ where $\Lambda_0$ is the value of the scalar potential at the dS minimum. The corresponding lifetime $\tau \sim 1/\Gamma$ is exponentially small as compared with the Poincar\'e recurrence time which is reassuring. Furthermore the transition from dS to an AdS is preferred over dS to dS and the CDL transition towards decompactification dominates the dS to dS transitions.

An interesting observation was made in \cite{Johnson:2008vn,Aguirre:2009tp} regarding the actual implementation of the bounce solution in a toy model version of flux compactifications with two dS vacua plus the runaway. The claim is that not only the decompactification transition is preferred but that the potential bounce solution connecting the two dS vacua necessarily follows into the runaway towards decompactification providing a potential  obstacle to implement the transition. Note however that contrary to the toy model in which both transitions  corresponded to the CDL type, in flux compactifications the flux transition is of the BT type whereas the decompactification is of the CDL type.

\newpage

\section{String Inflation}
\label{sec:Sinfla}

\subsection{Embedding Inflation in String Theory}

As described in Section \ref{SecCO}, inflation is the leading candidate theory for the origin of structure in the universe. This reason alone would justify attempts to embed inflation in string theory (see \cite{Quevedo:2002xw,Kallosh:2007ig,Burgess:2004jk,Burgess:2006nod,Burgess:2006uv,Cline:2006hu,HenryTye:2006uv,McAllister:2007bg,Copeland:2011dx,Cicoli:2011zz,Burgess:2011fa, Yamaguchi:2011kg, Burgess:2013sla,Baumann:2014nda,Silverstein:2016ggb,Nastase:2019mhe} for detailed reviews on different aspects of string cosmology and inflation in string theory). However, there are more specific reasons why a consistent theory of quantum gravity is necessary for a full understanding of the inflationary dynamics.

\subsubsection{Motivations for String Inflation}

There are several reasons to require a consistent embedding of inflation models within 
string theory. Below, we list those which we consider most relevant: 

\begin{itemize}

\item \emph{UV sensitivity of inflation}: As we discussed in Section \ref{sec:infla}, single-field inflation is driven by the dynamics of a scalar field with mass $M_{\rm inf}$ below the Hubble scale during inflation $H_{\rm inf}$. This can be see from the  second potential slow-roll parameter $\eta_V$ introduced in \eqref{eq:etaV}, which  can be expressed as
\begin{equation}
\setlength\fboxsep{0.25cm}
\setlength\fboxrule{0.4pt}
\boxed{
\eta_V\simeq \left|\frac{M_{\rm inf}^2}{H^2}\right|\ll 1\,,
\label{etaProb}
}
\end{equation}
where the condition $\eta_V \ll 1$ guarantees that inflation lasts for long enough to solve the horizon problem and flatness problems. However, in the absence of a protection mechanism based on (approximate) symmetries, quantum corrections are expected to make these scalars heavy, so that $\eta_V\simeq \mathcal{O}(1)$. 

There are two generic sources of $\mathcal{O}(1)$ corrections to the $\eta_V$-parameter. First, quantum effects tend to generate corrections to scalar masses of order the cut-off of the effective field theory $\Lambda$, $M_{\rm inf}\simeq \mathcal{O}(\Lambda)$. For controlled inflationary model building with $H_{\rm inf}\lesssim \Lambda$, one has from (\ref{etaProb}) $\eta_V \gtrsim 1$. Secondly, even if the scalar potential arising from relevant and marginal operators yields $\eta_V \ll 1$, irrelevant operators of the form
\begin{equation}
\setlength\fboxsep{0.25cm}
\setlength\fboxrule{0.4pt}
\boxed{
    \Delta V = c\, V(\phi)\left(\frac{\phi}{\Lambda}\right)^2\,,
}
\end{equation}
would produce a correction to $\eta_V$ of order 
\begin{equation}
\setlength\fboxsep{0.25cm}
\setlength\fboxrule{0.4pt}
\boxed{
    \Delta \eta_V = 2 c \left(\frac{M_{\rm Pl}}{\Lambda}\right)^2\quad\text{with}\quad \Lambda \lesssim M_{\rm Pl}\,,
}
\end{equation}
that would violate the slow-roll condition if $\Lambda \ll M_{\rm Pl}$, or if $c\simeq \mathcal{O}(1)$ for $\Lambda \simeq M_{\rm Pl}$. 

This is usually referred to as the $\eta_V$-problem in  single-field inflation. A solution to it, if not based on fine-tuning, requires the presence of symmetries which can be only  provided by a consistent UV embedding. These symmetries, even if not exact (as for the case of global symmetries in quantum gravity),  may be enough to explain why $c \ll 1$. This UV-sensitivity of inflation implies that inflationary model building 
can only be trusted with a knowledge of its UV completion. Note that a robust UV embedding is needed even if the $\eta_V$-problem is addressed by relying on fine-tuning, since one has to ensure that the underlying theory features enough freedom to allow $\eta_V$ to be 
tuned to a small value.

On the other hand, as we discussed in Section \ref{sec:multinf}  the $\eta_V$ parameter in multi-field inflation, \eqref{etaVmulti}, does not have to be small. Indeed, the masses of the inflatons can be larger than the Hubble scale $H_{\rm inf}$ during inflation without affecting slow-roll \cite{Chakraborty:2019dfh,Aragam:2021scu}\footnote{We discuss an example of this below. See Appendix A of \cite{Chakraborty:2019dfh} for  a list of examples in field theory, and \cite{Aragam:2021scu} for  more details and examples in supergravity.}. The problem of UV sensitivity of inflation in this case cannot be formulated in terms of the inflaton masses, i.e.~$\eta_V^m$ of Eq.~\eqref{etaVmulti}. 

\item \emph{Trans-Planckian field excursions and the tensor-to-scalar ratio}: We saw in Eq. (\ref{LythBound}), reproduced here, that the size of the tensor fluctuation can be related to the field displacement of the inflaton by,

\be
\setlength\fboxsep{0.25cm}
\setlength\fboxrule{0.4pt}
\boxed{
 \frac{\Delta\varphi}{\Mp} = \int^{N_{\rm hc}}_{N_{\rm end}}{ dN \,\sqrt{\frac{r}{8}} } \,.
 \label{Lyth2}
 }
 \ee
Large and observable values of the tensor-to-scalar ratio $r$ correspond to trans-Planckian field excursions of the inflaton $\varphi$. Indeed, the simplest field theory models of single-field  inflation, such as chaotic inflation, all lead to substantially trans-Planckian field excursions (see Eqs.~\eqref{eq:monoinf}-\eqref{eq:natuinf} in Sec.~\ref{sec:infla}).

Eq.~(\ref{Lyth2}) represents a rare example where the Planck scale -- the quantum gravity energy scale -- explicitly appears inside a quantity that is observable, not only in principle but also possibly in practice with current technology. By definition, a trans-Planckian inflationary field excursion requires the existence of a potential within the UV-complete theory that sustains an approximately de Sitter phase across a Planckian distance in field space. Whether such regions can exist can only be answered within a quantum gravity theory.  For any potential $V_{inf} (\varphi)$, generic non-renormalisable corrections with $\mc{O}(1)$ coefficients $\lambda_n$,
\be
\setlength\fboxsep{0.25cm}
\setlength\fboxrule{0.4pt}
\boxed{
V_{full}(\varphi) = V_{inf}(\varphi) + \sum \lambda_n \frac{\varphi^n}{\Mp^n},
\label{ghjt}
}
\ee
will destroy the flatness of trans-Planckian field excursions. Whereas the $\eta_V$ problem only required control over the quadratic term in the potential, flatness on trans-Planckian scales requires control over all terms in Eq. (\ref{ghjt}).  

Again, the multi-field case is more subtle. In that case, $\Delta\varphi$ corresponds to the multi-field {\em inflationary trajectory}. This may or  not, coincide with the {\em geodesic trajectory} in field space, which in general will be curved. 
Thus in the multi-field case, the potential can be very steep (indicating the possibility of heavy inflatons, as discussed before), while inflation proceeds along a suitable flat non-geodesic trajectory, i.e.~$\Omega/H\gtrsim 1$, along which $\epsilon_V^m, \eta_V^m\gtrsim 1$ (see Eqs.~\eqref{eq:w}, \eqref{epsimulti}, \eqref{etaVmulti})\footnote{Moreover, when more than one scalar field is active during inflation, entropic perturbations may be converted -- more or less effectively -- into adiabatic perturbations, introducing a  {\em transfer angle} $\Theta$ \cite{Wands:2002bn} in \eqref{Lyth2} through $r=16\,\epsilon\cos^2\Theta$.}.
 
Having a UV-complete theory of quantum gravity, such as string theory, thus allows to address 
questions about the structure of the potential and whether it supports geodesic (single-field) or non-geodesic  trans-Planckian field excursions. 
This is especially relevant given 
that many of the best known field theory models of inflation, such as chaotic inflation or natural inflation, require trans-Planckian field excursions. The difficulties of attaining flat potentials over such field excursions in string theory has been appreciated for a long time (e.g. see \cite{Banks:2003sx}) and has recently received renewed attention with the Swampland Distance Conjecture (see the discussion in section \ref{sec:Alt}).

\item \emph{UV constraints on model building}: Field theoretic models of inflation can be characterised by almost any kind of inflationary potential, with very different predictions for the main cosmological observables. However the requirement of a sensible embedding into string theory, based on explicit Calabi-Yau realisations with full moduli stabilisation, can strongly constrain the allowed shape of the inflationary potential.

\item \emph{Connection to observations}: Inflation represents a unique possibility to test high scale physics like string theory due to the fact that the inflation scale is expected to be not too far from the Planck scale. Moreover, string theory applications to inflation may lead to the discovery of features which are either generic in the string landscape, or alternatively impossible to achieve from string theory with control over the effective field theory (one case may be large primordial tensor modes together with low-energy supersymmetry). 
Such predictions can lead to sharper tests of string theory scenarios, in particular through studying the interplay between cosmology and particle physics within the same model. 

\item \emph{Initial conditions}: Embedding inflation into string theory can help to understand the delicate issue of initial conditions and the mechanism through which inflation started.

\item \emph{Reheating}: A complete understanding of reheating after the end of inflation requires the knowledge of all relevant degrees of freedom at inflationary energies to make sure that the inflaton energy density is transferred to the production of Standard Model degrees of freedom without too much energy being lost into hidden sector particles. To perform this analysis it is clearly crucial to have a robust UV embedding of inflation into string theory to control all the microscopic degrees of freedom.
\end{itemize}

\subsubsection{Requirements for String Inflation}

Besides the reasons to embed inflation into string theory, let us briefly discuss what are the main conditions that a perfect working model of string inflation should satisfy: 

\begin{itemize}
\item \emph{Moduli stabilisation}: All the closed and open string moduli should be stabilised. The scalar potential should feature a region suitable to drive inflation without any orthogonal runaway direction which would destabilise the inflationary dynamics. Ideally, the same scalar potential should also allow for a proper description of the late-time evolution of the universe, including both the reheating epoch and also either a de Sitter minimum or a quintessence direction. 

\item \emph{Hierarchies:} A controlled low-energy effective field theory should be characterised by the following mass hierarchy throughout the whole inflationary dynamics:
\bea
&&\label{eq:standardH}
M_{\rm inf}< H_{\rm inf} < M_{\rm KK} < M_s < M_{\rm Pl}\,, \qquad {\text{ standard hierarchy, single and multi-field}}\\
&&
H_{\rm inf} < M_{\rm inf}<M_{\rm KK} < M_s < M_{\rm Pl}\,, \qquad {\text{ multi-field inflation }}
\label{eq:multiH}
\eea
where $H_{\rm inf}$ is the Hubble scale during inflation and $M_{\rm inf}$ denotes the inflaton mass during inflation. In the multi-field case, it denotes the eigenvalues of the mass matrix ${\mathbb M}^a_{\,\,\,b}=\nabla^a\nabla_b V$ along the inflationary trajectory.  
In the standard hierarchy case \eqref{eq:standardH},   all the moduli with mass $m\gtrsim H_{\rm inf}$ are heavy and sit at their minimum during inflation. The moduli with a mass of order the Hubble scale during inflation or smaller, $m\lesssim H_{\rm inf}$, should be analysed in detail as they can affect the inflationary dynamics. In this sense, the cleanest situation is the single-field case where all the moduli different from the inflaton can be decoupled since they are heavier than the Hubble scale. Other scenarios are instead intrinsically multi-field and need a deeper analysis (one possible exception involves additional axion fields which always remain massless and decouple from the dynamics). Let us also stress that the requirement to realise inflation below the Kaluza-Klein and the string scale is just due to our poor understanding of the full dynamics involving stringy corrections but it does not need to be a regime preferred by Nature.

\item \emph{Computational control}: Any scalar potential receives quantum corrections of several kinds in different regions of the underlying parameter and field spaces. Hence any working model of accelerated expansion needs to have computational control over the effective field theory. This requires that the size of all cycles should be fixed above the string scale and both the $\alpha'$ and the string loop perturbative expansions should be under control. Leading and higher order non-perturbative effects should also be computable and the backreaction from fluxes and localised objects should also be suppressed, or properly included.

\item \emph{Calabi-Yau embedding}: In order to have a successful model of inflation from string theory, a string-inspired supergravity setup is clearly not enough. Ideally, one should build a globally consistent Calabi-Yau (or equivalent) compactification with a rigorous description of the underlying geometry and topology, and an explicit orientifold involution and brane setup. In this way one can have control over the underlying K\"ahler cone which determines the field space, and can check if the brane setup and the divisors have the right topology to generate the desired perturbative corrections to the K\"ahler potential and instanton contributions to the superpotential to freeze the moduli and generate the inflationary potential. Given that the real world does not feature just an initial epoch of accelerated expansion, the same Calabi-Yau compactification should allow for the realisation of a Standard Model-like sector with a non-Abelian gauge group and chiral matter, together with a viable reheating process to produce it after the end of inflation. 

\item \emph{Comparison with data}: An ideal working model of string inflation should clearly match all existing data about cosmological observables. The main data to fit are the scalar spectral index and the amplitude of the density perturbations, together with the present upper bound on tensor modes. Stringent bounds arise also from non-Gaussianities and the amplitude of isocurvature modes which become particularly relevant in multi-field models. In fact, fields which are lighter than the inflaton, like ultra-light axions typical of string compactifications, might behave during inflation just as spectator fields which acquire isocurvature fluctuations that might be in tension with data if their later contribution to dark matter is considerable. Moreover bounds from dark radiation and dark matter overproduction should be respected in the reheating and post-inflationary epoch. 

\item \emph{Novel predictions}: Ideally, a successful model will also produce novel predictions for upcoming experiments that are characteristic of the scenario and also distinguish it from other related models.

\end{itemize}

It is clear that no existing model in the literature satisfies all these requirements perfectly. However, it is useful to hold them all in mind as the ultimate aspirational target.

\subsection{Single-Field Inflation}

Let us start our discussion with the case where only one modulus remains light and can be identified with the inflaton, or where more fields are light but the cosmological predictions remain effectively those of single-field inflation. The most robust models are based on shift symmetries which protect the flatness of the inflationary potential and provide an elegant solution to the $\eta_V$-problem in single-field inflation. These shift symmetries are always approximate and can be either saxionic, in the case of K\"ahler moduli \cite{Burgess:2014tja,Burgess:2016owb}, or axionic, in the case of $C_0$, $C_2$, $C_4$ and complex structure axions \cite{Pajer:2013fsa}. Denoting schematically the relevant complex modulus as $T = \tau + {\rm i} \theta$, saxionic shift symmetries look like:
\begin{equation}
T \to T' = T + \alpha\, , \quad \alpha \in \mathbb{R}\qquad \Rightarrow \qquad \tau \to \tau' = \tau + \alpha\,,
\label{rhoshift}
\end{equation}
while axionic shifts take the form:
\begin{equation}
T \to T' = T+ {\rm i} \alpha\, , \quad\alpha \in \mathbb{R} \qquad \Rightarrow \qquad \theta \to \theta' = \theta + \alpha\,.
\label{axionshift}
\end{equation}
Depending on the effects which break these symmetries, the inflationary potential can take different forms. 

In the absence of a symmetry-based mechanism to solve the single-field $\eta_V$-problem, accelerated expansion can still be achieved through an appropriate fine-tuning of the underlying parameters which allows for cancellations among higher dimensional operators that would otherwise ruin the solution. Let us now present a broad overview of different single-field models of string inflation classifying them in terms of the mechanism exploited to realise inflation: based on either saxionic/axionic shift symmetries or fine-tuning.

\subsubsection{Exponential Potentials from Saxionic Shift Symmetries}

Due to the typical no-scale structure of type IIB flux compactifications (a structure to the effective field theory which is special and absent in `generic' theories) \cite{Cremmer:1983bf,Burgess:2020qsc}, which holds at tree-level and even at 1-loop order in type IIB, the K\"ahler moduli different from the overall volume mode enjoy an effective non-compact shift symmetry of the type (\ref{rhoshift}). In fact, all the K\"ahler moduli are flat at tree-level. Due to the extended no-scale structure which suppresses the contributions of string loops \cite{Cicoli:2007xp}, the first effect which generates a potential is at $\mathcal{O}(\alpha'^3)$ \cite{Becker:2002nn}. However, this higher derivative effect depends just on the overall volume $\mathcal{V}$, leaving all the other moduli flat at this order of approximation. Hence, all the K\"ahler moduli orthogonal to the volume mode are in principle promising inflaton candidates since they are leading order flat directions. 

Different effects can break their shift symmetry and generate the inflationary potential. These can be $\mathcal{O}(\alpha'^3)$ corrections at $(F$-${\rm term})^4$ level \cite{Ciupke:2015msa}, $\mathcal{O}(\alpha'^4 g_s^2)$ string loop effects \cite{Berg:2005ja,vonGersdorff:2005bf,Berg:2007wt,Cicoli:2008va}, or even non-perturbative corrections. In the end, their effect can be captured by exponential contributions of the form $V_0\,e^{\pm \phi/f}$, where $f$ is an effective decay constant whose value depends on the topology of the K\"ahler modulus and the nature of the symmetry breaking effect. Generically, $g_s$ and $\alpha'$ perturbative corrections are power-law in terms of the canonically unnormalised moduli $\tau$ \cite{Cicoli:2021rub}, and become exponential in terms of the canonically normalised field $\phi$ since $\tau = e^{\phi/f}$. On the other hand, non-perturbative corrections are exponential in terms of the original moduli $\tau$, and remain exponential after canonical normalisation since they are relevant just for blow-up and Wilson moduli whose canonical normalisation is power-law. Therefore, the resulting inflationary potential for the canonically normalised field $\phi$ in the plateau region which sustains inflation takes the schematic form (we ignore additional contributions which guarantee the presence of a minimum after the end of inflation):
\begin{equation}
\setlength\fboxsep{0.25cm}
\setlength\fboxrule{0.4pt}
\boxed{
V = V_0 \left(1-{\cal C}_0\,e^{-(\phi/f)^p}\right),
}
\label{Refeq}
\end{equation}
where ${\cal C}_0$ is a constant depending on the details of the model. 
Different K\"ahler moduli correspond to different values of the effective decay constant $f$. We list the main models proposed in the literature:

\subsubsection*{Fibre Inflation}

In this case the inflaton is the volume of a K3 or $T^4$ divisor and the Calabi-Yau is a fibration over a $\mathbb{P}^1$ base \cite{Cicoli:2011it}. The potential can be generated by various perturbative corrections: $\mathcal{O}(g_s^2 \alpha'^4)$ and $\mathcal{O}(g_s^4 \alpha'^4)$ string loops \cite{Cicoli:2008gp}; $\mathcal{O}(g_s^4 \alpha'^4)$ string loops and $\mathcal{O}(\alpha'^3)$ effects at $(F$-${\rm term})^4$ level \cite{Broy:2015zba}; $\mathcal{O}(g_s^2 \alpha'^4)$ string loops and $\mathcal{O}(\alpha'^3)$ corrections at $(F$-${\rm term})^4$ order \cite{Cicoli:2016chb}. 

In each case the effective decay constant is $\mathcal{O}(1)$ in Planck units. More precisely, the original model features a potential that in the inflationary region matches (\ref{Refeq}) with $f=M_{\rm Pl}/\sqrt{3}$, $p=1$ and ${\cal C}_0=4$. The total inflationary potential is of the form (setting $\Mp=1$):
\begin{equation}
\setlength\fboxsep{0.25cm}
\setlength\fboxrule{0.4pt}
\boxed{
V =V_0\left[3-4\,e^{-\frac{1}{\sqrt{3}}\phi} + e^{-\frac{4}{\sqrt{3}}\phi} + \delta \left(e^{\frac{2}{\sqrt{3}}\phi}-1\right)\right],
}
\end{equation}
where $\delta \propto g_s^4 \ll 1$. See Fig. \ref{Fig:Fiber} for a qualitative plot of this potential. This model is characterised by a trans-Planckian inflaton field range of order $\Delta\phi\simeq \mathcal{O}(5)\,M_{\rm Pl}$ \cite{Cicoli:2018tcq} which corresponds to $H_{\text{inf}}\simeq 5\times 10^{13}$ GeV. Ref. \cite{Cicoli:2020bao} performed a detailed analysis determining the values of the microscopic parameters which give the best fit to the most recent cosmological datasets, obtaining $n_s = 0.9696^{+0.0010}_{-0.0026}$ and $r = 0.00731^{+0.00026}_{-0.00072}$ at $68\%$ CL (for Planck 2018 temperature and polarisation data only) at $N_e\simeq 52$ efoldings (see \cite{Bhattacharya:2020gnk} for an analysis making making use of cobaya \cite{Torrado:2020dgo}
and modechord \cite{Handley:2015fda}). 

This construction is among the best developed string inflationary models since it features: moduli stabilisation, globally consistent Calabi-Yau embeddings with chiral matter on D7-branes \cite{Cicoli:2016xae,Cicoli:2017axo}, analysis of preheating effects \cite{Antusch:2017flz,Gu:2018akj}, study of perturbative reheating for Standard Model sectors 
on both D3-branes \cite{Cicoli:2022uqa} and D7-branes \cite{Cicoli:2018cgu}, analysis of the production of primordial black holes \cite{Cicoli:2018asa} and the associated generation of secondary gravity waves \cite{Cicoli:2022sih}, study of the behaviour of axionic isocurvature modes \cite{Cicoli:2018ccr,Cicoli:2019ulk,Cicoli:2021yhb,Cicoli:2021itv}, production of a CMB power loss at large angular scales \cite{Cicoli:2013oba, Cicoli:2014bja}.

\begin{figure}[t]
\begin{center}
\includegraphics[width=130mm,height=80mm]{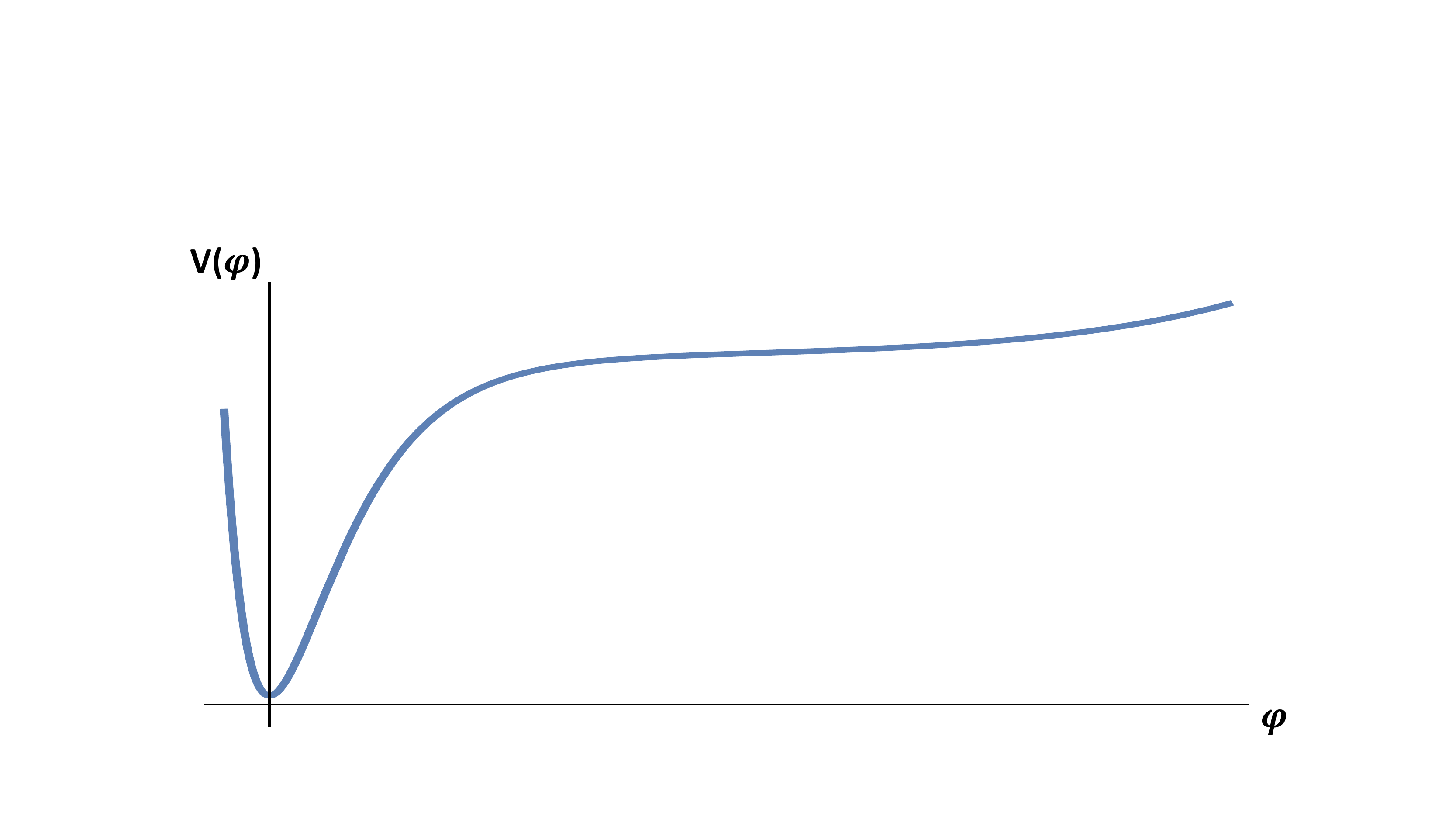} 
\caption{A typical scalar potential for a fibre modulus (with all other moduli stabilised), showing a relatively large plateau with field range of order $\Delta\phi\simeq \mathcal{O}(5)\,M_{\rm Pl}$. } 
\label{Fig:Fiber} 
\end{center}
\end{figure}

\subsubsection*{Blow-up Inflation}

In this model the inflaton is the volume of a diagonal del Pezzo divisor which resolves a point-like singularity. The potential is generated by either Euclidean D3-instantons or gaugino condensation on D7-branes, and the overall volume is kept stable during inflation via the presence of additional blow-up modes\footnote{See also \cite{Krippendorf:2009zza} for a purely supersymmetric realisation of blow-up and fibre inflation in a set-up that does not need to  include an uplifting term. } \cite{Conlon:2005jm}, or logarithmic loop corrections at $\mathcal{O}(g_s^2 \alpha'^3)$ \cite{Antoniadis:2019rkh}. This is a small-field model characterised by a sub-Planckian inflaton excursion that results in a very low Hubble scale, of order $H_{\rm inf}\simeq 5\times 10^8$ GeV and a negligibly small tensor-to-scalar ratio, $r\simeq 10^{-10}$. The potential is again of the form (\ref{Refeq}) with $p=4/3$ and an effective decay constant which scales as $f \simeq M_s \simeq M_{\rm Pl}/\sqrt{\mathcal{V}}$. In fact, the potential for the canonically normalised inflaton reads (setting again $M_{\rm Pl}=1$):
\begin{equation}
\setlength\fboxsep{0.25cm}
\setlength\fboxrule{0.4pt}
\boxed{
V = V_0\left(1-\lambda \,\phi^{4/3} \,e^{-\mu\,\phi^{4/3}}\right),
}
\end{equation}
where $\lambda\sim \mathcal{V}^{5/3}$ and $\mu\sim \mathcal{V}^{2/3}$. See Fig.~\ref{Fig:BlowUp} for a 2-dimensional plot of the trajectory of Blow-up Inflation in the ($\mathcal{V}$ mode, inflaton) plane. This potential can receive large loop or higher derivative corrections that might spoil its flatness. Thus these effects need to be absent by construction or be suppressed by small coefficients. A detailed numerical analysis of inflationary solutions in the full multi-field potential was performed in \cite{Blanco-Pillado:2009dmu}, finding a robust prediction for $n_s\sim 0.96$ for $N_e=60$. A generalisation of the model including also the axionic partner of the inflaton has been presented in \cite{Bond:2006nc}. Two-field generalisations where two of the K\"ahler moduli act as inflatons were presented in \cite{Yang:2008ns,Berglund:2009uf,Kawasaki:2010ux}. In particular, the model in \cite{Kawasaki:2010ux} turns out to be a double inflation model, producing a peak in the power spectrum at observationally interesting scales.

The post-inflationary evolution of this model has been studied in detail and determines the number of efoldings of inflation which, in turn, gives the prediction for the scalar spectral index: $n_s\simeq 1-2/N_e$ \cite{Martin:2013tda}. 
After the end of inflation preheating effects cause a violent non-perturbative production of inflaton self-quanta \cite{Barnaby:2009wr} whose perturbative decay leads to the formation of a thermal bath \cite{Cicoli:2010ha,Cicoli:2010yj,Allahverdi:2020uax}. If the Standard Model lives on D7-branes wrapping the inflaton cycle, this leads to $N_e\simeq 52$ and $n_s\simeq 0.961$. Notice however that when the visible sector is sequestered from the sources of supersymmetry breaking, this initial epoch of radiation dominance can end rather quickly due to the emergence of an early matter dominated era due to the oscillations of the overall volume mode, reducing the number of efoldings to $N_e\simeq 45$ \cite{Cicoli:2016olq}.\footnote{A reduction of the number of efoldings can also be due to a very long reheating epoch from the inflaton decay when the Standard Model lives on D7-branes wrapped around a local 4-cycle which does not intersect with the inflaton blow-up mode \cite{Cicoli:2022fzy}.} The corresponding prediction for the scalar spectral index would then be $n_s\simeq 0.955$ which is in slight tension with CMB Planck data\footnote{Constraints on K\"ahler inflation models with  WMAP7 where studied in \cite{Lee:2010tk}.}\cite{Bhattacharya:2017ysa,Bhattacharya:2017pws}. Let us finally mention that an explicit embedding of this model in a globally consistent Calabi-Yau compactification with a chiral visible sector on D3-branes at singularities has been given in \cite{Cicoli:2017shd}.

\begin{figure}[H]
\begin{center}
\includegraphics[width=120mm,height=85mm]{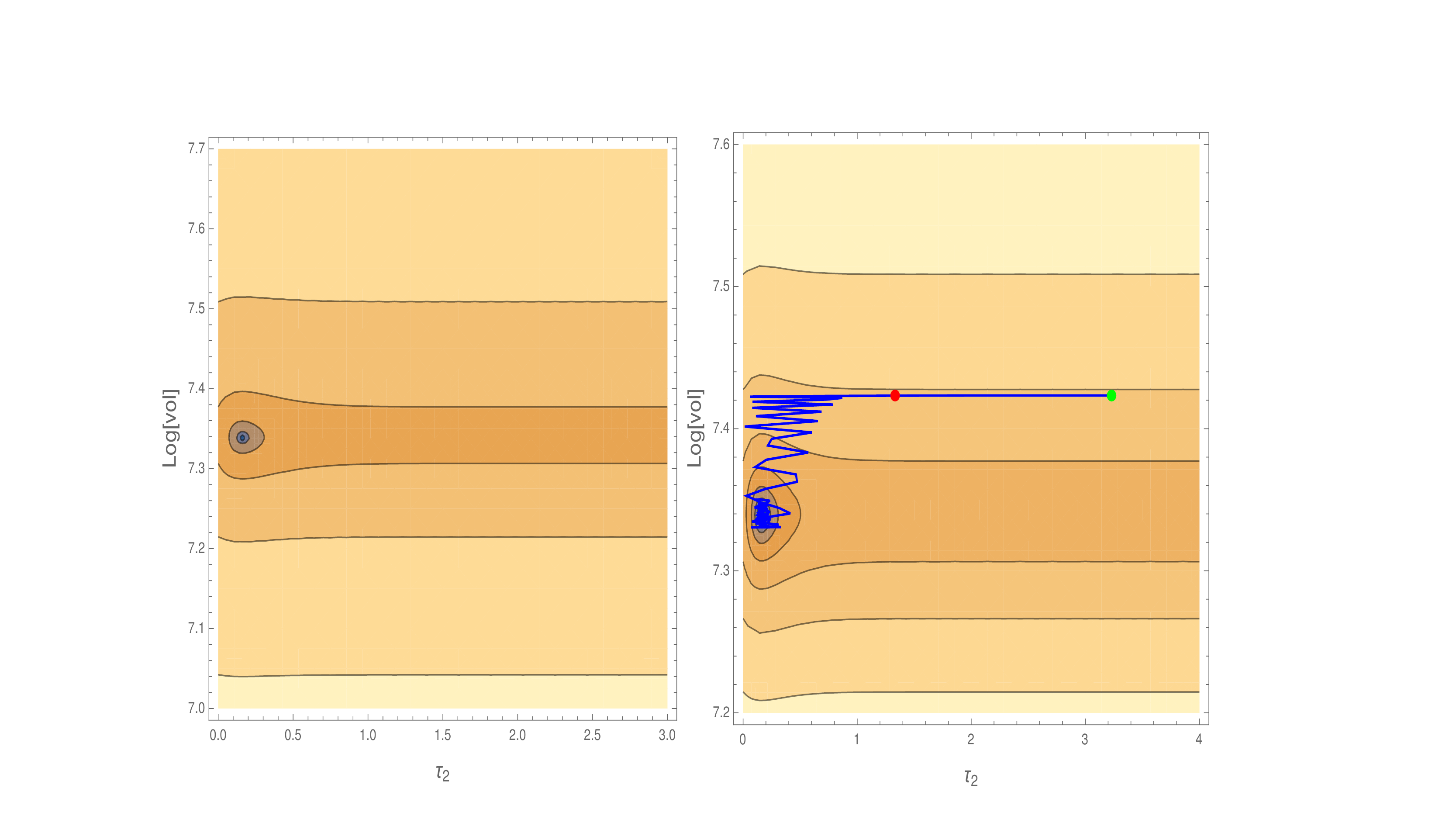} 
\caption{An example for the evolution trajectory of Blow-up Inflation in the ($\mathcal{V}$ mode, blow-up inflaton $\tau_2$) plane. The diagram shows level lines and the location of the global minimum. The trajectory of the inflationary field is illustrated with the green dot representing horizon exit and the red dot the end of inflation determined by $\epsilon=1$ (figure taken from \cite{Cicoli:2017shd}). The plot on the right hand side shows also the displacement of the $\mathcal{V}$ modulus during inflation.} \label{Fig:BlowUp} 
\end{center}
\end{figure}

\subsubsection*{Poly-instanton Inflation}

In this model \cite{Cicoli:2011ct} the inflaton is the volume of a so-called Wilson divisor, i.e.~a rigid divisor with a Wilson line \cite{Blumenhagen:2012kz}. This topological property implies that this divisor appears in the superpotential only via highly suppressed poly-instanton contributions \cite{Blumenhagen:2008ji} which make its potential naturally flat \cite{Cicoli:2011yy}. This potential is again of the form (\ref{Refeq}) with $p=1$ and an effective decay constant that scales as $f \simeq  M_p/\ln\mathcal{V}$. More precisely, the inflationary potential for this class of models reads:
\begin{equation}
\setlength\fboxsep{0.25cm}
\setlength\fboxrule{0.4pt}
\boxed{
V = V_0\left(1-\kappa \,\phi\,e^{-\kappa\phi}\right)\,,
}
\end{equation}
with $\kappa \simeq \ln\mathcal{V}$. The predictions of this model lie between the ones of Fibre and Blow-up Inflation. In fact, the tensor-to-scalar ratio is of order $r\simeq 10^{-5}$ together with $n_s\simeq 0.958$ at $N_e\simeq 52$. Note that explicit Calabi-Yau embeddings of this model with poly-instanton effects have been constructed in \cite{Blumenhagen:2012ue, Gao:2013hn,Lust:2013kt}.

\subsubsection{Trigonometric Potentials from Axionic Shift Symmetries}

Closed string axions enjoy continuous shift symmetries of the type (\ref{axionshift}) which are perturbatively exact. Non-perturbative effects break them down to discrete shift symmetries. These symmetry breaking effects scale as $V_0\,e^{- i \theta}$, and the resulting scalar potential for the canonically normalised field $\phi = f \theta$ looks like:
\begin{equation}
\setlength\fboxsep{0.25cm}
\setlength\fboxrule{0.4pt}
\boxed{
V = V_0 \left[1-\cos\left(\frac{\phi}{f}\right)\right].
}
\label{AxionPot}
\end{equation}
This potential is known to give rise to accelerated expansion only for $f>M_{\rm Pl}$ (see section \ref{subsIM}). However, this result can never be achieved with control over the effective field theory since the axion decay constant scales as $f\sim M_{\rm Pl}/\tau$, where $\tau$ denotes the volume of a divisor in string units, or more in general a combination of divisors \cite{Svrcek:2006yi,Cicoli:2012sz}. Given that a fundamental requirement to trust the low-energy action is $\tau> 1$, $f$ needs to be sub-Planckian. 

A similar line of reasoning arises from the weak gravity conjecture \cite{Arkani-Hamed:2006emk} applied to axions which results in $f S \lesssim M_{\rm Pl}$ where $S$ is the instanton action. Demanding $S>1$ to keep control over the instanton expansion, implies $f<M_{\rm Pl}$, and so no inflation driven from the potential (\ref{AxionPot}). 

Let us now briefly describe the two most promising and studied way-outs to this problem.

\subsubsection*{Natural Inflation from Alignment}

This model requires at least two axions whose decay constants, even if sub-Planckian, are aligned by fine-tuning. This generates an effective trans-Planckian decay constant for the lightest eigenmode \cite{Kim:2004rp,Choi:2014rja,Kappl:2014lra}. The scalar potential of this model looks like:
\begin{equation}
\setlength\fboxsep{0.25cm}
\setlength\fboxrule{0.4pt}
\boxed{
V = \Lambda_a  \left[1-\cos\left(a_1\frac{\phi_1}{f_1}+a_2\frac{\phi_2}{f_2}\right)\right]  
+ \Lambda_b \left[1-\cos\left(b_1\frac{\phi_1}{f_1}+b_2\frac{\phi_2}{f_2}\right)\right].
}
\end{equation}
In the limit where $a_1/a_2 \to b_1/b_2$, a linear combination of the two original axions (say $\tilde{\phi}$) is flat and its decay constant $\tilde{f}$ becomes infinite. For example, for $a_1=b_1=1$ one has:
\begin{equation}
\tilde{\phi}= \frac{f_2\phi_2-a_2 f_1\phi_1}{a_2^2 f_1^2 + f_2^2}\qquad\text{and}\qquad\tilde{f}=\frac{\sqrt{a_2^2 f_1^2 + f_2^2}}{|a_2-b_2|} \xrightarrow[a_2\to b_2]{} \infty.
\end{equation}
The predictions for the main cosmological observables depend crucially on the value of $\tilde{f}$. Taking again $N_e\simeq 52$ as benchmark point, $r\lesssim 0.1$ requires $\tilde{f}\lesssim 8$. For $\tilde{f}=8$ we obtain $n_s\simeq 0.960$ and $r\simeq 0.098$. Larger values of $\tilde{f}$ are in tension with present bounds on $r$, while smaller values of $\tilde{f}$ yield a scalar spectral index which tends to be too low. The effective field theory is also marginally under control and it is rather hard to build explicit Calabi-Yau orientifold models which are globally consistent due to the need to rely on large ranks for the condensing gauge groups. Interesting models have however been proposed using combinations of $C_0$, $C_2$ and $C_4$ axions \cite{Higaki:2014pja,Ben-Dayan:2014zsa,Long:2014dta,Gao:2014uha,Li:2014lpa,Ben-Dayan:2014lca, Abe:2014xja, Angus:2021jpr}. A survey on axion inflation in  type IIB string theory compactifications and the possibility of  enlarging the field range via alignment was performed in \cite{Long:2016jvd}. They determined an upper bound on the inflationary field range of $\Delta\phi \lesssim 0.3\Mp$.

\subsubsection*{N-flation}

This scenario involves several axions, each of them with a sub-Planckian decay constant, whose collective motion however results in an effective trans-Planckian decay constant \cite{Dimopoulos:2005ac}. The simplest version of this model features a scalar potential of the form:
\begin{equation}
\setlength\fboxsep{0.25cm}
\setlength\fboxrule{0.4pt}
\boxed{
V=\sum_{i=1}^N \Lambda \cos\left(\frac{\phi_i}{f}\right) \simeq \frac12 \left(\frac{\Lambda^2}{f}\right)^2 \sum_{i=1}^N \phi_i^2\,,
}
\end{equation}
where we have approximated the potential around the minimum. An initial displacement of each axion of order $f$ generates an overall displacement of the radial field $\rho= \sqrt{\sum_{i=1}^N \phi_i^2}$ of order $f_{\rm eff}=\sqrt{N}\,f$. Slow-roll inflation can then be easily achieved for $N\gg 1$. The predictions for the main cosmological observable are very similar to the ones of Natural Inflation: $n_s\simeq 0.96$ and $r\simeq 0.1$, with $r$ in tension with data \cite{Kim:2011jea}. Moreover, obtaining $\mathcal{O}(50)$ efoldings of inflation and matching the observed amplitude of scalar fluctuations requires a very large number of axions, $N\gtrsim 10^5$, and a relatively small Calabi-Yau volume $\mathcal{V}\simeq 10^3$. This implies that the stabilisation of the corresponding saxions at perturbative level is hardly under control \cite{Cicoli:2014sva}. In addition, all these light species can contribute to the renormalisation of the Planck mass.  Attempts to embed N-flation in string theory have been proposed in \cite{Easther:2005zr,Grimm:2007hs,Cicoli:2014sva,Grimm:2014vva}

\subsubsection{Power-law Potentials from Axionic Shift Symmetries}

Several scenarios corresponding to large field inflation have been proposed using string theoretical axions (historically, several of these were motivated by the claimed observation of primordial tensor modes by the BICEP collaboration \cite{BICEP}). In principle, these offer
the attractive possibility of both minimising fine tuning through an underlying shift symmetry while also offering the chance to compare with observations on relatively short time-scales.
 
\subsubsection*{Axion Monodromy}

Another scenario of large field inflation driven by closed string axions is Axion Monodromy (AM) where the axionic shift symmetry is broken at tree-level. In this case the resulting inflationary potential is power-law \cite{Silverstein:2008sg,McAllister:2008hb}. 

The original model is characterised by a linear potential, with two main scenarios depending on the nature of the axion: 
\begin{itemize}
\item \text{$B_2$-axion monodromy}: Inflation is driven by the $B_2$-axion $b$ whose potential originates from the reduction of the DBI action of a D5-brane wrapped around a 2-cycle $\Sigma_2$:
\begin{equation}
\setlength\fboxsep{0.25cm}
\setlength\fboxrule{0.4pt}
\boxed{
V (b) = V_0 \sqrt{{\rm Vol}(\Sigma_2) + b^2} \simeq \mu^3 \frac{\phi}{f}\,,
}
\end{equation}
where we have expanded for large values of $b$ and we have canonically normalised.

\item \text{$C_2$-axion monodromy}: Inflation is driven by the $C_2$-axion $c$ whose potential originates from the reduction of the DBI action of an NS5-brane wrapped around a 2-cycles $\Sigma_2$:
\begin{equation}
\setlength\fboxsep{0.25cm}
\setlength\fboxrule{0.4pt}
\boxed{
V (c) = V_0 \sqrt{{\rm Vol}(\Sigma_2) + g_s^2 c^2} \simeq \mu^3 \frac{\phi}{f}\,,
}
\label{VAxMon}
\end{equation}
where we have expanded for large values of $b$ and we have canonically normalised.
\end{itemize}
Note that, given that the $B_2$-axion appears in the K\"ahler potential $K = -3\ln\left(T+\bar{T}+\gamma b^2\right)$, axion monodromy models where the inflaton is $b$ are affected by the $\eta_V$-problem. This is not true for the case of the $C_2$-axion $c$, which is therefore the most promising candidate to drive axion monodromy. Note that this model is affected by backreaction effects \cite{Conlon:2011qp, Valenzuela:2016yny}, which may be addressed by using bifurcated throats \cite{Retolaza:2015sta}. 
See also \cite{Andriot:2015aza} for a no-go theorem on embedding  the original axion monodromy inflation mechanism of \cite{Silverstein:2008sg} in a concrete compactification.

In terms of observational consequences, this model yields $n_s\simeq 0.971$ and $r\simeq 0.083$ at $N_e\simeq 52$, with large tensor modes in tension with data. A potential way-out to reduce the tensor-to-scalar ratio relies on flattening effects which could make the inflationary potential shallower due to an appropriate interaction of the inflaton with heavy field which have been integrated out \cite{Dong:2010in}.    A qualitative picture of this mechanism is shown in Fig.~\ref{Fig:Flattening}. A further source of flattening, arising from fluxes, was analysed in \cite{Landete:2017amp} in the context of type IIB/F-theory flux compactifications with mobile 7-branes, which allowed for $0.14 \gtrsim r \gtrsim 0.04$.

Moreover, a cosine modulation of the scalar potential (\ref{VAxMon}) generated by non-perturbative effects, can yield oscillations in the scalar power-spectrum \cite{Flauger:2009ab,Peiris:2013opa} and resonant non-Gaussianities \cite{Hannestad:2009yx}. Depending on the size of the modulations, these can  change the background evolution allowing successful inflation for smaller field ranges in axion monodromy and sub-Planckian decay constants in a modulated axion inflation \cite{Parameswaran:2016qqq}. Modulated potentials also have interesting phenomenology such as the production of abundant primordial black holes, which can form all or part of the dark matter \cite{Ozsoy:2018flq}. More general expressions for the potential of axion monodromy are:
\begin{equation}
\setlength\fboxsep{0.25cm}
\setlength\fboxrule{0.4pt}
\boxed{
V = \mu^{4-p} \left(\frac{\phi}{f}\right)^p \,,
}
\end{equation}
where $p$ can be either $p=2/3$ \cite{Silverstein:2008sg} of $p=2$ \cite{Kaloper:2008fb,Berg:2009tg,Palti:2014kza}.

\begin{figure}[t]
\begin{center}
\includegraphics[width=140mm,height=80mm]{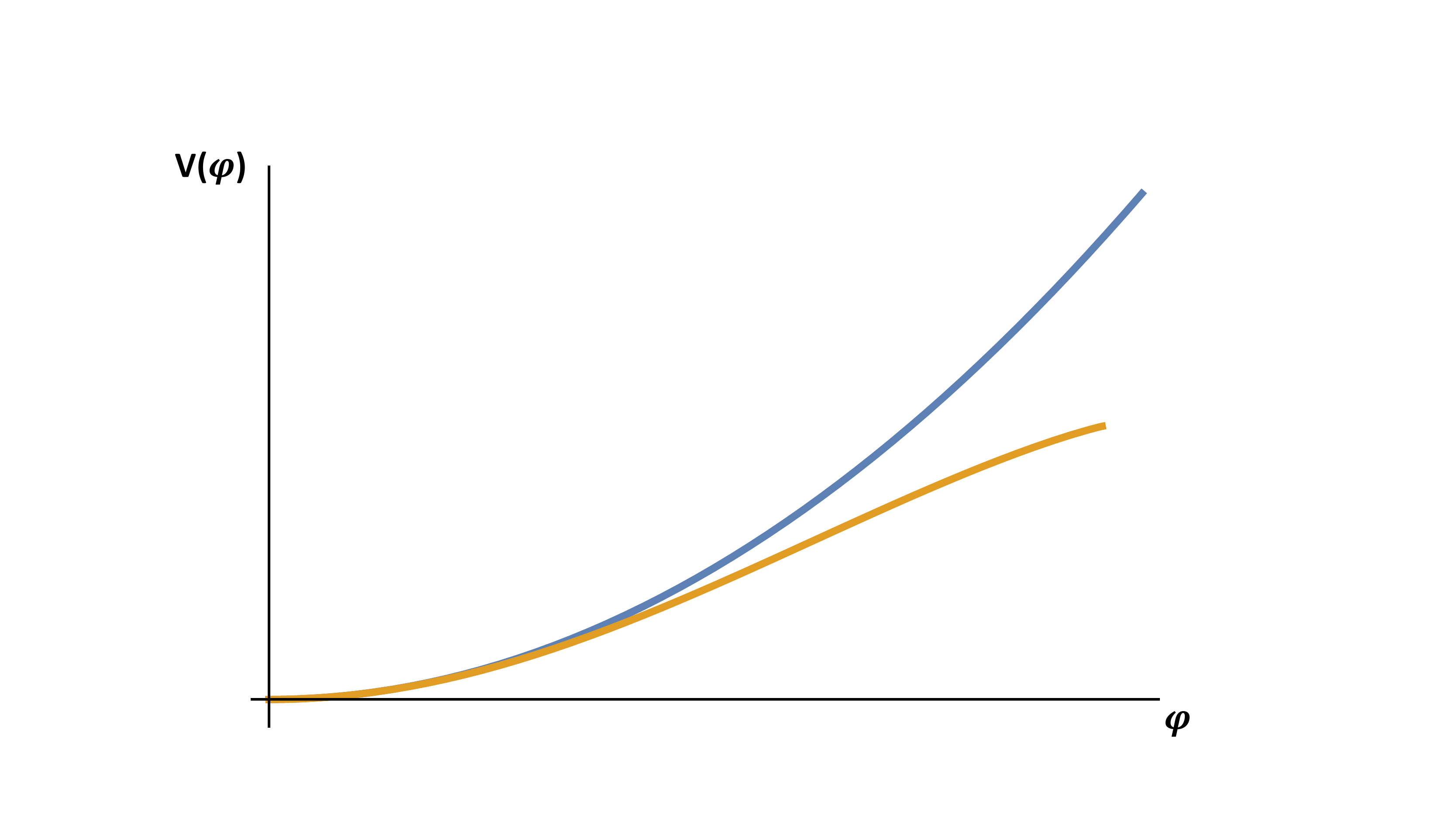} 
\caption{An example for a power law potential as appearing in Axion Monodromy or Alignment proposals. Here the simplest case of a quadratic potential is shown together with a flattening mechanism that modifies the corresponding potential after introducing couplings with heavy moduli fields.} 
\label{Fig:Flattening} 
\end{center}
\end{figure}

\subsubsection*{F-term Axion Monodromy}

An alternative axion monodromy mechanism was introduced in \cite{Marchesano:2014mla,Blumenhagen:2014gta,Hebecker:2014eua}. In this proposal, the idea is that the same background fluxes used for moduli stabilisation already generate a tree-level F-term scalar potential for the axion. The shift symmetry of the axion is broken by the background  fluxes to generate a non-oscillatory term in the potential, leading to diverse inflaton potentials, including linear large field behaviour, chaotic inflation, as well as potentials with even higher powers. Therefore, the potentials take the form 
\be
 V \sim A + B\, \theta^2 +C\, \theta \cos(\theta) +\dots \, ,
 \ee
 where $\theta$ is the axion  and $A, B$ depend on  other moduli and the fluxes, and the dots include further mixed terms, including sines and cosines multiplied by powers of $\theta$ \cite{Flauger:2014ana,Kobayashi:2015aaa,CaboBizet:2016uzv}. As mentioned before,  subleading (but sufficiently large) modulations can superimpose periodically steep cliffs and gentle plateaus onto the underlying potential. This can allow sufficient efolds of inflation with smaller field ranges lowering the tensor-to-scalar ratio, even achieving natural inflation with sub-Planckian axion decay constants \cite{Parameswaran:2016qqq,Kadota:2016jlw,Ozsoy:2018flq}. Realisations of this mechanism using different open and closed string axions can be found in \cite{Ibanez:2014kia,Franco:2014hsa,Hayashi:2014aua,Ibanez:2014swa,Garcia-Etxebarria:2014wla,Escobar:2015fda,Escobar:2015ckf,Hebecker:2015tzo,CaboBizet:2016uzv,Landete:2016cix}. 
The consistency of this new scheme with moduli stabilisation was analysed in \cite{Blumenhagen:2014nba,Hebecker:2014kva} and \cite{Blumenhagen:2015kja,Blumenhagen:2015qda,Blumenhagen:2015xpa} in the context of non-geometric flux compactifications.\footnote{See \cite{Damian:2018tlf} for a two field analysis of the models in \cite{Blumenhagen:2015kja}.} Further constraints have been analysed in \cite{Landete:2017amp,Kim:2018vgz}.

\subsubsection*{D7-Brane Inflation}

D7-deformation moduli enjoy an approximate shift symmetry at large complex structure which can be used to protect the inflaton potential. This can allow for stringy realisations of both large and small field models of inflation. In the large field case, the symmetry is broken by three-form fluxes which can develop a tunable scalar potential of the typical chaotic inflation form \cite{Hebecker:2014eua}:
\begin{equation}
\setlength\fboxsep{0.25cm}
\setlength\fboxrule{0.4pt}
\boxed{
V = \frac{m^2}{2}\,\phi^2\,.
}
\end{equation}
Such a model however predicts an amplitude of tensor modes above present observational bounds. 

On the other hand, small field models of inflation can be developed by focusing on systems of D7-branes wrapped around two different representative of the same family of 4-cycles which intersect over a 2-cycle with non-zero gauge flux. This gauge flux generates a force between the D7-branes that attracts them. This scenario is called Fluxbrane Inflation and gives rise to the following D-term inflationary potential for the relative position of the two D7-branes \cite{Hebecker:2011hk,Hebecker:2012aw,Arends:2014qca}:
\begin{equation}
\setlength\fboxsep{0.25cm}
\setlength\fboxrule{0.4pt}
\boxed{
V = V_0\left(1+\alpha\ln\left(\frac{\phi}{\phi_0}\right)\right),    
}
\end{equation}
with $\alpha\simeq g_{\rm YM}^2/(16\pi^2)\ll 1$. For our benchmark point $N_e\simeq 52$, this potential yields $n_s\simeq 1-1/N_e\simeq 0.981$ which is in slight tension with CMB data. Moreover the tensor-to-scalar ratio is $r\simeq 4\alpha/N_e \simeq 5\times 10^{-6}$ for $g_{\rm YM}\simeq 0.1$.

\subsubsection*{Wilson Line Inflation}

The T-dual picture of inflation models based on branes at angles \cite{Garcia-Bellido:2001lbk,Blumenhagen:2002ua,Gomez-Reino:2002yja} is Wilson Line Inflation, as branes intersecting at angles are dual to magnetised branes and brane deformation moduli are dual to Wilson line moduli. Ref \cite{Avgoustidis:2006zp} presented a model of inflation driven by Wilson line moduli including moduli stabilisation. The scalar potential in the small field regime takes the form:
\begin{equation}
\setlength\fboxsep{0.25cm}
\setlength\fboxrule{0.4pt}
\boxed{
V = A- \frac{B}{\phi^2}\,,
}
\end{equation}
leading to the following predictions for the two main cosmological observables at $N_e\simeq 52$: $n_s\simeq 0.971$ and $r\simeq 10^{-8}$. Note that larger values of the tensor-to-scalar ratio can be obtained in models of warped Wilson Line DBI Inflation \cite{Avgoustidis:2008zu,Kooner:2015rza}.

\subsubsection{Potentials in the Absence of Explicit Symmetries}

Several string inflationary models do not enjoy an underlying UV symmetry which protects the inflationary potential against large quantum correction. However, one can still build a working inflationary model if the system allows for enough tuning freedom to suppress all dangerous corrections to the inflationary potential. We shall now briefly discuss the main classes of string models of inflation which are not based on any symmetry.

\begin{figure}[t]
\begin{center}
\includegraphics[width=130mm,height=80mm]{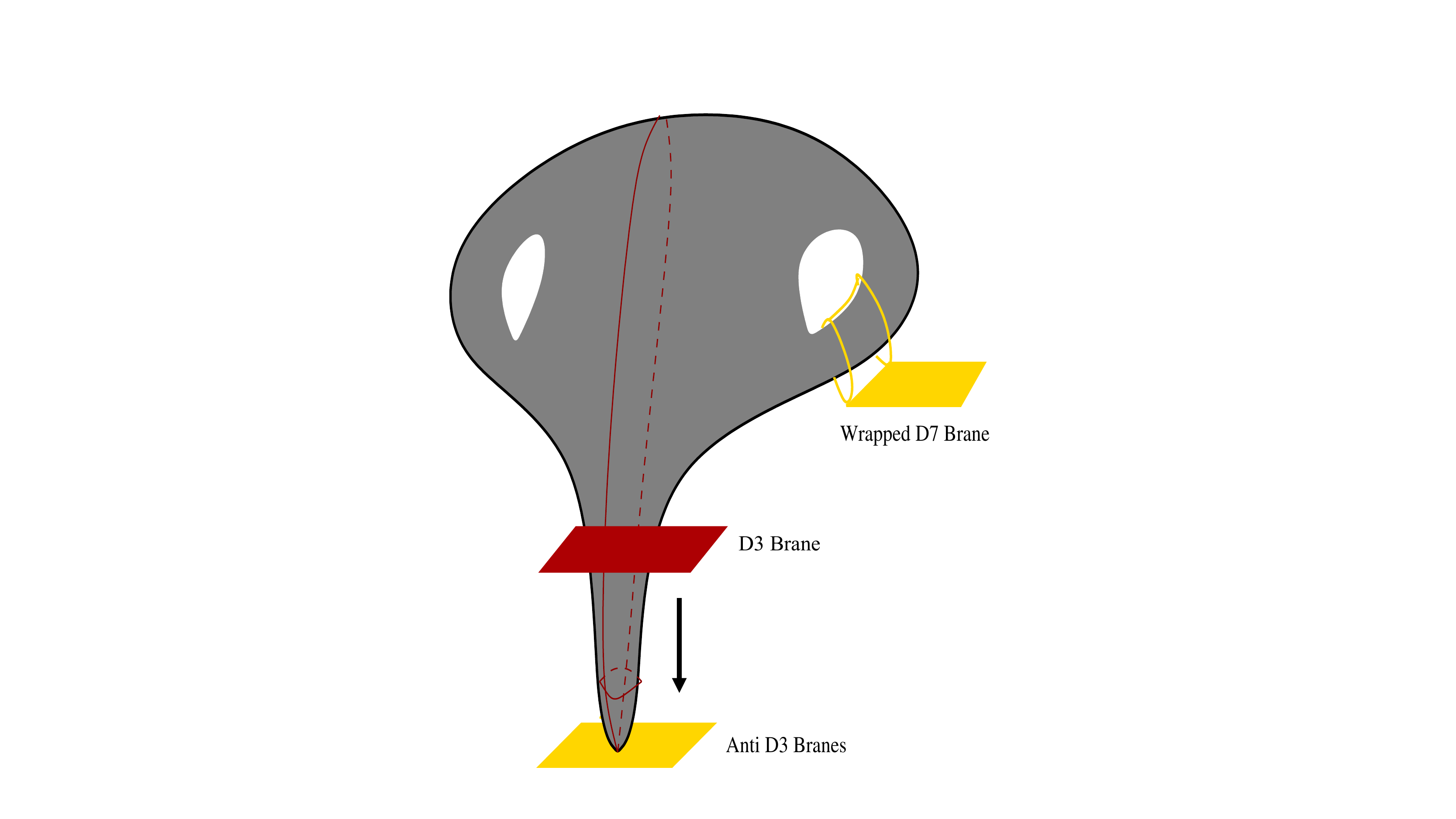} 
\caption{Warped brane-antibrane inflation. The set-up of KKLT is simply complemented with a moving D3 brane that is attracted to the anti-brane at the tip of a warped throat. In principle the warping allows for naturally small $\epsilon, \eta$ parameters. However, if, as in KKLT and LVS, moduli are fixed non-perturbatively then there are extra contributions to $\eta_V$ of order $\mathcal{O}(1)$ illustrating the single-field $\eta_V$ problem.} 
\label{Fig:KKLMMT} 
\end{center}
\end{figure}

\subsubsection*{{\rm D3}-$\overline{\rm D3}$ Inflation}

One of first attempts to realise inflation in string theory has used the separation between branes as the field which drives inflation \cite{Dvali:1998pa}. Since D-branes are BPS states their separation is a modulus with no potential. However if instead of two D-branes there are D-branes and anti D-branes there is a Coulomb attraction that gives rise to a scalar potential for the brane separation \cite{Burgess:2001fx,Dvali:2001fw}. More precisely, the model is based on a D3-$\overline{{\rm D3}}$ brane system where the inflaton is the open string modulus controlling the distance between the D3 and the $\overline{{\rm D3}}$ brane which move towards each other \cite{Burgess:2001fx,Dvali:2001fw}. For general studies of the dynamics of D-brane systems see e.g.~\cite{Burgess:2003qv, Burgess:2003mm, Burgess:2003tz, Easson:2007fz}. In \cite{Kachru:2003sx} these configurations were considered including warped compactifications in which warping can be used to get slow-roll (see Figure \ref{Fig:KKLMMT} for a cartoon of the warped brane-antibrane inflation set-up). The Coulomb potential for systems of D3-$\overline{{\rm D3}}$ branes in warped compactifications reads \cite{Kachru:2003sx}:
\begin{equation}
\setlength\fboxsep{0.25cm}
\setlength\fboxrule{0.4pt}
\boxed{
V = V_0 \left(1-\lambda\, \frac{V_0}{\phi^4}\right),
}
\label{Vbraneinfl}
\end{equation}
where the overall scale of the potential is set by the warped $\overline{{\rm D3}}$-brane tension: $V_0 \simeq T_3\,e^{4 A(r_{\rm IR})}$. Contrary to the original proposal in which the parameters of the potential do not give naturally small slow-roll parameters, 
the advantage of warping is that $V_0 \propto e^A$ can be as small as we want and both $\eta_V$ and $\epsilon$ can be as small as we want with $\epsilon \propto \eta_V^2 \ll \eta_V$. The predictions for the main cosmological observables evaluated at $N_e\simeq 52$ are $n_s\simeq 0.968$ and $r\simeq 7\times 10^{-8}$ \cite{Ma:2013xma}. 
This indicates that this is a small-field model,  in agreement with the microscopic bound on the field range for D3-brane inflation discussed in \cite{Baumann:2006cd}.  The field range can be slightly increased if  inflation is driven by the motion of a D5-brane wrapping an $2$-cycle \cite{Becker:2007ui}. In this model though, backreaction of the D5-brane needs to be checked. Furthermore, a D5-brane natural inflation model, was proposed in \cite{Kenton:2014gma}, where the D5-brane moved along an  angular direction in a warped resolved conifold \cite{PandoZayas:2000ctr,Klebanov:2007us}. The challenge in this model is to ensure that it is consistent with moduli stabilisation. 

In the original formulation of \cite{Burgess:2001fx,Dvali:2001fw} a crucial assumption of this model was that the dynamics which stabilises the K\"ahler moduli in the bulk does not alter the flatness of the potential (\ref{Vbraneinfl}) since at the time of the proposal there was no explicit scenario of moduli stabilisation. This is generically not the case since inflaton-dependent higher dimensional operators arise from both the tree-level K\"ahler potential and the prefactors of non-perturbative effects \cite{Kachru:2003sx}. A detailed study of how moduli stabilisation can be incorporated in this set-up has been reviewed in \cite{Baumann:2014nda} (crucial as including moduli stabilisation substantially changes the dynamics of the original proposal). See however a recent potential way-out exploiting perturbative stabilisation mechanism via RG effects \cite{Burgess:2022nbx} for which the warped inflation potential is at work even after moduli stabilisation.

One of the attractive points of D3-$\overline{{\rm D3}}$ inflation is the fact that the dimensionality of spacetime may play a role. In fact ref. \cite{Burgess:2001fx} proposed that in a gas of branes, D3-branes (for the type IIB case) are such that they can meet in 10 dimensions. This idea was further explored in \cite{Karch:2005yz, Durrer:2005nz}. Furthermore, the ending of brane inflation via the string tachyon is a stringy way to end inflation in a string theory realisation of hybrid inflation. The idea is that, while the branes approach each other, there is an open string state stretching between the two branes that gets lighter and lighter, at a critical distance becomes massless and, after that, becomes tachyonic. The end of inflation is the point where the tachyon reaches a minimum of its potential, which is essentially of the Mexican hat shape. This leads, in turn, to the prediction after inflation of cosmic strings (D1-branes) that may be one of the very few observable implications of 
string cosmology which is directly stringy in nature \cite{Sarangi:2002yt,Copeland:2003bj}.

D3-$\overline{{\rm D3}}$ inflation also offers the cleanest illustration of one of the more distinctive possibilities within \emph{string} inflation: the disappearance of the inflaton field at the end of inflation. The inflaton is the position modulus of the D3-brane, but after brane-antibrane annihilation this field is entirely removed from the effective field theory. This is in contrast to most field theory scenarios where, although the inflaton may decay, the field remains present in the theory.

\subsubsection*{Inflection Point Inflation}

Ref. \cite{Baumann:2007np,Baumann:2007ah,Krause:2007jk} have shown that the single-field $\eta_V$-problem of D3-$\overline{{\rm D3}}$ Inflation can be avoided since this system features enough tuning freedom to suppress dangerous Planck-suppressed higher dimensional operators that would generate corrections to the inflaton mass of order the Hubble constant during inflation.  

In the Kuperstein embedding, one has two contributions to the slow-roll $\eta_V$-parameter with opposite signs which can ensure $\eta_V\ll 1$, inducing inflation near an inflection point where the potential can be rewritten as an expansion as:
\begin{equation}
\setlength\fboxsep{0.25cm}
\setlength\fboxrule{0.4pt}
\boxed{
V \simeq V_0 \left(1+ \lambda_1 \frac{\phi}{M_{\rm Pl}} + \lambda_3 \frac{\phi^3}{M_{\rm Pl}^3} + \cdots \right) \, .
}
\end{equation}
This model is in slight tension with CMB observations since the prediction of the scalar spectral index depends on the total number of efoldings $N_e^{\rm tot}$. If $N_e^{\rm tot}\lesssim 120$, $n_s\gtrsim 1$, while $n_s\simeq 1-4/N_e$ only if $N_e^{\rm tot} \gg 2 N_e$. In this regime, $n_s\simeq 0.923$ for $N_e\simeq 52$ \cite{Linde:2007jn}, requiring a very large total number of efoldings which is however disfavoured by statistical considerations \cite{Agarwal:2011wm,McAllister:2012am}. Moreover, being a small-field model, the tensor-to-scalar ratio is negligible: $r\lesssim 10^{-6}$. For realisations
of inflection points in other settings see e.g \cite{Ben-Dayan:2013fva, Blanco-Pillado:2012uao, Maharana:2015saa}.

\subsubsection*{D3-D7 Inflation}

In cases without warping, inflation can occur in D3-D7 systems where the inflaton is a D3-brane position modulus whose D-term potential is generated by non-zero gauge fluxes on D7-branes which break supersymmetry. In order to stabilise the compact extra dimensions, the K\"ahler moduli can be frozen by gaugino condensation on D7-branes leading to extra contributions to the inflationary potential due to the mixing between K\"ahler and D3-position moduli in the K\"ahler potential. The final expression for the inflationary potential looks like \cite{Dasgupta:2002ew,Dasgupta:2004dw,Burgess:2008ir}: 
\begin{equation}
\setlength\fboxsep{0.25cm}
\setlength\fboxrule{0.4pt}
\boxed{
V = V_0 + A\,\ln\phi -\frac{m^2}{2}\, \phi^2+\frac{\lambda}{4}\,\phi^4\,.
}
\end{equation}
Defining $\alpha\equiv 2 m^2/V_0$ the prediction for the scalar spectral index is:
\begin{equation}
n_s=1-\alpha\left(1+\frac{1}{1-e^{-\alpha N_e}}\right)\simeq 1-\frac{1}{N_e} \quad\text{for}\quad \alpha \to 0\,.
\end{equation}
For $N_e\simeq 52$ and $\alpha\to 0$, this gives $n_s\simeq 0.981$ which is too blue. A value for $n_s$ compatible with Planck data ($n_s\simeq 0.966$) can be obtained for $\alpha\sim\mathcal{O}(0.01)$, leading however to a tension with current bounds on cosmic strings \cite{Haack:2008yb}. The tensor-to-scalar ratio turns out to be rather small once perturbativity and consistency with cosmic string bounds is required: $r\lesssim 10^{-6}$.

\subsubsection*{M5-brane inflation} 

In the context of M-theory compactifications \cite{Horava:1995qa,Horava:1996ma,Witten:1996mz}, an inflationary scenario was proposed, where the inflaton corresponds to the position of one or more M5-branes, moving along the interval. Inflation then comes to an  end as the M5-branes collide with and dissolve through small instanton transitions \cite{Buchbinder:2004nt,Becker:2005sg,Krause:2007jr}. The challenge in this scenario is to achieve a suitable inflationary potential, while simultaneously stabilising the geometric moduli. 

\subsubsection*{Racetrack Inflation}

Saddle point inflation can also be realised by an appropriate choice of the underlying parameters for a $C_4$-axion in the so-called Racetrack Inflation scenario which is characterised by a superpotential of the form \cite{Blanco-Pillado:2004aap,Blanco-Pillado:2006dgl}:
\begin{equation}
\setlength\fboxsep{0.25cm}
\setlength\fboxrule{0.4pt}
\boxed{
W = W_0 + A\,e^{-a T} + B\,e^{-b T}\,.
}
\end{equation}
Inflation takes place mainly along the axionic direction given by ${\rm Im}(T)$, however the saxion ${\rm Re}(T)$ also evolves. This model requires a contrived choice of underlying parameters with an inflationary trajectory which might be spoiled by quantum corrections (such as, for example, $\alpha'$ effects) that can either destabilise ${\rm Re}(T)$ or make the potential steeper along the ${\rm Im}(T)$-direction \cite{Greene:2005rn}. This is a small-field model which, in the original formulation of \cite{Blanco-Pillado:2004aap}, predicts $n_s\simeq 0.942$ and $r\lesssim 10^{-8}$ for our benchmark point $N_e\simeq 52$. 

\subsubsection*{Volume Modulus Inflation}

As we have already pointed out, the overall volume modulus of type IIB string compactifications $\mathcal{V}$ does not enjoy an approximate rescaling symmetry since the leading order $\alpha'^3$ effect that breaks the no-scale structure depends explicitly on $\mathcal{V}$. However, if enough tuning freedom is present, also the volume mode can also be used to drive inflation near an inflection point induced by at least 5 different contributions to its scalar potential. These can arise from $\alpha'$ effects, $g_s$ loops, higher F-terms, $\overline{D3}$-brane contribution and D-terms. The resulting scalar potential for the canonically normalised inflaton looks like \cite{Cicoli:2015wja} (see also \cite{Antoniadis:2020stf}):
\begin{equation}
\setlength\fboxsep{0.25cm}
\setlength\fboxrule{0.4pt}
\boxed{
V = V_0 \left(\kappa_{\alpha'}\,e^{-\sqrt{\frac{27}{2}}\phi}-\kappa_{g_s}\,e^{-\frac{10}{\sqrt{6}}\phi}-\kappa_{F^4}\,e^{-\frac{11}{\sqrt{6}}\phi}
+\kappa_{\bar{D3}}\,e^{-\sqrt{6}\phi}+\kappa_D\,e^{-\frac{8}{\sqrt{6}}\phi}\right),
}
\label{Vpot}
\end{equation}
where all the $\kappa$'s are flux-dependent $\mathcal{O}(1)$ coefficients. An example showing an inflection point around $\phi\simeq \mathcal{O}(7)$ and a late time minimum at $\phi\simeq \mathcal{O}(24)$ is shown in Fig. \ref{Fig:VolInfl}. Due to the large tuning freedom, this model can reproduce the observed value of the scalar spectral index but tensor modes are unobservable, as in any inflection point inflation model.

These models are particularly interesting to reconcile high scale inflation with low energy supersymmetry \cite{Conlon:2008cj}, as a potential solution to the tension pointed out in \cite{Kallosh:2004yh}. In fact, in any viable model the inflationary energy density should not exceed the height of the barrier towards decompactification, otherwise the volume mode would run-away to infinity:
\begin{equation}
H_{\rm inf}^2 M_{\rm Pl}^2 \lesssim V_{\rm barrier}\,.
\end{equation}
In KKLT and LVS models, this sets therefore the following bounds:
\begin{eqnarray}
V_{\rm barrier}^{({\rm KKLT})} &\sim& m_{3/2}^2 M_{\rm Pl}^2\quad\Rightarrow\quad H_{\rm inf}\lesssim m_{3/2}\,, \\
V_{\rm barrier}^{({\rm LVS})} &\sim& m_{3/2}^3 M_{\rm Pl}\quad\Rightarrow\quad H_{\rm inf} \lesssim m_{3/2} \sqrt{\frac{m_{3/2}}{M_{\rm Pl}}}\,.
\end{eqnarray}
Given that in a supergravity framework the mass of the supersymmetric partners is generically of order $m_{3/2}$, a high value of $H_{\rm inf}$ necessarily implies high scale supersymmetry. Two solutions rely on decoupling the height of the barrier from the soft terms $M_{\rm soft}$: ($i$) by tuning the scalar potential so that $V_{\rm barrier}$ becomes independent of $m_{3/2}$ as in racetrack models \cite{Kallosh:2004yh}, or ($ii$) by sequestering the sources of supersymmetry breaking from the visible sector so that $M_{\rm soft}\ll m_{3/2}$ \cite{Blumenhagen:2009gk,Aparicio:2014wxa}. Another possibility would instead be to exploit the fact that the gravitino mass is $\mathcal{V}$-dependent since $m_{3/2}^2 = e^K |W|^2$ with $K=-2\ln\mathcal{V}$. Hence, if inflation is driven by a rolling $\mathcal{V}$ mode with a potential similar to (\ref{Vpot}), the gravitino mass during inflation could be much higher than today. Being necessarily a small field model with a sub-Planckian field range, the tensor-to-scalar ratio is however very small, $r\lesssim 10^{-9}$, and so TeV-scale supersymmetry can be made compatible at most with $H_{\rm inf}\lesssim 10^{10}$ GeV, but not higher. 

\begin{figure}[t]
\begin{center}
\includegraphics[width=140mm,height=85mm]{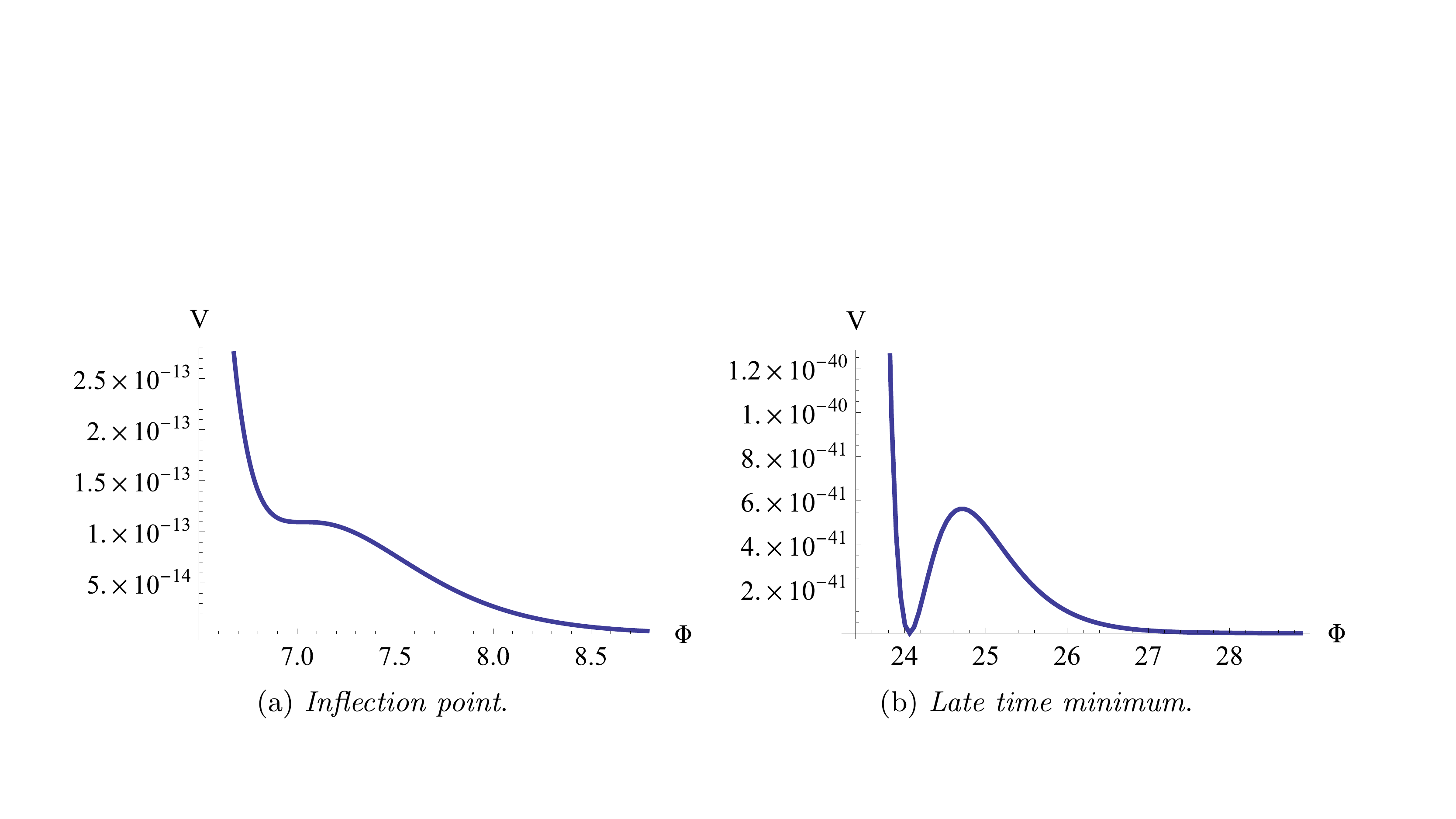} 
 \vskip -25pt
\caption{An example of  inflection point inflation for  volume modulus inflation  (figure taken from \cite{Cicoli:2015wja}). } 
\label{Fig:VolInfl} 
\end{center}
\end{figure}

\subsubsection*{DBI Inflation}

Accelerated expansion can also be achieved beyond the slow-roll approximation by exploiting the relativistic dynamics of spacetime-filling D-branes moving in a warped throat \cite{Alishahiha:2004eh,Silverstein:2003hf}. This is the so-called DBI inflation scenario characterised by non-canonical kinetic terms arising from the DBI action and a Lagrangian density of the form:
\begin{equation}
\setlength\fboxsep{0.25cm}
\setlength\fboxrule{0.4pt}
\boxed{
\mathcal{L} = - T(\phi) \left(\sqrt{1+\frac{(\partial \phi)^2}{T(\phi)}}-1\right)-V(\phi)\,,
\label{LDBI}
}
\end{equation}
where $\phi$ is the D-brane position modulus and $T(\phi)$ is the warped tension of the D-brane. The functional form of $T(\phi)$ and the potential depends on the dimensionality of the brane and determines the phenomenology of the scenario (see e.g.~\cite{Shandera:2006ax,Bean:2007hc,Ma:2013xma}). 

In the original set-up, a D3-brane moves relativistically in a warped throat. In this case, two scenarios can arise: the so-called UV-model where the brane moves from the UV end of the throat towards its tip \cite{Alishahiha:2004eh,Silverstein:2003hf}, and the IR-model where the brane moves from the tip of the throat to its UV end \cite{Chen:2004gc,Chen:2005ad}. The scalar potentials for each case read:
\begin{equation}
V = V_{\rm UV} \sim \phi^2\qquad\text{or}\qquad V = V_{\rm IR} = V_0-\frac{\beta}{2} H^2\phi^2\,,
\end{equation}
where $\beta$ is an $\mathcal{O}(1)$ positive constant. Accelerated expansion is obtained since the non-canonical form of the kinetic terms forces a speed limit on the motion of the D3-brane, under the assumption that quantum and backreaction effects do not modify the form of the Lagrangian (\ref{LDBI}). 
Besides D3-branes, DBI inflation driven by the relativistic motion of  D5- and D7-branes  in a warped throat has been studied in \cite{Kobayashi:2007hm}. Furthermore, just as in the non-relativistic case, also in this case, DBI inflation can be driven by warped Wilson lines, as studied in \cite{Avgoustidis:2008zu,Kooner:2015rza}.   

As discussed in \cite{Chen:2008hz,Baumann:2022mni},  realising DBI inflation in a consistent  string compactification is challenging. However, this scenario represents a key  example of an inflationary mechanism that relies on the symmetries of an ultraviolet theory. For this reason, it is interesting to defer the question of an explicit ultraviolet completion, and  investigate its rich phenomenology. It is worth noting that this scenario has motivated the observational search of specific cosmological signatures. Let us briefly summarise this. 

The DBI Lagrangian \eqref{LDBI} is a particular case of the more general Lagrangian corresponding to the so called $P(X, \phi)$ theories \cite{Armendariz-Picon:1999hyi,Garriga:1999vw,Chen:2006nt}, given by
\be
{\mathcal L} = P(X,\phi) \,,
\ee
where $X\equiv -\frac12(\partial\phi)^2$ and $P(X,\phi)$ is an arbitrary function of $X$ and $\phi$. Thus for DBI inflation we have 
\be
P(X,\phi) = -T(\phi)\left(\sqrt{1-\frac{2X}{T(\phi)}}-1\rp -V(\phi)\,.
\ee

The equation of motion for the Fourier modes of $\cal R$ \eqref{eq:curvatureR} is modified in DBI inflation to 
\be
\label{eq:Rdbi}
\setlength\fboxsep{0.25cm}
\setlength\fboxrule{0.4pt}
\boxed{
\cR''_k + 2\frac{z'}{z}\,\cR_k'+c^2_s\,k^2\,\cR_k=0\,,
}
\ee
where 
 \be
 c_s^2 = \frac{P_X}{P_X+2XP_{XX}} \,
  \ee
  is the sound speed, which is related to the ``Lorentz factor" as
  \be
  \gamma \equiv \lp 1-\frac{\dot\phi^2}{T(\phi)}\rp^{-1/2} = c_s^{-1} \,,
  \ee
   and thus differs from unity. The pump field is now given by  
$$z\equiv \frac{a\,\dot\varphi}{H\,c_s}\,,$$ 
 which satisfies
\be
\label{eq:pumpfdbi}
\setlength\fboxsep{0.25cm}
\setlength\fboxrule{0.4pt}
\boxed{
\frac{z'}{z} = aH\lp 1+\epsilon-\delta - s\rp\,,
}
\ee
where the new slow-roll parameter $s$ is associated to the change in the sound speed:
\be
s\equiv \frac{\dot c_s}{H\,c_s}\,.
\ee
The scalar power spectrum is  modified by $c_s$:
\be
\setlength\fboxsep{0.25cm}
\setlength\fboxrule{0.4pt}
\boxed{
\cP_\cR =  \frac{H^2}{8\pi^2\Mp^2 \,\epsilon\,c_s }\,\Bigg|_{k=aH}\,,
}
\ee
where all quantities are evaluated at {\em horizon crossing}, $k=aH$. On the other hand, the  amplitude of the tensor power spectrum is not modified with respect to the non-relativistic case \eqref{eq:PT}, since the speed of sound does not enter into the equation of motion for the tensor modes \eqref{eq:tensoreq}. 
The spectral tilt and tensor-to-scalar ratio are thus modified as
\be
\setlength\fboxsep{0.25cm}
\setlength\fboxrule{0.4pt}
\boxed{
n_s=1-2\epsilon -\eta -s\,,\qquad r = 16\,\epsilon\,c_s\,.
}
\ee
Let us note that the Lyth bound is modified by a factor of $c_sP_{\,X}$ in $P(X,  \phi)$ theories \cite{Baumann:2006cd}, which however is equal to one in the special case of DBI inflation, and therefore the correspondence between $\Delta\phi$ and $r$ is the same as in slow-roll inflation. Consequently, also in DBI inflation large tensors are not possible \cite{Baumann:2006cd,Lidsey:2007gq}. 

The new slow-roll parameter $s$ in the curvature perturbation equation \eqref{eq:Rdbi} offers an interesting new possibility  to enhance the curvature perturbation via the violation of its slow-roll condition \cite{Ozsoy:2018flq}. Indeed, during slow-roll,  $\epsilon, \delta, s \ll 1$, and we have the usual constant and decaying mode solutions to \eqref{eq:Rdbi} (see section \ref{sec:PrimF}). However, now slow-roll can be violated via large changes in $c_s$, that is $s>1$ in \eqref{eq:pumpfdbi}, and the decaying mode then becomes a growing mode, enhancing the scalar perturbations, potentially to sufficiently large values to produce abundant primordial black holes \cite{Ozsoy:2018flq}. A possible way to induce such a slow-roll violation is with features in the warp factor that is experienced by the moving  D-brane, e.g. due to duality cascades or annihilation of branes \cite{Hailu:2006uj,Miranda:2012rm}, or if the inflation-driving D-brane travels down a deformed double throat, caused by two separate stacks of D-branes or localised fluxes \cite{Franco:2005fd,Cascales:2005rj}, such that the warp factor experiences a strong decrease, inducing a large $s$.

One of the most interesting signatures of the DBI  mechanism, is the generation of equilateral non-Gaussianities  \cite{Alishahiha:2004eh}, whose amplitude is given by 
\be\label{fnlequil}
f_{\rm NL}^{\rm equil} = -\frac{35}{108}\lp\frac{1}{c_s^2} -1\rp\simeq-\frac{35}{108}\,\gamma^2\,.
\ee
The latest Planck constraints  $-73<f_{\rm NL}^{\rm equil}<21$ ($68\%$C.L.) \cite{Planck:2019kim} implies that $\gamma\lesssim 15$, which represents a strong constraint on the model. 

\subsubsection{Single-field String Inflation and Cosmological Observables}

After presenting a brief description of several examples of single-field models of string inflation, let us now summarise and compare their predictions for two main cosmological observables, the scalar spectral index $n_s$ and the tensor-to-scalar ratio $r$, evaluated at the benchmark point $N_e\simeq 52$. These predictions are listed in Tab. \ref{TabPredInfl}. 

\begin{table}[ht]
\centering 
\begin{tabular}{|c|c|c|} 
\hline
\cellcolor[gray]{0.9}{\bf String model} & \cellcolor[gray]{0.9} $\boldsymbol{n_s}$ & \cellcolor[gray]{0.9}$\boldsymbol{r}$  \\ 
\hline 
\hline
Fibre Inflation & $0.967$ & $0.007$  \\ 
\hline
Blow-up Inflation & $0.961$ & $10^{-10}$  \\
\hline
Poly-instanton Inflation & $0.958$ & $10^{-5}$  \\
\hline
Aligned Natural Inflation & $0.960$ & $0.098$  \\
\hline
$N$-Flation & $0.960$ & $0.13$ \\
\hline
Axion Monodromy & $0.971$ & $0.083$  \\  
\hline 
D7 Fluxbrane Inflation & $0.981$ & $5\times 10^{-6}$ \\
\hline
Wilson line Inflation & $0.971$ & $10^{-8}$  \\
\hline
D3-$\overline{\rm D3}$ Inflation & $0.968$ & $10^{-7}$  \\
\hline
Inflection Point Inflation & $0.923$ & $10^{-6}$  \\
\hline
D3-D7 Inflation & $0.981$ & $10^{-6}$  \\
\hline
Racetrack Inflation & $0.942$ & $10^{-8}$  \\  
\hline 
Volume Inflation & $0.965$ & $10^{-9}$  \\ 
\hline
DBI Inflation & $0.923$ & $10^{-7}$  \\
\hline 
\end{tabular}
\caption{Comparison among the predictions for the scalar spectral index and the tensor-to-scalar ratio of the main models of string inflation, evaluated as a benchmark point at $N_e\simeq 52$.}
\label{TabPredInfl}
\end{table}

Note that there is a relatively small number of inflaton candidates among all open and closed string moduli and most have been used in concrete proposals of string inflation. Note also that as per the scientific tradition, more than half of them are already in tension with the latest experimental bounds on $n_s$ and $r$. Models such as axion monodromy and fibre inflation will be further tested in the planned experiments for the next 5-10 years.

Let us stress that we focused just on a restricted list of single-field models which represent the most developed classes of {\it string} inflationary scenarios. A  broader ensemble of different models is present in the literature, even if most of them are just string-inspired, or supergravity-inspired, since they are based on ideas coming from string theory but are still lacking a solid stringy embedding or a detailed mechanism for moduli stabilisation. Just to name some of these examples, let us mention M-flation \cite{Ashoorioon:2009wa,Ashoorioon:2009sr,Ashoorioon:2011ki,Ashoorioon:2014jja}, $\alpha$-attractor models \cite{Kallosh:2013yoa,Kallosh:2013maa,Galante:2014ifa,Kallosh:2015lwa, Bhattacharya:2022akq}, sequestered inflation \cite{Kallosh:2021fvz,Kallosh:2021vcf}, axion inflation on a steep potential due to dissipation from gauge field production \cite{Anber:2009ua,Anber:2012du}, and chromonatural inflation \cite{Adshead:2012kp}.

\subsection{Multi-Field Inflation}
 
So far our discussion has been restricted to the case where the inflation proceeds along either a single direction -- such as  a closed string modulus, the radial direction of a D-brane moving in the 6-dimensional compact space, a single Wilson line, or a single combination of axions -- or with predictions that are effectively single-field, such as racetrack inflation. Indeed models are usually designed this way,  with all the non-inflaton fields sitting in their local minima as the inflaton rolls. This has the obvious advantage of simplicity, besides being effective in describing the primordial fluctuations, which are approximately scale invariant, statistically Gaussian, isotropic and homogeneous to high degree. 

Going beyond this simple picture, however, is not only well  motivated from an observational point of view, as future experiments may reveal interesting or unexpected physics (such as non-gaussianities, anisotropies, inhomogeneities), but also from a theoretical perspective. In particular, in string compactifications, moduli (spin-0) fields are ubiquitous, while spin-1 fields also enter in the process of moduli stabilisation (see Sec. \ref{sec:MS}). 

Thus a generic feature of string inflation models is that a significant number of  moduli and/or spin-1 fields, with a range of masses, may be  dynamically active during inflation. Their dynamics can thus contribute to the inflationary mechanism at the level of background or fluctuation evolution, and can leave imprints on the properties of scalar as well as tensor modes, for example by amplifying their spectra.  The resulting inflationary models can thus in general be quite complex, and they have  slowly started  to be explored in detail. We now summarise  recent results on multi-field dynamics and possible signatures in string theory (inspired) models of inflation.

\subsubsection{Multi-field D-brane inflation}

Our discussion above on D-brane inflation was restricted  to purely radial evolution of the D-brane. In general, a D$p$-brane can move in $(9-p)$ of the 6  compact dimensions. Indeed, the potential for the D-brane will in general depend on the angular directions as well as the radial one. Including these effects leads to multi-field models of D-brane inflation. 

\subsubsection*{Multi-field DBI inflation}

The phenomenology of DBI multi-field inflation has been comprehensively studied in \cite{Easson:2007dh,Huang:2007hh,Langlois:2008mn,Langlois:2008wt,Langlois:2008qf,Renaux-Petel:2009jdf,Gregory:2011cd,Emery:2013yua,Kidani:2014pka}
(for earlier work on multi-field inflationary perturbations see  \cite{Sasaki:1995aw,Gordon:2000hv,GrootNibbelink:2001qt}).\footnote{DBI inflation with $N$ D3-branes was studied in \cite{Ward:2007gs}.}
In the multi-field case, the Hubble friction can be assisted by a retarding force, e.g.~a centrifugal force in the two field case. This is the idea of {\em spinflation} \cite{Easson:2007dh}. Here, rather than rolling straight down the throat, the inflatons orbit towards the tip. Inflation then ends once
 the inflatons lose their angular momentum.  As shown in \cite{Easson:2007dh}, this prolongs inflation, although the number of e-foldings gained is very small, as angular momentum is redshifted away after only a few e-foldings.  This result is a consequence of the flat field space metric in this model and the dynamics changes if the field space has a non-trivial curvature, in which case angular motion can remain relevant throughout inflation \cite{Brown:2017osf}.

\subsubsection*{Cosmological perturbations for multi-field DBI inflation}

The cosmological perturbations in the two field DBI case are given by \cite{Langlois:2008mn,Langlois:2008wt,Langlois:2008qf}:
\begin{subequations}
 \begin{align}
     &\ddot Q_T + 3H\lp 1-s \rp \dot Q_T + \left(\frac{c_s\,k^2}{a^2} +m_T^2  \right) Q_T = 
\left(\Xi Q_N\right)^{\Large\dot{}} -\left( H\lp 2s-\epsilon\rp - \frac{c_s}{\dot \varphi}\left[\frac{f_T}{2c_sf^2}(1-c_s)^2 - V_T\right] \right) \Xi Q_N\,, \label{QT} \\
& \ddot Q_N + H\lp 3-s\rp \dot Q_N + \left(\frac{k^2}{a^2} +m_N^2 +\frac{\Xi^2}{c_s^2} \right) Q_N =- 
\frac{\dot \varphi}{\dot H} \frac{k^2}{a^2}\Xi \Psi,  \label{QN}
 \end{align}
\end{subequations}
where $Q_T=T_i Q^i $ and $Q_N=N_i Q^i$ are, respectively, the adiabatic (tangent) and entropy (normal) field fluctuations in spatially flat gauge; $\Psi$ is the Bardeen potential; $f(\phi^a) $ is the warped tension (equivalent to $1/T(\phi)$ in the single-field case \eqref{LDBI}) and $f_T$ and $f_N$ are, respectively, the tangent and normal projections of the derivative of the warped tension; and the coupling, $\Xi$, is given by
\begin{equation}\label{eq:Xi}
\Xi= -\frac{c_s(1-c_s)^2 \,f_N}{\dot \varphi f^2} -c_s(1+c_s^2)\,\Omega \,,
\end{equation}
with $\Omega$ the turning rate defined in \eqref{Omega}.
The adiabatic and entropic masses,  $m_T$ and $m_N$,  are given by 
\begin{subequations}
 \begin{align}
\label{amass}
\frac{m^2_T}{H^2}  &\equiv -\frac{3}{2}\eta- \frac{1}{4}\eta^2 -\frac{1}{2} \epsilon\eta+\frac32\eta \,s -\frac{1}{2}\frac{\dot\eta}{H} \,,\\
\label{emass}
\frac{m_N^2}{H^2} & \equiv c_s\frac{V_{NN}}{H^2} + \Mp^2\, \epsilon  \,{\mathbb R} - 
 w^2 - \frac{(2+c_s)(1-c_s)}{(1+c_s)}\frac{f_N}{H\,f}w\, \dot \varphi  - \frac{(1-c_s)^2f_{NN} }{2f^2H^2} 
 -\frac{(1-c_s)^3 f_N^2}{4(1+c_s)f^3H^2}\,.
\end{align}
\end{subequations}
Here, ${\mathbb R} $ is the field space curvature and $w$ is the dimensionless turning rate defined in Eq. \eqref{eq:w}. 
  
 DBI inflation is characterised by $c_s\ne1$.  For $c_s\ne1$,  when one or more (angular) fields are dynamical  during inflation, their quantum fluctuations lead to entropy perturbations, which propagate with the same speed of sound $c_s$ as the adiabatic mode. Moreover, assuming that the coupling, $\Xi$, between the adiabatic and entropy perturbations is very small, in the limit $c_s\ll 1$ the amplitude of the entropy perturbations is boosted relative to the adiabatic fluctuations, $Q_N\simeq Q_T/c_s$. 
In addition, the  amplitude of the bispectrum acquires a factor of $c_s^{-2}$:
\be
f_{\rm NL}^{\rm equil} \simeq -\frac{35}{108}c_s^{-2}\cos^2\Theta\,,
\ee
where $\Theta$ parameterises the transfer of entropy to adiabatic perturbations with $\Theta=0$ for no transfer and $\Theta =\pi/2$ for maximal. Thus $\Theta$ can in principle allow for larger values of $\gamma=1/c_s$ (see \eqref{fnlequil}). 

For more general situations, the dynamics (and phenomenology) of the perturbations will depend on the size of the coupling, $\Xi$, and the masses of the adiabatic and entropy perturbations. These in turn depend on the bending of the inflationary trajectory, $w$, the curvature of the field space, ${\mathbb R}$,  the warp factor, $f$, and its derivatives. 

\subsubsection*{Multi-field D3-brane inflation}

As we have mentioned above, a D$p$-brane can move in $(9-p)$ directions inside the internal 6-dimensional space. Therefore, for D3-brane inflation, besides the radial motion, there are 5 angular directions along which the brane can move. The two field  phenomenology of  warped D3-brane inflation   has been investigated  in  e.g.~\cite{Panda:2007ie,Chen:2008ada,Chen:2010qz}. On the other hand, the full 6-dimensional dynamics of warped D3-brane inflation has to be computed numerically and this has been done in \cite{Agarwal:2011wm,Dias:2012nf,McAllister:2012am,Hertog:2015zwh,Marzouk:2021tsz}. 

From these analyses it was found that in  trajectories of prolonged inflation, angular motion is relevant during the first few e-foldings of inflation, becoming  exponentially suppressed afterward. For models consistent with observable values of the spectral tilt, $n_s$ fell within the range $0.94\lesssim n_s\lesssim 1.10$, while the tensor-to-scalar ratio was in general very small $r\lesssim 10^{-12}$. Finally, cases with large non-Gaussianities were rare.

\subsubsection*{Multi-field D5-brane inflation}

A two field model of D5-brane inflation has recently been considered in \cite{Chakraborty:2019dfh}. This is a two field analysis of the model proposed in \cite{Kenton:2014gma}, where a D5-brane moves along a single angular direction of a warped resolved conifold, in a realisation  of natural inflation.  Interestingly, the two field dynamics of this model allow for two types of inflationary trajectories: an almost  geodesic one, where the masses of the two fields are lighter than $H$ realising the standard hierarchy \eqref{eq:standardH}, and a strongly non-geodesic one, where both fields are heavier than $H$, realising the `fat' hierarchy \eqref{eq:multiH} \cite{Chakraborty:2019dfh}. The potential takes the schematic form $V(r,\theta)=V(r)+W(r)\cos\theta$, while the field space metric depends non-trivially on the radial direction giving rise to a negatively curved space. An instantaneous axion decay constant can be defined as $\sqrt{g_{\theta\theta}(r)}$. The cosmological predictions of these two types of trajectories are very different and, in particular, a possible way to distinguish between them is via their predictions for non-Gaussianities. In the almost geodesic case, the cosmological predictions for the spectral tilt and tensor-to-scalar ratio are indistinguishable from the single-field natural inflation, but the non-Gaussianity parameter, $f_{\rm NL}$, is too large at ${\mathcal O}(10)$ and indeed in tension with the latest Planck constraints on the local non-Gaussianity parameter $f_{\rm NL} =-0.9\pm 5.1$ (68\%C.L.) \cite{Planck:2019kim}. In the strongly non-geodesic case, the predictions for $n_s$ and $r$ are modified as shown in figure \ref{fig:D5multi}, while $f_{\rm NL}$ is much smaller at  ${\mathcal O}({\rm few})$, and can be consistent with the Planck data.  These models suffer from the same theoretical challenges as the single-field case of \cite{Kenton:2014gma}, which we discussed above.

\begin{figure}[H]
\begin{center}
\includegraphics[width=.75\textwidth]{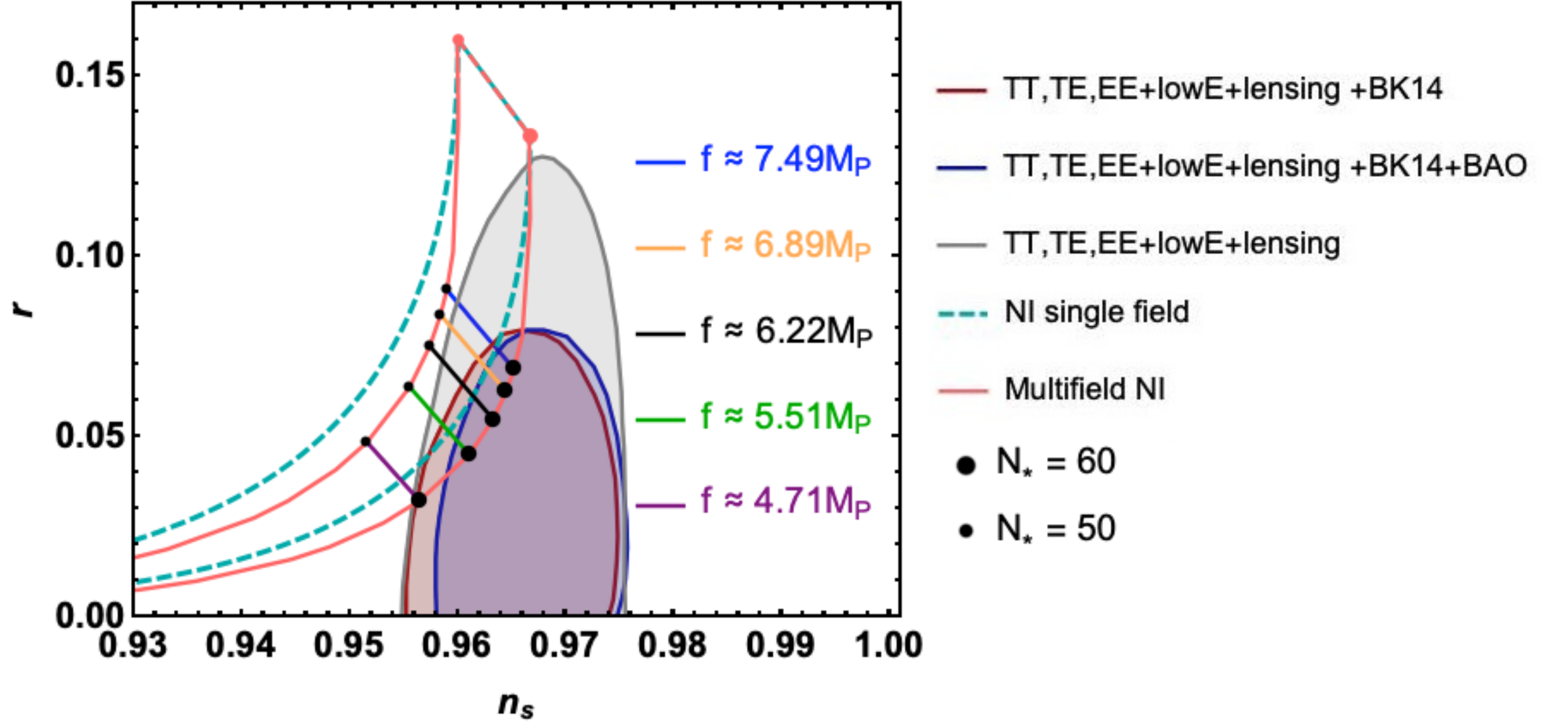} 
\caption{ The $(n_s, r)$ plane for the strongly non-geodesic D5-brane multi-field  inflation model discussed in the text. $f$ is an  instantaneous  axion decay constant, which depends on the parameters of the model. The single-field natural inflation predictions are indicated by the cyan dashed curve,
while the fat D5-brane predictions follow the continuous curve. This figure is taken from \cite{Chakraborty:2019dfh}, which can be consulted for more details.} 
\label{fig:D5multi} 
\end{center}
\end{figure}

\subsubsection{Closed string multi-field inflation and spectator fields}

Thus far, we considered D-brane multi-field inflation models, focusing on the $(9-p)$ (open string) moduli associated to the positions of the brane as it moves in the internal space. In these models, the (reasonable) assumption is that all other (closed string) moduli have been stabilised via some of the mechanisms discussed in Sec. \ref{sec:MS}. Fields with other spins are also usually taken to be  unimportant during the inflationary evolution. We now review multi-field models of inflation in the closed string sector, as well as models with spectator fields of  spin 0 and/or 1. 

\subsubsection*{Alternative sources for curvature fluctuations  }

One of the characteristics of  inflationary scenarios involving more than one fluctuating scalar degree of freedom is the presence of entropy perturbations besides the adiabatic one. 

Multi-field models then offer alternative mechanisms to generate the density perturbations after the end of inflation. Two generic mechanisms for achieving such post-inflationary isocurvature-to-adiabatic conversion are  the  {\em curvaton}  \cite{Linde:1996gt,Lyth:2001nq,Moroi:2001ct} and the {\em modulated reheating} \cite{Dvali:2003em,Kofman:2003nx,Dvali:2003ar} scenarios. In the former case, the adiabatic curvature perturbation is generated from the decay of a spectator field, the curvaton, after the end of inflation and significant non-Gaussianity may be generated. 
In the modulated reheating scenario on the other hand, perturbations  in one or more  spectator fields modulate  the decay rate of the inflation field which gives rise to reheating.  This then  converts fluctuations of the spectator fields to density fluctuations in the post-inflationary universe. The observed curvature perturbations can again be non-Gaussian.  

In \cite{Burgess:2010bz} a string theory realisation of the curvaton  scenario  was proposed  where the inflaton is a  blow-up modulus (see blow-up inflation above) and a  fiber modulus plays the role of curvaton field.   This scenario 
 allows for large local non-Gaussianity at a level ${\mathcal O}(10)$, which is  in tension with the latest Planck constraints, $f_{\rm NL} =-0.9\pm 5.1$ (68\%C.L.) \cite{Planck:2019kim}.

Similarly, in \cite{Cicoli:2012cy} a string theory realisation of the modulated reheating scenario was presented,  where inflation is driven by a fibre modulus (see fibre inflation above), while a blow-up mode acts as a modulating field. Interestingly, in this scenario, local non-Gaussianity at a level  ${\mathcal O}({\rm few})$ can be produced generically, which is in agreement with the  latest Planck constraints on $f_{\rm NL}$. 

Finally, it was argued in \cite{Dimopoulos:2006ms,Dimopoulos:2011ws} that a vector field could play the role of the curvaton field. Moreover, the contribution of a vector field to the curvature perturbation will in general be statistically anisotropic, as shown in \cite{Dimopoulos:2008yv}. 
The possibility of realising the vector curvaton scenario in D3-brane models of inflation was investigated in \cite{Dimopoulos:2011pe}. The vector curvaton was identified with the $U(1)$ gauge field that lives on the world volume of a D3-brane, while the inflaton sector could arise from the same brane or some other sector. 
The dilaton was considered as a spectator field that modulates the evolution of the vector field. Given the current hints towards statistical anisotropies in the power spectrum and bispectrum \cite{Planck:2015igc}, the vector curvaton represents an interesting possibility.

\subsubsection*{Spectator Chromonatural Inflation (SCNI)}

We have discussed how string theory compactifications involve several spin-0 closed and open string moduli, which can have interesting implications for inflation. 
Spin-1 fields are also present and may generate  interesting phenomenology. Above, we discussed a spin-1 vector field as a candidate to generate the adiabatic perturbations, potentially leading to detectable statistical anisotropies. 

The Chern-Simons (CS) coupling between a rolling axion and a non-Abelian $SU(2)$ gauge field,  $\chi\,F_{\mu\nu}^a\tilde F^{a\,\mu\nu}$, where $F=dA-g\, A\wedge A$, with $g$ the gauge coupling, was introduced in \cite{Adshead:2012kp} to `slow down' the axion via its associated friction term.\footnote{See \cite{Maleknejad:2011sq,Maleknejad:2011jw,Adshead:2012qe,Sheikh-Jabbari:2012tom} for the  related scenario of gauge-flation and its relation to chromonatural inflation.} In this way, natural inflation could  proceed even  with a sub-Planckian decay constant, $f<\Mp$ (that is, in a steeper potential). This scenario is called chromonatural inflation (CNI). Interestingly, in this scenario, fluctuations in the gauge field source tensor and scalar modes. This is because the gauge field  
$A_\mu=A_\mu^aT_a$, (with $T^a$ the $SU(2)$ generators), can have a spatially isotropic configuration given by $A^a_i=Q(t)\delta_i^a$ at the background level, while its  fluctuations give rise to tensorial perturbations from $\delta A_j^a \sim B_{ij}\delta_i^a$, sourcing the equations of motion for the tensors at the linear level:  $\Box h_{ij} =-16\pi G \Pi_{ij}$, where $\Pi_{ij}$ is the tensor part of the energy momentum tensor fluctuations $\delta T_{\mu\nu}$. 
In particular, the tensor modes experience a transient growth in one of their polarisations,  $h_{\pm} = h_{\pm}^{\rm vacuum} + h_{\pm}^{\rm source}$, 
enhancing the amplitude of the gravitational waves and  leading to the production of a chiral tensor spectrum, distinguishable from the tensor spectrum that arises in vanilla inflation scenarios. Thanks to the extra source of primordial gravitational waves (PGW), these can be produced at observable levels even with sub-Planckian field excursions, thus evading the Lyth bound \eqref{eq:Lythbd}. 
Further investigation has shown, however, that in CNI it is not possible to simultaneously satisfy the bounds on $r$ and those on the scalar spectral index $n_s$ \cite{Adshead:2013qp,Adshead:2013nka}. An interesting proposal to alleviate this problem was proposed in \cite{Dimastrogiovanni:2016fuu}. The idea is to separate the inflationary sector from the gauge-axion sector, which  acts only as a spectator, thus called spectator chromonatural inflation (SCNI) in \cite{Holland:2020jdh}.
A common feature of CNI and SCNI is the need for a large
axion-gauge CS coupling, $\lambda$:
\be
\frac{\lambda }{4f} \,\chi\,F_{\mu\nu}^A\tilde F^{A\,\mu\nu}\,,
\ee
where $F=dA-g\, A\wedge A$, with $g$ the gauge coupling and both axion and gauge field canonically normalised. 
A successful SCNI, leading to a consistent background evolution and a large enhancement of the PGWs to observable levels without excessive backreaction of the
gauge field fluctuations, requires indeed   
($i$) $\frac{\lambda}{f}\Mp\gtrsim 10^4$, leading to typical values of $\lambda\gtrsim {\mathcal O}(10^2)$, sub-Planckian decay constants $f\lesssim  {\mathcal O}(10^{-1})$, and ($ii$) small gauge couplings $g\lesssim  {\mathcal O}(10^{-2})$. Obtaining these values represents a non-trivial theoretical challenge as pointed out  in \cite{Agrawal:2018mkd,Holland:2020jdh,Bagherian:2022mau}. 

Axions and non-Abelian gauge fields are common ingredients in string theory compactifications, and thus it is natural to ask whether the SCNI model can be realised successfully in a UV complete theory and, if not, what are the main challenges. Some attempts to do this have appeared in the literature recently. In \cite{DallAgata:2018ybl}, an embedding of  CNI in ${\cal N}=1$ supergravity was presented. 

Although, as we have mentioned, CNI is observationally
unviable, one could in principle add an inflationary sector into the setup in \cite{DallAgata:2018ybl}, keeping the axion-gauge sector as spectators, to construct a model of SCNI. However, the viability and phenomenology of such a model will need to be carefully analysed when adding more fields.

Later, in \cite{McDonough:2018xzh}, an embedding of SCNI into a  string theory scenario was presented. 
The  model considers  gaugino condensation on magnetised D7-branes in type IIB CY orientifold compactifications, and the axion associated to the 2-form potential $C_2$ present
in the compactification (used in \cite{Long:2014dta,Ben-Dayan:2014lca} to realise single-field natural inflation in string theory). The inflationary sector is given by a model of blow-up modulus inflation within the LVS. 
An explicit construction was not presented, and importantly, the  backreaction of the gauge field tensor fluctuations on the background was not considered. 

In view  of these results, \cite{Holland:2020jdh} considered in detail the requirements for a successful realisation of the SCNI scenario in explicit  string theory setups. Specifically, as already  mentioned, the construction should give ($i$) a successful background evolution, ($ii$) a sufficiently large enhancement of the tensor fluctuations to  detectable levels by future experiments, and ($iii$) a controllable backreaction from the gauge field tensor fluctuations.  The inflationary sector was given by blow-up inflation in the LVS framework,\footnote{Though a realisation with fibre inflation was also discussed in \cite{Holland:2020jdh}.} which, if taken alone as a single-field inflation, would give a tensor-to-scalar ratio that is too small to be observationally relevant.  Embedding into a multi-field K\"ahler inflation model requires 3 K\"ahler moduli and, in order to realise  SCNI, one needs to moreover introduce a spectator sector. This requires a fourth K\"ahler modulus and gaugino condensation on a multiply-wrapped magnetised stack of $N$ D7-branes, whose gauge field fluctuations couple to a $C_2$ axion. The full moduli stabilisation and cosmological evolution
of the inflaton, as well as the spectator sector, was analysed in detail and thus it possible to  explicitly identify the necessary parameters and their values in order to achieve the 3 goals stated above. Specifically, these parameters are: the magnetic flux $m$ on the D7-brane stack, the degree of the condensing group $N$, and the wrapping number $n$. The typical values for these parameters to achieve a successful SCNI are $(m,N,n) \sim {\cal O} (10^4,10^5,25)$. For the fibre inflation case, these numbers are slightly improved, $(m,N,n) \sim {\cal O} (10^2,10^3,1)$, though fibre inflation allows for a much larger tensor-to-scalar ratio.

\subsubsection*{Multi-field Axion Monodromy}

Axion monodromy inflation represents an interesting scenario with a very rich phenomenology, in particular when considering multi-field extensions. Although there are no explicit string theory constructions of these, given their interesting phenomenology, we review here two  field theory models and a realisation in supergravity. 

\begin{enumerate}
\item[{\bf i.}] \underline{ Multiple axions Axion Monodromy}
Similar to N-flation, one possibility is to consider several axions whose  shift symmetry is broken at tree-level generating a leading power-law  term. This case was considered in \cite{Wenren:2014cga}, where it as shown that the spectral index is shifted red-wards from the single-field predictions. 

Later, in \cite{DAmico:2021vka,DAmico:2021fhz} the same generalisation was consider, with the distinctive feature that  inflation happens in two (or more) stages of monodromy inflation,  separated by non-inflating epochs, such as matter domination\footnote{Double inflation models have been considered in the past \cite{Silk:1986vc} to decouple
the spectrum on large and small scales. Models with multiple stages of inflation were called rollercoaster cosmology in \cite{DAmico:2020euu}.}. This model allows for a spectral index  which fits the CMB constraints, and $0.02 \lesssim r \lesssim 0.06$, which should be considered alongside the observational bound from BK-Planck 2020 $r\lesssim 0.3$   \cite{BICEP:2021xfz}. The authors also consider the possibility that the  first inflaton couples to a  $U(1)$ vector field,  producing  vectors near the end of the first stage of inflation, which in turn can source tensors during the intermediate matter epoch. These tensor modes turn out to be chiral and could be accessible to future gravitational wave experiments at different scales.

\item[{\bf ii.}] \underline{Axion-saxion Axion Monodromy}
While the models above focus on several axions, axions are usually coupled to their companion saxions, which are assumed to be stabilised in axionic inflation. 
However, the axion-saxion system can evolve cosmologically with very interesting effects. 
This was considered in \cite{Bhattacharya:2022fze}, which studied an ${\cal N}=1$ supergravity model with an axion-saxion system that evolves non-trivially, giving rise to several interesting effects: ($i$) the fields execute transient strong non-geodesic motion without the requirement of a large field space curvature\footnote{In \cite{Aragam:2021scu} it was shown that strong non-geodesic trajectories in supergravity seem to require large field space curvatures. }. This originates  from transient violations of slow-roll, $\eta\gtrsim1$, caused by the modulations  in the scalar potential. ($ii$) The non-trivial dynamics  lead to a large enhancement of the adiabatic power spectrum at small scales, providing the first concrete realisation of resonant features studied recently in the literature \cite{Fumagalli:2020nvq,Braglia:2020taf,Fumagalli:2021cel,Fumagalli:2021dtd}. These can lead to considerable production of light PBHs and a large and wide spectrum of induced GWs. The potential takes the simple form
\be\label{eq:VMAM}
\setlength\fboxsep{0.25cm}
\setlength\fboxrule{0.4pt}
\boxed{
V = \frac{M^2 }{\beta} \left(\rho^2 + \theta^2 + \frac{2\lambda}{M} \,e^{-b \rho} \left[\theta\,  \cos{(b \,\theta)}+  \rho\,  \sin{(b\, \theta)} + \frac{\lambda}{2M} \,e^{-b \rho}\right] \right)\,,}
\ee
where $\theta$ is the axion and $\rho$ the saxion, both of their  leading terms being quadratic\footnote{This potential arises from the following potentials: $\Mp^{-2}\,K = - \alpha \log[(\Phi + \bar \Phi)/\Mp -\beta S\bar S/\Mp^2]$ and
$W = S(M\Phi + i \lambda e^{- b\Phi} )$, where $S$ is a nilpotent superfield and $\Phi=\rho+ i \theta$ \cite{Ozsoy:2018flq,Bhattacharya:2022fze}.}. In Fig. \ref{fig:VMAM} we show the  
inflationary trajectory and in Fig.\ref{fig:psgw} we show the adiabatic   and GW spectra for a selection of parameters (see \cite{Bhattacharya:2022fze} for details). However, due to the large oscillations, the spectral index and tensor-to-scalar ratio at CMB scales have variations that violate current constraints. 

\begin{figure}[H]
\center{
\includegraphics[width=0.55\textwidth]{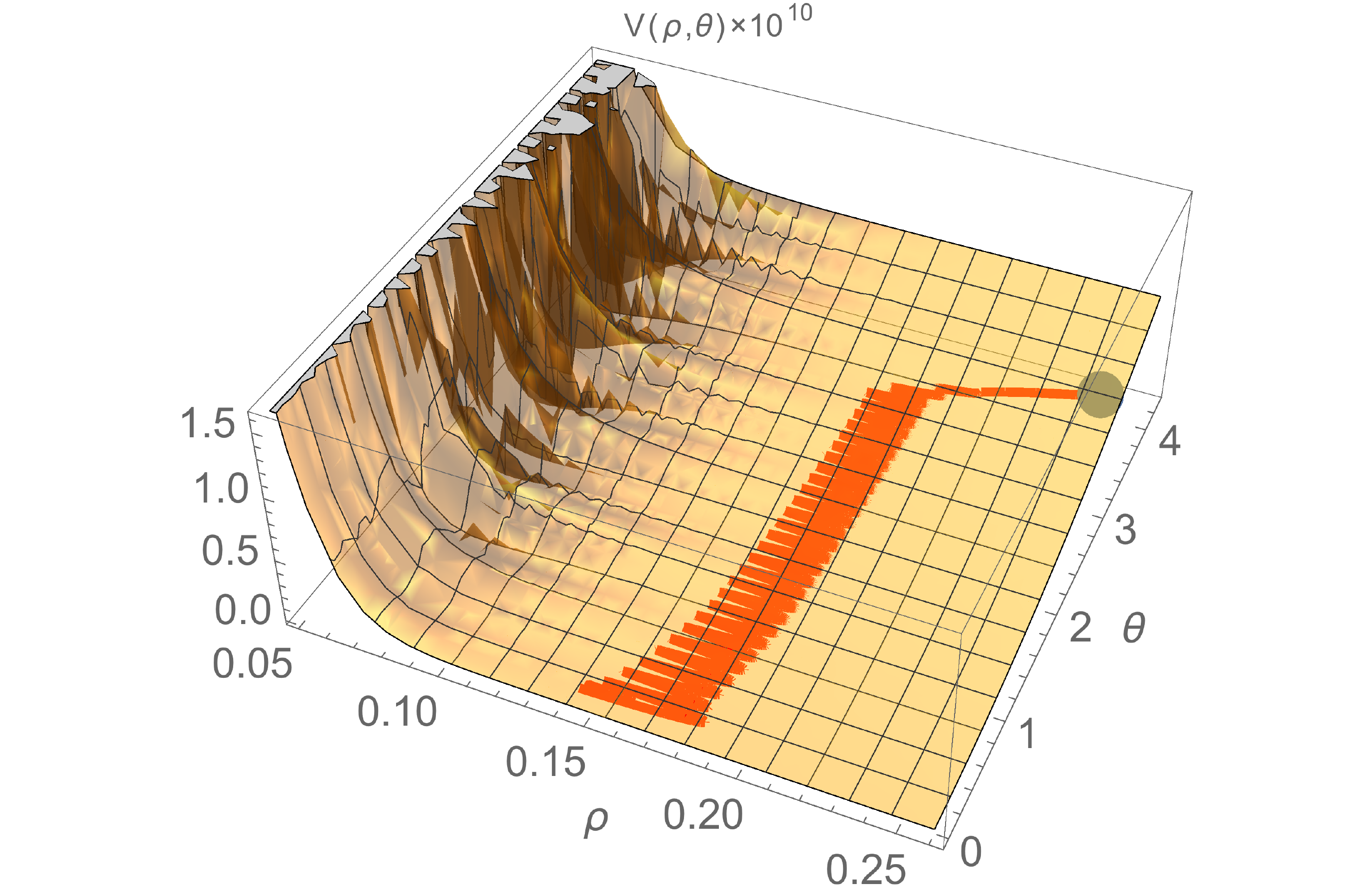}}
\caption{Inflationary trajectory of axion-saxion system as they move in the scalar potential \eqref{eq:VMAM} \cite{Bhattacharya:2022fze}.}
\label{fig:VMAM}
\end{figure}

\begin{figure}[H]
\center{
\includegraphics[width=0.49\textwidth]{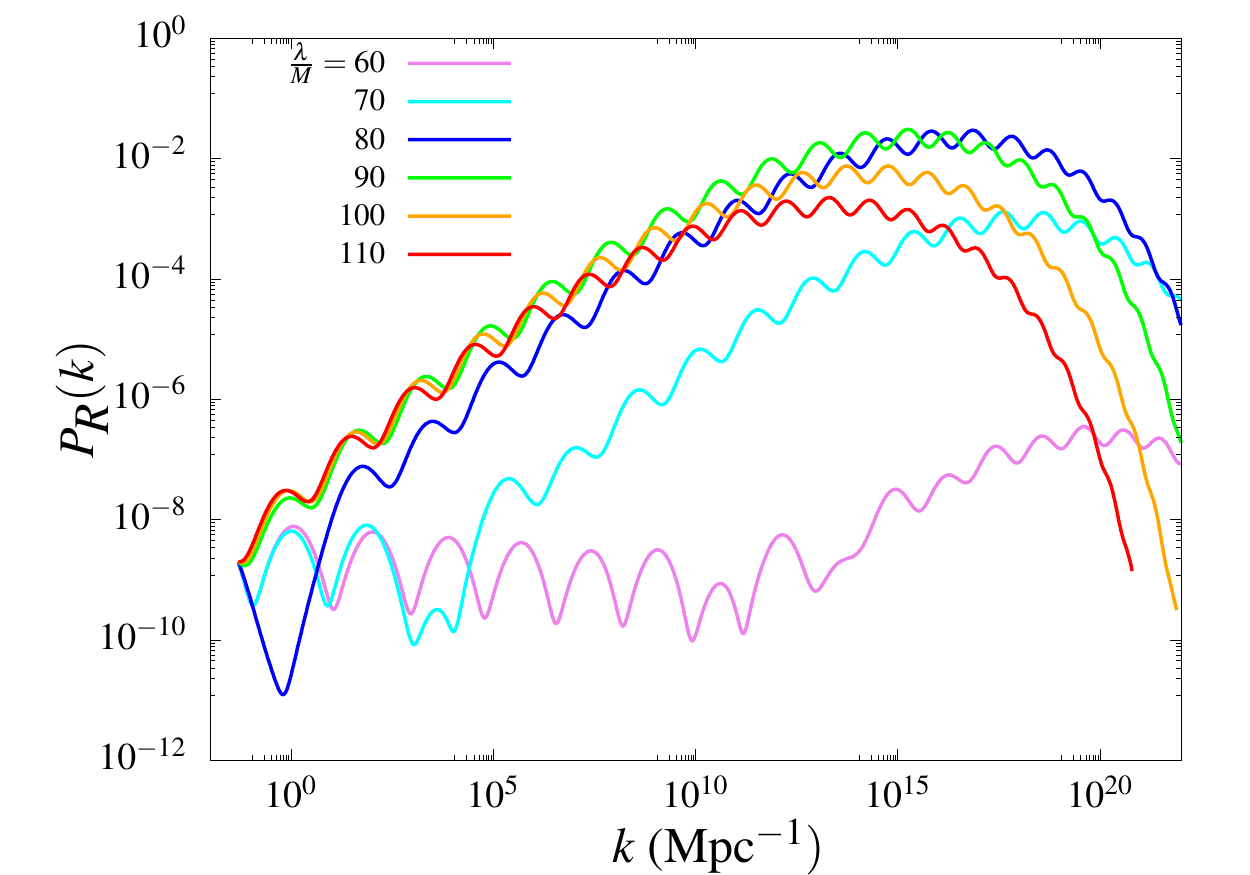}
\includegraphics[width=0.48\textwidth]{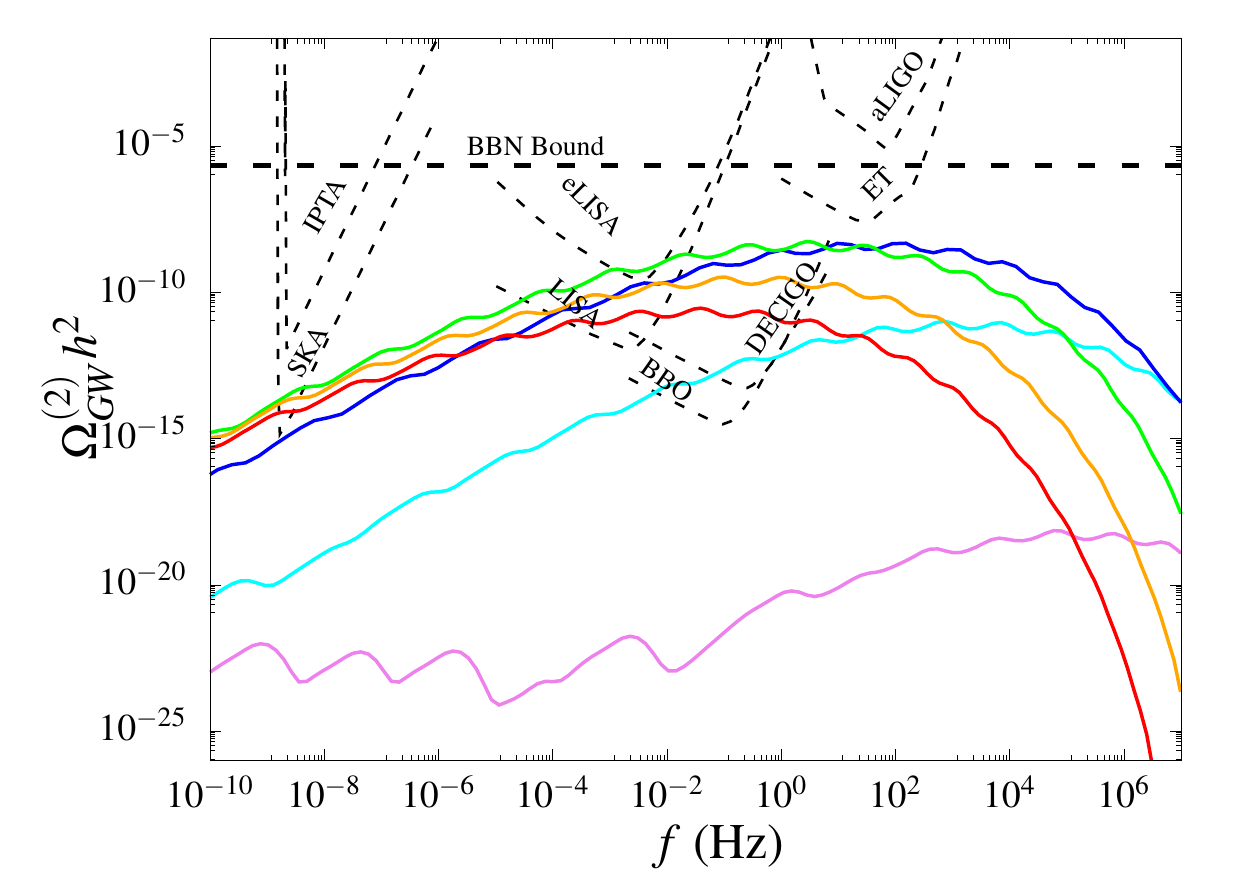}
}
\caption{Adiabatic (left) and GW (right)  spectra for a  selection  of parameters given in table 2 of \cite{Bhattacharya:2022fze}. The adiabatic power spectra are   computed using the code \texttt{PyTransport} \cite{Mulryne:2016mzv}. The left panel shows the variation of $P_{\mathcal{R}}(k)$ for different values of $\lambda /M$, with fixed $b=50$. The right panel shows the variation of $\Omega _{\rm GW}^{(2)}h^2$. 
}
\label{fig:psgw}
\end{figure}

\end{enumerate}

\newpage

\newpage

\section{Post-Inflation}
\label{reheating}

This section refers to physics that originates between the end of inflation and the start of the thermal Hot Big Bang. It begins with the universe still dominated by the vacuum energy of inflation, but now moving away from slow-roll as the inflationary epoch terminates. It ends as the universe settles into the Hot Big Bang: a radiation-dominated epoch with the energy density predominantly in relativistic thermalised Standard Model degrees of freedom. In this section, we focus on what happened between these two eras. This is not a comprehensive review of all aspects of cosmology in this epoch. Instead, we focus on those aspects where stringy physics is especially relevant. Readers interested in a more general treatment of the standard cosmology can consult e.g. \cite{Kolb:1990vq, Baumann:2022mni}, while an earlier discussion of aspects of moduli physics in this epoch is \cite{150207746} and a review of non-standard expansion histories is \cite{Allahverdi:2020bys}.
 
 While it is true that there exists a `standard' cosmological account of reheating, involving a rapid transfer of energy from inflationary degrees of freedom to relativistic Standard Model degrees of freedom, in string theory cosmologies there are no strong reasons to expect this standard account to hold. Although some aspects of the standard cosmology may be preserved in some string theory models, the standard cosmology may be modified in (at least) three ways. First, through the existence of large field displacements between the end of inflation and the final vacuum. Second, in there being no necessary relationship between the inflaton field and the field responsible for reheating. Third, through the expectation of a long moduli-dominated epoch in the universe culminating in moduli-driven reheating. These possibilities are illustrated in Fig. \ref{ReheatingCartoon}. In addition, UV complete string models may connect aspects of early universe and particle physics that otherwise appear uncorrelated.
 
\begin{figure}[ht]
\centering
\includegraphics[width = 0.9\textwidth]{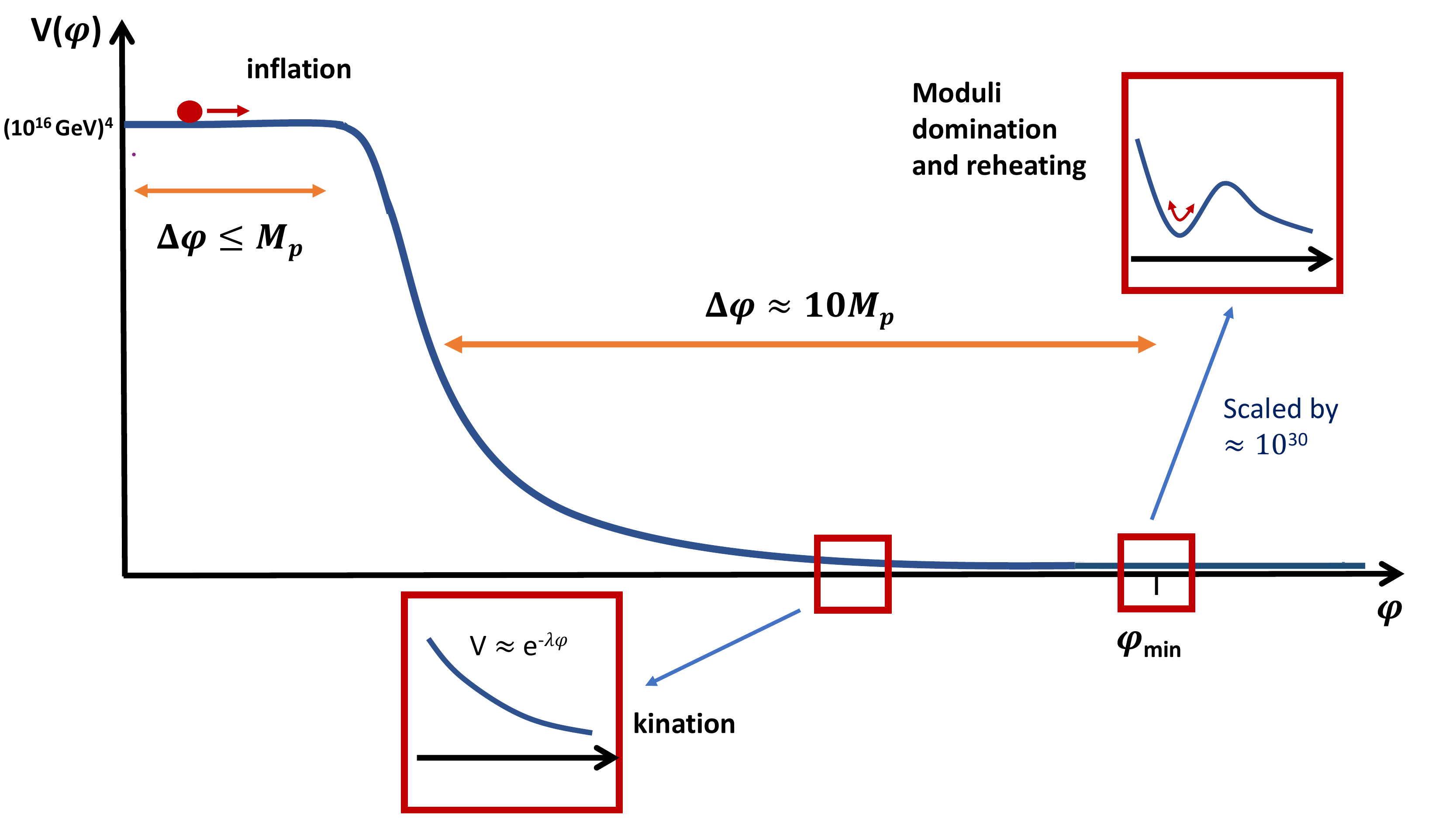} 
\caption{A cartoon of one way moduli and stringy physics can substantially modify the post-inflationary history of the universe. Following a period of inflation at relatively high energies, several epochs may occur prior to the start of the Hot Big Bang. We show here the case of a kination epoch followed by moduli domination leading to late reheating. Note the large range of scales that may arise in the scalar potential and the scalar 
field displacement. In particular, the barrier after the minimum may be $20$ (or more) orders of magnitude smaller than the energy scale during inflation ($V_{\rm barrier}\simeq 10^{-20} V_{\rm inf}$).}
\label{ReheatingCartoon}
\end{figure}

\subsection{The Standard Cosmology}

We start with a brief review of the `standard' account of post-inflationary cosmology. During the inflationary epoch, the universe was dominated by the vacuum energy density of a scalar field and the evolution of the universe was well approximated by
\begin{eqnarray}
H(t) & = & H_{\rm inf}\,,  \\
a(t) & = & a(t_{\rm init})\, e^{H_{\rm inf}(t - t_{\rm init})}\,,  \\
\rho(t) &=& V_{\rm inf} = 3 H_{\rm inf}^2 M_{\rm Pl}^2\,.
\end{eqnarray}
During inflation, the inflaton slowly rolls down a flat potential. As inflation ends, the inflaton starts rolling rapidly and oscillates about the minimum of the potential. The amplitude of the inflaton $\phi$ can be modelled by
\begin{equation}
\label{infdecay}
\setlength\fboxsep{0.25cm}
\setlength\fboxrule{0.4pt}
\boxed{
\ddot{\phi} + (3 H + \Gamma) \dot{\phi} = - m^2 \phi\,.
}
\end{equation}
Strictly, Eq. (\ref{infdecay}) is only correct if the inflaton is oscillating rapidly about its minimum.

The coherent oscillations of the inflaton are equivalent to a population of massive inflaton quanta. These quanta behave as massive particles and decay to the Standard Model
with a decay rate $\Gamma$ and with a lifetime $\tau \equiv \Gamma^{-1}$. For times $t \ll \tau$ the universe is matter-dominated and 
filled by $\phi$ particles. It evolves as
\begin{eqnarray}
a(t) & = & a(t_{\rm init}) \left( \frac{t}{t_{\rm init}} \right)^{2/3}\,, \\
\rho(t) & = & \frac{\rho_{\rm init}}{a^3}\,.
\end{eqnarray}
For times $t \gg \tau$, all $\phi$ particles have decayed, and the universe is radiation-dominated, leading to
\begin{eqnarray}
a(t) & = & a(t_{\rm init}) \left( \frac{t}{t_{\rm init}} \right)^{1/2}\,,  \\
\rho(t) & = & \frac{\rho_{\rm init}}{a^4}\,.
\end{eqnarray}
At this point, reheating has fully occurred and there is no energy density left in the inflaton field, with all energy transferred to the Standard Model. While nothing in the above specifies the lifetime $\tau$ which marks the transition into the Hot Big Bang, it is normally expected to be short, lasting for few or no e-folds in the scale factor.

In a full treatment, the process of reheating is time-dependent as decays are stochastic and governed by quantum mechanics.
However, an analytically simple approximation which captures most of the relevant physics is the \emph{instantaneous decay approximation}. In this approximation, all inflaton particles are assumed to decay at the same time $\tau$, at which $H = H_{\rm dec}$. As $H = 1/(2t)$ during radiation domination, $H_{\rm dec} = \Gamma/2$.

As the energy density of thermalised radiation is set as
\begin{eqnarray}
\rho_\gamma & = & \frac{\pi^2}{30} g_* T^4\,, \\
V & = & 3 H^2 M_{\rm Pl}^2\,,
\end{eqnarray}
we can evaluate the reheating temperature as $V_{\rm dec} = 3 H_{\rm dec}^2 M_{\rm Pl}^2$, to obtain
\begin{equation}
\setlength\fboxsep{0.25cm}
\setlength\fboxrule{0.4pt}
\boxed{
T_{\rm rh}^4 = \frac{90}{g_* \pi^2}\,H_{\rm dec}^2 M_{\rm Pl}^2\,.
}
\end{equation}
This formula defines the reheat temperature, corresponding to the last stage of significant entropy injection. The subsequent evolution of the universe is a radiation-dominated Hot Big Bang. 

The reheating temperature specifies the maximum temperature at which the formalism of a thermalised Hot Big Bang can be applied. The actual value of the reheating temperature is crucial for many questions in particle cosmology and astroparticle physics. For example, scenarios of thermal leptogenesis rely on reheating temperatures $T_{\rm rh} \gtrsim 10^{11}\,\rm{GeV}$, whereas many scenarios of dark matter production, such as thermal freeze-out, involve $T_{\rm rh} \gtrsim 1\,\rm{GeV}$. Nonetheless, the only \emph{strong} constraint is that reheating must occur prior to the time when nucleosynthesis commences ($t \sim 0.01\,\hbox{s}$, $T_{\rm BBN} \sim 1 \, \rm{MeV}$). Once the Hot Big Bang starts, the subsequent evolution of the universe is well approximated by
\begin{eqnarray}
a(t) & = & a(t_{\rm init}) \left( \frac{t}{t_{\rm init}} \right)^{\frac12}\,, \nonumber \\
\rho(t) & = & \frac{\pi^2}{30} g_* T^4\,.
\end{eqnarray}
Here $T$ denotes the thermalised temperature and $g_*$ the effective number of relativistic degrees of freedom
\begin{equation}
g_* = \sum_{\rm bosons} g_{\rm b}  + \frac78 \sum_{\rm fermions} g_{\rm f}\,.
\end{equation}

In the `standard' treatment of reheating, it is normally assumed that the inflaton will decay (and reheat the universe) via tree-level perturbative couplings to the Standard Model, e.g. 
\begin{equation}
y \phi \bar{Q} Q\,,
\label{infperd}
\end{equation}
where $y$ is the Yukawa coupling, analogous to the couplings of the Standard Model Higgs ${\cal H}$. For such perturbative decay channels, the decay rate is
\begin{equation}
\Gamma_\phi \sim \frac{y^2}{16 \pi} m_\phi\,.
\label{renormdecay}
\end{equation}
If the inflaton decays via perturbative and renormalisable channels, as in Eq. (\ref{infperd}) above, then
\begin{equation}
\setlength\fboxsep{0.25cm}
\setlength\fboxrule{0.4pt}
\boxed{
T_{\rm rh} \sim  \left( \frac{y^2}{16 \pi} m_\phi M_{\rm Pl} \right)^{\frac12}\,.
}
\end{equation}
For $m_\phi \gtrsim 1\,\hbox{TeV}$, the resulting reheat temperature is rather large; such large reheat temperatures are necessary for various Beyond-the-Standard-Model scenarios such as thermal leptogenesis.

In string cosmologies, we expect substantial modifications to this picture, both in terms of the physics of reheating and also in terms of the evolution of the fields between the end of inflation and the final minimum. The latter is chronologically earlier and so we consider it first. 
In doing so, we are compelled to consider a problem associated to a tension between inflationary and Standard Model energy scales.

\subsection{The Overshoot Problem}

As described in Sec. \ref{ModuliSection} the field space of string compactifications is described by moduli which can vary across many Planckian distances in their expectation values. There is no reason that the present day vacuum of the theory should occupy a similar location in moduli space now 
as it did during the inflationary epoch; indeed, they may be separated by many Planckian distances in field space. In that case, the theory has to move in moduli space from one location to the other. As with other similar points in this section, this is not a \emph{necessary} feature of models of string inflation; however, our focus here is on the novel behaviours that can occur in string cosmology, rather than the cases where a string cosmology replicates more standard scenarios.

Such a cosmological evolution can give rise to a dynamical problem. As detailed in Sec. \ref{sec:infla}, inflation is the leading candidate theory that can simultaneously explain both the large-scale homogeneity of the universe and the presence of small-scale inhomogeneities which have subsequently grown into galaxies and the other cosmic structures we see today. Inflation is characterised by an exponential growth in the scale factor of the universe driven by the vacuum energy of a scalar field. The rate of this growth arises from the size of the vacuum energy during inflation. While not yet known, it is naturally large. Expressed in terms of the tensor-to-scalar ratio $r$ which stage 4 CMB experiments hope to measure, it is
\begin{equation}
V_{\rm inf}^{1/4} \sim r^{1/4}\, 10^{16}\, \rm{GeV}\,.
\end{equation}
However, the quasi-de Sitter state experienced during inflation is, by definition, not the final vacuum and string theory/supergravity models of inflation typically end with rolling moduli.

During the inflationary epoch, we expect the characteristic scales to be large. In supergravity models, unless special and specific cancellations occur, the scale of the potential is $V \sim m_{3/2}^2 M_{\rm Pl}^2$. This follows from
\begin{equation}
V = e^K \left[ K^{i\bar{\jmath}} D_i W D_{\bar{\jmath}} \overline{W} - 3 \vert W \vert^2 \right]
\end{equation}
and $m_{3/2} = e^{K/2} \vert W \vert$ (in units where $M_{\rm Pl} = 1$). Based on this, for $r \gtrsim 10^{-12}$ (as holds in essentially all models of inflation) we expect the gravitino mass during inflation to satisfy $m_{3/2} \gtrsim 10^9 \, \rm{GeV}$.

The key point is that when inflation ends, the state of the universe is \emph{likely to involve rolling moduli fields, with potential energies far greater than any of those currently applicable in particle physics}. The cosmological evolution must take us from this state to the current vacuum, in the process dissipating all this early potential energy. 

The overshoot problem is the question of why, and how, this evolution ends up in the current vacuum, as opposed to simply running off to the 10-dimensional decompactification limit. In terms of the 4-dimensional effective potential, it is always the case that $V \to 0$ 
in the asymptotic limits where $g_s \to 0$ or $\mathcal{V}_s \to \infty$ at the boundary of moduli space. This can be seen in several ways, but the most intuitive is to note that since the string scale relates to the 4-dimensional Planck scale as (where $\mathcal{V}_s$ is the volume of the Calabi-Yau measured in units of $l_s^6$ in terms of the 10-dimensional string frame metric in the convention where string and Einstein frame metrics coincide in the vacuum\footnote{The relation between the metrics in string and Einstein frames in 10 dimensions is convention dependent and in general is given by  $G_{MN}^S = e^{\frac{\phi-\phi_0}{2}}G_{MN}^E$, the above choice of conventions corresponds to 
$\phi_0= \langle \phi \rangle$. Note that this differs from the convention used in the moduli stabilisation section.})
\begin{equation}
\label{convvvv}
\setlength\fboxsep{0.25cm}
\setlength\fboxrule{0.4pt}
\boxed{
M_s = \frac{g_s}{\sqrt{4\pi}}\frac{M_{\rm Pl}}{\sqrt{\mathcal{V}_s}}\,,
}
\end{equation}
then in 4-dimensional Einstein frame these asymptotic limits correspond to $M_s \to 0$. As all physical scales in string theory (including the scalar potential) depend on $M_s$, as $M_s \to 0$ it will always be the case that $V \to 0$.

If post-inflationary evolution starts with the universe having large positive vacuum energy in the deep interior of moduli space, why does it not just evolve so that it runs away all the way to the zero-potential decompactification limit? (see Figure \ref{Fig:Kination}).

As the current vacuum is a local minimum of the potential, it is surrounded by a barrier to decompactification. However, the height of this barrier is typically  $\sim m_{3/2}^2 M_{\rm Pl}^2$ (following from the general structure of the supergravity scalar potential), and many scenarios assume $m_{3/2}^{\rm now} \ll m_{3/2}^{\rm inf}$. A primary example where $m_{3/2}^{\rm now} \ll m_{3/2}^{\rm inf}$ is explicitly realised is when inflation is driven by the volume modulus close to an inflection point at small field values \cite{Conlon:2008cj, Cicoli:2015wja}. The height of the barrier is minuscule compared to the height of the original potential -- so how can the barrier trap the fields? To visualise this problem, one can imagine a rollercoaster released under gravity from a height of a hundred metres and which must stop at an endpoint where the track rises by one millimetre -- except the inflationary case involves a far greater disproportion in the relative potential heights.

The overshoot problem was first formulated by Brustein and Steinhardt \cite{Brustein:1992nk}. A key point of their analysis, which remains true today, is that stringy potentials are naturally extremely steep. To appreciate this, note that the two most important moduli are the dilaton ($S$) and volume modulus ($T$), as these incorporate $g_s$ and $\mathcal{V}$ and so control the asymptotic runaway limit to large volume and weak coupling. The K\"ahler potential is logarithmic in these moduli, typically
\begin{equation}
\setlength\fboxsep{0.25cm}
\setlength\fboxrule{0.4pt}
\boxed{
K = - \ln \left(S + \bar{S} \right) - 3 \ln \left( T + \overline{T} \right).
}
\end{equation}
Examining the metric $K_{S\bar{S}}$ or $K_{T\overline{T}}$, this implies that the \emph{canonically normalised} field is logarithmic in either the string coupling $g_s$ or the volume $\mathcal{V}$.

Any string-derived potential that is a simple power-law in the dilaton or volume -- as would arise from any perturbative weak coupling expansion in either $g_s$ or $\alpha'$, or also any potential whose natural scaling is $M_s^4$ -- will then correspond to a potential that is a runaway exponential in the canonically normalised field $\phi$, $V(\phi) \propto e^{-\lambda \phi}$. Note that while such a runaway exponential cannot itself have a minimum, it may nonetheless be a good approximation to the moduli potential over a large region of its field range, and so be a good approximation to the full potential.

Such an exponential potential
\begin{equation}
V(\phi) \propto e^{-\lambda \phi}
\end{equation}
is already rather steep. We note that in string theory `flat' exponentials do not occur; current evidence 
suggests that $\lambda > \sqrt{2}$ and accelerated expansion cannot occur (e.g. see \cite{Garg:2018zdg, Olguin-Trejo:2018zun, ValeixoBento:2020ujr, Cicoli:2021fsd, Rudelius:2022gbz, Calderon-Infante:2022nxb, Shiu:2023nph} for recent discussions).

Moreover, in many cases stringy potentials actually arise from non-perturbative terms in the superpotential: examples are gaugino condensation in racetrack models, or the non-perturbative effects in KKLT. For such models the potential in terms of the canonical field is instead a double exponential:
\begin{equation}
V(\phi) \propto e^{-\alpha e^{\beta \phi}}\,.
\end{equation}
Like a second slice of chocolate cake, this is perhaps too much of a good thing; manifestly, such double exponentials are exceptionally steep
and, plotted on a linear axis, resemble a vertical cliff -- which simply serves to re-emphasise the overshoot problem.

\subsection{The Road to a Solution: Rolling and Tracker Solutions}

The overshoot problem is a well-posed question about the immediate dynamics of post-inflationary string cosmology. It looks like a sharp and severe problem. However, it turns out there is a clear roadmap to a solution which also offers the attractive possibility of quasi-universal behaviour in the early universe (within the context of string compactifications). First, however, we describe the simplest solutions to the overshoot problem (but also ones which remove most of the interesting physics from it).

The overshoot problem exists when the scales present in inflation are substantially larger than the scales of the barrier preventing decompactification. If this is not the case, then there is clearly no overshoot problem. There are two easy ways to achieve this but both are somewhat artificial. The first is through models of extremely low scale inflation. If $V_{\rm inf} \sim \left( 1\, {\rm TeV} \right)^4$, then the inflationary potential is comparable to the weak scale and there is no overshooting problem to explain (for example, see \cite{hepph0103243, 09122324}).
However, it is difficult to embed such models into string theory as viable UV-complete examples of inflation.

The second `trivial' solution to the overshoot problem is to require that the scales of the post-inflationary vacuum are the same as that of the inflationary potential. The barrier height is then, automatically, comparable to the scale of the potential during inflation. In this case, there would be manifestly no overshoot problem. Examples of models utilising this approach are \cite{0411011, 0611183, 11124488}. The problem with this solution (and why we use the word `trivial' to refer to it) is that it ducks the question of where the weak scale (or indeed, any of the other particle physics hierarchies) actually comes from. While it is intellectually consistent to regard the weak scale as a pure accident of nature, arising from either a highly fine-tuned Standard Model or a similarly fortuitous cancellation resulting in a light gravitino mass (i.e. $m_{3/2}^2 \ll m_\phi^2$ from fine-tuning), this avoids all the deep `why' questions about the Standard Model and its energy scales. One can then reasonably  ask: why care about using string theory in the first place to understand the hierarchy problems of the Standard Model?

Another approach to mitigate the overshoot problem comes from cases where the Standard Model is 
sequestered from the sources of supersymmetry breaking, since in this case low-energy supersymmetry could be compatible 
with a large gravitino mass. However, such sequestering requires a particular realisation of the Standard Model with D3-branes at singularities and several sources of desequestering can ruin this picture e.g.~\cite{Conlon:2010ji, Berg:2012aq}. Moreover, even in the presence of sequestering, TeV-scale supersymmetry can be compatible with at most $m_{3/2}\sim 10^{11}$ GeV, which would still be 3 orders of magnitude smaller than a Hubble scale during inflation of order $H_{\rm inf}\sim 10^{14}$ GeV. Such scenarios would, therefore, likely still have some remnant overshoot problem to be solved.

A more attractive and more ambitious solution to the overshoot problem (compared to the above) comes from the ideas of kination epochs and tracker solutions. To see how these arise, we start with the equations of motion for an evolving scalar:
\begin{subequations}
\label{fgha1}
\begin{empheq}[box=\widefbox]{align}
\ddot{\phi} + 3H\dot{\phi} & =  - \frac{\partial V}{\partial \phi}\,, \\
H^2 & = \frac{1}{3 M_{\rm Pl}^2} \left( V(\phi) + \frac{\dot{\phi}^2}{2} \right).
\label{fgha2}
\end{empheq}
\end{subequations}
The Hubble friction term (proportional to $H$) implies that all sources of energy act as friction on a scalar rolling down a potential. If large enough, this braking friction will avoid the overshoot, by slowing down the scalar sufficiently so that it fails to climb over the barrier to decompactification. While there are contributions to Hubble friction from both the potential itself and any kinetic energy of the scalar field, the self-braking is not by itself enough to avoid overshoot.
\begin{center}
\begin{figure}
\includegraphics[width=14cm]{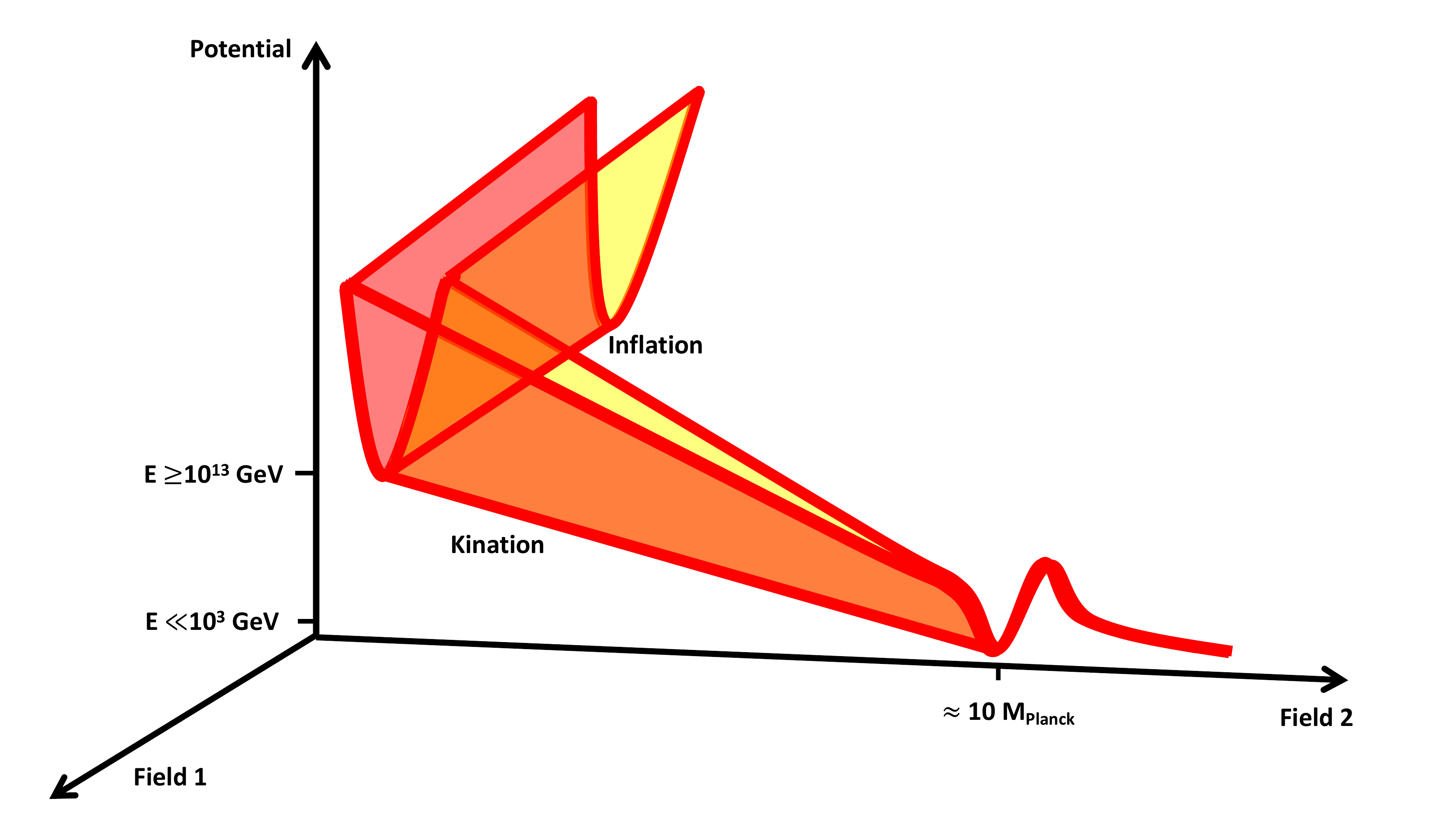}
\caption{A typical potential featuring high scale inflation and a low barrier toward decompactification which can potentially suffer from the overshoot problem. Figure by Elisa Quevedo.}
\end{figure}
\label{Fig:Kination}
\end{center}
In slow-roll inflation, the largest contribution to $H$ is the energy density in the potential. However, once a field starts rolling down an exponential (or even double-exponential) slope, the steepness of the potential implies that the kinetic energy of the inflaton rapidly becomes large. In particular, 
it dominates over the contributions of the potential density. The universe then enters a phase where its energy density is dominated by the kinetic energy of the scalar field -- a \emph{kination} phase (first named in \cite{hepph9606223}). For a recent review of kination, see \cite{211101150}.

In such a kination epoch, the scalar field equations of motion Eq.~(\ref{fgha1}) and (\ref{fgha2})
reduce to 
\begin{eqnarray}
\ddot{\phi} + 3H\dot{\phi} & =  & 0\,, \\
H^2 & = &  \frac{\dot{\phi}^2}{6 M_{\rm Pl}^2}\,.
\label{HEq}
\end{eqnarray}
This results in an equation for $\phi$:
\begin{equation}
M_{\rm Pl} \, \ddot{\phi} + \sqrt{\frac32}  \dot{\phi}^2 = 0\,,
\end{equation}
which is solved by 
\begin{equation}
\label{fieldkination}
\setlength\fboxsep{0.25cm}
\setlength\fboxrule{0.4pt}
\boxed{
\phi = \phi_{\rm init} + \sqrt{\frac23} M_{\rm Pl} \ln \left( \frac{t}{t_{\rm init}} \right).
}
\end{equation}
Here the initial condition has been set as $\phi(t_{\rm init}) = \phi_{\rm init}$. The residual integration constant has been fixed by requiring that a time coordinate of $t=0$ represents (at least formally) an initial singularity where the energy densities diverge. It is worth noting that during kination, the field moves through approximately one Planckian distance in field space each Hubble time. 
This is an interesting feature from the perspective of \emph{string} cosmology; such trans-Planckian field excursions are home territory for string theory and require a theory of quantum gravity to ensure adequate control of the effective field theory expansion over such large displacements. Any extended kination epoch, lasting for many Hubble times, will result in a field traversing a markedly trans-Planckian distance (such a kination epoch can also have an interpretation as a 10-dimensional Kasner solution \cite{Apers:2022cyl}).

The scale factor behaves as
\begin{equation}
a(t) \propto t^{1/3}\,,
\end{equation}
which follows immediately from $H^2 \equiv \left(\frac{\dot{a}(t)}{a(t)}\right)^2 = \frac{\dot{\phi}^2}{6 M_{\rm Pl}^2}$. During a kination epoch, the energy density  therefore drops off as
\begin{equation}
\setlength\fboxsep{0.25cm}
\setlength\fboxrule{0.4pt}
\boxed{
\rho_{\rm kin}(t) \propto a(t)^{-6}\,.
}
\end{equation}
By comparing with $\rho \propto a^{-3}$ or $\rho \propto a^{-4}$ (behaviours of matter and radiation domination), we see that kinetic energy dilutes much faster. This implies that during a fast-rolling kination phase, any initial sources of matter or radiation will -- over time -- catch up with the kination energy. At this point, their additional Hubble friction can effectively stop the evolution of the field (it becomes overdamped) until the energy densities of the universe have fallen sufficiently for the slope of the potential to become important again.

At this point, the evolution enters an attractor tracker solution. The `attractor' nature refers to the fact that many initial conditions converge onto the same solution. The `tracker' property refers to the fact that
fixed proportions of the energy density lie in each of potential energy, kinetic energy and radiation (or matter) \cite{Wetterich:1987fm, Ferreira:1997hj, Copeland:1997et}. The use of tracker solutions, and additional Hubble friction to avoid overshoot, goes back a long way (for example, see  \cite{hepth9805005, hepth0001112, hepph0010102, Barreiro:2005ua, hepth0408160, hepth0505098}).

We now describe the properties of the tracker solution (mostly following the analysis of \cite{Copeland:1997et}).
The existence of the tracker solution relies on the presence additional contributions to energy density that redshift slower than kinetic energy.
For a generic cosmic fluid with equation of state 
\begin{equation}
P= (\gamma - 1) \rho\,, \qquad \qquad \rho \sim a^{-3 \gamma}\,,
\end{equation}
and so a slower redshift than kinetic energy requires $\gamma < 2$. Both matter and radiation satisfy this condition. Given the high inflationary scales, there does not appear to be an obvious candidate for stable matter at the end of inflation (although, as possibilities, one could consider either primordial black holes or relatively heavy axions with $m_a < H$, which become non-relativistic shortly after the end of inflation).

Instead,  we focus on the relatively universal case of initial radiation, where $\rho_{\rm extra} = \rho_\gamma$ (note we use $\rho_\gamma$ to denote any form of radiation, not just photons). There are many good candidates for such radiation (for example, gravitons, axion-like particles or extra $U(1)$ gauge bosons).

The Friedmann equations are
\begin{eqnarray}
\dot{H} & = & - \frac{1}{2 M_{\rm Pl}^2}\Big(\rho_{\gamma}+P_{\gamma} + \dot{\phi}^2\Big) = - \frac{1}{2 M_{\rm Pl}^2} \Big( \gamma \rho_{\gamma}+ \dot{\phi}^2\Big)\,, \\
H^2 & = & \frac{1}{3 M_{\rm Pl}^2}\Big(\rho_{\gamma}+ \frac12\dot{\phi}^2+ V(\phi) \Big)\,,
\end{eqnarray}
with energy conservation set by
\begin{equation}
\dot{\rho}_{\gamma}= -3 H \big( \rho_{\gamma}+P_{\gamma} \big) = -3 H \gamma \rho_\gamma\,.
\end{equation}
The attractor nature is made manifest by transforming to the variables
\begin{equation}
x = \frac{\dot{\phi}}{M_{\rm Pl}} \frac{1}{\sqrt{6} H}\,, \quad \quad y = \sqrt{\frac{V(\phi)}{3}} \frac{1}{M_{\rm Pl} H}\,.
\end{equation}
These variables are equivalent to the fractional energy densities in kinetic and potential energy, $\Omega_k = x^2$ and $\Omega_p = y^2$, while the energy density in radiation is set by $\Omega_{\gamma}= 1- x^2 - y^2$.
The dynamical evolution then becomes
\begin{eqnarray}
  x'(N) &=& -3x - \frac{V'(\phi)}{V(\phi)}\sqrt{\frac32} y^2 + \frac32 x \big[ 2x^2 + \gamma(1-x^2-y^2) \big], \nonumber \\
  y'(N) &=& \frac{V'(\phi)}{V(\phi)}\sqrt{\frac32} xy + \frac32 y \big[ 2x^2 + \gamma(1-x^2-y^2) \big], \label{eq:xy2} \\
H'(N) &=& -\frac32 H (2x^2 + \gamma(1-x^2-y^2)), \nonumber \\
\phi'(N) &=& \sqrt{6} x, \nonumber 
\end{eqnarray}
where the time variable is $N = \ln a$.

One of the simplest possible potentials (which happily also holds for LVS and other models where the potential is power-law in either the dilaton or volume moduli) is where the potential can be approximated by a single (steep) exponential,\footnote{In the context of LVS, the single exponential is a good approximation while the field rolls down the slope. Near the minimum, other terms become important. The full potential takes the form
\begin{equation}
V(\phi)= V_0 \big( (1-\varepsilon \phi ^{3/2}) e^{- \lambda \phi} + \delta e^{- \sqrt{6} \phi }\big)\,. \nonumber
\end{equation}
Here the uplift parameter $\delta$ needs to be fine-tuned to achieve a dS vacuum at $\phi \sim \varepsilon^{-2/3}$. 
}
\begin{equation}
\setlength\fboxsep{0.25cm}
\setlength\fboxrule{0.4pt}
\boxed{
V = V_0\,\exp \left( - \lambda \frac{\phi}{M_{\rm Pl}} \right),
}
\end{equation}
so that $V'(\phi)/ V(\phi) = - \lambda/M_{\rm Pl}$. While in an LVS context, $\lambda = \sqrt{27/2}$, the precise value of $\lambda$ is not essential for the existence of the tracker solution. We regard $\phi = 0$ as corresponding to the central region of moduli space (with either $g_s \sim 1$ or $\mathcal{V} \sim 1$). The precise value of $V_0$ will depend on the details of the compactification, but for reasonable values of $W_0$ we expect it to be of order $M_{\rm Pl}^4$.
In this regime, the system \eqref{eq:xy2} is known to have a stable attractor solution where the scalar field and radiation have a fixed ratio of energy densities. The fixed point is characterised by
\begin{equation}
\setlength\fboxsep{0.25cm}
\setlength\fboxrule{0.4pt}
\boxed{
\Omega_k = x^2 = \frac{3}{2} \frac{\gamma^{2}}{\lambda^{2}}\quad \quad  \Omega_p = y^2 = \frac{3(2-\gamma) \gamma}{2 \lambda^{2}} \quad \quad  \Omega_{\gamma}= 1-x^2-y^2 = 1-\frac{3 \gamma}{\lambda^{2}}\,.
}
\end{equation}
If the attractor solution is obtained before the rolling field reaches the barrier, it will not overshoot.

The presence of such attractor solutions has several appealing features. First, it provides a natural \emph{mechanism} for solving the overshoot problem \cite{hepth9805005, hepth0001112, hepph0010102, Barreiro:2005ua, hepth0408160, hepth0505098, 07092810, Conlon:2008cj, 220700567, Alam:2022rtt} (for some alternative approaches, for which we do not have space for a full discussion, see \cite{08073216, 08104251, 10122187, 11124488}). The additional energy content provides a Hubble friction that slows down the rolling runaway field sufficiently to ensure that it does not overshoot the barrier to infinity, but ends in the desired low-energy vacuum. The extra friction provides a natural solution to the overshoot problem, and allows the true vacuum to be located with relative ease.

Second, the required \emph{ingredients} for this mechanism to work -- additional matter or radiation -- are not exotic and are generally present within string compactifications. To be more specific, some extra radiation will always be present at the end of inflation. Associated to the quasi-de-Sitter phase of inflation is a temperature,
\begin{equation}
T_{\rm dS} = \frac{H}{2 \pi}\,.
\end{equation}
The thermal bath associated to this temperature will lead to an energy density in radiation at the end of inflation given by
\begin{equation}
\label{dsenergy}
\rho_{\gamma, {\rm init}} = \frac{\pi^2}{30} g_* \left( \frac{H_{\rm inf}}{2 \pi} \right)^4\,.
\end{equation}
This is a `minimal' level of radiation, as its origin is effectively universal -- whatever the inflationary model, we expect it to be accompanied by thermal background radiation of this magnitude. Other, more model-dependent, sources of inflation include the perturbative conversion of the inflaton degrees of freedom into radiation (i.e.~the analogue of decays in a time-dependent background) or radiation from any cosmic strings that may be formed at the end of inflation, such as in models of brane inflation.

Finally, this mechanism also suggests that an extended period of kination may be a generic, or universal, feature of string cosmology in the early universe: it appears to be a common, if not universal, expectation of string cosmology that the evolution of the universe passes through a kination epoch and then a scaling solution. This would be appealing as it would simplify the search for any distinctive observational footprints from string cosmology: generic behaviour provides a clearer and cleaner target than scenarios where `anything goes'.

All that said, it is also worth enumerating the potential challenges and disadvantages associated to this solution to the overshoot problem. The first such challenge is associated to the large field displacements required. From Eq. (\ref{fieldkination}) we see that during a kination epoch, in every Hubble time the field passes through a Planckian distance in field space. A kination epoch that lasts for a significant number of Hubble times will involve a notably trans-Planckian field excursion. Furthermore, to reach the tracker solution it is likely that such an excursion is necessary; the tracker solution requires radiation to `catch up' with the kinetic energy. However, the nature of inflation is that the initial level of radiation will be highly sub-dominant. For example,  Eq. (\ref{dsenergy}) would give rise to
\begin{equation}
\Omega_{\gamma, {\rm init}} \sim \frac{H_{\rm inf}^2}{M_{\rm Pl}^2}\, .
\end{equation}
The lower the inflationary scale is, the smaller $\Omega_{\gamma, {\rm init}}$ is, and the longer the kination epoch needs to last in order for the radiation to catch up and lock onto the tracker solution.

However, such large field excursions are not without their problems. As reviewed in Sec. \ref{Sec:Swamp}, according to the Swampland Distance Conjecture, towers of states come down in mass whenever a field traverses a trans-Planckian distance $\Delta \phi > M_{\rm Pl}$ \cite{Ooguri:2018wrx},
\begin{equation}
\setlength\fboxsep{0.25cm}
\setlength\fboxrule{0.4pt}
\boxed{
M_i \propto e^{-\alpha \Delta \phi / M_{\rm Pl}}\,,
}
\end{equation}
where $\alpha$ is an $\mathcal{O}(1)$ number. Such towers of states may modify the low-energy effective field theory and lead to additional corrections to the Lagrangian, although note that the presence of a descending tower of states is not necessarily a problem: some examples, such as Kaluza-Klein states in the asymptotic large volume regime, are relatively well understood and do not necessarily lead to control issues.

The second, related, challenge is that most possible origins of initial seed radiation generate rather small amounts of radiation, at a level proportional to $\left(H_{\rm inf}/M_{\rm Pl}\right)^2$. For models with low inflationary scales, it would not be possible for fields to reach the tracker solution before they overshoot the barrier and are on their way towards decompactification. In practice, the tracker solution may be hard to locate and in this case, another mechanism to avoid overshoot would have to be found. When the mechanism works, it works well -- but the tracker requires rather specific conditions on the available field displacement in order for it to work.

\subsection{Moduli Domination}

However it happens, overshoot must be avoided in any cosmology that can describe our universe. Sooner or later, the theory must approach our own vacuum and gradually settle into it. It is here that another characteristic aspect of string cosmologies come into play: the expectation of a period of moduli domination. Why?

In expanding universes, matter and radiation redshift as
\begin{subequations}
 \begin{align}\label{matterrad}
    \rho_{\rm mat} & \propto  a(t)^{-3}\,, \\
     \rho_{\rm rad} & \propto  a(t)^{-4}\,,
 \end{align}
\end{subequations}
and so matter wins out over radiation.
Although both familiar and basic, Eq.~(\ref{matterrad}) implicitly contains one of the most important elements of \emph{string} cosmology. As discussed in Section \ref{ModuliSection}, moduli originate from higher-dimensional modes of the graviton and interact through gravitationally suppressed couplings. On dimensional grounds, the decay rates of such moduli are set as
\begin{equation}
\setlength\fboxsep{0.25cm}
\setlength\fboxrule{0.4pt}
\boxed{
\Gamma_\phi = \frac{\lambda}{16 \pi} \frac{m_\phi^3}{M_{\rm Pl}^2}\,,
}
\end{equation}
where $\lambda$ is a dimensionless $\mathcal{O}(1)$ constant, whereas particles with renormalisable perturbative decays have decay rates given by Eq. (\ref{renormdecay}). Compared to these, the lifetimes of the scalar moduli are enhanced by a factor of $\left(M_{\rm Pl}/m_\phi\right)^2$. Indeed, as the Planck scale is the silverback mountain gorilla of energy scales in physics, moduli also outlive other particles with non-renormalisable interactions suppressed by (merely) the GUT scale.

When heavy particles decay, their decay products are normally relativistic. With radiation redshifting as $\rho_{\rm rad} \sim a^{-4}$ and matter
redshifting as $\rho_{\rm mat} \sim a^{-3}$, the relativistic products from any `early' decays rapidly grow sub-dominant to any surviving 
matter present.
With the evolution of cosmic time, a universe crowded with particles inevitably becomes dominated by the longest-living, latest-decaying matter. As gravity is, both empirically and theoretically, the weakest force, this implies that it is a generic expectation of string compactifications that the universe will go through a stage where its energy density is dominated by the mass-energy of moduli particles, for which all interactions are non-renormalisable and suppressed by the Planck scale. 

This era of \emph{moduli domination} is one of the most generic and distinctive expectations of string cosmology, and it is one of the most notable ways in which string cosmology differs quite substantially from many field theory approaches to inflation where reheating is assumed to be driven by fields with couplings that are either renormalisable or, at least, suppressed by scales far lower than the Planck scale (see Fig.~\ref{fig:AlternativeHistory}). 

\begin{figure}[ht]
    \centering
    \includegraphics[width = 0.9\textwidth]{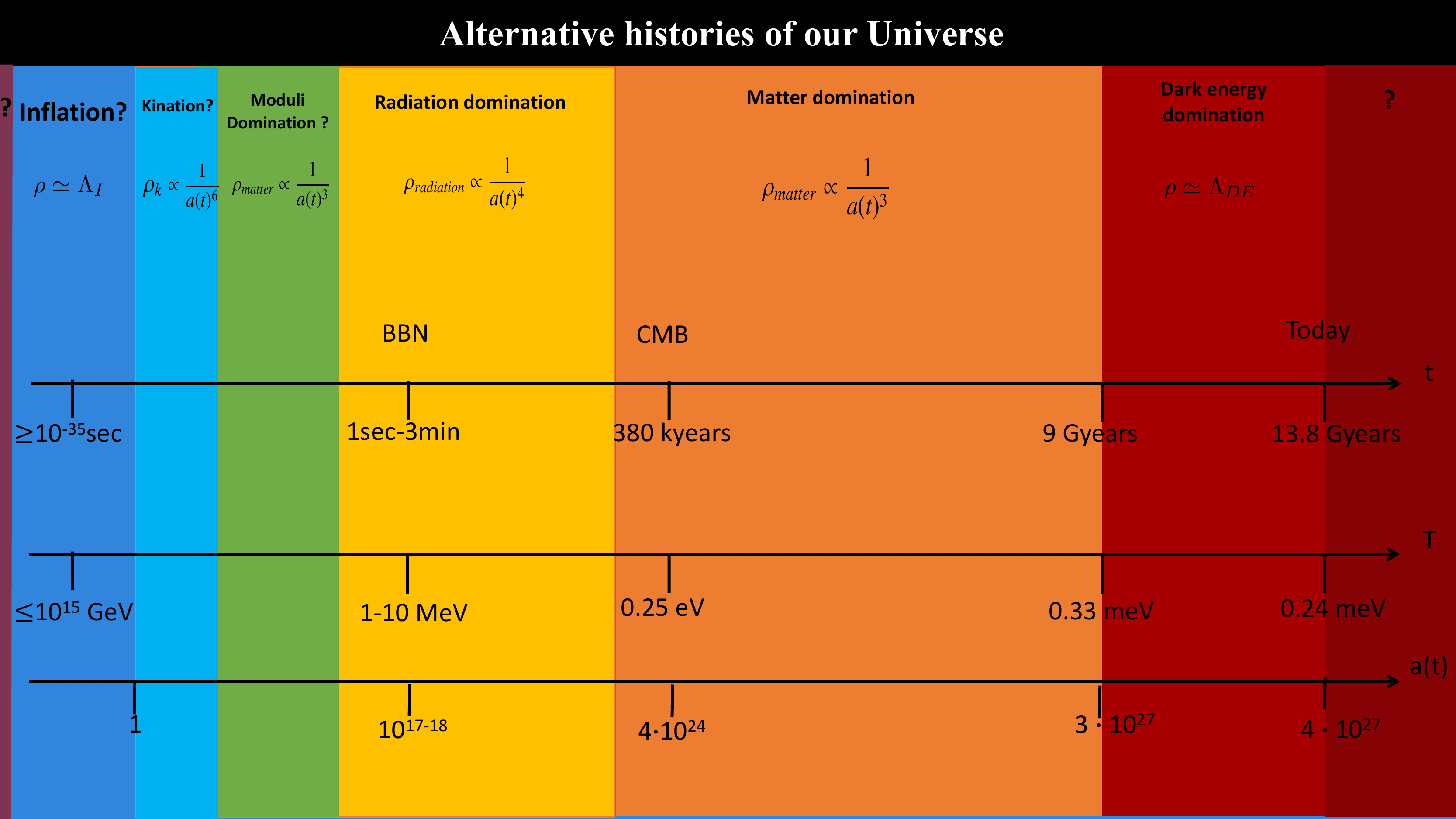} 
    \caption{Alternative histories of the universe including potential periods of kination and moduli domination.}
    \label{fig:AlternativeHistory}
\end{figure}

While not strictly unique to string theory (the key feature is the presence of massive scalars with gravitational-strength interactions), it represents a very different cosmological history to many Beyond-the-Standard-Model post-inflationary scenarios, which involve a rapid transfer of energy from the inflationary degrees of freedom into Standard Model particles.

Sometimes string theory is seen as an esoteric UV issue of little interest to hard-working practical cosmologists studying the universe one trillionth of a second after the Big Bang. It is, therefore, important to note that the cosmology of such standard field theory scenarios is unstable to the inclusion of a sector with only gravitationally coupled particles (i.e. moduli). As described above, as long as there is \emph{some} initial amplitude in the moduli fields, we expect this energy density to grow so that the universe passes through an epoch of moduli domination.

Naively, one may think it possible to avoid this by assuming that the inflaton is charged only under Standard Model degrees of freedom, 
such that all inflationary dynamics only involves a displacement in the inflaton field. The claim is that, in this case, there would be no amplitude in the moduli degrees of freedom or, put another way, the post-inflation moduli would not be displaced from their final minimum during inflation. However, in practice it is very hard to engineer this: in the context of \emph{any} effective Lagrangian with a UV completion in string theory, there will almost always be an initial displacement of the moduli from the final minimum, and thus some amplitude in the moduli field.  This is particularly so for the universal moduli -- the overall volume and the dilaton (see \cite{Cicoli:2016olq} for one explicit computation of the volume mode displacement during inflation).

Why? We illustrate this in the context of IIB compactifications, but the argument extends easily. The supergravity scalar potential
is (with $M_{\rm Pl} =1$)
\begin{equation}
\label{scalarpot}
V = e^K \left( K^{i \bar{\jmath}} D_i W D_{\bar{\jmath}} \overline{W} - 3 \vert W \vert^2 \right).
\end{equation}
The K\"ahler potential is
\begin{equation}
K = - 2 \ln \mathcal{V}(T + \overline{T}) - \ln \left( \int {\rm i} \Omega \wedge \overline{\Omega} \right) - \ln (S + \overline{S})\,,
\end{equation}
where $\mathcal{V}$ is the volume, $T$ the chiral superfield containing the volume modulus, and $S$ the superfield for the dilaton. As this K\"ahler potential directly depends on $\mathcal{V}$ and $S$, the presence of the $e^K$ factor implies the scalar potential (\ref{scalarpot}) will always depend \emph{explicitly} on these fields. \emph{Any} source of energy will give a contribution to the potential for these fields; this includes the inflationary contribution to the potential that is absent in the final minimum. This implies that, during the inflationary epoch, these fields are necessarily displaced from the final minimum: in terms of the moduli fields defined as oscillations about the final vacuum, such fields have non-zero amplitudes during inflation.

If arguments in terms of ${\cal N}=1$ supergravity Lagrangians sound somewhat abstract or specific to supergravity, there is a more physical way of putting the same point. The dilaton (string coupling) and volume modulus both directly enter the relationship between the string scale and the 4-dimensional Planck mass (where $\mathcal{V}_s$ the volume of the Calabi-Yau measured in units of $l_s^6$ in terms of the 10-dimensional string frame metric in the convention where string and Einstein frame metrics coincide in the vacuum i.e those described above equation (\ref{convvvv}))
\begin{equation}
\setlength\fboxsep{0.25cm}
\setlength\fboxrule{0.4pt}
\boxed{
M_s = \frac{g_s}{\sqrt{4\pi}} \frac{ M_{\rm Pl}}{\sqrt{\mathcal{V}_s}}\,.
}
\end{equation}
Just as in string theory the string mass $M_s$ is \emph{the} fundamental physical scale, so is $g_s$ \emph{the} fundamental measure of the intrinsic 
strength of quantum effects (including gauge couplings). Therefore, all physical scales (in particular, potential energies) depend on $M_s$ -- and so all potentials will depend on the string coupling and the volume modulus. From this, it immediately follows that the inflationary part of the potential acts a source term for the dilaton and volume modulus -- and so once this is removed, the expectation values of these fields will shift, implying that during inflation these fields are displaced from their final minimum.

We note here another potential important stringy difference to field theory scenarios of inflation. In field theory scenarios, inflation ends with a rolling inflaton. Although the inflaton ultimately decays so that only the curvature perturbation survives, the inflaton is still present at the end of
 inflation and the inflaton still exists as a potential excitation within the theory. In string theory, inflation may end with the inflation having already disappeared at the end of inflation. In examples like brane inflation, where inflation terminates with brane/anti-brane annihilation, the inflaton field (the brane position) has entirely vanished in the post-inflationary epoch. Nonetheless, the argument that such a universe should find its way to a moduli-dominated epoch is 
 unchanged\footnote{Reheating after brane/anti-brane inflation has a rich phenomenology (many of the features are
 intrinsically string theoretic). See e.g  \cite{Jones:2003da, Polchinski:2004ia, Barnaby:2004gg, Kofman:2005yz, Chialva:2005zy, Frey:2005jk}. }. 

The resulting era of \emph{moduli domination} is therefore generic and one of the most universal expectations in string cosmology. Moreover, when reheating does proceed only via non-renormalisable interactions suppressed by a factor of $M_{\rm Pl}$ (as for moduli), then `standard' expectations as to the reheating temperature are greatly modified. The reheating temperature now becomes
\begin{subequations}
\begin{empheq}[box=\widefbox]{align}
T_{\rm rh} & \simeq  \left( \frac{\alpha}{4 \pi} \right)^{\frac12} \left( \frac{m_\phi}{M_{\rm Pl}} \right)^{\frac12} m_\phi \\
& \simeq  1 \, \hbox{GeV} \, \left( \frac{m_\phi}{10^6 \, \hbox{GeV}} \right)^{3/2}.
\label{reheatmoduli}
\end{empheq}
\end{subequations}
In a stringy context, we expect the reheating temperature to be much lower than in field theory scenarios. Physics expected to take place in the early universe (e.g.~baryogenesis) must proceed via scenarios that can operate at these relatively low temperatures.

The fact that in string theory, reheating is expected to proceed via moduli decay (see Fig.~\ref{fig:ModuliDomination},) leads to various cosmological problems and/or opportunities 
that must be addressed. It should be emphasised again, though, that these problems are not specific to string theory; \emph{all consistent theories 
of the early universe must include gravity} and these questions arise in \emph{any} theory which includes in its spectrum scalar particles whose interactions are not stronger than gravitational strength (i.e. they are non-renormalisable and suppressed by 
explicit powers of $M_{\rm Pl}$): this issue cannot be rendered irrelevant by pretending it does not exist.

\begin{figure}[ht]
    \centering
    \includegraphics[width = 0.9\textwidth]{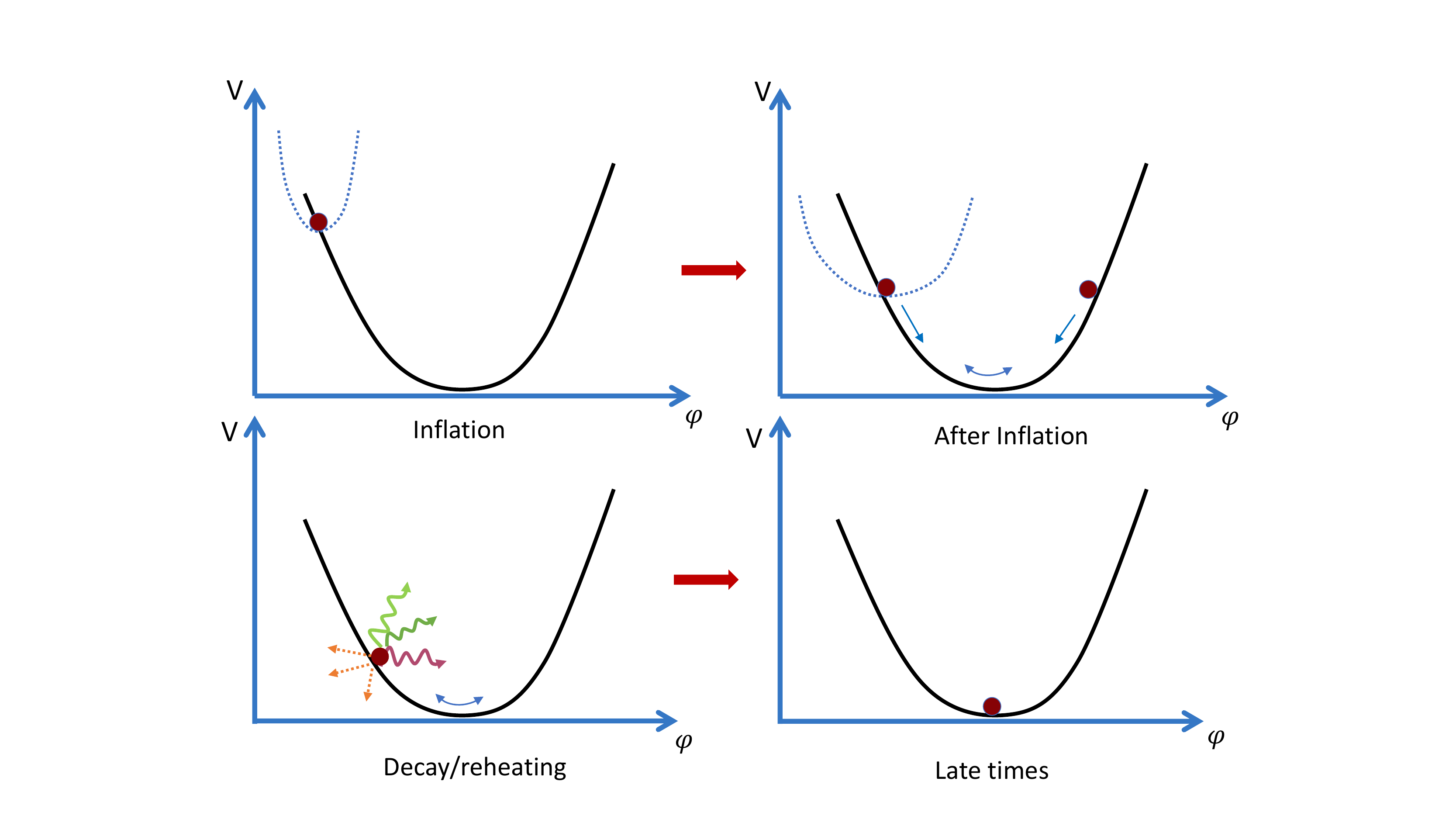} 
    \caption{Moduli domination. During inflation moduli fields which are not inflatons tend to be trapped at a point that does not correspond to their minimum. Only after inflation is finished do those moduli field start settling towards their minimum. Their coherent oscillations around this minimum come to dominate the energy density of the universe, behaving like matter domination, and particle production during this time is the source of reheating. This differs substantially from the standard picture of reheating from the inflaton right after inflation.}
    \label{fig:ModuliDomination}
\end{figure}

\subsection{Aspects of Moduli-Induced Reheating}

This epoch of moduli domination can affect various areas of early universe physics and
we now consider in more detail the implications of reheating driven by decays of moduli.

\subsubsection{The Cosmological Moduli Problem}

One of the best-known issues associated with moduli reheating is the \emph{cosmological moduli problem} (CMP) \cite{CoughlanRoss, hepph9308292, hepph9308325}. The $M_{\rm Pl}$-suppression in moduli interactions implies that they are long-lived. While the masses of moduli are model-dependent to some extent, in most cases these masses are comparable to the gravitino mass. When low-energy gravity-mediated supersymmetry is used to solve the hierarchy problem of the Standard Model, the soft terms are also normally comparable to the gravitino mass (note that for gauge-mediated models, $m_{3/2} \ll 1\, {\rm TeV}$, and this issue becomes more severe). For `generic' supergravity models, we then have
\begin{equation}
m_\phi \sim m_{3/2} \sim M_{\rm soft} \sim 1 \,{\rm TeV}\,,
\end{equation}
where $m_\phi$ denotes the modulus mass. For such mass scales, it
follows from Eq. (\ref{reheatmoduli}) that the resulting reheating temperature is $T_{\rm rh} \lesssim 1 \, {\rm MeV}$, insufficient for conventional nucleosynthesis. Generally, in any case where the lightest moduli have masses $m_\phi \lesssim 30\, \hbox{TeV}$, the period of moduli domination extends until a time period later than when Big Bang Nucleosynthesis (BBN) occurred. 
As observations of primordial element abundances make us highly confident that the standard picture of BBN is correct, such cosmologies are inconsistent with nature.

The above was first realised in the 1980s by Coughlan, Fischler, Kolb, Raby and Ross \cite{CoughlanRoss} for the supergravity Polonyi field and is the original cosmological moduli problem. It was subsequently re-emphasised and shown to be generic in the context of stringy scenarios of moduli stabilisation in \cite{hepph9308292, hepph9308325} and has existed since then as a standard reference point in discussions of string cosmology. The importance of this problem lies in its genericity: as long as moduli exist, they will tend to survive at low energies and tend to dominate the energy density of the universe. The original arguments relied on the assumption of low-energy supersymmetry to estimate their masses. However, the argument is more general than supersymmetry. Even if compactifications do not preserve supersymmetry, as long as the Kaluza-Klein scale is smaller than the Planck mass the moduli will survive at low energies. As moduli are the zero modes of Kaluza-Klein field, they only obtain masses through quantum effects. It is easy to estimate that their masses are at most of order $M_{KK}^2/M_{\rm Pl}$ \cite{Burgess:2010sy} which is much smaller than the KK and string scales as long as $M_{KK}<M_{\rm Pl}$.

The cosmological moduli problem is associated to the \emph{time} at which moduli decay, and through this, the Hubble scale (i.e. temperature) at which a conventional radiation-dominated Hot Big Bang is initiated. It may be useful here to distinguish here between a \emph{hard} cosmological moduli problem and a \emph{soft} cosmological moduli problem. 

The hard CMP is the requirement that $T_{\rm rh} > T_{\rm BBN}$: we \emph{know} that BBN occurred and so it is an absolute necessity that the reheat temperature is in excess of the temperatures required for BBN. There are several other aspects of physics we \emph{think} originated in the thermal Hot Big Bang, but where a precise answer is not known. Examples include mechanisms for baryogenesis and 
the origin of the dark matter relic abundance. Scenarios for these can require $T_{\rm rh} \gg T_{\rm BBN}$. As examples, scenarios of thermal leptogenesis can require $T_{\rm rh} \gtrsim 10^{11}\,{\rm GeV}$ \cite{Buchmuller:2005eh} whereas thermal freeze-out of WIMP dark matter tends to require $T_{\rm rh} \gtrsim m_{\rm DM}/20$, where $m_{\rm DM}$ is the dark matter mass. The soft CMP refers to the requirement that $T_{\rm rh}$ is sufficiently high to allow for all the other particle  physics, in addition to BBN, that we would like to occur in the Hot Big Bang. As the name suggests, this requirement is more nuanced and model-dependent, being tied to the actual physics used to generate (for example) the baryon asymmetry.

Since its original formulation, the importance of the CMP as a problem has waxed and waned. Most obviously, the `Problem' in the title originates from an assumption that supersymmetry would be present at TeV energy scales and would solve the hierarchy problem of the Standard Model. If low-energy supersymmetry is entirely absent -- or even if it simply ameliorates the hierarchy problem by appearing at 1000 TeV energy scales -- then the hard CMP loses its force. It is certainly true that, since the CMP was first proposed, the energy scales at which supersymmetry might plausibly be present have risen by an order of magnitude.

A further aspect is the development of explicit models of moduli stabilisation. The original CMP was formulated based on the properties of `generic' ${\cal N}=1$ 
supergravity scalar potentials. With the construction of actual scenarios of moduli stabilisation, it was realised that these `generic' expectations are often violated in actual models, which have additional structure compared to `generic' expectations. We give two examples.

The first example involves models based on non-perturbative stabilisation, where a gaugino condensate is balanced against either a constant term or another gaugino condensate. The standard examples of these are KKLT and racetrack models, as discussed in Sec.~\ref{ModuliSection}. In this case, there is a logarithmic enhancement in the masses of moduli relative to the gravitino mass, and also a logarithmic suppression in the mass of the soft terms relative to the gravitino mass \cite{hepth0411066, hepth0503216, hepph0702146}. More specifically
\begin{equation}
m_\phi \sim \ln \left( \frac{M_{\rm Pl}}{m_{3/2}} \right) m_{3/2} \sim \ln^2 \left( \frac{M_{\rm Pl}}{m_{3/2}} \right) M_{\rm soft}\,.
\label{kkltsoft}
\end{equation}
This implies that soft terms at $\sim 1 \, \rm{TeV}$ would actually be accompanied by moduli masses at 1000 TeV, greatly ameliorating the cosmological moduli problem.

The second example involves models which start with a no-scale structure (of which LVS is the principal example). At leading order (i.e. with exact no-scale as in GKP \cite{Giddings:2001yu}) a modulus remains massless despite the presence of both a finite gravitino mass and broken supersymmetry. In the full vacuum, this structure still survives 
approximately, with a modulus parametrically lighter than the gravitino mass:
\begin{equation}
m_\phi \ll m_{3/2}\,.
\label{noscalemass}
\end{equation}
The form and scale of the soft terms depends on the realisation of the Standard Model and the degree of sequestering present. Two possible scenarios have been identified \cite{Blumenhagen:2009gk, Aparicio:2014wxa, Reece:2015qbf}, building on earlier analyses of LVS soft terms in \cite{hepth0605141, hepth0609180, hepth0610129}. These lead to either  
\begin{equation}
m_{3/2} \sim \frac{M_{\rm Pl}}{\mathcal{V}} \gg m_\phi \sim M_{\rm soft} \sim \frac{M_{\rm Pl}}{\mathcal{V}^{3/2}}\,,
\label{lvssoft1}
\end{equation}
or
\begin{equation}
m_{3/2} \sim \frac{M_{\rm Pl}}{\mathcal{V}} \gg m_\phi  \sim \frac{M_{\rm Pl}}{\mathcal{V}^{3/2}} \gg M_{\rm soft} \sim \frac{M_{\rm Pl}}{\mathcal{V}^2}\,.
\label{lvssoft2}
\end{equation}
Irrespective of which case is realised, what Eqs. (\ref{kkltsoft}), (\ref{lvssoft1}) and (\ref{lvssoft2}) have in common is a pattern of masses very different to that assumed in the `generic' CMP.

Despite these results, and the general softening of the CMP in the absence of a discovery of weak scale supersymmetry, it persists as a question that must be addressed in all stringy cosmologies. Even if the hard CMP is not a problem as such, it is also a reminder that stringy cosmologies are expected to have lower reheating scales than in more conventional field theory models of the early universe.

\subsubsection{Dark Radiation Constraints on Moduli Decays}

The cosmological moduli problem is primarily associated with the \emph{time} at which moduli decay, and the danger that the transition to radiation domination occurs too late for conventional BBN epoch.

Other potential problems (or opportunities!) are associated to the decay products and branching ratios of moduli. As with the CMP, these problems can be formulated both  `generically' in terms of supergravity models without special structure in the couplings, and then refined in terms of the precise structures that arise within particular string compactifications. As with the CMP, the origin of all such problems is the requirement that reheating cannot lead to a universe that is significantly different than a radiation-dominated Hot Big Bang, in which almost all energy is contained in Standard Model degrees of freedom.

In particular, if significant amounts of energy density are injected into degrees of freedom that are decoupled from the Standard Model, this may lead to significant cosmological tensions. The simplest example of this is an over-production of \emph{dark radiation} -- a possible additional relativistic contribution to the energy density which is decoupled from the Standard Model degrees of freedom. While today, after ten billion years of matter and dark energy domination, such radiation would only contribute negligibly to the dynamics of the universe, in earlier epochs such radiation would significantly affect the expansion history. 

From observations of the CMB, we know that the early universe bounds the amount of additional `dark' radiation that can contribute to the energy density. This is conventionally expressed in terms of a bound on the number of effective neutrino species; this reflects the fact that, after they decouple, the neutrinos free-stream as relativistic particles decoupled from the other Standard Model degrees of freedom, and so behave (effectively) as dark radiation.
The magnitude of this limit can be expressed as \cite{Planck:2018vyg}:
\begin{equation}
\setlength\fboxsep{0.25cm}
\setlength\fboxrule{0.4pt}
\boxed{
\Delta N_{\rm eff} \lesssim 0.2\,.
}
\end{equation}
Here $N_{\rm eff}$ is the effective number of `extra' neutrino species beyond those of the Standard Model.

String compactifications contain many particles which could in principle constitute dark radiation, as they are both light and very weakly coupled to the Standard Model. One example is gauge bosons of unbroken dark sector gauge groups. These can arise either as hidden $U(1)$s present on `distant' branes within D-brane constructions or alternatively from $n$-form potentials reduced on internal $(n-1)$-cycles (e.g. $A^i_\mu = \int_{\Sigma_2^i} C_3$). Another type of particle with light and `protected' masses are chiral fermions belonging to matter sectors decoupled from the Standard Model. In models of branes at singularities, these can arise from matter originating on branes located at distant singularities. Here, the chirality protects the mass of the particles while the geometrical separation ensures the weakness of couplings to the Standard Model and ensures that such particles will not thermalise with Standard Model degrees of freedom.

However, in terms of dark radiation candidates within string theory, arguably the most appealing and universal particle candidates are axions or axion-like particles (ALPs) \cite{Svrcek:2006yi, Arvanitaki:2009fg, Cicoli:2012sz, Demirtas:2018akl}. Axions are $U(1)$-valued scalars: instead of the real line, they take values on a circle. They have a perturbative shift symmetry $a_i \to a_i + 2 \pi f_a$, which ensures that the potential remains flat to all orders in perturbation theory. Here $f_a$ is the \emph{axion decay constant} which specifies the radius of the $U(1)$ circle and also governs the strengths of axion-matter couplings. As non-perturbative corrections are exponentially suppressed, it is natural (depending on the precise details of the compactifications) for ALPs to remain effectively massless, and in particular sufficiently light so as to be a good candidate for dark radiation. The shift symmetry also implies that axions must couple via derivative interactions, which implies that their interactions with the Standard Model are suppressed by their decay constant $f_a \gg 1\, \rm{TeV}$. For all these reasons, energetic relativistic axions produced in the early universe are likely to persist as a relativistic background until the present day (forming what has been called a Cosmic Axion Background \cite{13053603, 210109287}).

We aim to formulate the generic problem succinctly. In string theory, reheating tends to be driven by moduli with gravitational-strength couplings. Certain moduli (for example, the overall volume modulus or the dilaton modulus) tend to couple to all sectors of the theory in a broadly democratic fashion. However, as there could easily be hundreds of hidden-sector particles, 
democratic decays would likely result in too much energy going into axions or other hidden sector modes. The decay modes 
\begin{equation}
\phi \to {\rm hidden}\,,
\end{equation} 
may produce dark radiation at a level far above the observable limits. As moduli-dominated reheating is generic in string theory, this constitutes a generic problem (or opportunity!) for string models of reheating. 

This problem of \emph{excessive dark radiation} is generic within string compactifications. If reheating originates from gravitationally-coupled sectors, it is hard to see why the fractional decay rate to non-Standard Model light hidden sectors should be close to zero: what makes the Standard Model sector special and preferred?

To formulate this problem more precisely, and move beyond generic statements, requires actual vacua in which we can determine the longest-lived modulus, its mass, its couplings and its decay modes. Although the `visible sector' in principle requires the actual Standard Model, even without a full Standard Model realisation in a compactification, toy proxies for a Standard Model can be used (for example, a D3-brane probing a Calabi-Yau singularity or a stack of D7-branes wrapping a divisor in the geometric regime) to allow for the calculation of decay modes to a `quasi-visible' sector.

As a consequence of its parametric scale separation, the Large Volume Scenario provides a clean environment in which the lightest modulus and its couplings can be determined. It is not surprising then, that in the context of LVS, this problem has been formulated especially sharply, starting originally with \cite{Cicoli:2012aq, 12083563} and then developed further on in 
\cite{12113755, 13041804, 13054128, Allahverdi:2014ppa, Angus:2014bia, 14036810, Cicoli:2015bpq, Cicoli:2018cgu, 210713383, Cicoli:2022fzy, Cicoli:2022uqa}. 

Focusing on LVS Calabi-Yau models where the overall volume is controlled by a single modulus, the lightest modulus is the overall volume modulus, which allows for a precise computation of the couplings and decay modes. These couplings are set by the relevant K\"ahler and superpotential terms. The superpotential terms drop out, as K\"ahler moduli such as the overall volume only appear in the superpotential non-perturbatively. The dependence of matter kinetic terms on the volume direction itself is given by 
\begin{equation}
K_{T\overline{T}} \propto \frac{1}{\mathcal{V}^{2/3}}\,.
\label{matterkahlermetric}
\end{equation}
This follows from the modulus K\"ahler potential $K = - 2 \ln \mathcal{V}$ and the requirement that the physical Yukawa couplings in local models are determined locally and so must be independent (at tree-level) of the overall volume \cite{hepth0609180}. The self-couplings of the volume multiplet can be found from the K\"ahler potential
\begin{equation}
K = - 3 \ln \left( T + \overline{T} \right),
\end{equation}
where $T$ is the superfield containing the volume modulus.
This is equivalent to $K = - 2 \ln \mathcal{V}$ when we restrict to only the overall volume mode. The resulting kinetic terms are
\begin{equation}
\mathcal{L}_{\rm kin} = \frac{3}{4 \tau_b^2} \partial_{\mu} \tau_b \partial^{\mu} \tau_b + \frac{3}{4 \tau_b^2} \partial_{\mu} a\, \partial^{\mu} a\,.
\label{Lvol}
\end{equation}
From Eq. (\ref{Lvol}) we can read off the couplings of the volume modulus to its own axion.

Assuming an MSSM-like local matter sector and no additional dark sectors, it was found in \cite{Cicoli:2012aq, 12083563} that there are \emph{two} dominant decay modes: the decay $\phi \to aa$ of the volume to the volume axion, and the decay $\phi \to H_1 H_2$ which proceeds via the Giudice-Masiero $\mu$-term. The coupling to the axion is universal and after canonical normalisation takes the form \cite{Cicoli:2012aq, 12083563}
\begin{equation}
\mathcal{L} = \frac34 \exp \left( -2 \sqrt{\frac23} \phi \right) \partial_{\mu} a \,\partial^{\mu} a\,,
\end{equation}
which gives a decay width
\begin{equation}
\setlength\fboxsep{0.25cm}
\setlength\fboxrule{0.4pt}
\boxed{
\Gamma_{\phi \to aa} = \frac{1}{48 \pi} \frac{m_\phi^3}{M_{\rm Pl}^2}\,.
}
\label{axionwidth}
\end{equation}
This axionic contribution to dark radiation is universal and unavoidable, as the volume axion is always present in the compactification, and at large volumes the couplings are entirely set by the tree-level K\"ahler potential. As well as these decays to the volume axion, the volume modulus can also decay to other light open- or closed-string axions present in the compactification \cite{210713383, Cicoli:2022uqa}.

Interestingly, all matter decay modes \emph{except} the one proceeding via a Giudice-Masiero term 
\begin{equation}
K = \dots + \left( \frac{Z H_u H_d}{\left( T_b + \overline{T}_b \right)} + {\rm c.c.} \right) + \dots 
\end{equation}
are  suppressed by an $\left(M_{\rm soft}/m_\phi\right)^2\ll 1$ factor, as the K\"ahler potential of Eq. (\ref{matterkahlermetric}) leads to a coupling
\begin{equation}
\phi ( {Q} \nabla^2 Q + Q \nabla^2 \bar{Q})\,,
\end{equation}
which is suppressed on-shell by a relative factor $\left(M_{\rm soft}/m_\phi\right)^2$ compared to modes like Eq. (\ref{axionwidth}). Note that in models where the visible sector is sequestered from the sources of supersymmetry breaking, also the decay of the volume mode into MSSM gauge bosons is suppressed since the tree-level gauge kinetic function is set by the dilaton. Meanwhile, the Giudice-Masiero term gives a decay width
\begin{equation}
\setlength\fboxsep{0.25cm}
\setlength\fboxrule{0.4pt}
\boxed{
\Gamma_{\phi \to H_u H_d} = \frac{2 Z^2}{48 \pi} \frac{m_\phi^3}{M_{\rm Pl}^2}\,,
}
\end{equation}
very similar to the axion decay width of Eq. (\ref{axionwidth}). 

Although some aspects are model-dependent (for example, the branching ratio to the Standard Model depends on the undetermined constant $Z$ and would be enhanced for models with extended Higgs sectors), this calculation illustrates the way in which moduli-dominated reheating is expected to produce dark radiation. Depending on perspective, this can be viewed either as a problem or an opportunity for such models. 

Another contribution to the volume mode decay into two Standard Model Higgs degrees of freedom $h$ comes from moduli-dependent loop corrections to the Higgs mass \cite{Cicoli:2022fzy}:
\begin{equation}
m_h^2 = M_{\rm soft}^2 \left[c_0 + c_{\rm loop} \ln\left(\frac{M_{\rm KK}}{m_{3/2}}\right)\right] 
\end{equation}
which, using $M_{\rm KK}\sim M_{\rm Pl}/\mathcal{V}^{2/3}$, induces the following couplings:
\begin{equation}
\mathcal{L}\supset -\sqrt{\frac32} \frac{m_h^2}{M_{\rm Pl}}\,\phi \,h^2 + \frac{c_{\rm loop}}{4}\sqrt{\frac32} \frac{M_{\rm soft}^2}{M_{\rm Pl}}\,\phi \,h^2\,.
\label{LoopCoupl}
\end{equation}
Clearly, the first coupling is irrelevant since it is suppressed by the Higgs mass. However the second coupling would give rise to a decay width that scales as 
\begin{equation}
\Gamma_{\phi \to hh} \sim \left(\frac{c_{\rm loop}}{Z}\right)^2 \left(\frac{M_{\rm soft}}{m_\phi}\right)^4\,\Gamma_{\phi \to H_u H_d}\,.
\end{equation}
For models without sequestering (which requires the Higgs mass to be fine-tuned to be much lighter than its `natural' value), $M_{\rm soft}\sim m_{3/2}\gg m_\phi$, $\Gamma_{\phi \to hh}$ dominates over $\Gamma_{\phi \to H_u H_d}$ and $\Gamma_{\phi \to aa}$, guaranteeing the absence of any dark radiation production. On the other hand, in sequestered models, $M_{\rm soft}\lesssim m_\phi$, implying that $\Gamma_{\phi \to hh}$ is subdominant with respect to $\Gamma_{\phi \to H_u H_d}$.

In a phenomenologically realistic scenario, it is obviously critical that the amount of dark radiation produced does not exceed the observational bounds. One approach to achieving this is to enhance the modulus couplings to the visible sector. For example, in sequestered models, one may assume that $Z \gg 1$ or that actual BSM physics involves a large number of Higgs sectors which enhances the visible sector branching ratio purely on multiplicity grounds. 

Although it is fair to describe LVS as the most developed scenario for precise studies of reheating and dark radiation production, similar questions of dark radiation production have also been studied for other scenarios of moduli-stabilisation, for example see \cite{ 13047987, 14072501, 150207746, 151207907,  190603025, 210803317, Baer:2023bbn}.

Our discussion has mostly treated the existence of dark radiation from moduli decay as a problem to be expunged. However, one can also regard it as an opportunity to be seized. The electromagnetic interactions of photons imply that the photon surface of last scattering is the CMB, which dates from a time 380,000 years after the Big Bang. For axion-like particles, the similar surface of last scattering extends deep into the primordial epoch of the universe, reaching far earlier than the epoch of BBN. Indeed, one would expect that a typical axion-like particle produced from modulus decay would propagate to the present day without interacting once. If it were possible to detect these particles, they would give a direct probe of the universe at an exceedingly early period. Such a surviving background of relativistic dark radiation cosmic axions was called a `Cosmic Axion Background' in \cite{13053603} and its phenomenology and detection possibilities studied in 
 \cite{13066518, 13104464, 13123947, 14065188, 14072501, 14114172, 150605334, 160208433, 210703420, Bhattacharya:2020zap, Banerjee:2022era}.

\subsubsection{Moduli-Induced Gravitino Problems}

As well as decays to axions or other massless particles, another potentially problematic channel for moduli decays is the decay mode to gravitini. If the lightest modulus has a mass $m_\phi > 2 m_{3/2}$, it can decay via the process $\phi \to 2 \psi_{3/2}$. If the modulus is gravitationally coupled and related to the superfield which broke supersymmetry (and thus generates the two spin-$\frac12$ components of the gravitino in the super-Higgs effect), we expect it couple to the gravitino modes and so have a notable branching ratio to gravitini. 

This is true in both KKLT and racetrack stabilisation. For the case of the hierarchically small gravitino masses required to address
the electroweak hierarchy problem,
\begin{equation}
m_\phi \sim \ln \left( \frac{M_{\rm Pl}}{m_{3/2}} \right) m_{3/2} \gg m_{3/2}\,,
\end{equation}
and so the gravitino can be treated as effectively massless compared to the modulus (as $m_{3/2} \sim m_\phi/30$). However, it is not true for LVS, where the inverse hierarchy $m_\phi \ll m_{3/2}$ applies.

The dangers of this decay channel were emphasised in \cite{hepph0602061, hepph0602081}, where it was dubbed the \emph{moduli-induced gravitino problem}. The dangers arise as the gravitini are long-lived and such long-lived gravitini can cause several problems. In particular, they can interfere with BBN if they live long enough so that they decay after the start of BBN with a late injection of energy (effectively reintroducing the modulus problem). Another problem can arise with the dark matter abundance. If the gravitini are stable, as in gauge-mediated scenarios, then the decay can result in over-production of dark matter. However, even if unstable then the decay can result in an overproduction of the neutralino LSP.

Building on the original work, further studies of the moduli-induced gravitino problem have been performed in \cite{hepph0603265, hepph0604140, hepph0604132, hepph0605297, hepph0607170, hepph0701042, hepph0703319, Heckman:2008jy, 09043773, 09082430, 12092583, 12104077, 13110052, 150402040, 160308399, 220106633}.

\subsubsection{Post-inflationary History and Predictions from Inflation}

As discussed in chapter \ref{SecCO}, in inflation the predictions for the spectral tilt and the tensor-to-scalar ratio are determined by the number of e-foldings $N_e$ between horizon exit of the CMB modes and the end of inflation. The quantity $N_e$ is determined by the so called `matching equation' which is the condition obtained by tracking the evolution of the energy density of the universe from the time of horizon exit of the CMB modes to the present times (see e.g. \cite{Planck:2013jfk}). Thus, inflationary predictions are sensitive to the post-inflationary history of the universe. As we have seen, in string models the presence of light moduli generically leads to a non-standard post-inflationary history involving epochs of moduli domination. This, in turn, affects the precise predictions and is crucial for obtaining the best fit regions for string inflationary models \cite{Easther:2013nga, Dutta:2014tya, Bhattacharya:2017ysa, Cicoli:2020bao, Bhattacharya:2020gnk, Neves:2020anh}. The analysis of \cite{Bhattacharya:2017ysa, Bhattacharya:2020gnk} integrates the CosmoMC analysis with Modechord \cite{Handley:2015fda} allowing for extraction of the best fit regions in the parameter space.

\subsubsection{Finite-temperature Corrections and Decompactification}

Reheating at the immediate end of inflation can produce an initial epoch of radiation domination with several particles in thermal equilibrium. 
Although this may subsequently be diluted by moduli,
this generates finite-temperature corrections which in 4-dimensional string models can be a dangerous source of decompactification, see e.g \cite{Buchmuller:2004xr, Buchmuller:2004tz, Danos:2008pv, Anguelova:2009ht, DiMarco:2019czi, Gallego:2020vbe,Alam:2022rtt}. This can be seen from the fact that thermal loops yield a moduli-dependent contribution to the scalar potential of the form (at 1- and 2-loop order):
\begin{equation}
\setlength\fboxsep{0.25cm}
\setlength\fboxrule{0.4pt}
\boxed{
V_T \sim T^4 g^2 + T^2 M_{\rm soft}^2\,,
\label{VT}
}
\end{equation}
where $g^2\sim \tau_{\rm SM}^{-1}$ is the Standard Model coupling set by the modulus supporting the stack of branes which reproduce the visible sector, and $M_{\rm soft}\lesssim m_{3/2}$ is the mass of the supersymmetric particles running in the loop (we ignored ordinary matter since it would give rise to a subdominant contribution). 

As we have already seen, the KKLT potential scales as $V_{\rm KKLT}\sim m_{3/2}^2 M_{\rm Pl}^2$ while the LVS potential scales as $V_{\rm LVS}\sim m_{3/2}^3 M_{\rm Pl}$. Given that $M_{\rm soft}\lesssim m_{3/2}$ and $T< M_{\rm Pl}$, the second term in (\ref{VT}) is always a small correction to the KKLT potential. This is true also in the LVS case since $T < M_{\rm KK} \sim M_{\rm Pl}/\mathcal{V}^{2/3}$ implies $T^2 M_{\rm soft}^2 < M_{\rm KK}^2 m_{3/2}^2 \sim M_{\rm Pl}^4/\mathcal{V}^{10/3} \ll V_{\rm LVS} \sim M_{\rm Pl}^4/\mathcal{V}^3$ for $\mathcal{V}\gg 1$. Hence the only dangerous term in (\ref{VT}) is the one proportional to $T^4$ which would cause a runaway towards decompactification unless $T^4 g^2$ is subdominant with respect to the zero-temperature moduli potential. This problem can be avoided if 
the initial reheating temperature is below a maximal temperature given in KKLT by \cite{Buchmuller:2004tz}: 
\begin{equation}
T_{\rm rh}<T_{\rm max}\sim \sqrt{m_{3/2} M_{\rm Pl}}\sim 10^{11}\,{\rm GeV}\quad\text{for}\quad m_{3/2}\sim 10\,{\rm TeV}
\end{equation}
and in LVS by \cite{Anguelova:2009ht}:
\begin{equation}
T_{\rm rh}<T_{\rm max}\sim \left(m_{3/2}^3 M_{\rm Pl}\right)^{1/4}\sim 10^8\,{\rm GeV}\quad\text{for}\quad m_{3/2}\sim 10\,{\rm TeV}\,.
\end{equation}

\subsubsection{Baryogenesis and Moduli-Domination}

As we have seen, 4-dimensional string models are characterised by late-time epochs of moduli domination which end when the moduli decay, leading to a reheating temperature of order $T_{\rm rh}\simeq 0.1\,m_\phi\sqrt{m_\phi/M_{\rm Pl}}$. The actual value of $T_{\rm rh}$ clearly depends on the lightest modulus mass $m_\phi$, which in string models is correlated with the scale of supersymmetry breaking. As an illustrative example, consider LVS models with sequestered supersymmetry breaking which can lead to TeV-scale supersymmetry without a cosmological moduli problem. In this case $m_\phi \sim M_{\rm soft}^{3/4} M_{\rm Pl}^{1/4}$. For $M_{\rm soft}\sim 1$ TeV, we obtain $m_\phi \sim\mathcal{O}(10^7)$ GeV which, in turn, leads to $T_{\rm rh}\sim \mathcal{O}(1-10)$ GeV. This reheating temperature is clearly too low to allow for standard scenarios of baryogenesis like thermal leptogenesis or EW baryogenesis which require higher reheating temperatures (and so much larger moduli masses, incompatible with any form of low-energy supersymmetry). 

If we require (relatively) low-energy supersymmetry, a reheating temperature around $T_{\rm rh}\sim \mathcal{O}(1-10)$ GeV necessitates a non-thermal mechanism for baryogenesis. One interesting option is given by Affleck-Dine baryogenesis \cite{Affleck:1984fy} that exploits MSSM D-flat directions which carry a net baryon number. If lifted by supersymmetry breaking effects, they can acquire a non-zero VEV during inflation. After inflation, they first oscillate around the minimum and then decay into fermions generating baryon asymmetry. Interestingly ref. \cite{Allahverdi:2016yws} has shown that in K\"ahler moduli inflation inflaton-dependent F-terms can induce tachyonic masses during inflation for these MSSM D-flat directions, creating the right initial conditions for Affleck-Dine baryogenesis. However, this mechanism has the problem of generating (in general) too much baryon asymmetry. Nevertheless, this can be avoided in string models where the initial baryon asymmetry is diluted by the decay of the lightest modulus which gives \cite{Kane:2011ih}:
\begin{equation}
\frac{n_B}{s} \simeq \frac{|A|}{m_\phi}\frac{T_{\rm rh}}{m_\phi}\left(\frac{\phi_0}{M_{\rm Pl}}\right)^2\,,
\end{equation}
where $A$ is the soft trilinear $A$-term, $\phi_0$ the initial modulus displacement and $n_b/s$ the baryon-to-entropy ratio. Interestingly, the values $|A|\sim 1$ TeV, $m_\phi\sim 10^6$ GeV, $T_{\rm rh}\sim 1$ GeV and $\phi_0\sim \mathcal{O}(0.1-1)\,M_{\rm Pl}$, can naturally reproduce the observed baryon-to-entropy ratio $n_B/s\sim 10^{-10}$.  Affleck-Dine baryogenesis after (string) axion inflation was studied in \cite{Akita:2017ecc}, where it was found the axion oscillations could produce sufficient baryon asymmetry just after inflation even without a soft-supersymmetry breaking $A$-term.

\subsection{Non-standard cosmologies from D-branes}
\label{ssec:DST} 

From our discussion so far, it is clear that string theory cosmology models strongly motivate non-standard cosmological evolution. In particular, the epoch between the end of inflation and the onset of BBN remains highly unconstrained by current observations. Such non-standard evolution can be driven by the scalar fields in the theory as discussed above (see Fig.~\ref{fig:AlternativeHistory}).

Besides kination and moduli domination, a different scenario arises when D-branes are present. For D-branes moving in the internal compact space, the scalar field(s) associated to the open string representing, e.g.~the radial position of a D-brane along a warped throat, couples {\em disformally} to  matter living on such a brane \cite{Koivisto:2013fta}. Indeed, the induced metric on the brane takes the form $\tilde g_{\mu\nu} = C(\phi) g_{\mu\nu} + D(\phi) \partial_\mu\phi \partial_\nu\phi$, which is a particular case of the more general relation between two metrics in scalar-tensor theories introduced by Bekenstein in \cite{Bekenstein:1992pj}. This non-trivial coupling between matter and the scalar field can change the cosmological expansion rate, $\tilde H$ in the early evolution of the universe with interesting and potentially testable implications\footnote{See \cite{Allahverdi:2020bys} for a review on implications of non-standard cosmologies.}. Interestingly, for a `pure disformal' modification of the expansion rate, which arises for $C=1$, $D=D_0=$const.~from (stacks of) D-branes moving in the internal space in the early universe, the modified expansion rate can change the standard prediction for the production of thermal dark matter, giving earlier freeze-out temperatures and larger cross-sections compared to the standard case  \cite{Dutta:2016htz,Dutta:2017fcn}. 
Moreover, such an early period of D-brane scalar-tensor domination may be tested by its effect on the stochastic primordial gravitational wave background as shown in \cite{Chowdhury:2022gdc}. 
Due to non-standard cosmological expansion, the amplitude of the PGW spectrum can be enhanced over a wide range of frequencies covering various sensitivity ranges of future GW experiments such as LISA \cite{amaroseoane2017laser}, DECIGO \cite{Seto:2001qf}, and the Einstein Telescope \cite{Maggiore:2019uih}. Specifically, a disformally dominated epoch is characterised by a peaked spectrum with a {\em broken power-law} profile, with slopes that depend on the scalar-tensor theory considered \cite{Chowdhury:2022gdc} (see Fig.~\ref{Fig:DisformalGW}).  Moreover, if the scalar potential for the D-brane scalar field  satisfies suitable conditions, this scalar field can naturally source an early period of {\em early dark energy} \cite{Chowdhury:2023} (see sec.~\ref{subs:EDE}). Disformal D-brane scalar tensor theories are based on symmetries appearing in D-brane models with  interesting and potentially testable phenomenological implications. However, to fully explore their cosmological implications, it remains to embed  these scenarios into fully developed string constructions with moduli stabilisation. 

\begin{figure}[H]
\begin{center}
\includegraphics[width=10.00cm]{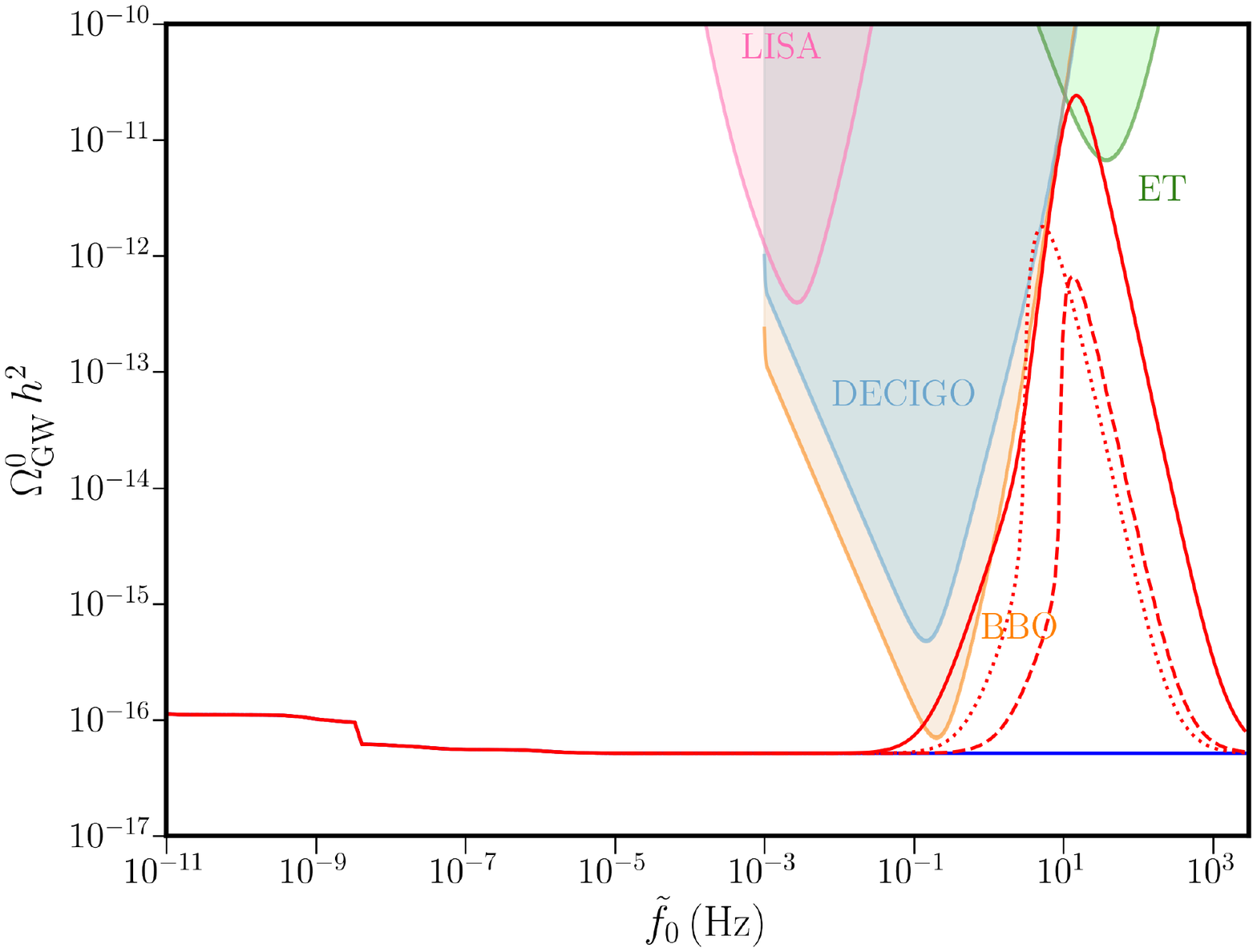}
\end{center}
\vskip -95pt
\caption{Gravitational wave spectrum for purely disformal D-brane scenarios with $C=1$, $D=D_0=4.822\times 10^{-37}{\rm GeV}^{-4}$, solid line. The vertical line indicates the frequency corresponding to the initial temperature ($\tilde T_i=10^{11}\,{\rm GeV}$). Figure taken from \cite{Chowdhury:2022gdc}, see this reference for further details.  (The dashed and dotted lines correspond to phenomenological cases discussed in \cite{Chowdhury:2022gdc}.)
}\label{Fig:DisformalGW}
\end{figure}

\subsection{Non-thermal Dark Matter from String Theory}

Post-inflationary eras of matter domination driven by late-time decaying string moduli have strong effects on dark matter. We briefly discuss here how moduli domination can modify the standard picture of the two most promising scenarios for dark matter: TeV-scale WIMPs and the QCD axion. We also comment on the possibility to realise fuzzy dark matter from string theory using ultra-light axions. Moreover we show how, in a given class of models, the underlying UV correlations between different aspects of string phenomenology can constrain the nature of dark matter. Finally we briefly comment on the string-inspired idea of dynamical dark matter.

\subsubsection{Non-thermal WIMPs}

Thermal WIMPs are arguably the most developed and best studied candidates for cold dark matter. In the standard paradigm they are assumed to be in thermal equilibrium in the early universe after inflation. Subsequently, dark matter drops out of thermal equilibrium and its abundance freezes out at a temperature of order $T_{\rm f} \sim m_{\rm DM}/20 \gtrsim 10$ GeV as annihilation becomes inefficient. This mechanism, even if it has been named the `WIMP miracle', requires a very specific annihilation cross section. For example, in the Minimal Supersymmetric Standard Model (MSSM), neutralino dark matter candidates typically give too much (for Bino-like neutralinos) or too little (for Higgsino/Wino-like neutralinos) relic density. 

However, from the string theory point of view, many standard thermal dark matter scenarios seem very hard to achieve since the generic late-time decay of long-lived moduli typically erases any previously produced dark matter relic density. Dark matter is instead 
produced non-thermally from the decay of the lightest modulus at $T_{\rm rh}$ \cite{Moroi:1999zb, Acharya:2008bk, Acharya:2009zt, Allahverdi:2013noa, 14091222, Allahverdi:2014ppa, Aparicio:2015sda, 150205406, 150804144, 151106768, Aparicio:2016qqb, Allahverdi:2018iod, Allahverdi:2020uax, 220106633, Cicoli:2022uqa}. The requirement to avoid any cosmological moduli problem and to obtain TeV-scale supersymmetry, typically yields stringy scenarios with $T_{\rm BBN} \sim \mathcal{O}(1)\,{\rm MeV} < T_{\rm rh}\sim \mathcal{O}(1)\,{\rm GeV} < T_{\rm f}\sim\mathcal{O}(10-100)\,{\rm GeV}$. 

Interestingly, non-thermal dark matter scenarios vastly enlarge the parameter space of particle physics models since the presence of an extra parameter, $T_{\rm rh}$, in the determination of the dark matter relic abundance, can allow portions of the parameter space which would be ruled out in the standard thermal case. Let us now analyse this point in more detail.

The abundance of any existing dark matter particle is diluted from the modulus decay by a factor which is at least of order $(T_{\rm f}/T_{\rm rh})^3$ that can easily be as large as $10^6$. Thus dark matter has to be produced from the modulus decay, which leads to a dark matter abundance of the form:
\begin{equation}
\frac{n_{\rm DM}}{s} = {\rm min}\left[\left(\frac{n_{\rm DM}}{s}\right)_{\rm obs}\frac{\langle\sigma_{\rm ann}v\rangle^{\rm th}_{\rm f}}{\langle\sigma_{\rm ann}v\rangle_{\rm f}}\left(\frac{T_{\rm f}}{T_{\rm rh}}\right),Y_\phi {\rm Br}_{\rm DM}\right],
\label{NonThDM}
\end{equation}
where $\langle\sigma_{\rm ann}v\rangle^{\rm th}_{\rm f}\simeq 3\times 10^{-26}\,{\rm cm}^2\,{\rm s}^{-1}$ is the value of the annihilation rate needed in the thermal scenario to reproduce the observed dark matter abundance:
\begin{equation}
\setlength\fboxsep{0.25cm}
\setlength\fboxrule{0.4pt}
\boxed{
\left(\frac{n_{\rm DM}}{s}\right)_{\rm obs} \simeq 5\times 10^{-11}\left(\frac{1\,{\rm GeV}}{m_{\rm DM}}\right),
}
\end{equation}
$Y_\phi \equiv 3 T_{\rm rh}/( 4 m_\phi)$ is the yield factor associated to the dilution due to the entropy released by the modulus decay, and ${\rm Br}_{\rm DM}$ denotes the branching ratio of the modulus decays into R-parity odd particles which decay to dark matter.

Depending on the annihilation cross section and the reheating temperature, two scenarios arise:

\begin{enumerate}
\item \textbf{Annihilation Scenario}:

In the \emph{Annihilation Scenario} the dark matter particles produced from the lightest modulus decay undergo some annihilation. This is described by the first term on the right-hand side of eq. (\ref{NonThDM}). This mechanism occurs when
\begin{equation}
\langle\sigma_{\rm ann}v\rangle_{\rm f}=\langle\sigma_{\rm ann}v\rangle^{\rm th}_{\rm f}\left(\frac{T_{\rm f}}{T_{\rm rh}}\right).
\end{equation}
Given that $T_{\rm rh} < T_{\rm f}$, the Annihilation Scenario can match the observed dark matter abundance only if $\langle\sigma_{\rm ann}v\rangle_{\rm f}>\langle\sigma_{\rm ann}v\rangle^{\rm th}_{\rm f}$ as in the case of Higgsino-like neutralinos which, due to their large annihilation cross section, tend instead to be underproduced in the standard thermal case. On the other hand, non-thermal production of Higgsino-like neutralinos from moduli decays can also yield the correct dark matter relic abundance for masses as low as a few hundreds of GeV \cite{Aparicio:2015sda, Aparicio:2016qqb}. 

\item \textbf{Branching Scenario}:

In the \emph{Branching Scenario} the final dark matter abundance is the same as the one produced from the lightest modulus decay since the residual annihilation of dark matter particles is inefficient. This mechanism is hence described by the second term on the right-hand side of eq. (\ref{NonThDM}). Clearly, this can happen if 
\begin{equation}
\langle\sigma_{\rm ann}v\rangle_{\rm f}<\langle\sigma_{\rm ann}v\rangle^{\rm th}_{\rm f}\left(\frac{T_{\rm f}}{T_{\rm rh}}\right).
\end{equation}
This condition is always satisfied when $\langle\sigma_{\rm ann}v\rangle_{\rm f}<\langle\sigma_{\rm ann}v\rangle^{\rm th}_{\rm f}$ as in the case of Bino-like neutralinos which instead tend to be overproduced in the standard thermal scenario due to the smallness of their annihilation cross section. Clearly, the Branching Scenario could be realised also for $\langle\sigma_{\rm ann}v\rangle_{\rm f}>\langle\sigma_{\rm ann}v\rangle^{\rm th}_{\rm f}$
if $T_{\rm f}/T_{\rm rh}$ is very large.
\end{enumerate}

\subsubsection{QCD axion}

First of all, let us stress that QCD axion models are UV sensitive (as they rely on a symmetry or periodicity to protect the axion potential against corrections raising the axion mass) and so any construction of these models can be trusted only by embedding them in a complete UV theory where it is known that the symmetry is maintained (e.g. string theory). 

In the context of field theory models of the QCD axion, where the axion arises as the phase of a scalar field that breaks a $U(1)_{PQ}$ symmetry, this UV sensitivity can be seen through the so-called \emph{axion quality problem} which is related to the fact that Planck-suppressed higher-dimensional operators with $\mathcal{O}(1)$ coefficients would make the QCD theta angle shift away from the vanishing value set by the standard QCD axion potential\cite{Kamionkowski:1992mf, Barr:1992qq, Holman:1992us}. This can be seen schematically as follows: in the simplest QCD axion model $\Phi =\rho \,e^{i \vartheta}$ is charged under the $U(1)_{\rm PQ}$ symmetry which is spontaneously broken by $\langle \rho\rangle = f_{\rm QCD} \neq 0$ and $\vartheta = a/f_{\rm QCD}$ is the pseudo-Goldstone boson playing the role of the QCD axion. Its potential is generated by QCD instantons and scales as
\begin{equation}
\setlength\fboxsep{0.25cm}
\setlength\fboxrule{0.4pt}
\boxed{
V \simeq \Lambda_{\rm QCD}^4 \left[1-\cos\left(\frac{a}{f_{\rm QCD}}\right)\right],
\label{VQCDax}
}
\end{equation}
which features clearly a minimum at $\langle a \rangle=0$, solving the strong CP problem. However higher-dimensional Planck-suppressed operators would tend to regenerate a non-zero effective QCD theta angle since
\begin{equation}
\setlength\fboxsep{0.25cm}
\setlength\fboxrule{0.4pt}
\boxed{
\Delta V \sim \frac{\Phi^n}{M_{\rm Pl}^{n-4}}\qquad\Rightarrow\qquad \Delta V \sim \frac{f_{\rm QCD}^n}{M_{\rm Pl}^{n-4}}\,\cos\left(\frac{n a}{f_{\rm QCD}}+\theta_n\right)
}
\end{equation}
would be sudominant with respect to (\ref{VQCDax}) for $f_{\rm QCD}\simeq 10^{12}$ GeV (the preferred value to reproduce the observed dark matter relic abundance without any tuning) only for $n\gtrsim 14$. 

One appealing aspect of string models of axions is that string theory naturally provides symmetries which forbid 
dangerous corrections to axion masses. When the QCD axion is realised as a closed string mode, it inherits geometric properties from the extra-dimensional compact manifold. The gauge symmetry associated to the higher-dimensional $p$-form whose reduction on an internal $p$-cycle gives the axionic mode descends to give the exact non-perturbative shift symmetry that protects the axion mass against large perturbative corrections. On the other hand, when the QCD axion instead occurs as the phase of an open string mode, this symmetry can be one of the effective global $U(1)$'s which are typical of D-brane constructions. The resulting symmetries are exact as discrete symmetries, but as continuous symmetries they are broken by non-perturbative effects. For recent discussions of the axion quality problem see  \cite{Dvali:2022fdv,Burgess:2023ifd}.

Another effect of string theory on axion physics is that moduli decays can dilute the QCD axion dark matter produced from the standard misalignment mechanism which leads to a relic abundance of the form:
\begin{equation}
\setlength\fboxsep{0.25cm}
\setlength\fboxrule{0.4pt}
\boxed{
\left(\frac{\Omega_{\rm QCD}}{\Omega_{\rm DM}}\right)\simeq \left(\frac{f_{\rm QCD}}{10^{12}\,{\rm GeV}}\right)^{7/6}\,\theta_{\rm in}^2\,.
\label{QCDabundance}
}
\end{equation}
Clearly, for natural $\mathcal{O}(1)$ values of the initial misalignment angle $\theta_{\rm in}$, the axion decay constant cannot be larger than $f_{\rm QCD}\lesssim 10^{12}$ GeV. However this upper bound can be pushed to larger values in the presence of a late-time epoch of moduli domination if the moduli decay after the formation of the QCD axion condensate. This can happen if the reheating temperature from the decay of the lightest modulus is below the scale of QCD strong dynamics, $1\,{\rm MeV}\lesssim T_{\rm rh }\lesssim \Lambda_{\rm QCD}\sim 200\,{\rm MeV}$. As shown in \cite{Kawasaki:1995vt, Fox:2004kb}, the new expression for the QCD axion relic abundance becomes:
\begin{equation}
\setlength\fboxsep{0.25cm}
\setlength\fboxrule{0.4pt}
\boxed{
\left(\frac{\Omega_{\rm QCD}}{\Omega_{\rm DM}}\right)\simeq 50
\left(\frac{T_{\rm rh}}{1\,{\rm MeV}}\right)\left(\frac{f_{\rm QCD}}{10^{16}\,{\rm GeV}}\right)^2 \theta_{\rm in}^2\,.
}
\end{equation}
Thus reheating temperatures close to BBN (as $T_{\rm rh}\sim 5$ MeV) can enlarge the typical axion window $10^9\,{\rm GeV}\lesssim f_{\rm QCD}\lesssim 10^{12}\,{\rm GeV}$, allowing for $f_{\rm QCD}\sim 5\times 10^{14}$ GeV without the need to tune the initial axion misalignment angle. This dilution mechanism is particularly important for models where the QCD axion decay constant is relatively large. A study of
 distributions of axion decay constants in the string landscape is \cite{Broeckel:2021dpz}.

\subsubsection{Fuzzy dark matter}

Another interesting scenario is fuzzy dark matter where dark matter is made of ultralight axion-like particles \cite{Hu:2000ke,Schive:2014dra,Hui:2016ltb,Hui:2021tkt}. The best candidate to realise fuzzy dark matter is an axion with mass around $10^{-22}$ eV and decay constant $f\sim 10^{16\div 17}$ GeV since the wave nature of such a particle can suppress kpc scale cusps in dark matter halos and reduce the abundance of low mass halos~\cite{Schive:2014hza,Schive:2014dra,Hui:2016ltb}. Regardless of the relevance of fuzzy dark matter to solve these observational problems (whose actual origin is under debate), it is interesting to focus on fuzzy dark matter since it has been claimed to arise naturally from string theory due to the smallness of its mass and the largeness of its decay constant. 

Ref. \cite{Cicoli:2021gss} analysed fuzzy dark matter from string theory studying how moduli stabilisation affects the masses and decay constants of different axions in type IIB compactifications. The result is that matching the whole observed dark matter abundance without tuning the axion initial misalignment angle is not a generic feature of 4-dimensional string models since it requires a slight violation of the weak gravity conjecture applied to axions. This can be easily seen as follows in the case of a single axion with Lagrangian: 
\begin{equation}
{\cal L}=\frac{1}{2}\,M_{\rm Pl}^2\,f^2(\partial \theta)^2-A\,M_{\rm Pl}^4\,e^{-S}\cos(\theta)\,,
\label{eq:AxionLagr}
\end{equation}
where $f$ is the axion decay constant and $S$ the instanton action. The axion mass for the physical axion $\phi=f\theta$ is then given by
\begin{equation}
m_\phi^2=A \, e^{-S}\,\frac{M_{\rm Pl}^4}{f^2}\,,
\label{eq:axionmassgen} 
\end{equation} 
which implies:
\begin{equation}
S\,f=-f\ln\left(\frac{m_\phi^2\,f^2}{A\, M_P^4} \right).
\label{eq:naivebound}
\end{equation}
Given that a GUT scale decay constant implies that the Peccei-Quinn symmetry is broken before inflation, the fuzzy dark matter abundance produced from the misalignment mechanism can be expressed as \cite{Cicoli:2012sz}:
\begin{equation}
\frac{\Omega_{\phi}h^2}{0.112}\simeq 2.2 \times \left(\frac{m_\phi}{10^{-22} \mbox{eV}}\right)^{1/2}
\left(\frac{f}{10^{17}\mbox{GeV}}\right)^2 \theta_{\rm mi}^2\,,
\label{eq:DMabundance}
\end{equation}
where $\theta_{\rm mi}\in [0,2\pi]$ is the initial misalignment angle. Assuming natural initial conditions, $\theta_{\rm mi}\sim\mathcal{O}(1)$, $100\%$ of dark matter is therefore reached for $m_\phi \simeq 10^{-22}$ eV and $f\simeq 10^{17}$ GeV. Substituting these values in (\ref{eq:naivebound}), for natural $\mathcal{O}(1)$ values of the prefactor $A$, one obtains $S\,f\simeq 10\,M_{\rm Pl}$ which requires a slight violation of the WGC bound $S\,f\lesssim M_{\rm Pl}$ \cite{Arkani-Hamed:2006emk, Alonso:2017avz,Hebecker:2018ofv}. 

Ref. \cite{Cicoli:2021gss} considered the case of $C_4$ axions, $C_2$ axions and thraxions \cite{Hebecker:2018yxs}. The results for $S\,f$ are summarised in Tab.~\ref{tab:closedaxions} and Fig.~\ref{fig:final_plot_bounds} shows the predictions confronted with present and forthcoming observations for an $\mathcal{O}(1)$ initial misalignment angle and different values of the microscopic parameters which lead to a controlled effective field theory. The outcome is that the best candidates to realise fuzzy dark matter in string theory are $C_2$ axions and thraxions, and to some extent also $C_4$ axions in certain limits.

\begin{table}[t!]
\begin{center}
		\centering
		\begin{tabular}{l | l c | c | }
\hline
		\cellcolor[gray]{0.9}  	Axion &  \cellcolor[gray]{0.9}    $S\,f/M_{\rm Pl}$          \\ [5pt]
			\hline\\ [-5pt]
			$C_0$ 	& $\lesssim 1/\sqrt{2}$      \\ [7pt]
			$B_2$	 & $\lesssim 1$  \\[7pt]
			$C_2$ 	   & $\left\{\begin{array}{ll}
				S_{\rm ED1}f\lesssim 1 \ \\
				S_{\rm ED3}f\lesssim \sqrt{g_s}\,\mathcal{V}^{1/3}
			\end{array}\right.$	  \\[15pt]
			$C_4$ 	    &  $\lesssim \sqrt{3/2}$ 	  \\[5pt]
			$C_{2,\text{thrax}}$ 	    & 
        $\lesssim \frac{3\pi M^3\sqrt{g_s} }{\mathcal{V}^{1/3}}$ \\[6pt]
			\hline
		\end{tabular}
\end{center}
		\caption{Bounds on $S\,f$ for different classes of closed string axions. $\mathcal{V}$ is the Calabi-Yau volume in Einstein frame, $g_s$ is the string coupling and $M$ is a flux quantum number in the throat. \label{tab:closedaxions}}
	\end{table}

\begin{figure}[ht]
    \centering
    \includegraphics[width = 0.9\textwidth]{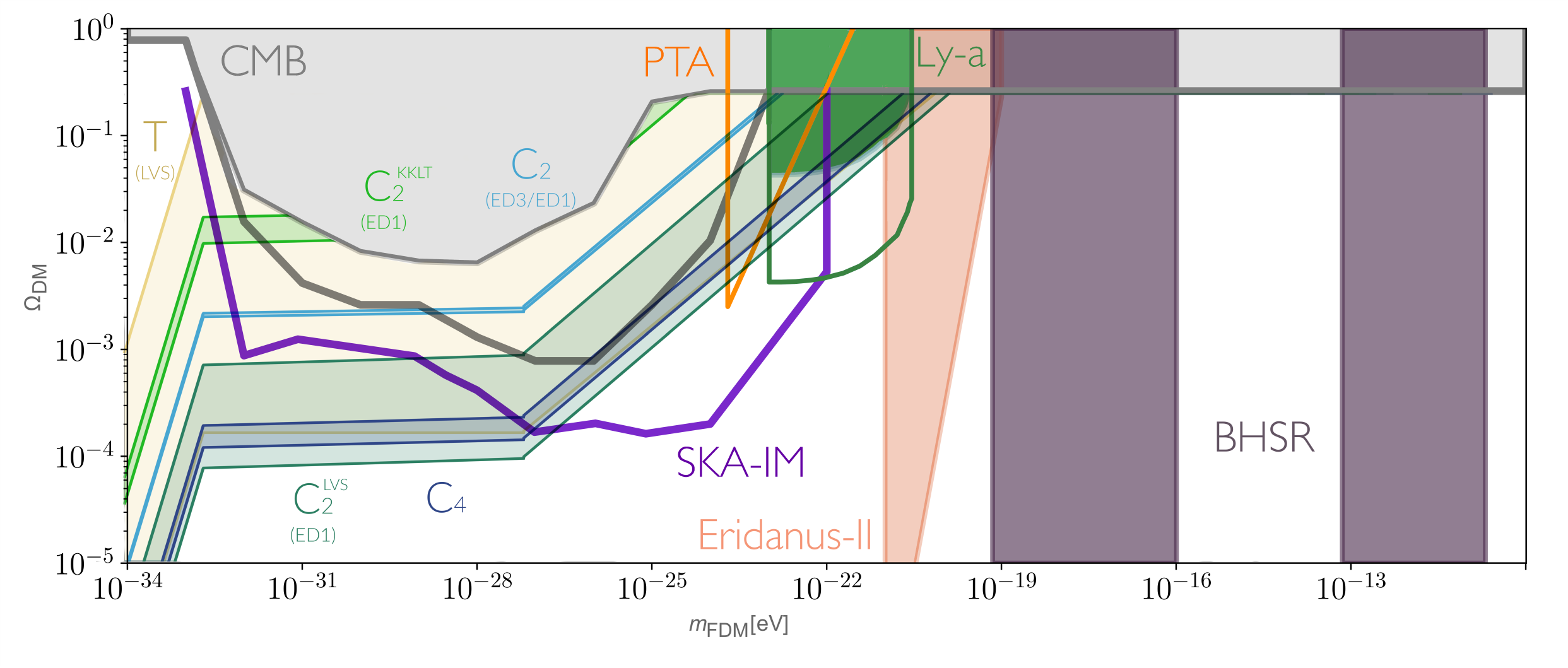} 
    \caption{Predictions for the mass and total dark abundance of $C_4$ axions (blue stripe), $C_2$ axions (light blue stripe for ED3/ED1 effects, dark/light green stripe for ED1 effects in LVS/KKLT), and thraxions in LVS (sand stripe). The results are compared to current and future experimental bounds. Figure taken from \cite{Cicoli:2021gss}.}
    \label{fig:final_plot_bounds}
\end{figure}

\subsubsection{Dark Matter and the Interplay between Inflation and Post-Inflation}

So far, we have discussed different aspects of string cosmology in a somewhat separate manner. However, one of the main features of any UV embedding is the presence of  correlations among different physical phenomena like inflation, supersymmetry breaking, reheating, dark radiation and dark matter. This non-trivial interplay can lead to distinct predictions in the context of certain scenarios. 

The implications of this interplay for most string models have not been explored in depth yet since this study requires a thorough control over moduli stabilisation, the underlying Calabi-Yau geometry and an explicit realisation of the Standard Model. The first attempts in this direction have been mainly performed within LVS string models of inflation where the inflaton is a K\"ahler modulus and the visible sector is realised either on D3-branes at singularities or on D7-branes wrapped around divisors in the geometric regime. We therefore focus on the LVS class of string inflationary models to illustrate how the interplay between different physical constraints can lead to precise predictions for the origin of dark matter.
\begin{itemize}
\item \textbf{Fibre Inflation with visible sector on D7-branes:}

We first mention Fibre Inflation \cite{Cicoli:2008gp} with the visible sector living on a stack of D7-branes wrapped around the inflaton divisor $\tau_{\rm fibre} = {\rm Re}(T_{\rm fibre})$ (similar considerations apply also to the case where the D7-branes wrap a blow-up mode). In this case, the overall volume is controlled by two divisors, $\mathcal{V}=\sqrt{\tau_{\rm fibre}}\tau_{\rm base}$. Matching the amplitude of the density perturbations fixes $\mathcal{V}\simeq 10^3$ which, in turn, gives a rather high inflationary scale, $H_{\rm inf}\sim M_{\rm Pl}/\mathcal{V}^{5/3}\sim 10^{13}$ GeV. Given that D7-branes do not lead to sequestering, the soft terms are of order the gravitino mass and tend to be very high, $M_{\rm soft}\sim m_{3/2} \sim M_{\rm Pl}/\mathcal{V} \sim 10^{15}$ GeV. Due to this high value, WIMP neutralino dark matter cannot work since it would lead to an overproduction both in the thermal and in the non-thermal case. Hence R-parity has to be broken and the lightest supersymmetric particle has to be unstable. Another potential dark matter candidate is the QCD axion which in this scenario is naturally reproduced by the closed string mode ${\rm Im}(T_{\rm fibre})= a_{\rm QCD}/f_{\rm QCD}$ with a large decay constant $f_{\rm QCD} \sim M_{\rm Pl}/\mathcal{V}^{2/3}\sim 10^{16}$ GeV. Since $f_{\rm QCD}>H_{\rm inf}$, the QCD axion is a flat direction during inflation and acquires large isocurvature fluctuations that are in tension with the present bound
\begin{equation}
H_{\rm inf}   \lesssim 10^{-5}\left(\frac{\Omega_{\rm DM}}{\Omega_{\rm QCD}}\right) \theta_{\rm in} f_{\rm QCD}\,,
\label{IsoBound}
\end{equation}
where $\theta_{\rm in}$ is the initial misalignment angle, $\Omega_{\rm DM}$ is the observed dark matter relic density and $\Omega_{\rm QCD}$ is the QCD axion abundance given by (\ref{QCDabundance}). For $f_{\rm QCD}\sim 10^{16}$ GeV, $\Omega_{\rm QCD}\sim \Omega_{\rm DM}$ for $\theta_{\rm in}\sim 0.01$. Substituting these values in (\ref{IsoBound}) would give $H_{\rm inf}\lesssim 10^9$ GeV which is clearly in contradiction with the fact that in Fibre Inflation $H_{\rm inf}\sim 10^{13}$ GeV. This implies that the QCD axion can form only a tiny fraction of the dark matter abundance. 

Another option is fuzzy dark matter from the ultralight ALP ${\rm Im}(T_{\rm base})= a_{\rm ALP}/f_{\rm ALP}$. However also this ALP has a very large decay constant $f_{\rm ALP} \sim M_{\rm Pl}/\mathcal{V}^{2/3}\sim 10^{16}\,{\rm GeV}> H_{\rm inf}$ which creates a tension with current isocurvature bounds. In fact, the ALP contribution to dark matter is given by (\ref{eq:DMabundance}) which can give $\Omega_{\rm ALP}\sim \Omega_{\rm DM}$ for $m_{\rm ALP}\sim 10^{-20}\,$ eV and $\theta_{\rm in}\sim \pi$. Substituting these values in (\ref{IsoBound}) would give $H_{\rm inf}\lesssim 10^{11}$ GeV which is again in contradiction with $H_{\rm inf}\sim 10^{13}$ GeV for Fibre Inflation. Hence also fuzzy ALP dark matter can at most account for $1\%$ of dark matter. 

The remaining options to realise dark matter in Fibre Inflation with visible sector on D7-branes seems therefore to be dark glueballs on a hidden sector wrapping a blow-up mode \cite{Halverson:2016nfq, Halverson:2018olu}, or primordial black holes (PBH). Ref. \cite{Cicoli:2018asa} has shown that the inflationary potential features enough tuning freedom to induce a near inflection point where an epoch of ultra slow-roll can enhance the amplitude of scalar fluctuations. This leads to PBH formation at horizon reentry in a way compatible with PBHs accounting for the whole dark matter abundance. Interestingly, this mechanism yields also the production of secondary gravity waves which might be observable in future interferometers \cite{Cicoli:2022sih}. 

Note that this scenario features two potentially dangerous ultra-light axions, the QCD axion ${\rm Im}(T_{\rm fibre})$ and the ALP ${\rm Im}(T_{\rm base})$, that could cause dark radiation overproduction from the inflaton decay. This is however not the case since the dominant branching ratio is the one associated to the inflation decay into the visible sector due to decays to gauge bosons \cite{Cicoli:2018cgu} and to Higgses via the enhanced loop-induced coupling in (\ref{LoopCoupl}). Hence in this model $N_{\rm eff}$ is in practice indistinguishable from its Standard Model value. Moreover, the reheating temperature from the inflation decay is $T_{\rm rh}\simeq 10^{12}$ GeV, leading to $N_e\simeq 53$ e-foldings of inflation required to solve the horizon problem.

\item \textbf{Fibre Inflation with visible sector on D3-branes:}

Let us now consider Fibre Inflation with the Standard Model realised on D3 branes at a singularity obtained by shrinking down an exceptional del Pezzo divisor $T_{\rm dP}$. This case has been studied in \cite{Cicoli:2022uqa}. Even if supersymmetry breaking is sequestered, the soft terms are still too high to allow for either thermal or non-thermal dark matter neutralino since $M_{\rm soft}\gtrsim m_{3/2}^2/M_{\rm Pl} \sim M_{\rm Pl}/\mathcal{V}^2 \sim 10^{11}$ GeV. The model features two potential candidates for fuzzy dark matter, given by the two ultra-light ALPs ${\rm Im}(T_{\rm fibre})$ and ${\rm Im}(T_{\rm base})$. However their decay constant is again rather high, $f_{\rm ALP}\sim M_{\rm Pl}/\mathcal{V}^{2/3}\sim 10^{16}$ GeV, implying that isocurvature bounds prevent them to form $100\%$ of dark matter. 

A promising dark matter candidate in this case is given instead by the QCD axion realised as the phase $\vartheta$ of an open string mode $\varphi = \rho\,e^{i\vartheta}$ living on the D3-brane stack at the del Pezzo singularity. Its decay constant is set by the VEV of $\rho$ which appears in the D-term potential:
\begin{equation}
V_D \simeq g^2 \left(\rho^2-\xi\right)^2 \qquad \Rightarrow \qquad \langle\rho^2\rangle = \xi \sim \frac{\tau_{\rm dP}}{\mathcal{V}}\,.
\label{Dstab}
\end{equation}
This relation fixes a combination of $\rho$ and $\tau_{\rm dP}$ which is given mainly by the del Pezzo divisor whose axionic partner is eaten up by an anomalous $U(1)$ \cite{Cicoli:2013cha}. The radial part of $\varphi$ is then stabilised by supersymmetry breaking contributions which scale schematically as:
\begin{equation}
V \simeq \pm \,m_0^2 \,\rho^2 + A \,\rho^3\,,
\end{equation}
where $m_0$ is the soft scalar mass and $A$ the soft trilinear coupling. If $\varphi$ is non-tachyonic, $\langle\rho\rangle=0$ which, in turn, fixes $\tau_{\rm dP}=0$ at the singularity without any axion left over. However, if $\varphi$ is tachyonic, its radial part acquires a non-zero VEV of order:
\begin{equation}
f_{\rm QCD} = \langle\rho\rangle \sim \frac{m_0^2}{A} \sim M_{\rm soft}\gtrsim 10^{11}\,{\rm GeV}\,.
\label{rhoVEV}
\end{equation}
This mass scale lies exactly in the right window for QCD axion dark matter. Moreover, given that $f_{\rm QCD} \sim 10^{11}\,{\rm GeV}< H_{\rm inf}\sim 10^{13}\,{\rm GeV}$, the Peccei-Quinn symmetry is unbroken during inflation, and so the axion is not constrained by any isocurvature bound. Note that, substituting (\ref{rhoVEV}) in (\ref{Dstab}) would give $\tau_{\rm dP}\sim \mathcal{V}^{-3}\ll 1$, with the blow-up mode still in the singular regime. 

In this model, reheating is driven by the perturbative decay of the inflaton fibre modulus $\tau_{\rm fibre}$ whose main decay channels are into the open string QCD axion $\vartheta$, the two closed string ALPs ${\rm Im}(T_{\rm fibre})$ and ${\rm Im}(T_{\rm base})$, and the Higgses of the MSSM. The 3 axionic degrees of freedom contribute to extra dark radiation which can be avoided only in the presence of a relatively large Giudice-Masiero coupling $Z\gtrsim 3$ since the loop-induced inflaton-Higgs coupling (\ref{LoopCoupl}) is in this case subdominant due to sequestering \cite{Cicoli:2022uqa}. The relevant contribution to the K\"ahler potential is:
\begin{equation}
K \supset Z\,\frac{H_u H_d}{\tau_{\rm base}^\lambda \tau_{\rm fibre}^{(1-\lambda)}}\,,
\label{GMFibre}
\end{equation}
which can induce an inflaton-Higgs coupling only if $\lambda\neq 1/3$, as suggested by explicit toroidal computations \cite{Aparicio:2008wh}, otherwise the denominator in (\ref{GMFibre}) would just be a function of the overall volume, implying an effective decoupling of the inflaton from the Higgs \cite{Angus:2014bia}. The final reheating temperature is around $T_{\rm rh}\sim 10^{10}$ GeV which requires $N_e\simeq 52$ e-foldings of inflation. 

\item \textbf{K\"ahler Moduli Inflation with visible sector on D7-branes:}

Let us now consider K\"ahler Moduli Inflation where the Hubble scale during inflation is lower, $H_{\rm inf}\sim 10^8$ GeV. We focus on cases where the Standard Model lives on D7-branes which can wrap the inflaton divisor or another blow-up mode which does not intersect with the inflaton \cite{Cicoli:2010ha}. Let us consider 3 cases separately:

\begin{enumerate}
\item \textbf{Inflaton cycle not wrapped by any D7-stack}: This case has been studied in \cite{Cicoli:2022fzy}. The inflaton blow-up mode $\tau_{\rm inf}$ is wrapped just by a Euclidean D3-brane instanton and the Standard Model lives on D7-branes wrapping a blow-up mode $\tau_{\rm SM}$ which does not intersect with $\tau_{\rm inf}$. The closed string axion ${\rm Im}(T_{\rm SM}$) plays the role of the QCD axion with a decay constant set by the string scale, $f_{\rm QCD}\sim M_s\sim M_{\rm Pl}/\sqrt{\mathcal{V}}\sim 10^{15}$ GeV, since $\mathcal{V}\sim 10^6$ in K\"ahler Moduli Inflation. In this case the QCD axion is a viable dark matter candidate which can satisfy present isocurvature bounds. In fact, from (\ref{QCDabundance}) we realise that $\Omega_{\rm QCD}\sim \Omega_{\rm DM}$ for $\theta_{\rm in}\sim 0.02$. Substituting this result in (\ref{IsoBound}), implies $H_{\rm inf}\lesssim 2\times 10^8$ GeV which is marginally in agreement with the inflationary scale of K\"ahler moduli inflation. Note again that neutralino dark matter would not work due to the high scale of the soft terms, $M_{\rm soft}\sim M_{\rm Pl}/\mathcal{V}\sim 10^{11}$ GeV. 

Let us now focus on dark radiation production. Even if the volume mode is the lightest modulus, it decays before the inflaton for two reasons: $(i)$ its loop-enhanced coupling to the Higgs (\ref{LoopCoupl}) which is the dominant coupling since $M_{\rm soft}\sim m_{3/2}$ for D7-branes; ($ii$) the inflaton geometric separation from the Standard Model suppresses its coupling to visible sector gauge bosons. Therefore reheating is driven by the decay of the inflaton modulus whose main decay channels are into: the volume mode $\tau_{\rm big}$ and its axionic partner ${\rm Im}(T_{\rm big})$, the blow-up mode supporting the Standard Model $\tau_{\rm SM}$ and its axionic partner ${\rm Im}(T_{\rm SM})$, and visible sector gauge bosons $A$. Subsequently, $\tau_{\rm big}$ decays into $A$ and ${\rm Im}(T_{\rm big})$, and $\tau_{\rm SM}$ decays into $A$ and the QCD axion ${\rm Im}(T_{\rm SM})$. The final prediction for extra dark radiation is very precise, $\Delta N_{\rm eff}\simeq 0.14$, and within current observational bounds. 

\item \textbf{Inflaton cycle wrapped by the SM D7-stack}: This case has been considered in \cite{Barnaby:2009wr,Cicoli:2010ha,Cicoli:2016olq}. The SM lives on D7-branes wrapped around the inflaton divisor $\tau_{\rm inf}$ whose axionic partner ${\rm Im}(T_{\rm inf})$ can be the QCD axion with $f_{\rm QCD}\sim 10^{15}$ GeV. This closed string QCD axion can form $100\%$ of dark matter in agreement with present isocurvature bounds, similarly to the previous case. As far as reheating is concerned, $\tau_{\rm inf}$ decays before the volume mode $\tau_{\rm big}$ which decays mainly into Standard Model Higgs degrees of freedom via the loop-enhanced coupling (\ref{LoopCoupl}). This guarantees the absence of extra dark radiation. Neutralinos need again to be unstable otherwise they would overproduce dark matter. 

\item \textbf{Inflaton cycle wrapped by a hidden D7-stack}: This case has been studied in \cite{Allahverdi:2020uax}. The inflaton divisor $\tau_{\rm inf}$ is wrapped by a hidden D7-stack and the Standard Model is built on D7-branes wrapping another blow-up mode $\tau_{\rm SM}$ which does not intersect with $\tau_{\rm inf}$. This case is particularly interesting since it can lead to the correct abundance of super-heavy neutralinos with mass of order $m_{\rm DM}\sim M_{\rm soft}\sim m_{3/2}\sim 10^{10-11}$ GeV thanks to two effects: ($i$) the initial production of neutralinos from the inflaton decay is tiny since the inflaton decays mainly into hidden sector degrees of freedom due to the geometric separation between $\tau_{\rm inf}$ and $\tau_{\rm SM}$; ($ii)$ the decay of the volume mode dilutes the neutralinos produced from the inflaton decay without being able to reproduce them since $m_{\tau_{\rm big}}<m_{\rm DM}$. Such a super-heavy dark matter candidate is particularly interesting since statistical studies of the supersymmetry breaking scale in the landscape seem to prefer higher scales of supersymmetry breaking \cite{Denef:2004cf, Broeckel:2020fdz} (see however the recent results obtained in \cite{Cicoli:2022chj}). Moreover, if exponentially suppressed R-parity violating couplings are induced by non-perturbative effects, the decay of a dark matter particle with $m_{\rm DM}\sim 10^{10-11}$ GeV could explain the recent observation of ultra-high-energy neutrinos by ANITA \cite{Heurtier:2019git, Dudas:2020sbq}. Dark radiation overproduction is again avoided thanks to the loop-enhanced volume-Higgs coupling (\ref{LoopCoupl}). 
\end{enumerate}

\item \textbf{K\"ahler Moduli Inflation with visible sector on D3-branes:} 

In K\"ahler Moduli Inflation the visible sector can also be realised on D3-branes at singularities with sequestered supersymmetry breaking. This model has been studied in \cite{Cicoli:2010yj, Cicoli:2012aq, 12083563, Allahverdi:2013noa, Allahverdi:2014ppa, Cicoli:2015bpq}. In this case the soft terms can be around the TeV scale, $M_{\rm soft}\sim M_{\rm Pl}/\mathcal{V}^2\sim \mathcal{O}(1-10)$ TeV for $\mathcal{V}\sim 10^7$, and so neutralinos are promising dark matter candidates. Their production mechanism is however non-thermal since reheating is driven by the volume mode decay with $T_{\rm rh}\sim\mathcal{O}(1)$ GeV. Dark radiation overproduction from the volume mode decay into ultra-light bulk axions can be avoided for relatively large values of the Giudice-Masiero coupling $Z\gtrsim 2$. The decay of the volume mode dilutes standard thermal neutralinos since $T_{\rm rh}< T_{\rm f}\sim m_{\rm DM}/20$ for $m_{\rm DM}\gtrsim \mathcal{O}(10)$ GeV. The dark matter production mechanism is the \emph{Annihilation scenario} and the WIMP is a Higgsino-like neutralino with $m_{\rm DM}\sim 300\,{\rm GeV}$ for the cMSSM \cite{Aparicio:2015sda} and, more in general, $300 \,{\rm GeV}\lesssim m_{\rm DM}\lesssim 850\,{\rm GeV}$ for any supersymmetric model with a Higgsino LSP \cite{Aparicio:2016qqb}.
\end{itemize}

An interesting observation emerges from this classification of dark matter candidates in LVS inflationary models: standard dark matter particles like TeV-scale WIMPs (even if non-thermal) and a QCD axion with an intermediate scale decay constant correlate with sequestered supersymmetry breaking and a slight tension with dark radiation overproduction (which requires a Giudice-Masiero coupling larger than unity), while non-standard dark matter candidates like primordial black holes, superheavy WIMPs or a QCD axion with a large decay constant correlate with no sequestering and in practice no dark radiation production due to the loop-enhanced volume coupling to Higgses.

A crucial task for string cosmology is to extend this investigation more in general to understand how the combination of UV correlations and different requirements like matching the observed amplitude of primordial density perturbations and dark matter abundance, isocurvature and dark radiation bounds together with supersymmetry breaking patterns, can constrain classes of models and single out different dark matter candidates.

\subsubsection{Dynamical Dark Matter}

The infinite towers of states that live in the string bulk interact with the Standard Model only gravitationally, and therefore are dark matter as far as the physics on the Standard Model brane is concerned. These states (which include string oscillator states, KK states, winding states, heavy moduli, etc.) are unstable, and so they do not behave as standard WIMP-like dark matter with a lifetime that exceeds the age of the universe. However, they can lead to a string-inspired alternative framework for dark matter physics, called `dynamical dark matter' where there are many towers of decaying bulk states throughout the history of the universe \cite{Dienes:2011ja, Dienes:2011sa, Dienes:2016vei}. This dark matter framework does not violate phenomenological constraints on the dark sector as long as the cosmological abundances of these states are properly balanced against their lifetimes.   

\subsection{Oscillons and Moduli Stars}

The ubiquity of scalar fields in string theory makes it natural to consider the implications of moduli domination as previously discussed. There is another generic aspect of scalar field potentials that may be very relevant in string theory, namely the possibility of having non-trivial solitonic configurations from moduli fields. Derrick's theorem states quite generally that energetic arguments make it impossible to obtain localised time-independent solutions of the Klein-Gordon equation. However, time-dependent solutions are allowed. In the case of a complex scalar field, the $U(1)$ symmetry makes it possible to find stable stationary solutions with the phase of the scalar being linear in time:  the {\it Q-Balls} found by Coleman are the standard examples of non-topological solitons for which  the solution is supported by the self-interactions of the 
scalar field.\footnote{Non-topological solitons differ from standard topological solitons, like kinks, vortices or skyrmions, where stability is guaranteed by the conservation of a topological charge. For non-topological solitons, the conserved charge is rather a Noether charge.} The non-linearities of the field equations prevent the fast dispersion of these localised configurations, giving rise to stable or long-lived compact objects that, if produced in the early universe, may lead to potentially observable effects such as gravitational waves. If the system is supported by gravity rather than the self-interactions of the scalar field,  then the configuration is known as a {\it boson star} or {\it oscillaton} and is a solution of the Einstein-Klein-Gordon equations.  

Contrary to standard stars in which pressure from internal nuclear fusion compensates the action of gravity to sustain the system, in boson stars this role is played by the Heisenberg uncertainty principle,  $\Delta x \Delta p\geq \hbar/2$ with $\Delta p\simeq m$ for a boson of mass $m$ and $\Delta x\simeq R_{\rm min}$ with $R_{\rm min}$ the minimum radius, $R\geq 1/m$ then prevents the system from collapsing. In this case there is a maximum mass $M_{\rm max}$ determined from the expression for the Schwarzschild radius $R_S=2GM$ so $M_{\rm max}\simeq M_{\rm Pl}^2/m$. This is different from fermionic stars for which $R_{\rm min}\simeq M_{\rm Pl}/m_{\rm f}^2$ and $M_{\rm max}\simeq M_{\rm Pl}^2/m_{\rm f}^3$ with $m_{\rm f}$ the mass of the fermion. Since the typical mass of a boson star is much smaller than that of a fermion star they are usually called {\it mini-boson stars}. For particular cases in which the scalar self-interactions are relevant (like in a quartic scalar potential) the maximum mass of a boson star would be of the same order as a fermion star. In this case the star is simply called a boson star. In this recent work \cite{Fox:2023aat}, the authors derived bounds on the parameters of  axion-like particles  based on features of axion stars.

For real scalar fields, long-lived quasi stable solutions have also been found, known as {\it oscillons}, depending on the self-interactions of the scalar field (see Fig.~\ref{fig:Oscillons}). If gravity is responsible for sustaining the compact system then these solutions are called {\it oscillatons}. See Tab.~\ref{tab:boson_stars} for a summary. For a recent review of oscillons and oscillatons including the many references to the original work and other reviews, see \cite{Visinelli:2021uve}.

\begin{table}
\begin{center}
\centering
\begin{tabular}{ | c | c | c |  }
\hline
\cellcolor[gray]{0.9}  {\bf Scalar} &  \cellcolor[gray]{0.9} {\bf Without gravity } &  \cellcolor[gray]{0.9} {\bf With Gravity}   \\
\hline \hline 
{Complex}   & Q-Balls  & (Mini) Boson Stars   \\
\hline
{Real}   & Oscillons  & Oscillatons  \\
\hline
\end{tabular}
\end{center} 
\caption {Summary of different solitonic configurations from moduli fields.}
\label{tab:boson_stars}
\end{table}

Let us start briefly reviewing the conditions for oscillon formation of a field $\phi$ with potential $V(\phi)$. The field starts oscillating  at a value $\phi_{\rm init}$ around its minimum at $\phi_{\rm fin}$. The Fourier modes of the perturbations $\delta\phi_k$ are determined by the equation
\begin{equation}
\setlength\fboxsep{0.25cm}
\setlength\fboxrule{0.4pt}
\boxed{
\delta{\ddot\phi_k}+3H\delta\dot\phi_k+\left(\frac{k^2}{a^2(t)}+V''(\phi(t)) \right) \delta\phi_k=0\,.
}
\end{equation}
If the perturbations $\delta\phi$ grow enough as the field oscillates around its minimum such that the non-linearities of the potential become important and the potential is shallower than quadratic (meaning that the interaction terms are attractive and compete with the mass term) then oscillons can be formed.

Typically this happens for the standard quartic potential that we can write in the form:
\begin{equation}
V(\phi)= m^2\Lambda^2\left[\frac12\left(\frac{\phi}{\Lambda}\right)^2+\frac{1}{3!}\frac{\lambda\Lambda}{m^2} \left(\frac{\phi}{\Lambda}\right)^3+\frac{1}{4!}\frac{g\Lambda^2}{m^2}\left(\frac{\phi}{\Lambda}\right)^4 +\cdots\right].
\end{equation}
Written in this way we identify $\phi \simeq \Lambda$ as the scale for which the interaction terms compete with the mass term. This means $\lambda\sim m^2/\Lambda, g\sim m^2/\Lambda^2$. In the simple symmetric case $\lambda=0$  the potential is shallower than quadratic if $g<0$, which means an attractive interaction. If the field starts oscillating at a value of $\phi\sim \Lambda$  the non-linearities in the potential allow for the formation of oscillons of mass $M\simeq \Lambda^2/m$\footnote{Note that the mass of the corresponding star is inversely proportional to the mass of the underlying particle, an interesting duality, and therefore the range of masses covers many orders of magnitude, from microscopic objects to macroscopic ones. Their stability is related to the stability of the corresponding particle.} and size $R\simeq 1/m$. Comparing this radius with the Schwarzschild radius $R_S\simeq M/M_{\rm Pl}^2$ we can see that gravity becomes relevant for $R\simeq R_S$ which implies $\Lambda\simeq M_{\rm Pl}$.

\begin{figure}[ht]
    \centering
    \includegraphics[width = 0.7\textwidth]{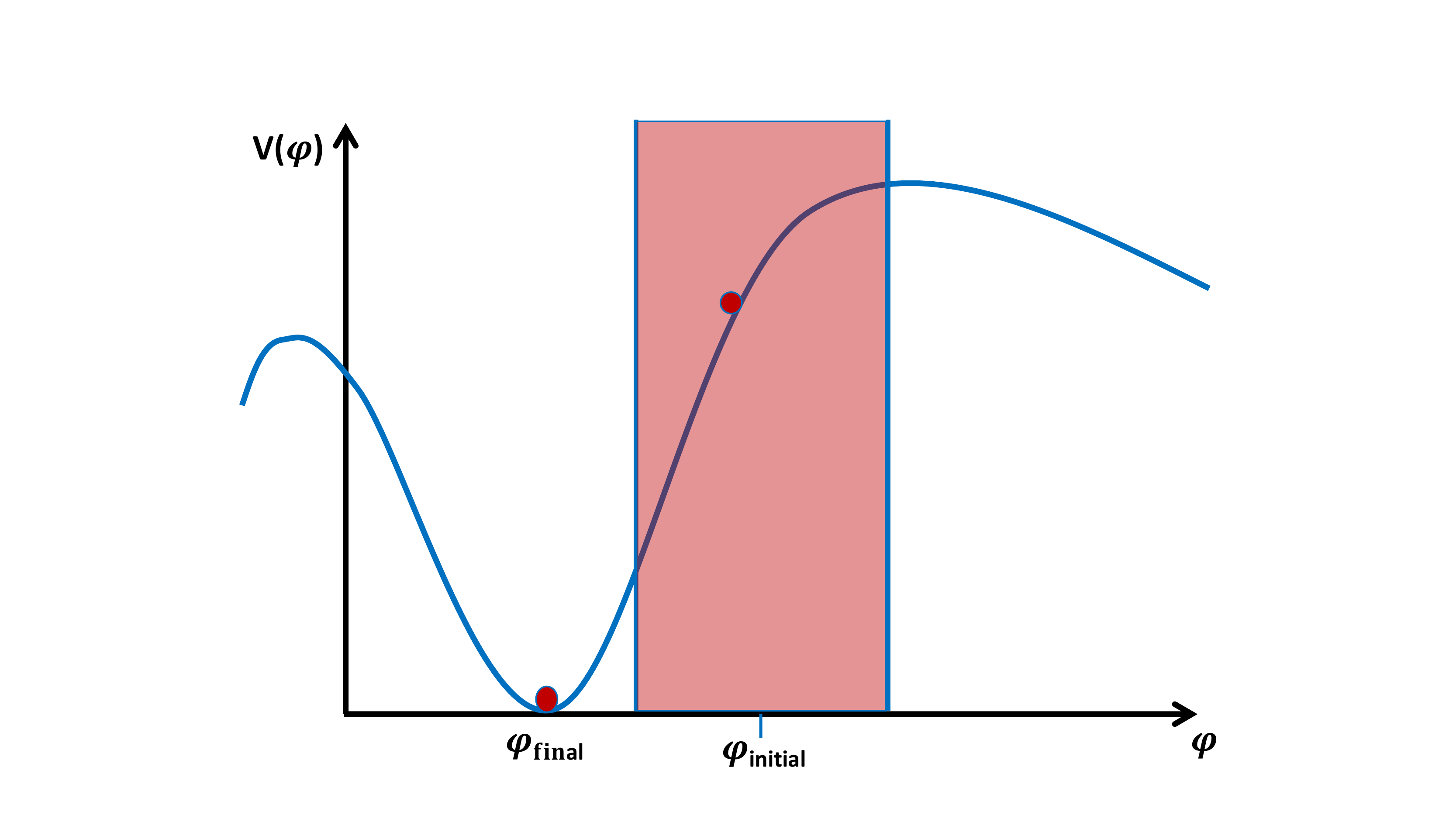} 
    \caption{Non-linear effects giving rise to oscillons.}
    \label{fig:Oscillons}
\end{figure}

Neglecting gravity ($\Lambda \ll M_{\rm Pl}$) oscillons can be formed  by {\it tachyonic reheating} with tachyonic oscillations for which the homogeneous field starts oscillating in the region where $V''<0$ (since the term tachyonic) and then the modes for which $\frac{k^2}{a^2(t)}+V''(\phi(t))<0$ will grow exponentially as it can be seen from the equation for the perturbations.

A second source for exponential growth of perturbations is {\it parametric resonance} for which the perturbation frequency $\omega^2_k=k^2/a^2+V''$ varies non-adiabatically ($|\dot\omega_k/\omega_k^2|<<1 $ is violated).

It is natural to ask if the scalar potentials computed from string compactifications such as KKLT and LVS models allow for the existence of oscillons and/or oscillatons. This has been done for oscillons in \cite{Antusch:2017flz} where it was found that oscillons can be formed in KKLT scenarios, as well as for blow-up modes in the LVS scenario (see Fig.~\ref{fig:Oscillons2}). For KKLT the mechanism is parametric resonance whereas for blow-up modes the mechanism is tachyonic oscillations. Large moduli, like the volume modulus or fibre moduli do not give rise to oscillons. For KKLT and blow-up modes the spectrum for gravitational waves was computed and found to be substantially different in both cases. In principle, gravitational waves (GW) could allow us to {\it hear the shape and size of the extra dimensions} by measuring the GW spectrum. However, the frequencies obtained naturally fall in the Giga Hertz regime, far beyond the reach of Earth interferometers such as LIGO and VIRGO and any future experiments which probe frequencies below the kilo Hertz (and also outside the range of LISA \cite{amaroseoane2017laser} or future space interferometers which will probe even smaller frequencies). 
\begin{figure}[ht]
    \centering
    \includegraphics[width = 0.7\textwidth]{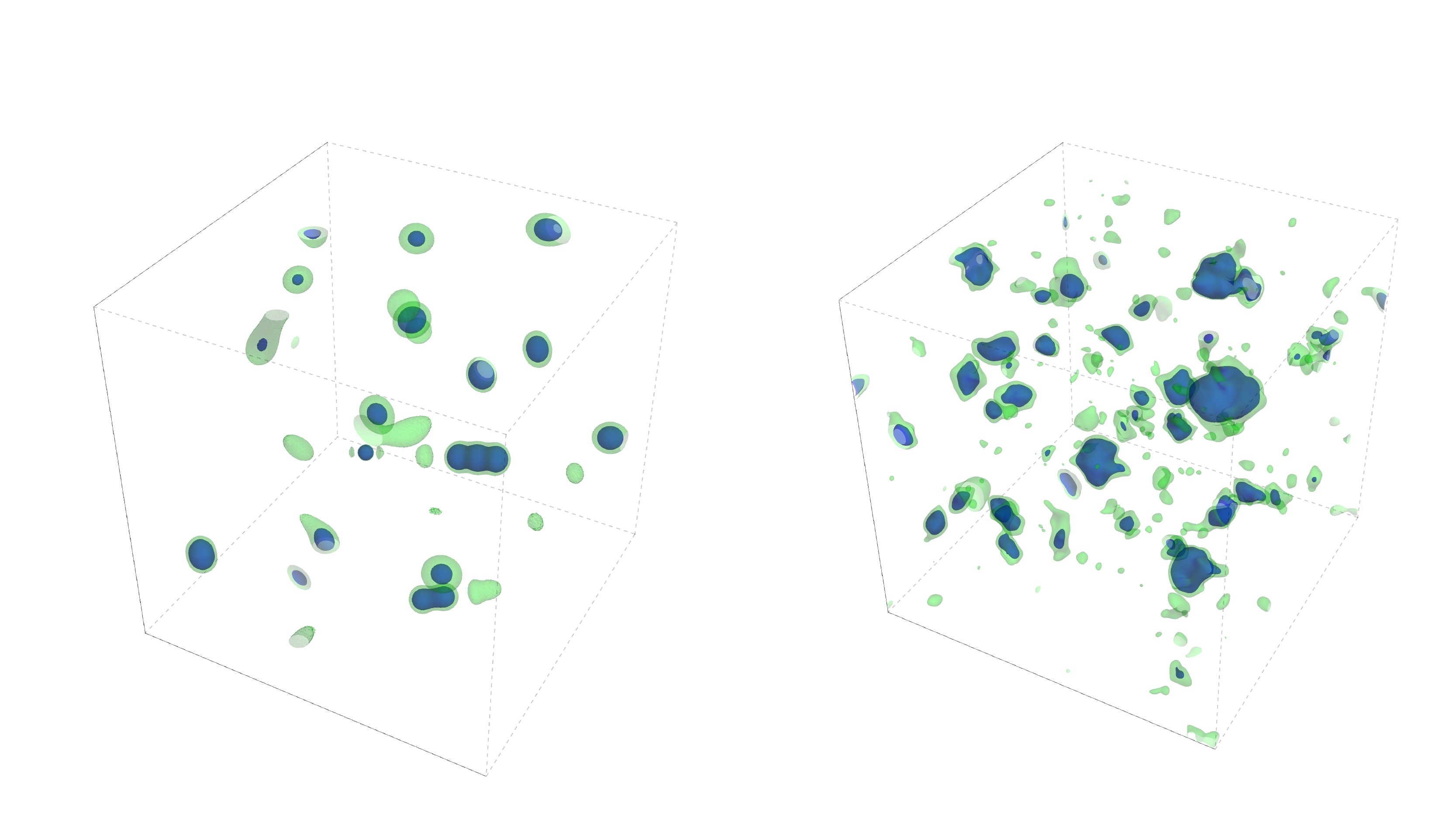} 
    \caption{3D modelling of oscillons for KKLT and blow-up in LVS.}
    \label{fig:Oscillons2}
\end{figure}

Such stringy oscillons are one of a large number of potential sources of gravitational waves of ultra high frequencies (UHF-GWs) at MHz range and above (see Fig.~\ref{fig:Oscillons3} for an example in KKLT and LVS models). Other sources are cosmic strings, phase transitions, preheating, boson stars, etc. Essentially, every model beyond the Standard Model may predict sources of GWs at high frequencies, with higher-energy processes giving higher frequencies (a rough rule of thumb is that GUT scale energies $10^{17}$ GeV correspond to GHz frequencies). Furthermore, usually the higher the frequency the smaller the potential experiment to detect them. However, the required sensibility (measured by the strain of the spectrum) increases with energies and therefore it is more challenging to detect them.  On the other hand, contrary to lower frequencies, there are no standard astrophysical sources at such frequencies and 
therefore any observation would imply either exotic astrophysics or a stochastic background of GWs with cosmological origin, hinting at physics beyond the Standard Model and early universe cosmology \cite{Ringwald:2020ist,Muia:2023wru}. This opens-up an interesting new challenge towards devising ways to search for UHF-GWs  (for a general discussion see \cite{Aggarwal:2020olq}; it is worth noticing that, even though there are many sources of UHF-GWs, see Fig.~\ref{fig:UHFGW},  it was the study of string cosmology that gave rise to this initiative).
\begin{figure}[ht]
    \centering
    \includegraphics[width = 0.8\textwidth]{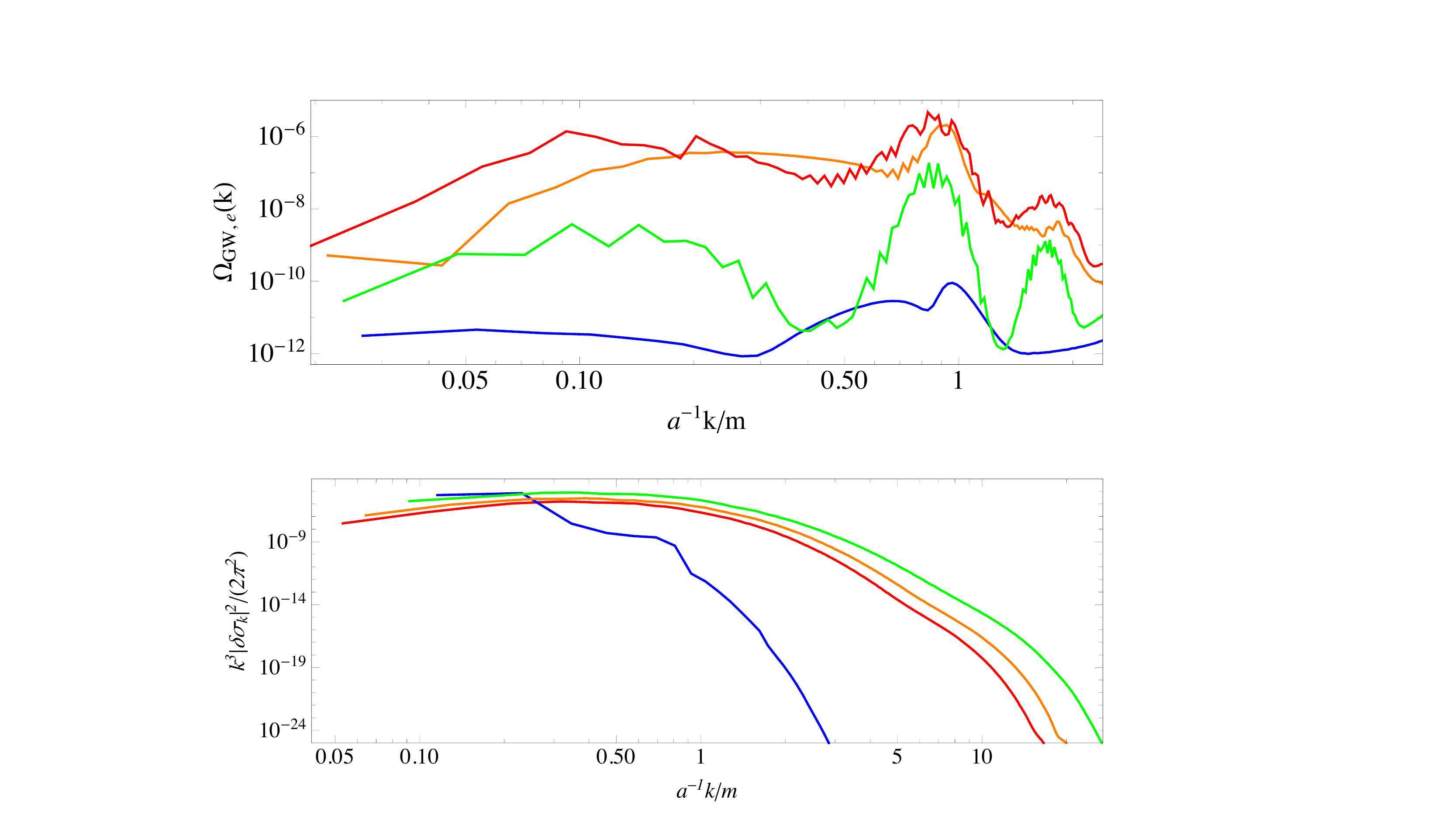} 
    \caption{Spectrum of GWs for KKLT and blow-up modes in  LVS at different moments in time \cite{Antusch:2017flz}. Even though  in both cases the gravitational waves are in the Giga Hertz range, the spectrum is different for different fields and scenarios. In principle potential observation of high frequency gravitational waves can differentiate among different scenarios.  }
    \label{fig:Oscillons3}
\end{figure}

For boson stars/oscillatons there is also the possibility that the inhomogeneities collapse to black holes. A full study of the Einstein-Klein-Gordon equations needs to be studied and recently developed codes for numerical relativity, such as GRChombo \cite{Andrade:2021rbd}, have been used to explore potentials like KKLT and LVS but also potentials appearing in axion monodromy. The growth in the energy density may lead in some cases to gravitational collapse and the production of primordial black holes. This is an interesting avenue of string cosmology that is only starting to be developed \cite{Krippendorf:2018tei, Muia:2019coe, Nazari:2020fmk}.

\begin{figure}[H]
    \centering
    \includegraphics[width = 1.1\textwidth]{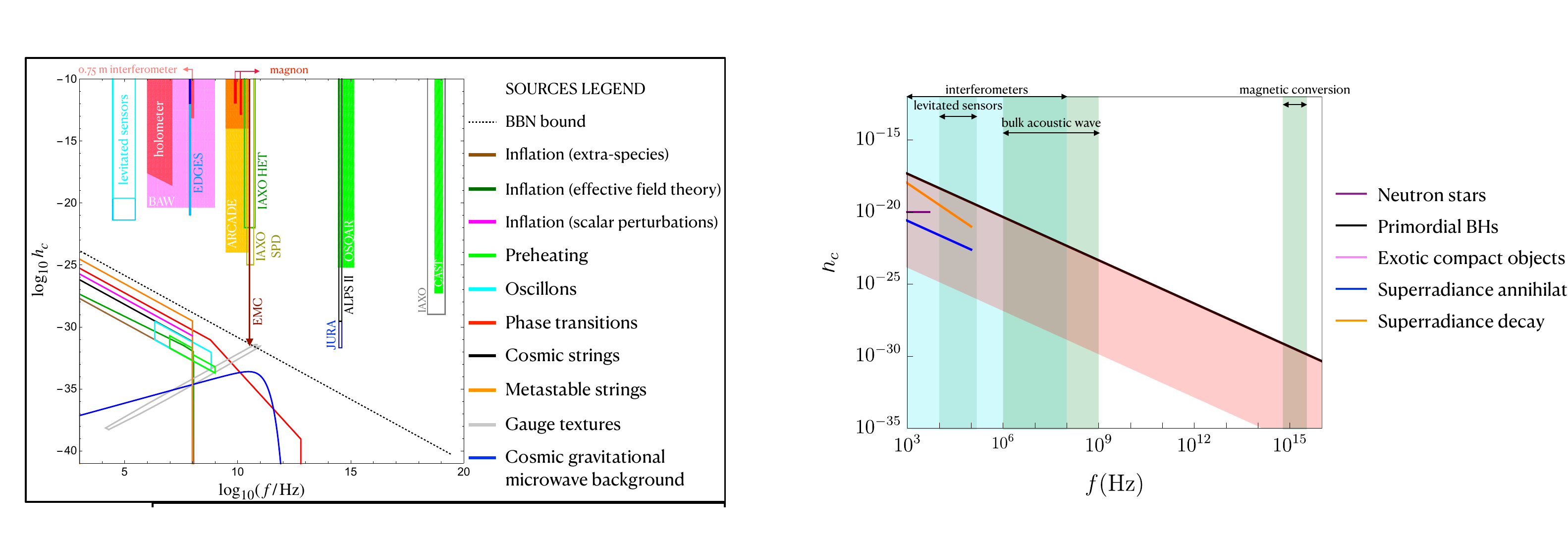} 
    \caption{Different sources of high frequency gravitational waves and the experimental constraints. Taken from \cite{Aggarwal:2020olq}.}
    \label{fig:UHFGW}
\end{figure}

\subsection{Cosmic Strings and Superstrings}

String theory is normally regarded as a theory of small scales and early times. However, there is one potential enormous exception to this: the possibility of a network of fundamental cosmic superstrings (or D1-strings) that stretch across the sky. For a long time, cosmic strings were regarded alongside inflation as one of the leading candidates to explain the origin of structure; in a field theory they have a natural origin as topological defects formed during processes of symmetry breaking in the early universe (as first described by Kibble in the 1970's \cite{Kibble:1976sj}).\footnote{One of the authors (JC) would like to note here that the town he lives in, Didcot, has streets named after Dirac, Higgs and Kibble.} Cosmic string were also appealing as a possible source of structure in the universe \cite{Zeldovich:1980gh, Vilenkin:1981iu}. Although observations of acoustic peaks in the CMB subsequently disfavoured cosmic strings as the primary source of structure in the universe, cosmic string networks could still exist at a subdominant level (CMB bounds on cosmic strings are described in \cite{Planck:2013mgr}).

Cosmic strings are, by convention, parametrised by a tension $G \mu$ (with $c=1$ and $G$ Newton's constant; although we normally use $\Mp$ in this review, here we conform to the standard convention for cosmic strings). 
Cosmic strings lose energy through radiation of gravitational waves and other light particles, forming a cosmic string network characterised by a scaling solution. These scaling solutions maintain a constant fraction of energy density in string relative to the overall universe during both matter and radiation epochs, with
\begin{equation}
\setlength\fboxsep{0.25cm}
\setlength\fboxrule{0.4pt}
\boxed{
\rho_{string} = \lambda \, G \mu \, \rho_{total},
}
\end{equation}
where $\lambda$ is an $\mathcal{O}(1 - 10)$ constant whose precise values depends on 
the detailed description of the network, and the possible energy loss channels. This is a complicated numerical problem (e.g. see \cite{Hindmarsh:1994re} for an older review and \cite{Gorghetto:2018myk} for more recent work). Observational bounds from potential modifications of the CMB power spectrum constrain
\begin{equation}
\setlength\fboxsep{0.25cm}
\setlength\fboxrule{0.4pt}
\boxed{
G \mu \lesssim 10^{-7}.
}
\end{equation}
The possibility of a cosmic string network consisting of fundamental strings was first considered in \cite{Witten:1984eb} with a negative conclusion: observational bounds on the tension of cosmic strings are incompatible with $ m_s \sim \Mp$ (and hence $\mu \sim G^{-1}$), and in the heterotic models then in vogue it is not possible to decouple the string and the Planck scales.

This conclusion has had to be revisited with the development of scenarios in which the fundamental string scale can be much less than the 4-dimensional Planck scale (such as in warped compactifications or in large volume compactifications such as LVS). For such models, there is no kinematic difficulty with the constraint $G \mu \lesssim 10^{-7}$: this bound is automatically satisfied in phenomenologically appealing versions of these scenarios.

Although this greatly ameliorates the kinematic constraints on the existence of networks of fundamental cosmic superstrings, one still needs a dynamical production mechanism. 
The most appealing such mechanism is brane-antibrane annihilation at the end of a period of brane inflation, which can lead to the formation of both D1-branes (i.e. D-strings) and fundamental strings, with $G\mu$ in the interesting range 
\be
10^{-12}\lesssim G\mu \lesssim 10^{-6}\,,
\ee
as discussed in \cite{Sarangi:2002yt, Jones:2003da}.
Cosmic string networks can only survive if the strings are themselves stable and do not immediately fragment or decay; a detailed analysis of such stability questions for cosmic superstrings is \cite{Copeland:2003bj}.
Cosmic strings are appealing both as a possible subdominant contribution to the energy density of the universe and 
possibly as  the sole available opportunity for the direct discovery of  macroscopic fundamental strings. 
One of the most promising routes by which cosmic (super)strings may be discovered is via gravitational radiation emitted from cusps, kinks or junctions, which may be detectable by future experiments, such as LISA \cite{amaroseoane2017laser}, and may be able to  detect loops with tensions as low as $G\mu \gtrsim 10^{-13}$ \cite{Damour:2000wa,Damour:2001bk,Damour:2004kw,Siemens:2006vk,Polchinski:2007qc}.  
Cosmic strings from a fundamental origin on the other hand, may be distinguishable from gauge theory cosmic strings by their network properties. 
For example, the reconnection probability $p$ of intersecting F- or D-strings might be much smaller than one, in contrast with ordinary strings, which have $p=1$ \cite{Jackson:2004zg}. Other effects due to (warped) extra dimensions on string evolution can help distinguish between standard and fundamental strings (see e.g.~\cite{Avgoustidis:2004zt,Blanco-Pillado:2005uwc,Avgoustidis:2007ju,OCallaghan:2010mtk,OCallaghan:2010jrm,Avgoustidis:2012vc}).
%
For further details, we refer the reader to the excellent dedicated reviews of cosmic strings and their properties \cite{Hindmarsh:1994re, Polchinski:2004ia, Copeland:2009ga}.

\newpage

\section{Dark Energy}
\label{sec:DE}

As we discussed in Sec. \ref{SecCO}, one of the most striking discoveries from the last century was the observational discovery of late-time cosmic acceleration from Type IA Supernovae (SN Ia) \cite{Riess:1998cb,Perlmutter:1998np}. This discovery represents one of the major puzzles of modern physics; its cause is generally dubbed {\em dark energy}, whose fundamental nature is still a mystery. According to observations, about $70\%$ of the energy density of the universe today consists of this unknown dark energy component. This has also been confirmed by other observations -- e.g. through Cosmic Microwave Background (CMB) \cite{WMAP:2003elm,Planck:2018vyg} or Baryon Acoustic Oscillations (BAO) \cite{SDSS:2005xqv} (however, see \cite{Sarkar:2007cx} for a dissenting view). 

Dark energy is characterised by an equation of state (see Sec.~\ref{SecCO})
\begin{equation}
\setlength\fboxsep{0.25cm}
\setlength\fboxrule{0.4pt}
\boxed{
w_{\rm DE} = p_{\rm DE}/\rho_{\rm DE} \,, \quad {\text{with}} \quad w_{\rm DE}< -1/3.
}
\end{equation}
Present observations suggest that the current energy density of dark energy is
$\rho_{{\rm DE}, 0} \sim 10^{-120}\,M_{\rm Pl}^4$, with an equation of state parameter given today by \cite{Planck:2018vyg}:
\begin{equation}
\setlength\fboxsep{0.25cm}
\setlength\fboxrule{0.4pt}
\boxed{
w_{{\rm DE}, 0} = -1.03 \pm 0.03 \,.
}
\end{equation}
 
The simplest candidate for dark energy, consistent with current data and used in the $\Lambda$CDM cosmological model, is a pure cosmological constant, $\Lambda$,  with $w_{\rm DE} =-1$. The cosmological constant can arise from the vacuum energy in particle physics, but the naive theoretical expectation for this is about $120$ orders of magnitude larger than the observed value \cite{Weinberg1}. 
To explain the extremely fine-tuned value of the cosmological constant, one possibility is to appeal to anthropic arguments  \cite{Weinberg2,Garriga:1999bf}, whose recent resurgence is motivated by the string theory landscape \cite{Douglas:2006es}. 

An alternative possibility is that the cosmological constant actually vanishes for reasons yet to be understood, calling for an alternative mechanism to explain the origin of dark energy.  The observations that constrain the value of $w_{\rm DE}$ today to be close to minus one (the value for a 
pure cosmological constant) say relatively little about its time evolution (see \cite{Planck:2018vyg} for constraints on a time-dependent equation of state parameter, using the phenomenological parameterisation $w = w_0 + (1-a)w_a$). So we can consider a situation in which the dark energy equation of state parameter changes with time, similarly to what happens during early universe inflation. As scalar fields naturally arise in supergravity and string theory, there are plenty of potential candidates for dark energy. Indeed, as we will see, many of the ideas discussed in Sec. \ref{sec:infla} to explain cosmic inflation from string theory can also be applied to the late-time cosmic acceleration. On the one hand, as only around a single e-folding of accelerated expansion is needed for dark energy, compared to $60$ e-foldings for inflation, constructing models of dark energy is easier. On the other hand, as dark energy is active today, it is constrained more strongly by  experiments and observations as well as by theoretical challenges as discussed in Sec.~\ref{sec:quint}.

We now consider recent developments using these two approaches to explain the current accelerated expansion of the universe within the context of string theory and string-inspired constructions.\footnote{For recent reviews see \cite{Berglund:2022qsb, Dutta:2021bih}.}

\subsection{Dark Energy as Vacuum Energy}

In this section we review the cosmological constant problem and the order of magnitude estimates of the vacuum energy.  We then describe how the string theory multiverse, using eternal inflation to populate it, has the potential to solve the cosmological constant problem, providing a (controversial) framework in which the fine-tuning of the vacuum energy is explained via anthropic arguments. Finally, we discuss the main questions that this paradigm leaves open.

\subsubsection{The cosmological constant problem}

The modern formulation of the cosmological constant problem actually raises two questions (see  \cite{Weinberg1, Carroll:2000fy, Padmanabhan:2002ji, Martin:2012bt, Li:2012dt, Burgess:2013ara,Padilla:2015aaa} for some reviews and \cite{Straumann:2002tv} for a historical account):
\begin{itemize}
\item Why is the cosmological constant so small but non-zero? 
\item The Coincidence Problem: why is the cosmological constant today comparable to the matter energy density today?
\end{itemize}
Phase transitions (as illustrated for bubble nucleation in figure \ref{fig:BubbleNucleation}) through the cosmological history would have changed the total vacuum energy, so any solution to the problem needs to ensure that the vacuum energy is suppressed throughout.  However, what makes the cosmological constant problem most challenging is that it is a low-energy problem, which can be posed at scales which we believe we understand very well. For instance, summing up vacuum-loop diagrams to a cutoff $M\sim 100$ GeV -- a reasonable scale given the success of the Standard Model in accelerator experiments -- the electron contributes:
\begin{equation}
\rho_e \sim \mathcal{O}\left(M^4\right) + \mathcal{O}\left( M^2 m_e^2 \right) + \mathcal{O}\left( m_e^4\ln \frac{M}{m_e}\right) \,,
\end{equation}
which is already around $55$ orders of magnitude too large. A recent estimate of the total vacuum energy from the Standard Model of particle physics, using a Lorentz invariant renormalisation scheme, gives a value for the energy density today of $\rho_\Lambda \sim -3.2 \times 10^8$ GeV$^4$ \cite{Koksma:2011cq,Martin:2012bt}. It is important to note that the gravitational effect of quantum loops has been experimentally observed \emph{in matter}; they are known to contribute to the inertial mass of particles via the Lamb shift and the nuclear electrostatic energy, where the equivalence principle has been verified to at least one part in a million \cite{Polchinski:2006gy}. Nonetheless, cosmological constraints on the vacuum energy would seem to indicate that quantum loops \emph{in the vacuum} do not gravitate. Although the length scale of the cosmological constant problem is huge, $H_0^{\;-1}$, as it occurs precisely at the point that quantum physics meets gravity, we might hope that a better understanding of string theory will offer a solution.

\begin{figure}[ht]
\centering
\includegraphics[width = 0.9\textwidth]{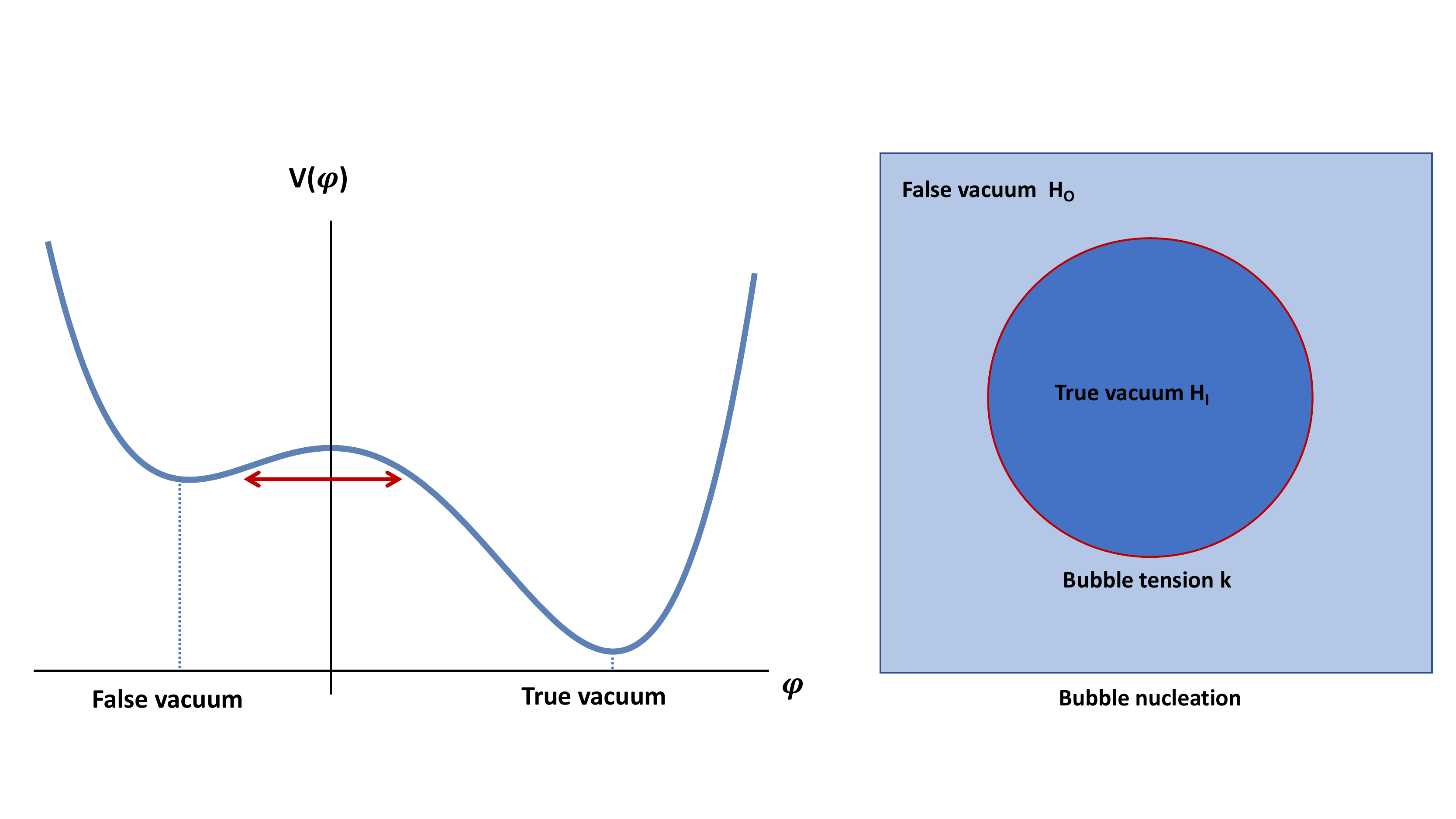} 
\caption{Bubble nucleation for a transition from a false to a true vacuum. The opposite transition is also allowed as long as the two spaces are dS.}
\label{fig:BubbleNucleation}
\end{figure}

\subsubsection{The anthropic principle and the string theory landscape}
\label{ssec:landscape}

Even without any microscopic understanding to hand, our very existence suggests that the cancellation of vacuum energy has to occur. If the vacuum energy were positive and much larger than the observed value, the growth of structure would have ceased too early preventing the formation of galaxies. If it were much larger and negative, the universe would have already collapsed in a Big Crunch. Remarkably, Weinberg \cite{Weinberg2} arrived at the \emph{prediction} that the vacuum energy must be small and of the same order as the matter energy density using this anthropic argument with bound:\footnote{Although note that cosmological constants much larger than the observed value were already known to be excluded.}
\begin{equation}
\setlength\fboxsep{0.25cm}
\setlength\fboxrule{0.4pt}
\boxed{
-10^{-123} M_{\rm Pl}^4 \lesssim \rho_\Lambda \lesssim 3 \times 10^{-121} M_{\rm Pl}^4 \,. 
}
\label{E:Weinwin}
\end{equation}
If one assumes that observers require galaxies, and that observers typically evolve soon after galaxy formation, then the \emph{Why Now?} problem is also solved, as allowing the vacuum energy to be as large as possible whilst allowing for galaxy formation places it in line with the matter energy density directly after galaxy formation.  For the argument to be complete, one also requires a theory to produce a multitude of vacua, including ones with anthropically viable total vacuum energies, and a mechanism to populate them all. This is what the string theory landscape \cite{Susskind:2003kw} is proposed to provide.

As reviewed in Sec. 3, string theory has a (probably finite \cite{Douglas:2003um, Ashok:2003gk} but) enormous number of solutions corresponding to 4-dimensional spacetimes at low energies, which we call the \emph{string theory landscape}. Each solution has a number of distinct contributions to the vacuum energy.  The value of the moduli potential energy discussed in Sec. 3 (which incorporates classical and leading order perturbative and/or non-perturbative effects) is one such contribution. To these should be added subleading quantum corrections in both the $g_s$ and $\alpha'$ expansions, amongst which those at $\mathcal{O}(\alpha'^0)$ incorporate standard field theory loops. By considering a string-inspired simplified model of such a setup, Bousso and Polchinski \cite{Bousso:2000xa} argued that the string theory landscape accommodates a discrete set of total vacuum energies that are sufficiently densely packed in Planck units to include the observed dark energy, and that moreover, all the corresponding vacua can be populated via an eternal inflation driven by non-perturbative bubble nucleation.

The discreteness of the distribution of vacuum energies arises from the topological nature of the input parameters for string compactifications -- flux quanta, D-brane and other localised source numbers, and an internal manifold characterised by its Hodge numbers -- which all contribute directly and indirectly to the vacuum energy.  Building on earlier work by Abbott \cite{Abbott:1984qf} and Brown and Teitelboim \cite{Brown:1987dd,Brown:1988kg}, Bousso and Polchinski \cite{Bousso:2000xa} (see also \cite{Bousso:2007gp} for a review) considered in particular the vacuum energy contribution from a set of $J$ 4-form background fluxes. In analogy to the electromagnetic field, each 4-form field-strength, $F_{(i)}$, is quantised in units of its source membrane's `electric'-charge $q_i$, $F_{(i)}^{mnpq} = n_i q_i \epsilon^{mnpq}$, and gives a positive vacuum energy contribution
\begin{equation}
\rho_{\rm 4-form} = \frac12 \sum_{i=1}^J n_i^2 q_i^2 \,.
\end{equation}
Crucially, the background 4-form fluxes are unstable to non-perturbative tunnelling effects described by Euclidean instantons, in analogy to how an electric field between two oppositely charged parallel capacitor plates discharges via Schwinger pair creation of electron and positrons, in the case of a 4-form, the depletion occurs via the spontaneous appearance of spherical membranes. Inside the membrane, the associated field strength is lowered by one unit of membrane charge: $n_iq_i \rightarrow (n_i-1)q_i$, with a subsequent reduction in energy, $(n_i-\frac12)q_i^2$ balanced by the initial mass of the membrane, which then expands outwards at the speed of light; see Fig. \ref{fig:BubbleNucleation}.  If one separates the total vacuum energy into the 4-form flux contributions and all the rest (which we may call the `bare' cosmological constant),
\begin{equation}
\rho_{\Lambda} = \lambda + \frac12 \sum_{i=1}^J n_i^2 q_i^2\,,
\end{equation}
it takes an exponentially long time for the
the background fluxes to deplete and so their vacuum energy contribution could eventually cancel an order one bare cosmological constant, $\lambda<0$, to give a total $\rho_\Lambda$ within the Weinberg window (\ref{E:Weinwin}).  This fine cancellation requires:
\begin{equation}
2|\lambda| \lesssim \sum_{i=1}^J n_i^2 q_i^2 \lesssim 2(|\lambda| + 10^{-118})\,.
\end{equation}
That this is possible for charges not much less than one relies on the fact that there are multiple distinct 4-form fluxes and associated membranes (see Fig. 1 from \cite{Bousso:2000xa}). These arise from M/string-theory compactifications as high-dimensional forms and branes, such as 5-branes wrapping distinct 3-cycles in the internal space. For example, an internal manifold with $\sim 500$ 3-cycles, with flux numbers ranging up to say $9$, would give $\sim 10^{500}$ different vacua for the construction. 

Whilst the Bousso-Polchinski toy model captures much of the physics of the landscape, realistic string constructions require further considerations \cite{Denef:2007pq}: moduli have to be stabilised; fluxes backreact on the geometry so that the bare cosmological constant $\lambda$ and charges $q_i$ themselves depend on the flux integers, leaving the cosmological constant to vary in unpredictable ways; the types and numbers of fluxes and branes that can be included are constrained by charge conservation or tadpole cancellation.  More sophisticated studies of the distributions of supersymmetric and non-supersymmetric Calabi-Yau flux compactifications of various string theories were carried out in \cite{Denef:2004ze, Douglas:2003um, Ashok:2003gk, Dienes:2006ut}. A recent development in this direction is the incorporation of K\"ahler moduli stabilisation, thereby ensuring that the sampling is only over the minima of the moduli potential (see e.g. \cite{Broeckel:2020fdz, Cicoli:2022chj} and references therein). Constructions with as many as $10^{272\,000}$ vacua have been proposed \cite{Taylor:2015xtz}, while F-theory models of dark energy have been proposed in \cite{Heckman:2019dsj, Heckman:2018mxl}.

In summary, string theory plausibly contains a landscape of vacua with a dense spectrum of vacuum energies that includes the observed vacuum energy, and a mechanism to populate them via membrane nucleation. The cosmological history of this scenario assumes that the universe starts with a large positive vacuum energy, expanding exponentially as dS space.  Eventually, somewhere within the universe, a membrane bubble will nucleate within which there is a lower vacuum energy.  This bubble will grow, but not as fast as the initial vacuum expands, the latter providing an ambient `multiverse'. Thus a process of eternal inflation ensues, with bubble nucleation continuing inside and outside existing bubbles, each bubble containing a long-lived open FRW universe and each jump in vacuum energy being not much smaller than one in Planck units, allowing for the creation of hot and dense universes. Eventually, one such bubble nucleation will create an anthropic universe, with the vacuum energy lowered to within the Weinberg range (\ref{E:Weinwin}) where the Big Bang begins.  See Fig. \ref{fig:Multiverse}.

\begin{figure}[ht]
    \centering
    \includegraphics[width = 0.99\textwidth]{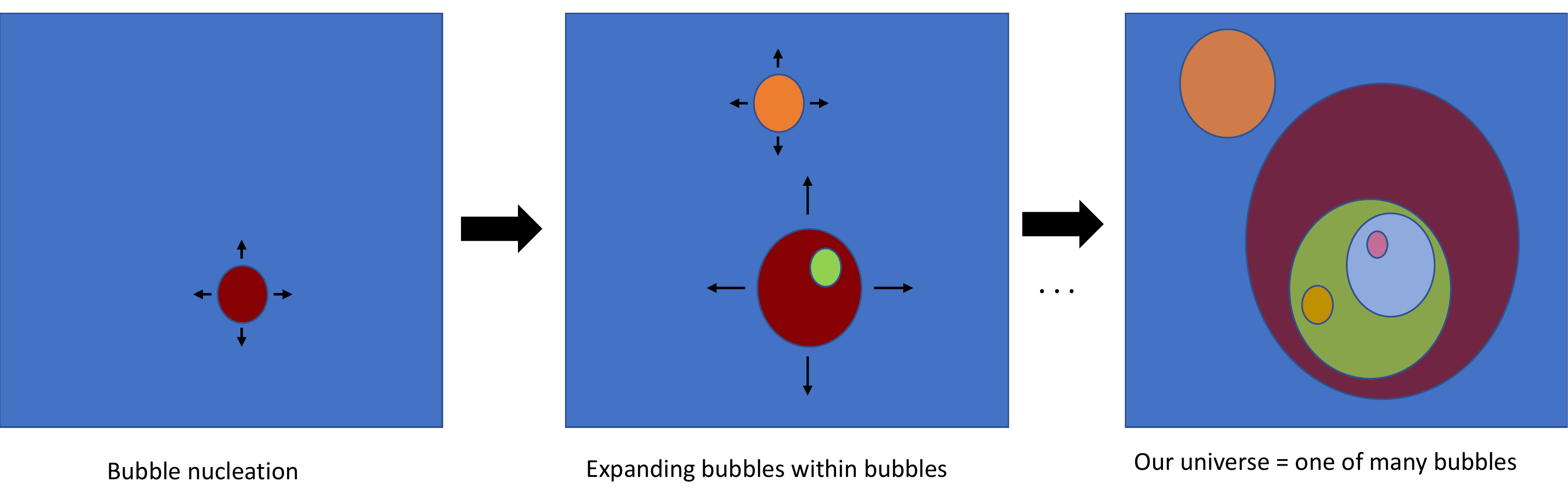} 
    \caption{Multiverse obtained by the process of bubble nucleation due to vacuum transitions from one dS vacuum to another with different cosmological constant.}
    \label{fig:Multiverse}
\end{figure}

\subsubsection{Open questions on the landscape}

The proposal of the string theory multiverse is deeply controversial. If true, it would motivates a paradigm shift in answer to the cosmological constant problem -- the value of the cosmological constant is environmental and our very existence implies it must happen to fall within the Weinberg range in our patch of the multiverse. In this case, there seems no reason to expect any further microscopic explanation at play. However, there remains work to be done.

Essential to the string theory landscape is an understanding of moduli stabilisation, reviewed in Sec. \ref{sec:MS}. So far, there are few attempts at achieving simultaneously moduli stabilisation and the Standard Model of particle physics (see \cite{Cicoli:2021dhg} for some recent progress). A complete model must (of course) include the vacuum energy of the Standard Model and dark matter amongst all the different contributions to be anthropically cancelled. 

To be more concrete, however, consider the simplified type IIB moduli stabilisation scenarios of KKLT and LVS. It is important to recognise that, although we can aim to make well-controlled dS vacua and -- remarkably -- even compute to good precision the vacuum energy in the weak coupling, weak curvature expansions, we cannot hope -- with current technology -- to obtain an explicit construction with fine-tuned vacuum energy $\mathcal{O}(10^{-120})$. The interplay between various classical and quantum effects are argued to give rise to metastable dS vacua, with the effective parameters such as $W_0$, $A$ and $a$ in $W =W_0+ A \, e^{-aT}$ (see Sec.~\ref{sec:MS}), ultimately depending on topological numbers such as flux integers and numbers of branes or instantons. The first obstacle arises from the difficulty to determine the exact dependence of parameters like $W_0$ and $A$ on the complex structure moduli and the explicit stabilisation of these moduli in terms of the discrete microscopic parameters. Moreover, even if discretely adjusting these integers allowed the effective parameters to be finely-tuned to give an anthropically viable metastable vacuum at weak coupling and large volume, the fine-tuning would typically be spoilt by higher order $g_s$ and $\alpha'$ corrections to the vacuum energy. To illustrate this point with a very simple toy model, consider a one-modulus system with canonical kinetic term and a scalar potential using an expansion in small $\phi$:
\begin{equation}
V(\phi) = \left(a_0 + a_2 \phi^2 + a_3 \phi^3 + a_4 \phi^4 + \dots\right)  \,.
\end{equation}
Assuming some mild hierarchy in the parameters $|a_3| \gg |a_i|$, $i\neq 3$, and $a_2 <0$, induces a metastable minimum $\langle \phi \rangle = -\frac{2 a_2}{a_3} \ll 1$ consistent with the expansion. If we assume that the leading coefficients are fine-tuned such that 
$$a_0-\frac{4}{27}\left(\frac{|a_2|^3}{a_3^2} \right) \sim 10^{-120},$$ 
then without further fine-tuning the higher order contributions $\mathcal{O}\left( \phi^4 \right)$ to the vacuum energy will dominate. Thus it is the full vacuum energy, including Standard Model/dark matter contributions and all $g_s$ and $\alpha'$ corrections -- right up to the order where the corrections are suppressed down to $10^{-120}\,M_{\rm Pl}^4$ without fine-tuning --  that has to be anthropically fine-tuned. This arguably weakens the significance of the challenges in uplifting the leading order AdS vacua of moduli stabilisation scenarios to dS (see \cite{deAlwis:2021zab}), although notice that the Standard Model contributions to the vacuum energy were found to be negative in \cite{Koksma:2011cq,Martin:2012bt}. Nevertheless, the key that the string landscape offers is the realisation that flux vacua lead to a finely-spaced distribution of vacuum energies over a suitably large range; even if the other contributions move this distribution up or down in energy, one may still expect vacuum energies of order of the observed dark energy amongst the final vacua.

Weinberg's anthropic window was derived by assuming that all other Standard Model parameters are fixed to their observed values. However, the string landscape suggests that all of these parameters would similarly vary from vacuum to vacuum.  For example, if effective field theory parameters are such that primordial density perturbations are larger or grow faster, anthropic arguments would require a larger positive vacuum energy. It is not yet clear how to use this framework to make predictions. Any statistical approach (see \cite{Denef:2004ze, Douglas:2003um, Broeckel:2020fdz, Broeckel:2021dpz, Cicoli:2022chj} and \cite{Kumar:2006tn} for a pedagogical review), by searching for those properties which have probability $\sim\mathcal{O}(1)$ in the landscape (see \cite{Page:2007bt, Hartle:2007zv} for discussions for and against our typicality) or identifying distinct low-energy features that are strongly correlated, would require an understanding of the measure on the space of vacua. Such a measure would need to include consideration of the dynamical mechanism that populates the landscape, which may prefer some vacua over others. This connects to the measure problem of eternal inflation, where bubbles of open universes with infinite spatial extent, reproduced infinitely many times, pose ambiguities in how to regularise \cite{Linde:1993xx, Guth:2000ka}.  

For some recent progress on searching the string landscape for models with phenomenological features using machine learning techniques see \cite{Cole:2021nnt}. Another strategy is to understand which classes of effective field theory can be consistently embedded into the string theory landscape (i.e. have a consistent UV completion into quantum gravity), and which classes instead lie in the swampland with no UV completion incorporating gravity \cite{Vafa:2005ui}. In fact, there are several conceptual challenges posed by any quantum theory on dS spacetime \cite{Witten:2001kn, Banks:2012hx, Maltz:2016iaw, Dvali:2018jhn}, most notably the absence of a well-defined S-matrix. Together with the technical challenges in achieving explicit, well-controlled moduli stabilisation with metastable dS vacua, this has led to the provocative conjecture that long-lived metastable dS vacua may actually lie in the swampland \cite{Garg:2018reu,Ooguri:2018wrx}. Although `may' does a lot of work in this last sentence, even if such a conjecture were true, it may still be of course reasonable to expect that our observed universe is sufficiently short-lived to stay in the landscape.

The landscape's approach to dark energy has led to several general proposals for combining it with the physics of vacuum transitions, a period of inflation, bubble collisions, etc. In particular, the general claim  that CDL vacuum transitions give rise to open universes after bubble nucleation has given rise to a holographic scenario for eternal inflation. See for instance \cite{Freivogel:2006xu,Sekino:2009kv}. For recent discussions to address the vacuum selection in terms of self-organised criticality see \cite{Kartvelishvili:2020thd,Giudice:2021viw}.

\subsection{Dynamical Dark Energy/Quintessence in String Theory}

In the previous section we discussed dS vacua in the string theory landscape as a possible explanation to present day accelerated expansion of the universe. However, given the debates around this proposal and observational prospects of measuring the dark energy equation of state and its time dependence, it is important to consider alternatives. The main alternative to a vacuum energy is a scalar field that is slow-rolling at positive energies, thus driving the accelerated expansion. This scenario goes under the generic name of {\em quintessence} \cite{Ratra:1987rm,Caldwell:1997ii} and much of the physics is similar to that of cosmic inflation, discussed in Sec. 4.
During quintessence, the equation of state of dark energy changes with time, similarly to inflationary cosmology, though initially Hubble friction can keep the field frozen yielding an effective cosmological constant. Quintessence scenarios with exponential or inverse power-law potentials are particularly attractive because they lead to equations of motion with attractor behaviour \cite{Ratra:1987rm} such that the present-day accelerated expansion is independent of the initial conditions. Moreover, these attractor solutions are usually scaling solutions such that the energy density scales as a power of the scale factor, and -- for the inverse power-law potentials -- the scalar field energy density can scale more slowly than the background fluid, allowing it to dominate eventually and drive the accelerated expansion \cite{Liddle:1998xm}.
 
Scalar fields abound in string theory and so it might seem rather natural to expect that one of them is at present rolling towards its minimum at a finite or infinite value. Among the scalar fields present in string theory, we can identify several potential candidates for quintessence: closed and open string string moduli, axions and runaway moduli. Moreover, scenarios where more than one scalar field drive the present day acceleration are also possible, as well as models of coupled dark sectors, where the dark energy and dark matter sectors are coupled. Time-dependent compactifications of the 10/11-dimensional supergravities that descend from string theory are also known to include accelerating cosmologies.

After reviewing the major challenges in the quintessence scenario, we will discuss each candidate in turn.

\subsubsection{Challenges for quintessence}
\label{sec:ChallengesQ}

Any quintessence scenario must meet a number of theoretical and observational challenges. Some of these are common to all quintessence models, irrespective of any string theory embedding (see \cite{Kolda:1998wq} for a review of phenomenological scenarios):
\begin{itemize}
\item \emph{Cosmological Constant Problem} -- quintessence scenarios start by assuming that some unknown mechanism fixes the background vacuum energy to zero, on top of which the quintessence dynamics plays out to drive the accelerated expansion. The symmetries that are known to do this job, e.g. supersymmetry or conformal symmetry, only work down to scales much higher than the observed vacuum energy, around the TeV scale, where they must be broken, although one exception may be the approximate shift symmetry associated with a pseudo-Goldstone boson \cite{Weinberg:1972fn}. The cosmological constant problem is the big elephant in the room of any quintessence construction. 

\item \emph{Radiative corrections to quintessence mass} -- for quintessence to drive an accelerated expansion, the quintessence scalar mass must be sufficiently light; $m_q \lesssim H_0$ with today's Hubble constant $H_0 \sim 10^{-33}$ eV.  On the other hand, as a scalar field, the quintessence mass is sensitive to quantum loops and -- in a similar way to the Higgs boson -- would be driven up to the UV cutoff. Again, symmetries like supersymmetry and conformal symmetry could help only down to TeV scales.  

\item \emph{`Why Now?' problem} -- Given the different scaling properties of radiation, matter and dark energy densities with the cosmological scale factor, $a(t)$, the current epoch -- in which all three of the energy densities are of the same order -- seems very special. Why does there exist an epoch in which all three densities are comparable, and why do we happen to live in this epoch? In quintessence models, the scalar field equation of motion may admit `tracker' solutions, in which the pressure and energy density in the scalar field tracks that of the dominant energy density \cite{Zlatev:1998tr, Steinhardt:1999nw}. Moreover, these tracker solutions may be late-time attractors, and hence be reached independently of the initial conditions.  However, in order for dark energy to come to dominate, the evolution must depart from the tracker solution so that the field ends up locked at an approximately constant value. To achieve this at the right time would seem to require some overall fine-tuning.

\item \emph{Fifth forces} -- Being a light boson, the quintessence field mediates long-range fifth forces. Current constraints \cite{Adelberger:2003zx} indicate that any scalar with mass less than around the meV scale must have weaker than Planckian couplings to the Standard Model. This would appear to rule out using one of the universal string moduli, such as the volume modulus or dilaton, as the quintessence field since these moduli couple to all fields with Planckian strength after Weyl rescaling to the Einstein frame. Note that even if such fields were to couple universally to matter, they would not simply lead to a renormalisation of Newton's constant as the nature of a spin-0 force is distinct from the nature of a spin-2 force. Non-universal moduli typically have couplings to the Standard Model that violate the equivalence principle, and such couplings are phenomenologically constrained to be at least a factor $10^{-11}$ weaker than gravity \cite{Damour:2010rp}. 
    
Another proposal is that fifth forces are screened in high ambient matter densities such as close to the Earth, due to non-minimal couplings to matter and/or certain derivative or non-derivative self-interactions \cite{Vainshtein:1972sx, Khoury:2003aq, Feldman:2006wg, Hinterbichler:2010es} (see \cite{Brax:2010gi, Hinterbichler:2010wu} for work towards embedding these mechanisms into string theory).  
    
Finally, pseudoscalars such as string axions, coupling as they do via derivative axial-current interactions of the form $\partial_\mu \theta (\bar{\psi} \gamma^\mu \gamma_5 \psi)$, need spin-polarised sources in order to be detectable via fifth forces and so evade fifth-force experiments using macroscopic bodies (searches for new mass-spin couplings are reviewed in \cite{Marsh:2015xka, Workman:2022ynf} with the strongest constraints coming from stellar cooling \cite{Raffelt:2012sp}, still far from the parameter space of quintessence).  In fact, axions can also be used to help screen scalar fifth forces \cite{Burgess:2021qti,Brax:2022vlf}; the non-linear target-space interactions between axions and saxions that typically appear in string constructions might convert would-be dilaton profiles to axion profiles, which can then be probed only by axion-matter couplings rather than dilaton-matter couplings. 

\item \emph{Time variation of fundamental constants} --  if quintessence is a string modulus that sets the visible sector gauge kinetic functions, Yukawa couplings or Planck mass, then its rolling would lead to unobserved time-variation of the fundamental constants. Similarly to fifth forces, this disfavours closed string moduli like the volume or dilaton as quintessence.  The problem may be reduced by using a local modulus geometrically sequestered from the visible sector. 

\end{itemize}
Quintessence models must also satisfy observational constraints, not just on the equation of state $w$ but also on compatibility with local $H_0$ measurements (for example, see
\cite{Colgain:2019joh, Banerjee:2020xcn, Lee:2022cyh, Heisenberg:2022lob, Heisenberg:2022gqk,Colgain:2022rxy} for recent discussions).

In addition to these overarching questions common to all quintessence scenarios, there are also a number of issues specific to string theoretic models:
\begin{itemize}
\item \emph{Moduli stabilisation problem} -- even supposing a light ($m_q \lesssim H_0 \sim 10^{-60}\,M_{\rm Pl}$), slowly-rolling modulus can be identified, all other moduli must be safely stabilised in a way compatible with the string mass above the TeV-scale (see \cite{Hebecker:2019csg} for a recent discussion). Given constraints from fifth-forces and time variation of fundamental constants, the stabilised moduli must include the overall volume modulus and the dilaton with $m \gtrsim 10^{-30}\,M_{\rm Pl}$, which increases to $m \gtrsim 10^{-14}\,M_{\rm Pl}$ when the cosmological moduli problem is taken into account. Possible quintessence candidates are then ratios of K\"ahler moduli or blow-up moduli that (hardly) affect the overall volume, complex structure moduli or axions. Moreover $M_{\rm soft} \gtrsim 10^{-15}\,M_{\rm Pl}$ from the absence of sparticles at the LHC, and $M_{\rm KK} \gtrsim 10^{-30}\,M_{\rm Pl}$ from tests of Newton's inverse square law (see \cite{Hebecker:2019csg} for further discussion in the context of the Large Volume Scenario, where the volume modulus may be used to suppress scales but is then also too light itself).

These phenomenological requirements imply a large hierarchy between the potential which stabilises the volume modulus $\mathcal{V}$, $V_0(\mathcal{V})$, and the one for the quintessence field $\phi$, $V_1(\mathcal{V},\phi)$ (recall the volume modulus couples to everything) \cite{Cicoli:2021skd}. In fact, the total dark energy potential should look like
\begin{equation}
V_{\rm DE}\simeq V_0(\mathcal{V}_{\rm min})+V_1(\mathcal{V}_{\rm min}, \phi)\simeq V_1(\mathcal{V}_{\rm min}, \phi)\,,
\end{equation}
where $\mathcal{V}$ needs to stabilised in a near-Minkowski vacuum, $V_0(\mathcal{V}_{\rm min}) \sim 0$, since $V_1$ is a tiny correction with respect to $V_0$ to guarantee that the mass of the volume mode is at least above the meV-scale while $m_q \sim 10^{-32}$ eV, and so a leading order stabilisation in an AdS vacuum would remain AdS even after adding $V_1$. Thus we require
\begin{equation}
\frac{V_1(\mathcal{V}_{\rm min}, \phi)}{V_0(\mathcal{V}_{\rm max})} \sim \left(\frac{m_q}{m_{\mathcal{V}}}\right)^2\lesssim 10^{-60}\,,
\label{QuintHierarchy}
\end{equation}
where $V_0(\mathcal{V}_{\rm max})$ is the value of $V_0$ at the potential barrier towards decompactification. Note that this is a huge hierarchy, which can hardly be obtained if both $V_0$ and $V_1$ are generated by perturbative corrections since it would require values of $\mathcal{V}$ too large to ensure $M_s\gtrsim 1$ TeV. As an illustrative example consider LVS models where $\mathcal{V}$ is fixed by $\mathcal{O}(\alpha'^3)$ effects which therefore set the size of $V_0$. If $\phi$ is a fibre modulus different from the inflaton, its potential would be generated by $\mathcal{O}(g_s^2 \alpha'^4)$ string loops which would set the size of $V_1$, yielding:
\begin{equation}
\frac{V_1}{V_0} \sim \frac{V_{g_s^2 \alpha'^4}}{V_{\alpha'^3}} \sim \frac{1}{\mathcal{V}^{1/3}}\lesssim 10^{-60}\qquad \Leftrightarrow\qquad 
\mathcal{V}\gtrsim 10^{180}\,,
\end{equation}
which would imply an extremely low string scale $M_s \sim M_{\rm Pl}/\sqrt{\mathcal{V}}\ll 1$ TeV. This shows that constructing a scalar potential with the hierarchy of scales as in (\ref{QuintHierarchy}) is a challenge. Note that in models where $\mathcal{V}$ does not evolve from inflation to today, this hierarchy could be even larger since preventing volume destabilisation during inflation (a quintessence version of the Kallosh-Linde problem) requires $\left(V_1/V_0\right)\lesssim \left(H_0/H_{\rm inf}\right)^2$. Depending on the inflationary scale, this yields $10^{-108}\lesssim \left(V_1/V_0\right)\lesssim 10^{-36}$ \cite{Cicoli:2021skd}. As pointed out in \cite{Cicoli:2021skd}, a natural way to achieve this huge hierarchy could be to fix $\mathcal{V}$ by perturbative effects, and then use axions as quintessence fields since their potential is generated by exponentially suppressed non-perturbative effects. Being axions, they also naturally avoid fifth-force problems, and their shift symmetry guarantees the radiative stability of the quintessence field mass. 
   
\item \emph{F-term problem} -- This is a rephrasing of the cosmological constant problem for quintessence models in the context of LVS moduli stabilisation \cite{Hebecker:2019csg}. The supersymmetry breaking in the Standard Model sector, say localised on some brane configuration, gives a large positive contribution to the scalar potential, $\delta V_{sb} \sim M^2 M_{\rm soft}^2 \sim 10^{-60}\,M_{\rm Pl}^4$, where $M$ is the mediation scale with $M\gtrsim 10^{-15}\,M_{\rm Pl}$. Clearly, in the absence of a source to cancel this large contribution, the vacuum energy would be way beyond the dark energy scale (see \cite{Cicoli:2012tz} for a scenario in which this large vacuum energy is supposed to finely cancel against contributions from the backreaction of non-supersymmetric visible sector branes). More generally, after supersymmetry breaking, the mass of the quintessence field would typically be of order the gravitino mass (see \cite{Chiang:2018jdg} and \cite{Cicoli:2018kdo} for some ways to evade this).
    
\item \emph{Sequestering in supergravity} -- Within the context of 4-dimensional $N=1$ supergravity, the usual low-energy description of string compactifications, it is difficult to suppress couplings between the quintessence field and Standard Model degrees of freedom such as the Higgs \cite{Denef:2018etk, Cicoli:2018kdo}. The scalar potential has to take the form 
\begin{equation}
V(\Phi, \chi)=e^{K(\Phi, \chi)}(|D_\chi W|^2 + |D_\Phi W|^2 - 3|W|^2),
\end{equation}
with $\Phi$ the quintessence superfield and $\chi$ denoting (collectively) the matter superfields. Assuming a maximal decoupling with K\"ahler potential and superpotential taking the form 
\begin{equation}
K=K_q(\Phi)+K_m(\chi),
\end{equation}
and 
\begin{equation}
W=W_q(\Phi) + W_m(\chi),
\end{equation}
and moreover taking a canonical normalisation $K_q = \Phi \bar{\Phi}$ and e.g. $W_q=\frac{2}{\beta}\sqrt{\Lambda} e^{-\frac12 \beta \Phi}$ so that \begin{equation}
V\sim \Lambda e^{-\beta \phi},
\end{equation}
one obtains couplings of the form 
\begin{equation}
\delta V \sim \frac{\bar{\Phi}_0}{M_{\rm Pl}^2} \delta\Phi |D_\chi W|^2, \qquad \delta \mathcal{L} \sim \frac{\bar{\Phi}_0}{M_{\rm Pl}^2} \delta\Phi \mathcal{L}_{\rm fermion}, \qquad \delta \mathcal{L} \sim -\frac{3T(G)}{16\pi^2} \frac{\Phi_0}{M_{\rm Pl}^2} \delta \Phi \mathcal{L}_{\rm gauge},
\end{equation}
(where $T(G)$ is a group theory number). 

Even if we choose $\Phi_0=0$ to evade present fifth forces constraints, at some point in the past we would have had $\Phi_0 \sim \mathcal{O}(1)M_{\rm Pl}$, giving Planck suppressed linear couplings between the quintessence field and the Standard Model for which there are strong bounds (see \cite{Chiang:2018jdg} for a recent example of quintessence model in supergravity). 

The most obvious way to try to achieve sequestering is to consider quintessence as an open string modulus geometrically separated from an open string realisation of the Standard Model, for which the K\"ahler potential takes the form 
\begin{equation}
K = -3M_{\rm Pl}^2 \ln \left( 1 - \frac{f(\Phi, \bar\Phi)}{3 M_{\rm Pl}^2} - \frac{g(\chi, \bar\chi)}{3M_{\rm Pl}^2} \right).
\end{equation}
It is not yet clear if sufficient coupling suppression can be achieved (see \cite{Conlon:2007gk, Cicoli:2010ha, Cicoli:2012cy, Abel:2008ai, Burgess:2008ri, Goodsell:2009xc, Cicoli:2011yh, Gan:2023wnp} for some stringy mechanisms of kinetic mixing, \cite{Berg:2010ha} for a summary of the challenges, \cite{Aparicio:2014wxa} for some proposed solutions and \cite{Acharya:2018deu} for some recent progress including a lower bound on the volume modulus to suppress the couplings between the Standard Model and other K\"ahler moduli).  
\end{itemize}

\subsubsection{Runaway quintessence -- single field} 
\label{S:singlefieldq}

The vast moduli sector of string compactifications offers the ideal place to look for quintessence candidates. Given the ubiquity of runaway moduli in string compactifications (see discussion on the Dine-Seiberg problem in Sec. 3), one may expect runaway quintessence scenarios to be easily realised, without the need for delicate interplay between various not-so-under-control ingredients as in dS vacua. As string moduli usually correspond to low energy couplings or expansion parameters, by setting up moduli to be slowly-rolling close to their asymptotic regime, parametric control\footnote{The expansion can be made arbitrarily good by making the expansion parameter arbitrarily small.} of expansions could be ensured with, moreover, a naturally suppressed vacuum energy.  Phenomenological bounds on the parameter $\lambda$ for runaway scalar potentials of the form $V(\phi) \sim V_0\,e^{-\lambda\phi}$ were found to be $\lambda \lesssim 1.02$ and $\lambda \lesssim 0.6$ at 3$\sigma$ in \cite{Akrami:2018ylq} and \cite{Agrawal:2018own}, respectively.

As mentioned above, the first candidate runaway moduli that one may think of are the overall volume and the dilaton since they are model independent; rolling towards infinite volume or zero coupling, respectively, in the asymptotic limit. Indeed, the dilaton was used early on as a quintessence candidate in \cite{Gasperini:2001pc,Damour:2002mi}, although ref. \cite{Cicoli:2021fsd} has recently shown that runaways for the volume mode and the dilaton at tree-level, where one could in principle achieve parametric control over all approximations, are never flat enough to drive an epoch of accelerated expansion. Nevertheless, the main difficulty with these moduli is that they couple to all matter fields. The strong bounds on fifth-forces and varying constants thus make these options untenable. On the other hand, if the runaway direction is a local modulus in a hidden sector it may plausibly avoid constraints from fifth forces and time variation of fundamental constants. Yet as we now discuss, even before worrying about these phenomenological challenges, it turns out to be intriguingly difficult to identify runaway directions in string theory that are sufficiently flat to source the required acceleration \cite{Olguin-Tejo:2018pfq,Bento:2020fxj}.  

For example, consider for simplicity a moduli stabilisation scenario in which all but one of the moduli are stabilised as heavy fields in a supersymmetric Minkowski vacuum. The remaining supersymmetric flat direction $\Phi$ is protected by non-renormalisation theorems \cite{Dine:1986vd}, but is ultimately lifted by non-perturbative effects. If this is a bulk modulus, it will typically have a K\"ahler potential and superpotential of the form:
\begin{equation}
\setlength\fboxsep{0.25cm}
\setlength\fboxrule{0.4pt}
\boxed{
K = -p\ln(\Phi + \bar{\Phi}) \qquad\textrm{and}\qquad W = A\, e^{-a\Phi} 
\label{E:DSrunaway}
}
\end{equation}
for some constants $p, A, a$. The resulting scalar potential for the saxion $\phi = {\rm Re}(\Phi)$, with runaway towards $\phi \rightarrow \infty$ is plotted in Fig. \ref{F:runaways} for $p=1,3$, and it is easily shown that the slow-roll parameter diverges at the tail:
\begin{equation}
\setlength\fboxsep{0.25cm}
\setlength\fboxrule{0.4pt}
\boxed{
\epsilon_V \rightarrow \frac{4}{p} a^2 \phi^2 \qquad\text{as}\qquad \phi \rightarrow \infty,
}
\end{equation}
and thus this potential is always too steep to drive an accelerated expansion \cite{Olguin-Tejo:2018pfq}.  

\begin{figure}[ht]
\centering
\includegraphics[width = 0.45\textwidth]{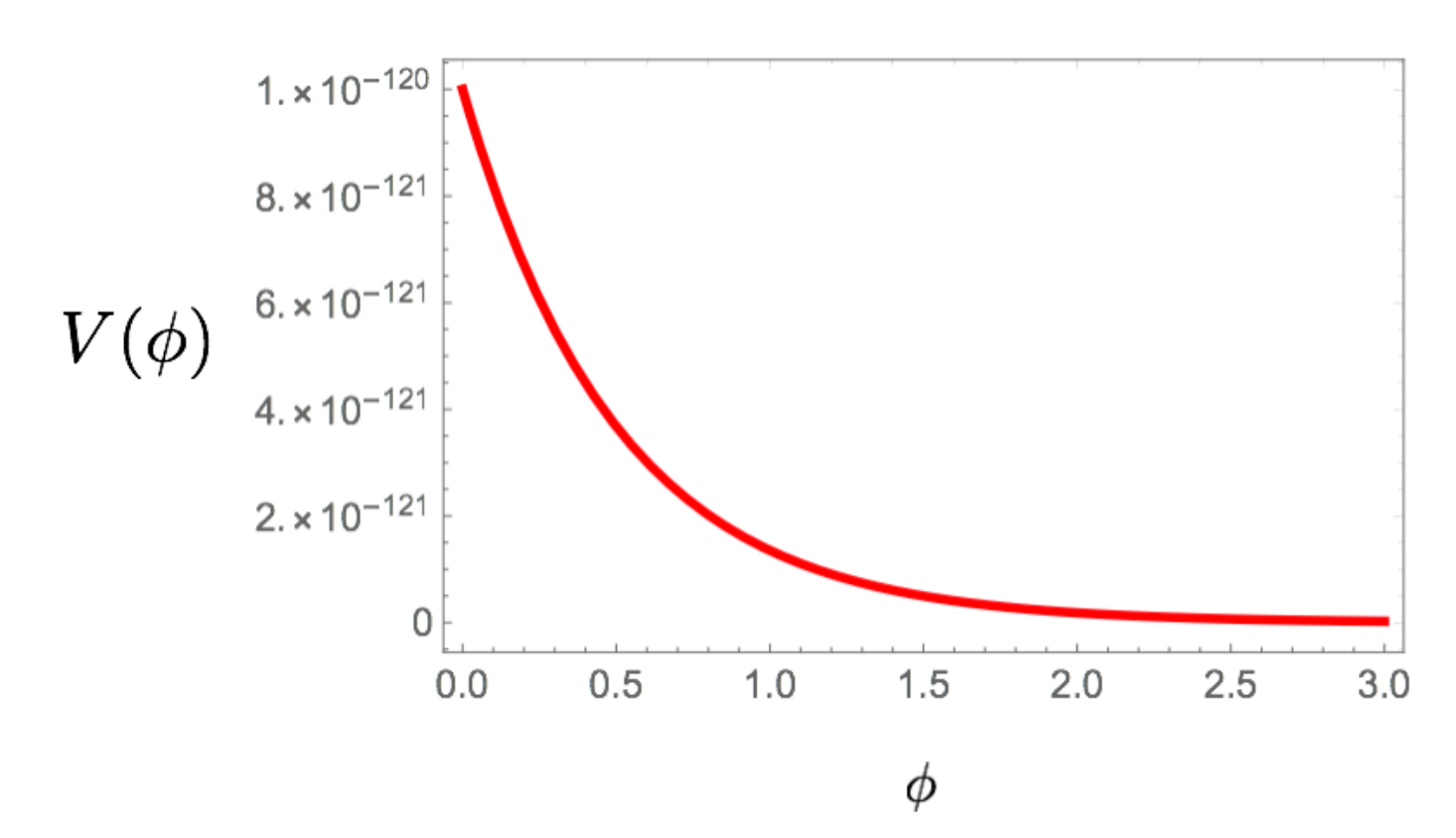} \qquad\includegraphics[width = 0.4\textwidth]{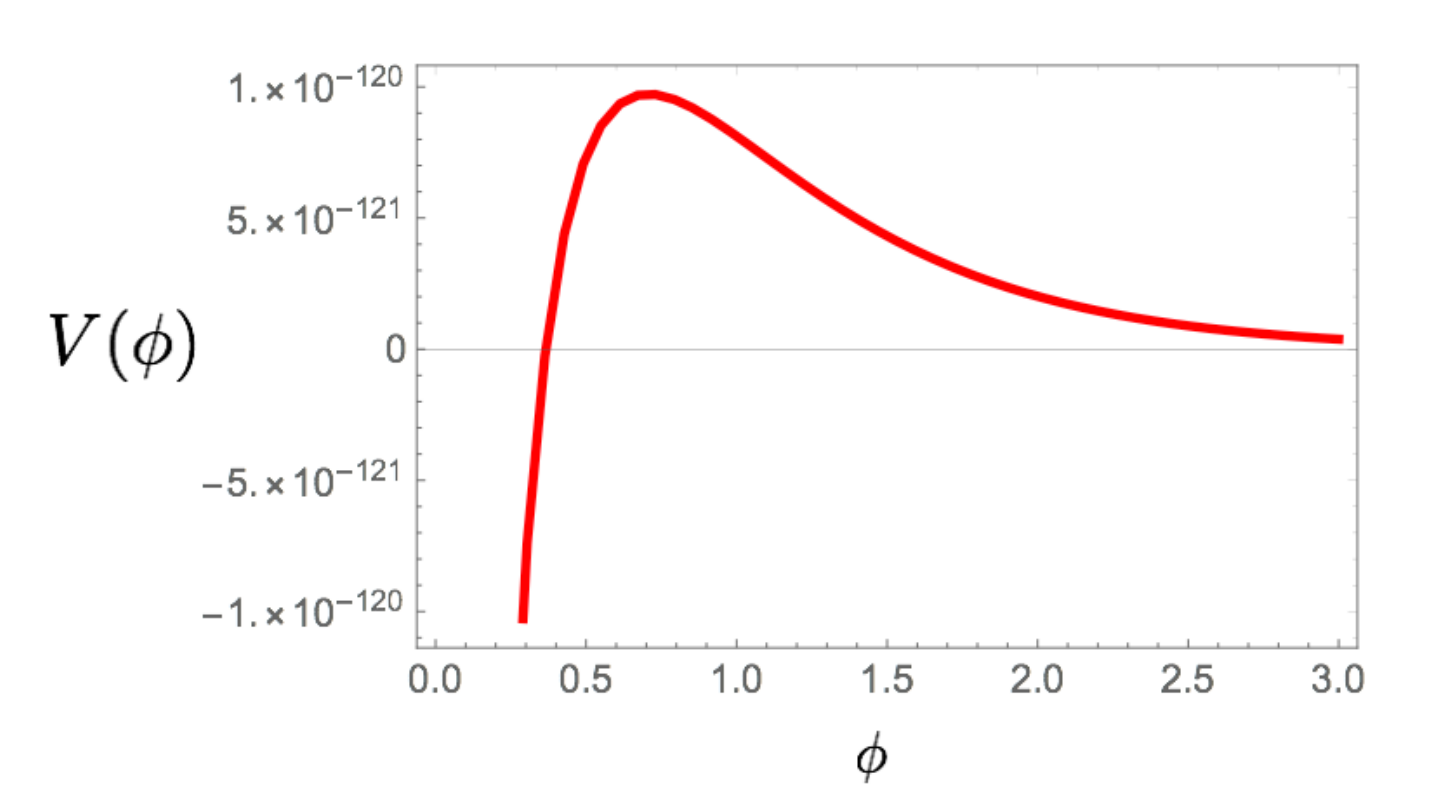}
\caption{Typical runaway potentials: the scalar potentials derived from eq. (\ref{E:DSrunaway}) for $p=3$ (left-hand side) and $p=1$ (right-hand side).}
\label{F:runaways}
\end{figure}

In fact, it has been shown using $N=1$ supergravity that all the typical string moduli -- whether bulk or local, whether lifted by perturbative or non-perturbative superpotentials -- have potentials that satisfy either $\epsilon_V > 1$ or $\rho_V < 0$ and so cannot drive an accelerated expansion, see Tab. \ref{quintrunnogos} \cite{Bento:2020fxj}. These results have been extended in \cite{Cicoli:2021fsd} considering no-scale models in type IIB, IIA and heterotic string compactifications. The limitations from string-inspired 4-dimensional $N=1$ supergravity on obtaining slow-roll potentials are reminiscent of similar results from \cite{Hellerman:2001yi} which show that, for a class of supersymmetric theories, it is impossible to relax into an asymptotic zero-energy supersymmetric minimum whilst accelerating (see also \cite{Rudelius:2021azq}). The theories considered are those with a single field with exponential potential $V\sim e^{-c\phi/M_{\rm Pl}}$; indeed, if the potential is to have a zero-energy supersymmetric vacuum and support $w_{\rm DE}={\rm constant}$ quintessence-like evolution, then it must have this form asymptotically, with $|c|< \sqrt{2}$. Assuming that the dynamics is indeed towards a zero-energy supersymmetric minimum at $\phi \rightarrow \infty$, Ref. \cite{Hellerman:2001yi} shows how 4-dimensional $N=1$ supersymmetry implies that having $V>0$ requires $c > \sqrt{6}$ and thus excludes slow-roll.

\begin{table}[ht]
\begin{center}
\centering
\begin{tabular}{| c | c | c | c | c |}
\hline
{\cellcolor[gray]{0.9}$K$} & {\cellcolor[gray]{0.9}$W$} & {\cellcolor[gray]{0.9}$V>0 \quad \epsilon_V<1$} \\
\hline
\hline
$-p\ln(\Phi+\bar{\Phi})$ & $W_0 + A\,e^{-a\Phi}$ & no-go \\
\hline
$-p\ln(\Phi+\bar{\Phi})$ & $W_0 + A\,\Phi^n$ & no-go \\
\hline
$k_0 + \frac{|\Phi|^{2p}}{k_1}$ & $W_0 + A\,e^{-a\Phi}$  & no-go \\
\hline
$k_0 + \frac{|\Phi|^{2p}}{k_1}$ & $W_0 + A\,\Phi^n$ & no-go except for $p=n$ \\
\hline
$k_0 + \frac{(\Phi+\bar{\Phi})^{2p}}{k_1}$ & $W_0 + A\,e^{-a\Phi}$  & no-go \\
\hline
$k_0 + \frac{(\Phi+\bar{\Phi})^{2p}}{k_1}$ & $W_0 + A\,\Phi^n$ & no-go except for $p=n$ \\
\hline
\end{tabular}
\end{center} 
\caption{Summary of no-go theorems for string inspired supergravity models of single-field runaway quintessence, and parameter points that evade them. The K\"ahler potentials correspond, respectively, to bulk moduli, fibre moduli and blow-up moduli, and the superpotentials correspond to the flat direction $\Phi$ being lifted, respectively, by a leading non-perturbative effect or a leading perturbative one.  No-scale scenarios, having $K=-3\ln(\Phi + \bar{\Phi})$ and $W$ independent of $\Phi$, are included in the case that no-scale-breaking occurs via non-perturbative corrections to $W$.}  
\label{quintrunnogos}
\end{table}  

Note that we have -- as is usual in quintessence and as is entirely unsatisfactory in string theory -- left aside the cosmological constant problem and the fact that supersymmetry in the Standard Model sector must be broken at least at TeV scales which is clearly a challenge for supersymmetric vacua where the supersymmetry breaking scale would naturally be set by the dark energy scale.  At the same time, the difficulties in obtaining runaway directions that sustain accelerated expansion seem to extend to non-supersymmetric setups. The runaway potentials 
associated with NS-NS tadpoles in non-supersymmetric string theories are also too steep for quintessence, taking the form $V=\Lambda\, e^{-\gamma \phi}$ with $\gamma=\frac52, \frac32, \frac32$ in the Einstein frame, for the SO(16)$\times$SO(16) theory, the orientifold $USp(32)$ Sugimoto model, and the type 0B' model, respectively (see \cite{Basile:2020mpt} for an extension of dS no-go theorems to non-supersymmetric string theories).

These no-go theorems against single-field runaway quintessence support the conjecture against dS vacua, which are indeed best motivated at the asymptotic boundary of moduli space \cite{Ooguri:2018wrx}. However, as we will discuss in Sec. \ref{S:multiq}, there may be paths to 
evade these no-go results by considering multifield scenarios.

\subsubsection{Hilltop quintessence}

Alongside runaway moduli, which are ubiquitous in string compactifications, dS maxima are also very easy to find. For example, the Dine-Seiberg runaway potential in eq. (\ref{E:DSrunaway}) for $p=1$ is shown in Fig. \ref{F:runaways} to have a dS hilltop. In \cite{Olguin-Tejo:2018pfq}, the possibility of sourcing quintessence at the hilltop was explored (see \cite{Dutta:2008qn} for a phenomenological analysis of hilltop quintessence). If the modulus starts close to the hilltop, it can remain frozen there by Hubble friction for much of the cosmological history, first sourcing an effective cosmological constant and then turning into a rolling quintessence field with observable consequences. The parameters could be chosen to match the observed dark energy, consistent with the refined dS swampland conjecture (see Secs.~\ref{S:DEswamp} and \ref{Sec:Swamp} below) and with sub-Planckian field displacements. Quantum fluctuations $\Delta \varphi \sim H/(2\pi)$ leave the field within the viable window close to the hilltop until $H \sim 0.01\,M_{\rm Pl}$. Although there is no need for fine-tuning in the Lagrangian parameters, the fine-tuning of initial conditions is difficult to motivate, perhaps resorting to some anthropic danger in starting from a point in field space that leads into a steep runaway.

Of course this is just a scenario, and even if motivation for the initial conditions could be found, work would need to be done to embed it in a fully fledged string construction with moduli stabilisation and control over subleading corrections; in particular it may be a challenge to achieve sufficient sequestering to hide fifth forces, higher order instanton corrections may not be suppressed, and -- as is usual in quintessence scenarios -- the cosmological constant problem and vacuum contributions from the susy-breaking visible sector have not been addressed. Further attention to the hilltop scenario was given in \cite{Cicoli:2021skd}, using the dS maximum for the volume modulus in the KKLT scenario, where it was found that matching with the observed dark energy would require an unacceptably light gravitino and light volume modulus.

\subsubsection{Axions as quintessence}

Axions are arguably the most attractive quintessence candidates \cite{Kaloper:2008qs, Panda:2010uq, Choi:1999xn,Cicoli:2021skd} as ($i$) being pseudo-scalars they evade the most stringent spin-independent fifth force bounds; ($ii$) they enjoy approximate shift symmetries which restrict the allowed couplings and protect the axion mass and potential energy density, which are otherwise UV sensitive quantities; ($iii$) their potential is generated by exponentially suppressed non-perturbative effects which can naturally reproduce the required hierarchy between the dark energy potential and the potential which fixes the volume modulus at leading order (for example via perturbative effects which break supersymmetry spontaneously) guaranteeing $M_{\rm soft}\gtrsim 1$ TeV and $m_{\mathcal{V}}\gtrsim 1$ meV in a way compatible with $m_q \sim 10^{-32}$ eV. 

From the string theory point of view, a rather generic prediction is the existence of a 
large number of axions \cite{Green:1987mn, Banks:1996ea, Banks:2002sd, Svrcek:2006yi, Conlon:2006tq}  -- sometimes called an \emph{axiverse} \cite{Arvanitaki:2009fg} -- arising from the KK reduction of higher-dimensional form fields on the topological cycles of the compactification space. As an internal space can easily have $\mathcal{O}(100)$ distinct cycles, there can be $\mathcal{O}(100)$ axions. Typically, the axion directions will be perturbatively flat due to shift symmetries descending from higher-dimensional gauge symmetries, and which may subsequently be lifted by non-perturbative effects that give masses $\sim \mathcal{O}(e^{-\tau}M_{\rm Pl})$ with $\tau$ the partner saxion that measures the size of the corresponding cycle. 

In principle, the range of axion masses present can cover a wide region all the way down to quintessence-like masses $H_0 \sim 10^{-33}$ eV and beyond, which could include dark energy candidates as well as possible ultra-light axion dark matter with $m \sim 10^{-22}$ eV \cite{Hui:2016ltb, Cicoli:2021gss}. However, note that this would require the presence of many non-perturbative effects in the compactification, all with different origins and of different magnitudes, with the corresponding saxions fixed at higher masses (to avoid cosmological problems). See \cite{Marsh:2015xka} for a review on axion cosmology and \cite{Krippendorf:2018tei} for a discussion on possible macroscopic compact objects from ultralight string axions. For an analysis studying the coupled evolution with dark matter see \cite{Kumar:2013oda}.

One specific realisation of this idea can be found in the LVS scenario \cite{Cicoli:2021skd}. At leading order and considering an appropriate uplifting sector, $\mathcal{V}$ is fixed by $\mathcal{O}(\alpha'^3)$ effects in a Minkowski vacuum where supersymmetry is spontaneously broken. At this level of approximation, the axionic partner of the volume modulus is a flat direction due to its shift symmetry that is perturbatively exact. Hence the mass scale of $\mathcal{V}$ and the supersymmetry breaking scale can be safely decoupled from $H_0$. The volume axion is lifted by tiny non-perturbative corrections that are exponentially suppressed in a power of the exponentially large volume, and so can naturally reproduce the dark energy scale since 
\begin{equation}
\setlength\fboxsep{0.25cm}
\setlength\fboxrule{0.4pt}
\boxed{
V_{\rm DE} \simeq e^{-\sqrt{\frac32} \frac{M_{\rm Pl}}{f}}\,M_{\rm Pl}^4\left[1-\cos\left(\frac{\varphi}{f}\right)\right]\,,
}
\label{Vax}
\end{equation}
where $\varphi$ is the canonically normalised volume axion with decay constant:
\begin{equation}
f = \sqrt{\frac32}\frac{N\,M_{\rm Pl}}{2\pi\mathcal{V}^{2/3}} \,   
\end{equation}
where $N=1$ for a Euclidean D3-brane instanton while $N$ is the rank of the condensing gauge group for gaugino condensation on D7-branes. Reproducing $V_{\rm DE}\sim 10^{-120}\,M_{\rm Pl}^4$ from (\ref{Vax}) requires $M_{\rm Pl}/f\sim 300$, and so moderately large volumes, $\mathcal{V} \sim 10^3 - 10^4$, which are however still large enough to trust the effective field theory. Moreover $M_s\sim 10^{16}$ GeV, $m_{3/2}\sim 10^{14}$ GeV and $m_{\mathcal{V}}\sim 10^{12}$ GeV, together with $m_q \sim 10^{-32}$ eV. 

Let us stress that any axionic field whose saxion partner is stabilised by perturbative effects will have such doubly exponentially suppressed masses. For example, whereas blow-up moduli tend to be stabilised as for their axionic partners non-perturbatively, leading to masses for saxions and axions of the same order, the saxions associated with fibre moduli tend to be stabilised perturbatively, and thus end up being heavier than their axion partners, allowing for an EFT where the saxions can be integrated out to leave an ultra-light axions to drive quintessence (see \cite{Emelin:2018igk} for a discussion on how, otherwise, lifting axion away from its minimum can lead to a steep runaway instability in the saxion direction).  

Unfortunately, as we have already discussed, it is not sufficient to have a light scalar field to drive a period of accelerated expansion; the scalar potential must be sufficiently flat. This is not the case for the potential (\ref{Vax}) since the axion decay constant is sub-Planckian, $f\sim M_{\rm Pl}/300$, while a sufficiently shallow potential would require a super-Planckian field displacement and axion decay constant \cite{Freese:1990rb}. Note that this is not just a consequence of matching the correct dark energy scale, but also a condition to trust the effective field theory. In fact, axion decay constants typically scale as $f \sim M_{\rm Pl}/\tau$ \cite{Svrcek:2006hf, Svrcek:2006yi, Arvanitaki:2009fg, Cicoli:2012sz}. Hence $\tau \gtrsim 1$ implies that axion decay constants are always sub-Planckian. Indeed super-Planckian decay constants are very difficult to obtain within string theory, and recent discussions in the context of the string swampland even suggest that they may be forbidden in quantum gravity via axionic versions of the weak gravity conjecture \cite{Klaewer:2016kiy, Blumenhagen:2017cxt, Palti:2017elp, Cicoli:2018tcq, Cicoli:2021gss}.

However, there are other possibilities for axions to drive an accelerated expansion without a super-Planckian decay constant: 
\begin{itemize}
\item \textbf{Axion hilltop quintessence:} for a single axion field, by fine-tuning the initial conditions sufficiently close to its hilltop, an accelerated expansion can be achieved \cite{Kamionkowski:2014zda, Cicoli:2021skd}. Even if one might think that not much tuning is necessary to achieve less than $1$ e-folding of late-time accelerated expansion, the initial conditions that allow for quintessence are likely to be destroyed by inflation. In fact, during inflation the axion is an ultra-light field which acquires quantum fluctuations of order $\Delta\varphi\sim H_{\rm inf}/(2\pi)$. The initial vicinity to the maximum is determined by the value of the axion decay constant. For example, for $f\sim 0.02\,M_{\rm Pl}$ which is roughly the largest possible value of $f$ compatible with a large volume expansion, this distance in Planck units has to be smaller than $10^{-18}$ \cite{Cicoli:2021skd}, imposing $H_{\rm inf}\lesssim 10^{-18}\,M_{\rm Pl}\sim 1$ GeV. For smaller values of $f$, the tuning gets even worse.

\item \textbf{Axion alignment quintessence:} for two or more light axionic fields, an alignment mechanism \cite{Kim:2004rp, Dimopoulos:2005ac, Shiu:2015xda, Cicoli:2014sva} may generate an effective super-Planckian decay constant out of two or more sub-Planckian decay constants (see \cite{Cicoli:2018kdo} for other possibilities for axion quintessence, which rely on the presence of a dS minimum). The lightest axion could play the role of dark energy while the axion which remains heavy could potentially account for dark matter.
\end{itemize}

\subsubsection{Branes, extra dimensions and string symmetries}
\label{S:DEbranes}

As a key constraint on quintessence candidates is always their interactions with Standard Model matter resulting in fifth forces, a natural place to look for string theory quintessence candidates is from hidden sector D-branes, as they may be coupled to visible sector D-branes with weaker-than-Planckian couplings.  

\paragraph{k-essence from branes}

Refs. \cite{Martin:2008xw, Gumjudpai:2009uy} explored whether the DBI action could give rise to quintessence attractor tracker solutions for the scalar field representing the position of the brane. This starts with a string-inspired, phenomenological action:
\begin{equation}
\setlength\fboxsep{0.25cm}
\setlength\fboxrule{0.4pt}
\boxed{
S_{\rm DBI}= -\int d^4x \; a^3(t) \left(T(\phi)W(\phi)\sqrt{1-\frac{\dot{\phi}^2}{T(\phi)}}-T(\phi)-\tilde{V}(\phi)\right), \label{E:DBIessence}
}
\end{equation}
where $T(\phi)$ is the warped tension of the brane and $W(\phi)$ and $\tilde{V}(\phi)$ are potential terms, the first coming from the nature of the D-brane stack (e.g. supersymmetry breaking effects, non-Abelian sectors, worldvolume fluxes) and the second from possible interactions with the bulk and other brane stacks.  This gives rise to an equation of state:
\begin{equation}
w_\phi = \frac{T(\phi)(\gamma-W(\phi))-\gamma \tilde{V}(\phi)}{\gamma T(\phi)(\gamma W(\phi)-1)+\gamma\tilde{V}(\phi)}
\end{equation}
with $\gamma = \left( 1-\frac{\dot{\phi}^2}{T(\phi)}\right)^{-\frac12}$. The simplest case is of a D3-brane moving in a 5-dimensional AdS space, describing the mid-region of a warped throat, $T(\phi) \propto \phi^4$ and $W(\phi)=1$. Assuming moreover that $\tilde{V}(\phi) \propto T(\phi)$ is justified as it follows from $\gamma=\,$constant, which in turn is an attractor scaling solution. Then $w_\phi=2w+1>1$, with $w$ the equation of state of the background fluid, so that $\phi$ scales faster than the background fluid and can never come to dominate the universe \cite{Martin:2008xw}.  More general forms for the functions characterising the DBI action, which may or may not derive from string theory, allows for viable k-essence models \cite{Gumjudpai:2009uy}, but only for super-Planckian field excursions.  Also, although  small scales for the vacuum energy and quintessence mass can be arranged, their robustness against quantum corrections is not yet addressed, 

\paragraph{Fine-tuning, branes and supersymmetry breaking}

Branes and extra dimensions potentially allow for symmetries which can help address the fine-tuning problems in dark energy in non-trivial ways. For example, supersymmetry may be badly broken in the visible sector branes, say around TeV scale, thus explaining the absence of superpartners for the Standard Model, whilst the large localised vacuum energy may curve the extra dimensions rather than the branes themselves.  Meanwhile, supersymmetry breaking in the bulk could be suppressed with respect to the visible sector branes by the Planck scale, leading to a corresponding suppression in the final vacuum energy. 

This idea was explored in the Supersymmetric Large Extra Dimensions (SLED) scenario \cite{Aghababaie:2003wz, Burgess:2004ib}, where two large extra dimensions explain the hierarchy problem \cite{Arkani-Hamed:1998jmv, Antoniadis:1998ig} and bring the string scale down to around TeV, whilst the KK scale and gravitino mass are suppressed to around meV such that bulk contributions to the vacuum energy go as $M_{\rm KK}^2 m_{3/2}^2 \sim {\rm meV}^4$. An interesting stringy embedding of the SLED scenario with anisotropic moduli stabilisation that leads to $2$ extra dimensions much larger than the other $4$ has been presented in \cite{Cicoli:2011yy}. In SLED, dark energy emerged as a quintessence scenario using the volume modulus associated with the large extra dimensions, which develops a logarithmic slow-roll potential \cite{Albrecht:2001cp, Albrecht:2001xt}.  

A related scenario based on \cite{Cicoli:2011yy} is \cite{Cicoli:2012tz}, which considers large extra dimensions in the context of LVS. Here, the quintessence field is a fibre modulus which has weaker-than-Planckian couplings to a visible sector that is localised on a blow-up modulus having no intersection with the quintessence fibre. As already mentioned, having a quintessence sector which is geometrically separated from the visible sector helps in suppressing dangerous fifth forces and time-variation of coupling constants. The dark energy potential is generated by tiny poly-instanton corrections to the superpotential and the model shares several features with those of a typical SLED model: $2$ exponentially large extra dimensions, a gravitino mass of order the cosmological constant scale and TeV-scale gravity.

For these proposals to really work, it is necessary to control the loops on the branes, where supersymmetry is only non-linearly realised; their contributions to the vacuum energy would need to be absorbed by backreaction onto the extra dimensional curvature, but they may instead lead to high curvature on the brane or instabilities and runaway solutions.  

Progress on these questions may be facilitated by recently developed tools in coupling non-supersymmetric matter to supergravity \cite{Komargodski:2009rz, Bergshoeff:2015tra, Dudas:2015eha, DallAgata:2015zxp, Schillo:2015ssx, Parameswaran:2020ukp}. This has been initiated in \cite{Burgess:2021juk}, where evidence was found that the supergravity form of the action -- and a small splitting in the gravity supermultiplet --  is stable against integrating out heavy, non-supersymmetric particles, thanks to the interplay with auxiliary fields associated with gravity, the goldstino and other supermultiplets in the supersymmetric gravity sector.  Having a gravity sector that is supersymmetric down to low energies, coupled to a visible sector where supersymmetry is non-linearly realised, is also motivated by the fact that gravity would have the weakest couplings to any supersymmetry breaking sector \cite{Arkani-Hamed:2006emk}. Cosmological bounds on the light gravitini and moduli that would arise with enhanced supersymmetry in the gravity sector are discussed in \cite{Kawasaki:2008qe, Feng:2010gw, Coughlan:1983ci, Banks:1993en, deCarlos:1993wie, Conlon:2007gk}.

How supersymmetry may help in the UV stability of dark energy models such as quintessence has been pursued further in the scenario of `yoga dark energy' \cite{Burgess:2021obw}. This proposes that a very supersymmetric gravity sector, combined with an accidental approximate scale invariance  and a `relaxon' scalar field \cite{Graham:2019bfu} with $m \lesssim m_e$ that dynamically reduces the leading non-gravitational vacuum energy, might explain the cosmological constant problem and the observed dark energy.  The supersymmetric gravity sector could be realised by brane supersymmetry breaking \cite{Sugimoto:1999tx, Dudas:2000nv, Antoniadis:1999xk, Kallosh:2014wsa, Kallosh:2016aep, GarciadelMoral:2017vnz, Cribiori:2019hod}, whilst accidental approximate scale invariance is a generic property of low-energy string vacua \cite{Berg:2005ja, Cicoli:2007xp, Cicoli:2021rub} and its interplay with supersymmetry is studied in \cite{Burgess:2020qsc}. The string interpretation of the relaxon remains to be explored in detail, as well as how well the scenario stands up to naturalness and phenomenological conditions.  

Interestingly, the UV completion \cite{Cribiori:2020sct} of brane supersymmetry breaking is not the standard supersymmetry but rather a so-called misaligned supersymmetry \cite{Dienes:1994np, Cribiori:2021txm}. The possibility that string theoretic modular invariance and misaligned symmetry may help with the cosmological constant problem has been discussed in \cite{Dienes:2001se}.  Although no complete realisation of this idea has been found, there exists non-supersymmetric setups that have an exponentially suppressed vacuum energy to two-loops \cite{Kachru:1998hd, Kachru:1998pg, Abel:2015oxa, Abel:2017rch}, thanks to Bose-Fermi cancellations between the massless field degeneracies in a non-supersymmetric visible sector and a non-supersymmetric hidden sector.

\subsubsection{Dark energy and the swampland}
\label{S:DEswamp}

The swampland aims to map out the set of effective field theories that can be ultraviolet completed consistently with quantum gravity (see Sec.~\ref{Sec:Swamp} below). The dS swampland conjecture \cite{Obied:2018sgi, Garg:2018reu, Ooguri:2018wrx} proposes that the scalar potential in any consistent effective field theory must satisfy either $|\nabla V| \gtrsim \frac{c}{M_{\rm Pl}}V$ or $\textrm{min}(\nabla_i \nabla_j V)\lesssim -\frac{c'^2}{M_{\rm Pl}^2} V$ where $c,c'>0$ are $\mathcal{O}(1)$ universal constants, ruling out even metastable dS vacua. Support for the dS conjecture is found in the parametrically asymptotically controlled regime of string theory via the swampland distance conjecture and Bousso's covariant entropy bound \cite{Ooguri:2018wrx} (it is important to note that the candidate string dS vacua discussed in Sec. 3 rely on numerical control rather than parametric control; see also \cite{Dvali:2018jhn, Dvali:2018fqu} for other quantum gravity arguments against long-lived dS and \cite{Rudelius:2019cfh} for some motivation via the problems of eternal inflation). In this regime, the strong dS conjecture \cite{Rudelius:2021azq} states that the strong energy condition should be satisfied at late times, implying $c=\sqrt{2}$, which would also forbid asymptotic accelerated expansion. 

A solution to the dark energy problem seems therefore to lie in a region of moduli space without full parametric control over the effective field theory, i.e.~where one cannot make approximations arbitrarily good. dS vacua or quintessence solutions could however be obtained with numerical control thanks to underlying parameters, like $W_0\ll 1$ in KKLT or $\mathcal{V}^{-1}\sim e^{-1/g_s}\ll 1$ in LVS, which can be made very small, even if not arbitrarily small (since the number of moduli and the D3 tadpole set a lower bound on $W_0$ and $g_s$). Note moreover that quintessence model building features the same challenges as the construction of dS vacua with however additional constraints coming from fifth forces, radiative stability of the quintessence mass, and huge hierarchies in the moduli stabilising scalar potential \cite{Cicoli:2021skd}. Hence, from this perspective, quintessence, instead of looking like a viable alternative to dS, seems to be under even less control than dS vacua. This consideration raises therefore some doubts on the validity of the dS conjecture in regions of the moduli space with numerical, instead of parametric, control due to the observational evidence of a present epoch of accelerated expansion and the current lack of robust dark energy alternatives to dS and quintessence.

Allowing for effective field theories with quintessence-like configurations, the trans-Planckian censorship conjecture \cite{Bedroya:2019snp} proposes that the possibility of sub-Planckian scale fluctuations redshifting until they cross the horizon and classically freeze out is unphysical, and therefore any epoch of super-luminal expansion must have a finite lifetime $\Gamma < H \ln H$. In asymptotic regions of field space, the trans-Planckian censorship conjecture implies the dS swampland conjecture with $c = \sqrt{2/3}$, however, deep in the interior of field space the former is compatible with metastable dS so long as it is sufficiently unstable quantum mechanically.

Ref. \cite{Agrawal:2018own} considered the implications of the swampland dS conjecture and the swampland distance conjecture and concluded that they would be in tension with early universe cosmic inflation: CMB observations bound the single-field inflationary slow-roll parameter $\epsilon < 0.0044$, which in turn bound the order one constant in the dS conjecture $c<0.094$. Moreover, any future detection of a tensor-to-scalar ratio of order $r \sim 0.01$ would indicate an inflaton field excursion of $\Delta \phi \sim 2\,M_{\rm Pl}$. 
In \cite{Han:2018yrk}  the authors coupled the vanilla exponential quintessence potential to the Higgs sector, and found that this coupling helps in addressing the electroweak vacuum instability problem. Moreover, they obtained the bound  $c>0.35$, consistent with current constraints $c\lesssim 0.6$ \cite{Agrawal:2018own,Akrami:2018ylq,Heisenberg:2018yae}.

Whether or not the swampland conjectures turn out to be true, it is interesting to consider to what new ideas they may lead. Ref. \cite{Montero:2022prj} uses the distance/duality conjecture, the smallness of dark energy, and current (albeit rather model dependent) observational constraints on extra dimensions \cite{Workman:2022ynf, Hannestad:2003yd}, to argue that the universe is in an asymptotic region of field space with a single large extra dimension -- named the `dark dimension' -- of size $l \sim \mu m \sim ({\rm meV})^{-1}$ along with a fundamental gravity scale $M \sim 10^{10}$ GeV. Whereas the brane scenarios discussed in Sec. \ref{S:DEbranes} use large extra dimensions to bring fundamental gravity down to the TeV-scale and to lower the supersymmetry breaking scale in the gravity sector, here the large extra dimension is motivated by the expectation that the one-loop vacuum energy goes as $M_{\rm KK}^4\sim ({\rm meV})^4$ for a tower of light states starting at $M_{\rm KK}$. It goes without saying that finding a viable string embedding with moduli stabilisation of the large dark dimension scenario is a difficult task. Moreover, it remains an open question how an accelerated cosmology is obtained in the asymptotic region of field space. 

Nevertheless, if dark energy is sourced by a rolling scalar field, the asymptotically exponentially light tower of states implied by the swampland distance conjecture could play the role of dark matter which continuously becomes lighter as the dark energy field rolls down its potential  \cite{Agrawal:2019dlm}. This so-called `fading dark matter' scenario may help to address the tensions in the late-time measurements of $H_0$ and $\sigma_8$ compared to the values inferred from early-times using the $\Lambda$CDM model. Alternatively, the tower of states may correspond to the massive KK gravitons, universally coupled to the Standard Model and produced as dark matter as the Standard Model sector cools down \cite{Gonzalo:2022jac}. Ref. \cite{Anchordoqui:2022txe} observes that a mesoscopic large extra dimension slows down the rate of Hawking radiation for black holes, thus prolonging the survival of low-mass primordial black holes, which can then compose all of dark matter. The correspondence between a 5-dimensional primordial black hole and 5-dimensional massive KK modes interpretation of dark matter was discussed in \cite{Anchordoqui:2022tgp}.

\subsubsection{Multi-field quintessence} 
\label{S:multiq}

Multi-field quintessence models can help overcome several of the problems suffered by single-field models, and are also well-motivated by string theory given the large numbers of scalar fields that arise in string compactifications, with non-trivial target-space geometries. Similarly to hybrid inflation, multi-field models provide mechanisms to exit the accelerated expansion. They also provide new avenues to achieve accelerated expansions with steep potentials, via non-geodesic trajectories or gradient flows. And sufficiently shallow potentials may be found in regions of weak couplings and parametric control, where competing terms in the multifield asymptotic limit can allow for geodesic trajectories along gradient flows, which represent shallow valleys in the multi-dimensional moduli space. We will now discuss these various proposals in more detail.

\paragraph{Hybrid quintessence for a graceful exit:}

The conceptual problems presented by an eternal dS, e.g. the absence of a well-defined S-matrix, are shared by quintessence models \cite{Hellerman:2001yi, Fischler:2001yj} unless there is a mechanism to end quintessence at some time in the future. Similarly to hybrid inflation \cite{Linde:1993cn}, this can be achieved in a multi-field hybrid quintessence model. A stringy realisation of this idea is given by quintessence from $2$ D3-branes separated by some large distance, $r$, and at some relative angle, $\theta$ \cite{Halyo:2001fb}. The relative angle breaks supersymmetry and generates a tracker quintessence potential for the $2$ fields: $V(\phi,\theta) = \theta^2 \frac{M_s^8}{M_{\rm Pl}^2 \phi^4}$. To obtain the observed dark energy scale, $\rho \sim 10^{-120}M_{\rm Pl}^4$, with $\theta, \phi \sim M_{\rm Pl}$, requires $M_s \sim$ TeV and hence $2$ large extra dimensions.  In the early universe, domination by radiation or matter ensures that $H>m_\theta$, so that $\theta$ is frozen at a constant VEV $\mathcal{O}(M_{\rm Pl})$.  If $\rho_\phi$ is greater than the energy density of the tracker solution, $\phi$ rolls quickly down its potential until it freezes at some VEV $\mathcal{O}(M_{\rm Pl})$, due to the large redshift of kinetic energy, by which time $\rho_\phi$ is much less than the tracker energy density. At this point both $\theta$ and $\phi$ behave as frozen quintessence with $w\approx -1$, and eventually the quintessence comes to dominate the universe. Later, $H$ falls below $m_\theta$, $\theta$ begins to roll, the accelerated expansion ceases, and the potential eventually vanishes as $\theta$ settles at its minimum, yielding to another era of matter domination. Although this scenario solves the problem of a well-defined S-matrix, it requires super-Planckian field displacements to achieve a shallow enough potential to drive the accelerated expansion as well as a moduli stabilisation scenario that results in two large extra dimensions.

\paragraph{Non-geodesic trajectories in string constructions:}

Just as for inflation, having multiple fields provides new avenues to achieve accelerated expansion with steep potentials via curved non-geodesic trajectories in the multi-dimensional target-space. String compactifications do give rise to non-linear sigma models in their 4-dimensional $N=1$ supergravity descriptions, providing in principle interesting target-space geometries for non-geodesic behaviour.\footnote{However, as shown in \cite{Aragam:2021scu}, non-geodesic trajectories in supergravity seem to require very large field space curvatures.} Moreover, non-geodesic trajectories can be longer than the geodesic distances that determine the masses of towers of states \cite{Landete:2018kqf, Hebecker:2017lxm}. 

Various effective field theory multi-field models have been proposed that are consistent with observational data and the swampland conjectures \cite{Sonner:2006yn, vandeBruck:2009gp,  Brown:2017osf, Russo:2018akp, Achucarro:2018vey, Cicoli:2020cfj, Cicoli:2020noz, Akrami:2020zfz}. Ref. \cite{Brinkmann:2022oxy} considers instead string-inspired multi-field models composed of saxion-axion pairs with the kinetic couplings and scalar potential expected for either closed string universal moduli or non-universal moduli such as blow-up modes. A possible setting for the universal moduli is \cite{Saltman:2004sn, Gallego:2017dvd}; these involve type IIB flux compactifications with all the complex structure moduli and axio-dilaton stabilised, leaving a single K\"ahler modulus with $K=-p\ln(T+\bar{T})$ and $V=V_0/(T+\bar{T})^p$ and $p=3$. Other string settings corresponding to different values of $p$ are outlined in Tab. \ref{T:multiunimodq}. A possible setting to consider candidate blow-up modes is type IIB orientifold flux compactifications with internal Calabi-Yau of the `weak Swiss cheese' form \cite{Cicoli:2018tcq}, assuming just one universal modulus, $\tau_b$, and one blow-up mode, $\tau_s$, with $\tau_s \ll \tau_b$, for which $K=-2\ln{\mathcal{V}} = -3\ln \tau_b + 2 \left(\tau_s/\tau_b\right)^{3/2}$. 

This yields specific polynomial kinetic couplings and scalar potential.  For the string-motivated couplings and potential discussed, and assuming that the dark energy epoch is entered from an epoch of matter domination as in our universe, Ref. \cite{Brinkmann:2022oxy} found that -- although multi-field accelerated cosmologies could easily be found -- none passed through the current observed values for $(\Omega_{\rm DE}, w_{\rm DE})\approx(0.7,-1)$ (starting from these observed values and working backwards, it was found that the observed values could be reached via initial conditions of kinetic domination). It remains an open question whether observationally viable models could be found in other multi-field string setups, with different couplings and potentials and/or more than two fields.  

\begin{table}[htp]
\begin{center}
\begin{tabular}{l c c c c c }\hline
\cellcolor[gray]{0.9} $p$ & \cellcolor[gray]{0.9} $X$ & \cellcolor[gray]{0.9} Theory & \cellcolor[gray]{0.9} Sources &   \cellcolor[gray]{0.9} $\mathcal{M}_{\rm internal}$ & \cellcolor[gray]{0.9} References \\ [5pt]
\hline
$1$ & $S = e^{-\varphi} + {\rm i}\, a$ & Heterotic & --- & $\mathrm{SU}(3)$ str. & \cite{Font:1990nt} 
\\ [5pt]
$2$ & $T_2 = {\rm Vol}(\Sigma_4^{(2)}) + {\rm i} \int_{\Sigma_4^{(2)}} C_{(4)}$ & Type IIB & D3/D7, O3/O7 & K3-fibered $\mathrm{CY}_3$  & \cite{Cicoli:2011it,Cicoli:2016xae,Cicoli:2017axo} 
\\[5pt]
$3$ & $T = {\rm Vol}(\Sigma_4) + {\rm i} \int_{\Sigma_4} C_{(4)}$ & Type IIB & D3/O3 & $\mathrm{CY}_3$  & \cite{Saltman:2004sn} 
\\[5pt]
$7$ & $Z = {\rm Vol}(\Sigma_3) +{\rm i}\int_{\Sigma_3} A_{(3)}$ & M-theory & KK6/KKO6 & $\mathrm{G}_2$ str. & \cite{Blaback:2019zig}
\\[5pt]
\bottomrule
\end{tabular}
\end{center}
\caption{Examples of string constructions having a 4-dimensional saxion-axion system with $K=-p\ln(\Phi+\bar{\Phi})$ and $V=e^K V_0$, which yields an exponential kinetic coupling and exponential runaway potential in the canonically normalised saxion. The two-field models arise after all other moduli present in the given setup are fixed. Table from \cite{Brinkmann:2022oxy}.}
\label{T:multiunimodq}
\end{table}

\paragraph{Gradient flows in string constructions:}

Ref.~\cite{Calderon-Infante:2022nxb} shows that the no-go theorems for single-field runaway quintessence in 4-dimensional $N=1$ supergravity can be evaded by considering saxionic multi-field models. Again focusing on regions of the moduli space at infinite field distance, which give parametric control as the weak coupling parameter in the relevant perturbative expansion goes to zero in that limit, the multi-field trajectories found turn out to be geodesic ones.  Asymptotically, the trajectories are gradient flows, completely determined by the shape of the potential and the field space metric: these flows are parameterised by $\lambda$ with $\dot{\phi}^k(\lambda) = -\mathcal{F}(\lambda)\partial^k V(\lambda)$ where the smooth positive function $\mathcal{F}(\lambda)$ takes care of reparameterisation invariance. The important point is that, with multiple fields, different terms can compete with each other in the potential even in the asymptotic limit. Consider for example the 4-dimensional $N=1$ scalar potentials arising from F-theory compactified on a Calabi-Yau fourfold with $G_4$-flux. Assume also that the K\"ahler modulus, of which the superpotential is independent, allows for a no-scale cancellation leaving a positive-definite scalar potential $V_{\rm NS} = e^K K^{i\bar{\jmath}}D_i W \overline{D_j W}$. Then, the asymptotic limit when approaching an infinite distance singularity in complex structure moduli space, keeping the overall volume $\mathcal{V}$ constant and assuming that partner axions have been stabilised or remain as flat directions, takes the form:
\begin{equation}
V_{\rm NS} = \sum_{\bf{l} \in \mathcal{E}} A_{\bf{l}} \prod_{k=1}^n \left(\phi^k\right)^{l_k},
\end{equation}
for $n$ saxions $\phi^k$, $\mathcal{E} \subset \mathbb{Z}^n$ and $A_{\bf{l}} \in \mathbb{R}$. The powers $l_k$ are constrained and classified by the framework of asymptotic Hodge theory, and overall, coefficients are such that $V\geq 0$ asymptotically and $V\rightarrow 0$ along at least one trajectory towards infinity. Consider for simplicity Calabi-Yau fourfolds with $2$ complex structure moduli only. Now, for flux choices such that the asymptotic potential has a single dominant term, e.g. 
\begin{equation}
V_{\rm NS}=f_4^2\,\frac{1}{\phi^1\phi^2}
\end{equation}
with $\vec{\phi}(\lambda)=(\alpha \lambda^3, \lambda)$, the situation is similar to the single-field case and the potential is too steep to drive an accelerated expansion. However, there are also flux choices that allow more than one term to compete in the asymptotic limit, e.g. 
\begin{equation}
V_{\rm NS} = f_2^2 \,\frac{\phi^2}{\phi^1}+h_0^2\frac{\phi^1}{\left(\phi^2\right)^3},
\end{equation}
with $\vec{\phi}(\lambda)=\left(\frac{\sqrt{5}}{3}\vline\frac{f_2}{h_0}\vline \lambda^2, \lambda\right)$. It can be shown that the dS coefficient for such an asymptotic gradient flow is $\frac{|\nabla V_{\rm NS}|}{V_{\rm NS}} = \sqrt{\frac27}$, which is sufficiently small to allow for an accelerated expansion.  

It is important to note that the potentials just written have ignored the K\"ahler moduli, in particular the universal volume modulus. Unless the volume modulus is stabilised, it will also contribute to the dS coefficient, rendering the potential once again too steep for an accelerated expansion; at the same time, once the volume is stabilised, the no-scale cancellation, assumed in order to ensure a positive-definite scalar potential, needs to be revisited.  

\subsubsection{Coupled dark sector models} 

Another way to overcome difficulties in finding accelerating cosmologies in string theory are scenarios in which the dark energy and dark matter sectors are coupled. Although there are strong constraints on dark energy interacting with the visible sector, from solar-system and table-top tests of gravity, there are no such constraints on dark energy interacting with dark matter. Given the rich hidden sectors offered by string constructions, coupled dark sectors are well-motivated and also come with interesting phenomenological signatures (see \cite{Bolotin:2013jpa, Wang:2016lxa} for a review into phenomenological models of interacting dark energy/dark matter).  

\paragraph{DBI-essence with disformally coupled dark matter:}

One such scenario arises from D-branes, where, as in DBI quintessence, dark energy arises from open strings representing the radial position of a D-brane along a warped throat, and now dark matter arises from open strings representing matter on the same hidden sector brane \cite{Koivisto:2013fta}.  The action for the dark energy scalar takes the form (\ref{E:DBIessence}) with $T(\phi)=h^{-1}(\phi)$, the inverse warp factor, and $W(\phi)=1$. The action for the dark matter, originating at low energies from $N$ particles on the moving D3-brane, e.g. from vector fields which have acquired St\"uckelberg masses $m_i$, takes the form:
\begin{equation}
S_{\rm DM} = -\sum_{i=1}^N\int d^4x\, m_i \, \sqrt{-\bar{g}_{\mu\nu}\dot{x}_i^\mu \dot{x}_i^\nu}\delta^{(4)}(x_i(\tau)-x_i),
\end{equation}
where $\bar{g}_{\mu\nu}=h^{-1/2}g_{\mu\nu}+h^{1/2}\partial_\mu x^m \partial_\nu x^m g_{mn}$ is the pull-back of the 10-dimensional metric onto the brane, $x^M(\xi^\mu)$ are the embedding of the brane worldvolume into the spacetime, and we have chosen the static gauge $\xi^\mu=x^\mu$.  The energy density that follows from $S_{\rm DM}$ is:
\begin{equation}
\rho = \sum_{i=1}^N m_i \delta^{(4)}(x_i(\tau)-x_i)\left(\frac{1}{T_3 h(\phi)}\right)^\frac14\sqrt{\frac{\dot{x}^2}{g}}\left(1-h(\phi),(u^\mu\partial_\mu\phi)^2\right)^{-\frac12}
\end{equation}
with $u^\mu = \frac{\dot{x}_\mu}{\sqrt{-\dot{x}^2}}$, from which one obtains a (so-called `disformal' \cite{Bekenstein:1992pj}) coupling with the dark energy scalar, $\phi$.  This non-trivial coupling leads to a non-conservation of dark matter energy density and a non-standard time-evolution as the universe expands.  

Consider two concrete examples, namely ($i$) an AdS warp factor, $h(\phi) \sim \phi^{-4}$, corresponding to a D-brane in the mid-region of a Klebanov-Strassler throat together with a mass term for the scalar potential, $\tilde{V}(\phi) \sim \phi^2$ and ($ii$) a constant warp factor, $h(\phi)\sim$ constant, corresponding to a D-brane in the bulk or close to the tip of the throat, and an inverse power law potential, $\tilde{V}(\phi) \sim \phi^{-2}$. 

 In both cases, a dynamical systems analysis reveals a late-time accelerating scaling expansion, with the universe becoming asymptotically devoid of all matter whilst maintaining a constant rate of acceleration. The acceleration may ultimately end, however, depending on what happens when the dark D-brane eventually reaches the tip of the throat. Note that the usual slow-roll conditions and Hubble scale mass for the dark energy scalar, $\phi$, is not necessary for this acceleration, thanks to the non-canonical kinetic term in the DBI action. The observed dark energy scale is obtained for $m_\phi \sim 10^{-60} M_{\rm Pl}^2$, which can be achieved for a D-brane close the tip of the warp throat (although the cosmological constant problem, as usual, has not been addressed). The consequent interchanges between dark energy and dark matter suggests that their energy densities should be around the same order, providing a solution to the coincidence problem. As well as phenomenological questions on the implications of coupled dark sectors for structure formation, it also remains necessary to embed the interesting D-brane couplings and potential into a fully fledged string compactifications with stabilised moduli.

\paragraph{Dark matter assisted dark energy}

As well as explaining the coincidence problem, coupling dark energy to dark matter can help to achieve accelerated expansion in ways consistent with quantum gravity expectations; with a graceful exit from the dS epoch, thus allowing S-matrices to be consistently defined \cite{Gomez:2022}. 

Motivated by the prevalence of hidden sectors in string theory, the thermal dark energy scenario \cite{Hardy:2019apu} (cf. thermal inflation \cite{Lyth:1995ka}) proposes the existence of a light hidden sector that is still in internal thermal equilibrium in the present-day universe, thus constituting dark radiation. A light hidden scalar $\phi$ -- which can be a modulus or matter field -- can then be stabilised in a temperature-dependent metastable dS minimum, away from its true AdS or Minkowski minimum, thanks to its interactions with bosons $\chi^i$ and fermions $\psi^a$ in the thermal bath. For example, with interaction terms $\lambda_i \phi^2 \chi^i \chi^i$ and $y_a \phi \psi^a \bar{\psi}^a$, and hidden sector temperature $T_h$ much greater than the masses $m_{\chi^i}, m_{\psi^a}$ in the thermal bath, finite temperature effects contribute to the scalar potential for $\phi$ as:
\begin{equation}
V(\phi, T_h) = V(\phi,0) + b T_h^2 \phi^2\,,
\end{equation}
where the constant $b$ depends on the interaction couplings, $\lambda_i, y_a$.  Consider e.g. $\phi$ with a Higgs-like zero-temperature potential $V(\phi,0) = \lambda(\phi^2-\phi_0^2)^2$, with zero-temperature minimum at $\phi=\phi_0$ giving Minkowski space and $m_\phi=2\sqrt{\lambda}\phi_0$. Then, for sufficiently high hidden sector temperatures $T_h > T_c= \sqrt{\frac{2\lambda}{b}}\phi_0$, finite temperature effects stabilise $\phi$ at $\phi=0$, at a dS minimum with vacuum energy $\rho_{\rm DE}=\lambda \phi_0^4$. Phenomenologically viable regions of parameter space can be found for scalar masses $\lesssim \mu$eV, which are much larger than usual quintessence scales.  The metastable dS survives until hidden sector temperatures fall below $T_c$, around which time there will typically be a first-order phase transition to the true global Minkowski minimum, thus allowing a well-defined S-matrix.

A related scenario is `locked dark energy' \cite{Axenides:2004kb} (cf. locked inflation \cite{Dvali:2003vv}). Starting with the same Higgs-like zero-temperature potential for $\phi$ and hidden sector quartic interaction $\lambda_\chi \phi^2 \chi^2$, one considers a region of parameter space where now $m_\chi \gtrsim H_0$, such that $\chi$ behaves as a dark matter field undergoing coherent oscillations, that is, fuzzy dark matter. Similar to the finite temperature contributions in the thermal dark energy case, the background amplitude in $\chi$ can drive $\phi$ to the maximum of its potential $\phi=0$, this happens when $\chi > \chi_c = \sqrt{\frac{2\lambda}{\lambda_\chi}}\phi_0$. Through the oscillation, $\chi$ spends some time with value $\lesssim \chi_c$; assuming an oscillation frequency $\omega_\chi = m_\chi$ this is given by $\Delta t \sim \frac{\chi_c}{m_\chi \bar{\chi}}$ with $\bar{\chi}$ the amplitude of the oscillations. So long as this time is less than the time scale of $\phi$'s tachyonic instability, $(m_\phi)^{-1}$ with $m_\phi$ the tachyonic mass, $\phi$ stays locked in a metastable dS. As the oscillations in $\chi$ are damped by the expansion of the universe, their amplitude decreases with time; the vacuum energy in $\phi$ comes to dominate the universe and drive an accelerated expansion, until, eventually, the amplitude in $\chi$ falls below the critical value $\chi_c$. Then $\phi$ rolls down to its true Minkowski minimum and the epoch of accelerated expansion ends, providing resolution to the S-matrix problem.  Phenomenological regions of parameter space have $m_\chi \lesssim 10$ MeV (so that it has not yet decayed) and, again, $m_\phi \lesssim \mu$eV.

The thermal and locked dark energy scenarios are string-inspired models, but it remains to embed them in fully fledged string constructions with moduli stabilisation, and so identify their phenomenologically viable parameter spaces.

\subsubsection{Time-dependent compactifications}

As per our discussion in Sec. 3, for any higher dimensional theory that satisfies the strong energy condition (i.e. has a stress energy tensor that, together with Einstein's equation, implies $R_{00} \leq 0$), there cannot be a time-independent compactification that has 4-dimensional accelerated expansion (and in fact this extends to massive IIA, which does not satisfy the strong energy condition) \cite{Gibbons:1984kp, Maldacena:2000mw}.  

However, time-dependent compactifications evade this no-go theorem \cite{Townsend:2003fx} and can give rise to transient periods of acceleration \cite{Townsend:2003fx, Ohta:2003pu, Ohta:2003ie, Ohta:2004wk, Roy:2003nd, Emparan:2003gg, Gutperle:2003kc}, corresponding to S-brane (`space-like' brane) solutions \cite{Gutperle:2002ai, Ohta:2003pu, Emparan:2003gg}. Indeed, any compactification that has a semi-positive definite scalar potential in the low-energy effective field theory can generate an accelerated expansion by starting with initial conditions such that the scalar field rolls up its potential with friction: eventually the field reaches a maximum, $\dot{\phi}=0$ and at this point there is an accelerated expansion if the potential dominates over other fluids present (see \cite{Townsend:2003qv} for a review). E.g. for a flat compactification with fluxes or compactification on a (compact) hyperbolic space, $H/\Gamma$ with $\Gamma$ a freely acting orbifold, the scalar potential takes the form $V(\phi)=\Lambda\, e^{-a\phi}$ with $\phi$ a geometric modulus. Thus, as $\phi$ evolves, the compactification is time-dependent. Despite this time-dependence of the compact space, the 4-dimensional Newton's `constant' is actually constant in the Einstein frame arrived at via a time-dependent Weyl rescaling; note that not all coupling constants, however, will be time-independent in this frame. Although there is no no-go theorem against compactifications with late-time accelerating cosmologies \cite{Russo:2018akp}, all explicit solutions found in 10/11-dimensional supergravities, which are low energy effective field theories of string theory, have either acceleration for some transient period that ends in deceleration or acceleration that tends to zero asymptotically; thus they have no cosmological horizon and no problem in defining an S-matrix (see, however, \cite{Dasgupta:2019gcd}, which argues for dS solutions in type IIB/M-theory with time-dependent compact manifolds and fluxes, in the presence of local and non-local quantum corrections).

Ref. \cite{Marconnet:2022fmx} contains a recent discussion on time-dependent compactifications of type IIA supergravity on Einstein, Einstein-K\"ahler and Calabi-Yau manifolds with fluxes, which suggests that cosmologies with either (semi-)eternal acceleration, or parametric control on the number of e-foldings, is generic; interestingly, all solutions found therein have negative spatial 4-dimensional curvature (an open universe). The stability of these solutions against small perturbations is still to be verified. Ref. \cite{Russo:2022pgo} discusses transient cosmological acceleration in a system with two axio-dilatons including one axion flat direction, which is a consistent truncation of maximal massive supergravity theories arising from string theory; these solutions correspond to flat FRW cosmologies with power-law scale factor, some including a transient dS-like epoch with $w\approx -1$.

Although these time-dependent compactifications with transient accelerated expansion have some regions of parameter space allowing to match the current observed values for energy density and equation of state $(\Omega_{\rm DE},w_{\rm DE})\sim(0.7, -1)$, there are a number of open issues. A suitable choice of frame yields a constant 4-dimensional Newton's constant, but the phenomenological viability of the time-dependence in other couplings needs to be verified \cite{Russo:2018akp,Townsend:2021wrs}. Matching the observed dark energy density requires too large a compact space, thus invalidating a 4-dimensional effective description \cite{Gutperle:2003kc}. A proper understanding of S-branes, with strong string coupling, is needed.  Furthermore, the scenario needs to be embedded into a full-fledged string construction that includes matter and moduli stabilisation. 

\subsubsection{Early dark energy}\label{subs:EDE}

So far we have discussed {\em late-time} cosmological acceleration in the context of string theory models. Recently, it has also been proposed to introduce an {\em early-time} dark energy epoch \cite{Poulin:2018dzj,Poulin:2018cxd}, motivated by the persistent discrepancy between the value of the cosmic expansion rate today, $H_0$, determined from direct measurements of distance and redshift, and  its value inferred from    the standard $\Lambda$CDM model using CMB measurements (for recent reviews on the $H_0$ discrepancies and phenomenological solutions see \cite{DiValentino:2021izs, Schoneberg:2021qvd}).
The Early Dark Energy (EDE) proposal postulates that there was a form of energy density, which contributes $\sim10\%$ of the total energy density of the universe briefly before recombination $z>3000$ and then  decays faster than radiation, so that it leaves the late-time evolution of the universe unchanged. This implies that  the expansion rate $H(t)$ is increased shortly before the formation of the CMB, which raises the estimated value of $H_0$ based on CMB data.
  
A phenomenological EDE model is given by a scalar field, $\varphi$, with potential (see \cite{Hardy:2019apu, Niedermann:2021ijp} for some alternatives)
\begin{equation}
\setlength\fboxsep{0.25cm}
\setlength\fboxrule{0.4pt}
\boxed{
\label{eq:VEDE}
V_{\rm EDE} = V_0\left[ 1- \cos \left(\frac{\varphi}{f}\right) \right]^n \,,
}
\end{equation}
where $V_0\sim {\rm eV}^4$, such that the energy density is comparable to that of the universe prior to recombination. If the initial field value is $\varphi/f\sim \pi$,  the field is initially frozen and behaves like a cosmological constant. Later, the field rolls-down its potential and begins to oscillate about its minimum with an equation of state $\omega_{\rm EDE} = (n-1)/(n+1)$, so that its energy density decays as $a\propto \rho^{-6n/(n+1)}$. Therefore, its energy density decays faster than radiation for $n>2$. The best fit values for $(n, f)$ are $n=3$ and $f\sim 0.2\, M_{\rm Pl}$ \cite{Poulin:2018cxd,McDonough:2021pdg}. 

Although the proposal is relatively simple, and successful in relaxing the Hubble tension \cite{Schoneberg:2021qvd}, it presents challenges, such as reproducing the energy scale $V_0\sim {\rm eV}^4$ and the UV origin of the potential \eqref{eq:VEDE}. For an axionic potential, the periodic terms in \eqref{eq:VEDE} can be seen as the leading terms in an instanton expansion. Therefore, the value $n=3$ would correspond to a delicate balance of the terms in an instanton expansion \cite{Rudelius:2022gyu}. Although most of the work on this proposal has been phenomenological, a proposal to realise the EDE potential \eqref{eq:VEDE} in supergravity and string theory has been put forward in \cite{McDonough:2022pku}, using racetrack-like type of superpotentials with suitable coefficients to ensure the required behaviour. 

A more detailed analysis of EDE from type IIB string theory has been recently performed in \cite{Cicoli:2023qri} which identified $C_2$ axions in LVS models as promising candidates to realize EDE from string theory with full closed string moduli stabilisation in a controlled effective field theory. The best working example of \cite{Cicoli:2023qri} is a typical Swiss-cheese LVS model with an orientifold odd modulus $G=\bar{S} b+{\rm i}\,c$ (where $b$ and $c$ are the axions arising respectively from the reduction of $B_2$ and $C_2$ on the orientifold odd 4-cycle) which mixes with the big modulus $T_b$. The superpotential takes the form
\begin{equation}
W = W_{\rm LVS }+A_1\, e^{-a (T_b+\mathfrak{f} G)}+A_2\, e^{-a(T_b+2 \mathfrak{f} G)}+A_3\, e^{-a (T_b+3 \mathfrak{f} G)}\,,
\label{WC2LVS}
\end{equation}
where $W_{\rm LVS}$ is the standard LVS superpotential and $a=2\pi/M$. The last 3 terms in (\ref{WC2LVS}) are generated by gaugino condensation on D7-branes with non-zero world-volume fluxes $\mathfrak{f}_k=k \mathfrak{f}$ ($k=1,2,3$) that yield the following EDE potential (after fixing ${\rm Im}(T_b)=b=0$):
\begin{equation}
V_{\rm EDE}= V_0\left[\frac52 -\frac{15}{4}  \cos\left(a \mathfrak{f}\, c \right)+ \frac32 \cos\left(2 a \mathfrak{f}\, c \right)
- \frac14 \cos\left(3a\mathfrak{f}\, c \right) \right],
\label{eq:VEDEaux}
\end{equation}
with $A_1 = -15 A/4$, $A_2 = 3A/2$, $A_3=-A/4$. After expressing the $C_2$ axion $c$ in terms of the canonically normalised field $\varphi$ with decay constant $f\sim 0.2 \sqrt{g_s}\,M\,\mathcal{V}^{-1/3}$, eq. (\ref{eq:VEDEaux}) reproduces the required EDE potential (\ref{eq:VEDE}) with $n=3$. Writing the relation between $f$ and the non-perturbative action $S$ as $f S\sim \lambda\,M_{\rm Pl}$, the EDE scale $V_0$ can be written schematically as:
\begin{equation}
V_0 = \Lambda\,e^{-S}\,M_{\rm Pl}^4 \simeq \Lambda\,e^{- \lambda M_{\rm Pl}/f}\,M_{\rm Pl}^4 \simeq \Lambda\,e^{- 5 \lambda}\,M_{\rm Pl}^4\quad \text{for}\quad f \simeq 0.2\,M_{\rm Pl}\,,
\label{KeyRel}
\end{equation}
which can reproduce $V_0 \sim 10^{-108}\,M_{\rm Pl}^4$ without tuning $\Lambda$ only if $\lambda \gg 1$. As summarised in Tab. \ref{tab:closedaxions}, ref. \cite{Cicoli:2021gss} found that $\lambda\sim\mathcal{O}(1)$ for $C_4$ axions with ED3/D7 non-perturbative effects and $C_2$ axions with fluxed ED1/D5 non-perturbative effects, while $\lambda\sim \sqrt{g_s}\,\mathcal{V}^{1/3}\gg 1$ for $C_2$ axions with fluxed ED3/D7 non-perturbative effects. Hence, this singles out this last case as the only one which can realise EDE without excessive fine-tuning. In fact, in this case the EDE scale and decay constant can be matched for $A\sim |W_0|\sim\mathcal{O}(1)$, $g_s\sim\mathcal{O}(0.1)$, $M\sim \mathcal{O}(100)$ (implying that fluxed ED3-instantons with $M=1$ would not give the right $f$ for $\mathcal{V}\gg 1$) and $\mathcal{V}\sim \mathcal{O}(5\times 10^5)$ which correlate with $m_{3/2}\sim\mathcal{O}(10^{13})$ GeV and $m_{\mathcal{V}}\sim \mathcal{O}(10^{10})$ GeV \cite{Cicoli:2023qri}.

\newpage

\section{Alternatives}
\label{sec:Alt}

So far we have followed the main highway of string cosmology, taking as a guideline the most standard and successful approach to the early universe, namely the inflationary universe. But, even this is limited in scope. String theory's standing as the prime candidate for a fundamental theory of Nature implies that
at some point it should be able to address the earliest moments of the Universe and the questions of what happened before inflation. 
Even more interestingly, it may give rise to totally different scenarios of the early universe which manifest an intrinsic stringy structure of matter and 
provide alternatives to the inflationary scenario. 

Over the years, several attempts have been proposed to explore this important possibility. It is essentially impossible to give a full review of 
all these proposals and we content ourselves with providing a short overview of the main ideas put forward in each case, 
referring to other reviews where these ideas are discussed in more detail \cite{Gasperini:2002bn,Lehners:2008vx,Brandenberger:2008nx,Battefeld:2005av,Brandenberger:2016vhg,McFadden:2009fg,Nastase:2019rsn}. 
In particular, we follow closely the summary presented in \cite{Quevedo:2002xw} and update the original presentations there.

Before discussing concrete proposals, we point out general areas where string theory and cosmology ought to intersect:

\begin{enumerate}
\item {\it Big-bang singularity.} Within the current understanding of string theory it is not possible to address the earliest point in the history of the universe. It is widely believed that before reaching the spacetime singularity, even spacetime itself could become an emergent quantity where a fully-fledged non-perturbative formulation of string theory (or quantum gravity) will be needed. For
various approaches to address this question see e.g \cite{Damour:2000wm, Damour:2002et, Cornalba:2003kd, Hubeny:2004cn,Fischler:1998st,  Craps:2006yb, Berkooz:2007nm, KalyanaRama:1997xt, Englert:2007qb, Barbon:2015ria, Engelhardt:2015gta} (we will discuss
some in detail in what follows).

\item{\it Hagedorn phase.} Since the early days of string theory it has been known that there is a critical temperature known as Hagedorn's temperature, $T_H$. 
As the number of string states increases exponentially with energy, the partition function
\be
Z={\rm Tr} e^{-\beta H},
\ee 
with $\beta=1/T$ and $H$ the Hamiltonian, diverges at a finite temperature $T_H=\frac{1}{4\pi\sqrt{\alpha'}}$. This indicates that going back in time from the present the Universe will reach a point in which the stringy nature will dominate and at the temperature $T_H$ a phase transition may occur. The precise nature of this phase is not known but the existence of a critical temperature clearly indicates that the stringy nature of the early universe would manifest itself at finite times.
  
  \item{\it Vacuum transitions.} The work by Coleman and collaborators to address quantum tunelling transitions among de-Sitter, anti-de-Sitter and Minkowski spacetimes could mark the beginning and end of our universe in a multiverse scenario. String theory should be able to provide a proper formulation to address these questions in a full quantum theory of gravity.
  
\item{\it The wave function of the universe.} The attempts to describe a wave function of the universe and the creation of the universe from nothing, as pioneered by Vilenkin, Hartle and Hawking, were developed using a semiclassical approach to gravity. String theory, being a UV complete theory, should be able to provide a proper formulation of these proposals.

\end{enumerate} 

Addressing these questions is important for string theory (and also any other proposal of quantum gravity), and may or may not lead to an inflationary universe. It is important to hold these questions in mind and consider with an open mind possible alternatives to inflation, following where the science leads but also not being contrarian for its own sake.

\subsection{Time-dependent String Backgrounds}

Formulating string theory in time-dependent backgrounds remains an open question. However, we can find with more ease time-dependent backgrounds 
of the low-energy ten-dimensional field theory for the different string theories. In these solutions, the metric, dilaton and antisymmetric tensors may be time-dependent with cosmological implications. A natural question is to search for solutions in which 4-dimensions expand and the other ones are either 
static or expand much more slowly, in order to address the fact that we only observe four large dimensions.

For this, we first consider 
a low energy string effective action including the metric $g_{MN}$, 
the dilaton $\varphi$ and the NS-NS antisymmetric tensor of the bosonic and heterotic strings $B_{MN}$. In an arbitrary number of dimensions $D=d+1$
the bosonic action takes the form (in units where $l_s = 1$):
\be
\setlength\fboxsep{0.25cm}
\setlength\fboxrule{0.4pt}
\boxed{
S\ = \ \int d^{D}x\sqrt{-g}\ e^{-\varphi}\left( R + \partial_M\varphi
\partial^M\varphi -\frac{1}{12} H_{MNP} H^{MNP}+\cdots \right),
}
\ee
where  $H_3=dB_2$. 

Two simple classes of solutions can be described directly, including only the dilaton and metric time dependence with vanishing torsion $H_{MNP}=0$:
\begin{enumerate}
\item {\it Linear Dilaton}. 
In the string frame there is a general solution of $\varphi=A\,t$, with $A$ a constant, and  metric $g_{MN}=\eta_{MN}$ \cite{Myers:1987fv,Antoniadis:1988vi}. Going to the Einstein frame, the solutions allow for a cosmological interpretation with
metric $ds_D^2=-dt^2+t^2 dx_{\small D-1}^2$ which may be given a cosmological interpretation with linearly expanding dimensions.
\item{\it Rolling toroidal radii}. 
A simple solution with vanishing curvature is to assume that all the spatial dimensions are tori with time-dependent radii $R_i(t)$ \cite{Mueller:1989in}. For the critical dimension, in the string frame:
\be
\setlength\fboxsep{0.25cm}
\setlength\fboxrule{0.4pt}
\boxed{
R_i(t)\propto  t^{p_i}\qquad e^{-\varphi(t)}\propto  t^p, \qquad \sum_i p_i^2=1, \qquad \sum_i p_i=p.
\label{muellers}
}
\ee
For different values of the constants $p_i$, this can generate either expanding or static solutions, but without any preference for the physical case $p_i>0, i=1,2,3$ and $p_i=0$ for $i>3$. The solution in Eq. (\ref{muellers}) is recognisable as a generalised Kasner solution, extended to include a rolling dilaton.
\end{enumerate}
Even though these are not realistic cosmological solutions, they were the first steps in the study of time-dependent solutions of string theory. Over the years further solutions have been studied with interesting cosmological interpretations.\footnote{It is important to keep in mind that four dimensional cosmology experienced by observers can be an effective one, as is clearly exhibited in  mirage cosmology
\cite{Kehagias:1999vr}}
 We will discuss some of them later on. In particular, solutions generalising the original chaotic solutions of 4D gravity from Belinsky, Lifshitz and Kalathnikov \cite{Belinsky:1970ew,Belinsky:1982pk}  have been much studied with interesting mathematical structure \cite{Damour:2000wm,Damour:2002et,Belinski:2017fas}. For a recent discussion of general classes of cosmological solutions see \cite{Hohm:2019jgu,Russo:2022pgo}. An interesting concrete study of  cosmological string theory backgrounds in two dimensions for which  three and four point functions can be explicitly computed has recently been done in \cite{Rodriguez:2023kkl}.

\subsection{String/Brane Gas Cosmology}

One of the first approaches to string cosmology  was the Brandenberger-Vafa scenario \cite{Brandenberger:1988aj} (see also \cite{Tseytlin:1991xk}, for reviews and recent discussions see \cite{Battefeld:2005av,Brandenberger:2006vv}). The idea is based on the simplest manifestation of T-duality in toroidal compactifications with radius $R$:
\be
\setlength\fboxsep{0.25cm}
\setlength\fboxrule{0.4pt}
\boxed{
M^2\ = \  \frac{n^2}{4R^2} + m^2 R^2 + N_L + N_R -2\,,
}
\ee
in units of the inverse string tension $\alpha'=1/2$. The integers $n$ and
 $m$ give the quantised momentum in the 
circle and the winding number of the string in the circle, $N_{L,R}$ are the 
left and right oscillator numbers respectively. 
We can easily see that the spectrum is invariant under the simultaneous exchange of $R\leftrightarrow 1/(2R)$
and winding and momentum modes, $n\leftrightarrow m$, with a self-dual radius of $R_c=1/\sqrt{2}$. This is  the simplest manifestation of $T$-duality in string theory.

Brandenberger and Vafa explored the possible implications of this duality in cosmology and made a few interesting remarks:

\begin{itemize}
\item
The concept of distance is different for radii greater and smaller than the self-dual radius $R_c$. For radii greater than $R_c$, distance can be defined in the standard way as the dual of the conjugate momenta. But below the critical radius, the winding models play the role of momenta and it is the conjugate dual of the winding modes that properly define distance,
Supposing this duality to hold in general, it is then possible to avoid distances smaller than the self-dual radius as the physics at those distances is equivalent to the physics at distances greater than the self-dual radius. In this sense, this concept of minimal length offers a way to avoid the big-bang singularity.
\item
Winding modes, and the tension associated to them, suggest that energetics can favour a small value of the corresponding circles. If winding modes annihilate with modes of the opposite winding, the corresponding circle can grow as large as possible and so essentially de-compactify. $p$-dimensional objects generically intersect in at most $2p+1$ dimensions. Therefore, winding strings ($p=1$) can annihilate in three spatial dimensions and allow these dimensions to become large, offering a potential explanation of the fact that we live in $3+1$ dimensions. This proposal is one of the very few concrete proposals to explain one of the most clear experimental properties of our observed universe, namely that it has three large dimensions and that the other ones, if they exist, are somehow hidden to us.\footnote{As mentioned in the brane inflation section, there are other proposals to address the dimensionality of spacetime \cite{Burgess:2001fx,Karch:2005yz, Durrer:2005nz}. For an independent  early proposal see \cite{Gibbons:1986uu}. } Some numerical studies have been made of this proposal, with mixed results (see for instance \cite{Sakellariadou:1995vk,Greene:2009gp,Greene:2012sa}).

\item
Independent of the dimensionality issue, this proposal emphasises that the current approach to cosmology in the FLRW model should be expanded such that the stress-energy tensor contribution to the density and pressure of the universe  should be that of a gas of strings and branes entering into the Einstein's equations used to study early universe cosmology. Extrapolating back in time from the present we reach higher and higher temperatures and at the critical Hagedorn temperature $T_H$ there is an expected  phase transition to a new phase dominated by a gas of  strings and branes \cite{Battefeld:2005av,Brandenberger:2006vv, Alexander:2000xv, Easther:2002qk, Easther:2003dd}. This is 
illustrated in figure \ref{Hagedorn}. Even though the nature of this phase is not known its cosmological implications may be possible to explore. 
\end{itemize}

\begin{figure}[t]
\begin{center}
\includegraphics[width=170mm,height=90mm]{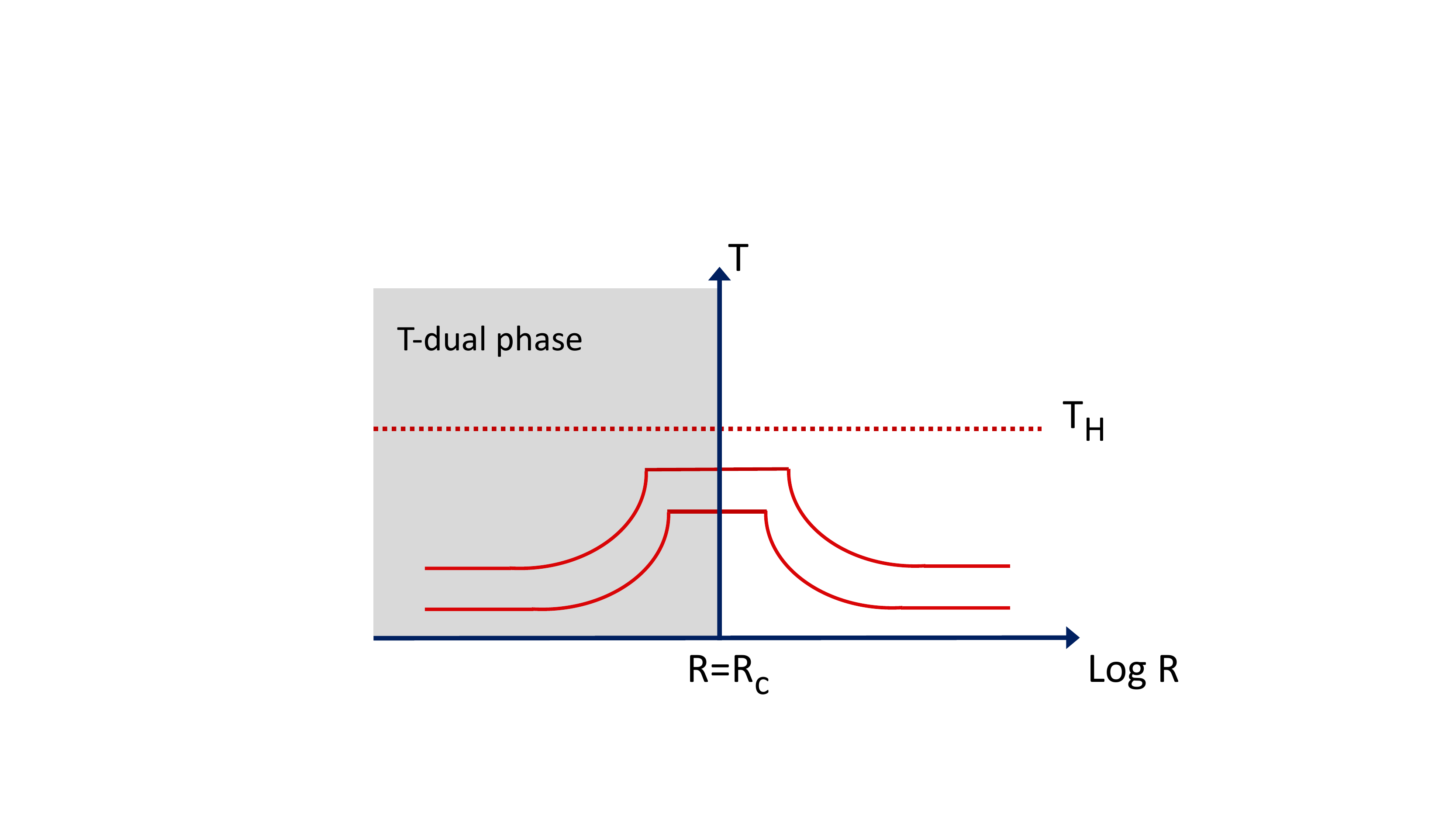} 
 \vskip -25pt
\caption{Temperature $T$ against the radius of the extra dimensions $R$ in string gas cosmology illustrating the Hagedorn temperature as a limiting temperature in string theory and the two dual phases of T-duality separated at the self dual point $R_c$.} \label{Hagedorn} 
\end{center}
\end{figure}

Although these ideas are very attractive, they have been mostly formulated in  simplistic cases for which all of the dimensions are circles. They have not been formulated on realistic set-ups in which chiral matter like in the Standard Model is present. A further issue is that the intuition that strings under tension causes cycles to shrink only holds under the assumption that the dilaton has been stabilised, e.g. by some standard mechanism such as gaugino condensation. This is manifest in 4-dimensional Einstein frame, where the radion is a linear combination of the string frame dilaton and radion.

The lack of understanding of the Hagedorn phase is also a major obstacle at the moment. However, these ideas may eventually be applicable in more realistic frameworks and it is worth to keep them in mind. In particular, after the emergence of the brane-world scenario, generalisation of these ideas may add to the arguments for the critical dimensionality of spacetime. For example, for D4 branes ($p=4$) to avoid annihilating each other they need to live in a universe with dimension greater than $2p+1=9$. This may put a bound on the brane dimensionality to be smaller than four \cite{Burgess:2001fx,Durrer:2005nz}.  A more detailed study for IIB string theory singles out  D3 and D7 branes \cite{Karch:2005yz} which are precisely the branes which host the Standard Model in F-theory and other local D-brane constructions of the Standard Model. For a realisation of inflation making use of the Hagedorn phase see \cite{Abel:2003jh}.

In the regimes where effective field theories are applicable, there also remains the standard challenge of implementing the scenario in realistic set-ups including moduli stabilisation (for a study of a possible realisation in type IIB string theory, see \cite{Frey:2005jk}).

\subsection{Pre Big-Bang Cosmology}

In the 1990s Veneziano and collaborators went beyond the  Brandenberger-Vafa  proposal by considering the possibility of
 $T$ duality in cosmological backgrounds much closer to  the FRW type. For an ansatz  of the type:
$ds^2= -dt^2 + \sum_{i=1}^d a_i^2(t)\ dx_i^2$
it can be seen that $T$ duality is a symmetry of the 
equations of motion acting as:
\be
\setlength\fboxsep{0.25cm}
\setlength\fboxrule{0.4pt}
\boxed{
a_i(t)\rightarrow \frac{1}{a_i(t)},
 \qquad \varphi\rightarrow \varphi - 2 \sum_i\log a_i.
 }
\ee
Since $a_i(t)$ represent the scale factor, as in FRW, this 
has been named {\it scale factor duality} \cite{Veneziano:1991ek}. Thus we can see that expanding and
contracting universes are related by this symmetry.

Gasperini and Veneziano combined this  symmetry with the standard time-reversal symmetry:
$a(t)\leftrightarrow a(-t)$, to allow for a possibility of considering cosmology before  $t=0$, in which the Hubble parameter increases instead of decreases.
Without duality, the symmetry under $t\rightarrow -t$ would send
$H(t)\rightarrow -H(-t)$ but, combining
 this with scale factor duality, it provides four different sign 
combinations for $ H(t)$. 
If the universe at late times is decelerating, $H$ would be a decreasing
 monotonic function of time for `positive' $t$, but a combination of duality
 and the 
$t\rightarrow -t$ transformation can give rise to  $H(-t) = H(t)$ so 
that this
 function can be even, as shown in figure \ref{PreBigBang}. They therefore proposed a scenario in which the universe accelerates from negative times
 towards the big-bang and then decelerates after the big-bang. The 
acceleration would 
indicate a period of inflation before the big-bang without the need of an
 scalar
 potential. This scenario is called {\it Pre Big-Bang Cosmology} \cite{Gasperini:1992em,Gasperini:2002bn,Gasperini:2007vw}, see also \cite{Tseytlin:1991wr}. 

A concrete solution for this system corresponds to the isotropic case
$a_i=a_j\equiv a(t)$ for which:
\be
a(t)= t^{1/\sqrt{d}}\qquad t>0\ , 
\ee
with a constant dilaton. For this solution, $H(t)\sim 1/t$ decreases monotonically 
with time. By applying the transformation $t\rightarrow -t$ and duality we
 can generate the four different branches of solutions:
\be
a(t)=t^{\pm1/\sqrt{d}}\qquad t>0, \qquad a(t)= \left(-t\right)^{\pm1/\sqrt{d}}\qquad t<0, \label{veneziano}
\ee
with 
$\varphi_\pm (\pm t)\ = \ (\pm\sqrt{d} - 1) \log(\pm t)$.

\begin{figure}[t]
\begin{center}
\includegraphics[width=160mm,height=70mm]{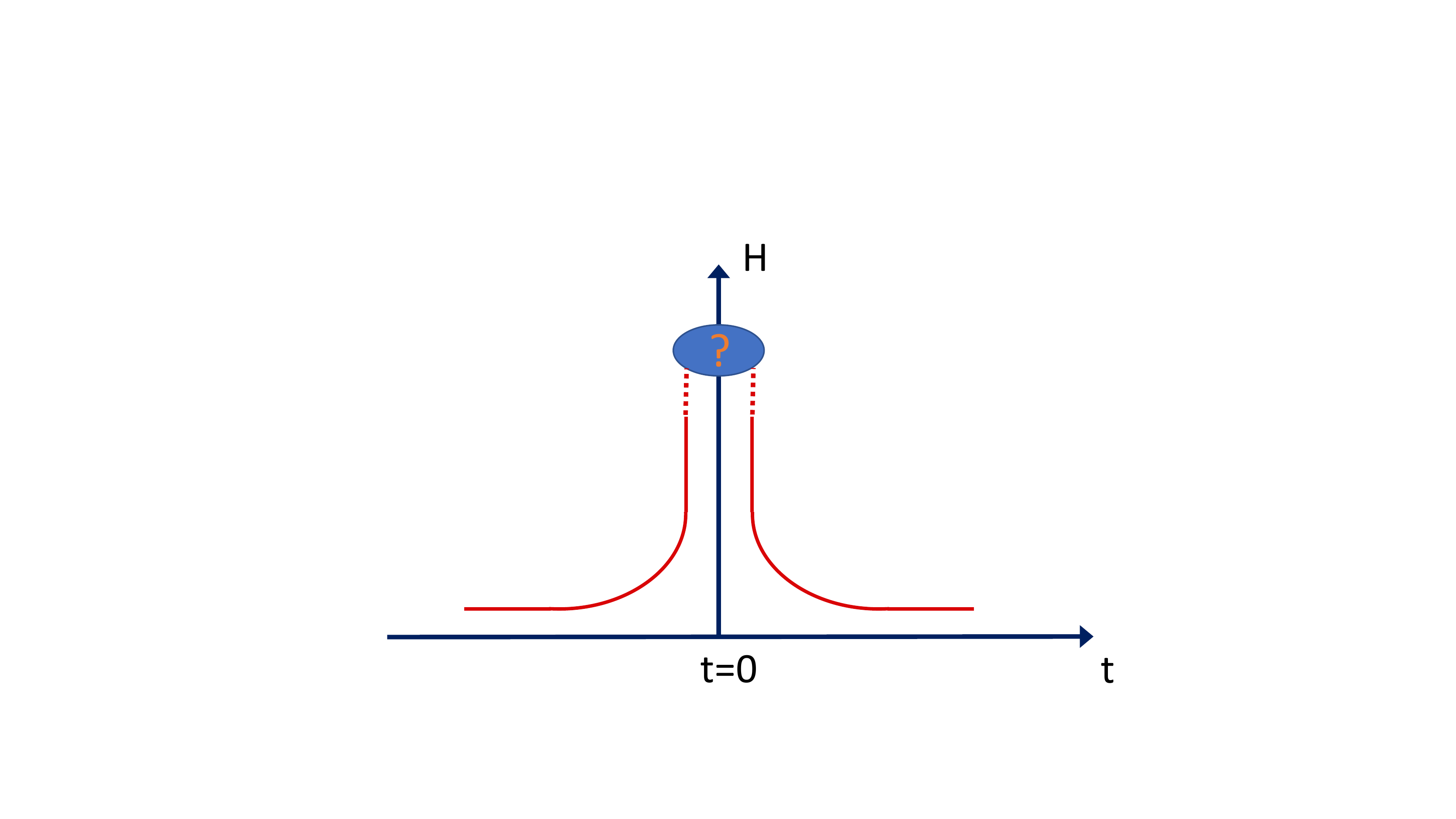} 
\caption{Possible realisation of the pre big-bang scenario with the past and future regions expected to match at the singularity with strong coupling and large curvature.} \label{PreBigBang} 
\end{center}
\end{figure}

In the two branches for which  $H>0$, the universe expands, which provides an 
interesting realisation of the pre big-bang scenario, as illustrated in figure \ref{PreBigBang}.
The solutions are such that there is  a singularity at $t=0$ but also in this
 region the dilaton blows up implying strong string coupling. 
The hope is that non-perturbative string effects provide a smooth 
matching between these two branches. Since the weak 
coupling perturbative string vacuum appears as a natural initial condition 
in the pre big-bang era, the scenario consists of an empty cold
 universe in the infinite past which expands in an accelerated way 
towards a region of higher curvature. Eventually, it approaches the region of 
strong coupling and large curvature, which is assumed will match smoothly
 to the post big-bang branch in which the universe continues expanding but now with
decelerated expansion.

 The spectrum of density perturbations in this scenario has been estimated, and claimed not to contradict the recent 
observations. The scenario also
provides testable differences compared to inflation via the tensor 
perturbations, which could be put to test in future searches for gravitational waves.

While this scenario has very interesting features, it has also been subject to criticism for several reasons. First, as
 the original authors pointed out,
 the main problem to understand is the graceful exit question,
namely how to pass smoothly from the pre to post Big-Bang period. As this requires describing the big-bang singularity, this is a major challenge. 
Close to the Big-Bang the perturbative treatment of 
string theory does not hold since the dilaton and the curvature increase,
 implying strong string coupling. Therefore, there is no concrete way to
 address this issue within the framework where the theory is formulated. 
Another important
 problem is the fact that the moduli are neglected from this analysis and 
there has to be a mechanism that stabilises the extra dimensions. 
Furthermore, the scale factor duality symmetry that motivated the scenario is not
 clearly realised
in  more realistic settings with nontrivial matter content and the fact that the 
dilaton will eventually be fixed by non-perturbative effects may change the 
setting of the scenario. 

On the other hand, this represents an explicit proposal for the early universe with interesting string inputs which  may prove useful in a
more realistic scenario. Furthermore, the study of this scenario has led to interesting potential signatures from gravitational waves and has triggered
much activity in this direction. After the detection of gravitational waves and the future progress in this direction, these preliminary studies of gravitational waves
from a string theory perspective may prove very useful for experimental searches and may be a useful guideline for alternative proposals.

\subsection{Ekpyrotic/Cyclic Scenario}

The ekpyrotic scenario (illustrated in figure \ref{Fig:HW}) was developed in the early 2000s by Khoury, Ovrut, Steinhardt and Turok and has presented itself as an interesting alternative to the inflationary universe, drawing its original inspiration from string theory \cite{Khoury:2001wf,Khoury:2001bz,Steinhardt:2002ih, Lehners:2008vx}.

The proposal was first formulated within a particular string theory scenario, namely the 11-dimensional formalism of Horava and Witten with one of the dimensions compactified in an interval $I$ (understood as a $Z_2$ orbifold of a circle $I=S^1/{\bf Z}_2$). 
The endpoints of the interval correspond to two parallel 10-dimensional spaces or {\it end of the world branes}  
which are the `fixed points' of the orbifold, with each hosting $E_8$ gauge theories, providing a strong coupling version of the heterotic string.
 Further compactification on a six-dimensional Calabi-Yau manifold then leaves two 
4D worlds at the ends of the interval in the 5D bulk. In principle, quasi-realistic
 models can be obtained from this approach, mostly using the topological
 properties of Calabi-Yau manifolds. It turns out that besides the end
 of the (interval) 
world branes, which we may also refer to as boundary  branes,
 there are also five-dimensional branes in these compactifications (M-branes) that are not restricted to
live at the fixed points and can move through the bulk.
 These are called bulk branes in order to differentiate them from the boundary branes. Overall, these
 configurations are reminiscent of models with both D-branes at orbifold singularities and also mobile branes.

\begin{figure}[t]
\begin{center}
\includegraphics[width=150mm,height=90mm]{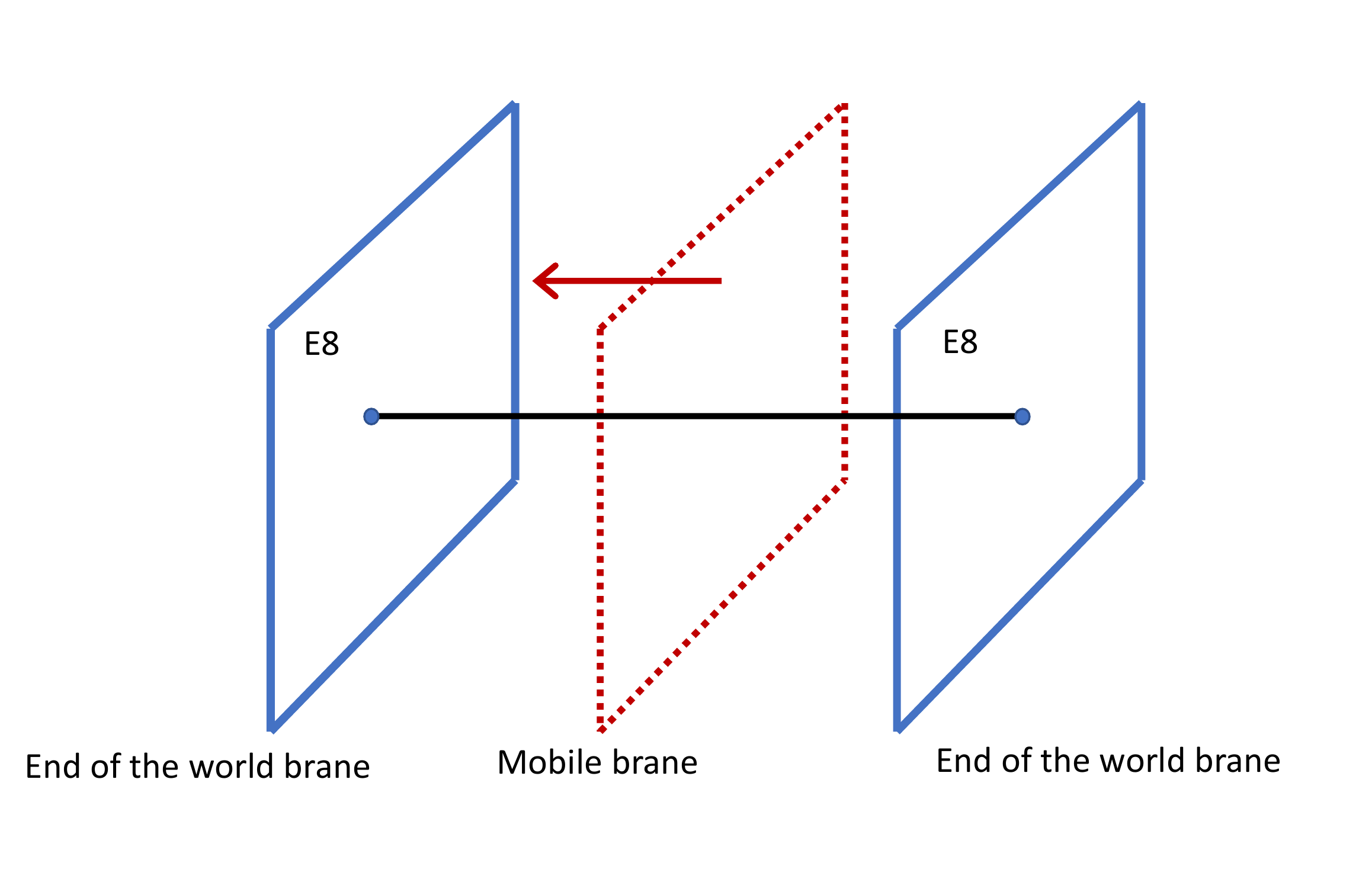} 
\caption{The Horava Witten scenario. 
Two surfaces or end of the world branes, each at the end of 
the interval, provide chiral matter and possibly interesting cosmology as proposed in the ekpyrotic/cyclic scenario. Ripples on the mobile brane may be the source of density perturbations.} \label{Fig:HW} 
\end{center}
\end{figure}

Khoury {\it et al.~} also made an interesting proposal regarding the collision of branes, not to obtain inflation as in the brane inflation discussed in section \ref{sec:infla}, but as an alternative to inflation. The original idea
was to assume that a (hidden) bulk brane going from one boundary of the interval to the other end, would collide with the second (visible) boundary 
brane and produce the Big-Bang. The bulk brane would be almost BPS -- by which it was 
meant that it is essentially parallel to the boundary branes --  and move freely and slowly from one end of the interval to the other.
Small quantum fluctuations  induce some ripples on this brane which, when colliding with the visible end of the world 
brane, would produce the density fluctuations measured in the CMB. There is 
no need of an inflation potential for this. A potential of the type 
$-e^{-\alpha Y}$, with $Y$ is the separation of the branes, was proposed, although not derived, describing  the attraction of the branes. 
The 5D metric is taken with a warp factor that implies that the
motion is
 from smaller to larger curvature across the interval and therefore the scale 
factor depends on the position of the brane in the interval. 

This proposal has received several critics The first involves 
the standard problems solved by inflation. The horizon and flatness
problems require the  branes to be almost exactly parallel before collision,
which may require a fine-tuning of initial conditions. Although relics such as
monopoles will not be present if the collision temperature is low
enough this argument needs to be properly quantified. There is also no general natural
dilution, as in the exponential expansion of inflation, making the solutions of these problems more
difficult in general.  The issue of fine tuning of
 the initial conditions in the ekpyrotic scenario has been widely debated.

However, the most prominent difficulty of this scenario is the following:
in a 4D description, $\dot a<0$ before the collision, while it is expected that $\dot a>0$ after the collision,
 which requires a transition from contraction to expansion, without, in principle, crossing a 
singularity. This is a problem because it violates the null energy condition.

Let us briefly review this argument. Consider gravity coupled to a scalar field: (setting $\kappa_5=1$) 
\be
{\cal L}\ = \ \sqrt{-g}\left( R-\frac{1}{2} 
\partial_\mu\phi\partial^\mu\phi-V(\phi)\right),
\ee
The energy density and pressure are given by:
\be
\rho\ = \ \frac{1}{2}\dot\phi^2 + V, \, \qquad p\ = \ \frac{1}{2}\dot\phi^2 - V .
\ee
Therefore, Einstein's equations imply:
\be
\setlength\fboxsep{0.25cm}
\setlength\fboxrule{0.4pt}
\boxed{
\dot H =  -\frac{1}{2}\left(\rho+p\right) = 
-\frac{1}{2}\dot\phi^2 \leq \ 0.
}
\ee
This implies that $H$ is  monotonically decreasing and 
we cannot go from contraction ($H<0$) to expansion ($H>0$).

This problem motivated a second version of this scenario 
 which does not include the mobile brane but considers the collision of the two end-of-the-world branes. In this case, there is 
a singularity at the moment of collision, since the size of the fifth dimension reduces to zero, 
 which, in principle, could allow a transition from 
contraction to expansion. The singularity happens only in the extra dimension because the scale 
factors of the branes remain finite during the process. After the collision, the two branes separate again and the scale factor increases (see Fig. \ref{F:ekpyvar}).
\begin{figure}[t]
\begin{center}
\vskip -10pt
\includegraphics[width=140mm,height=90mm]{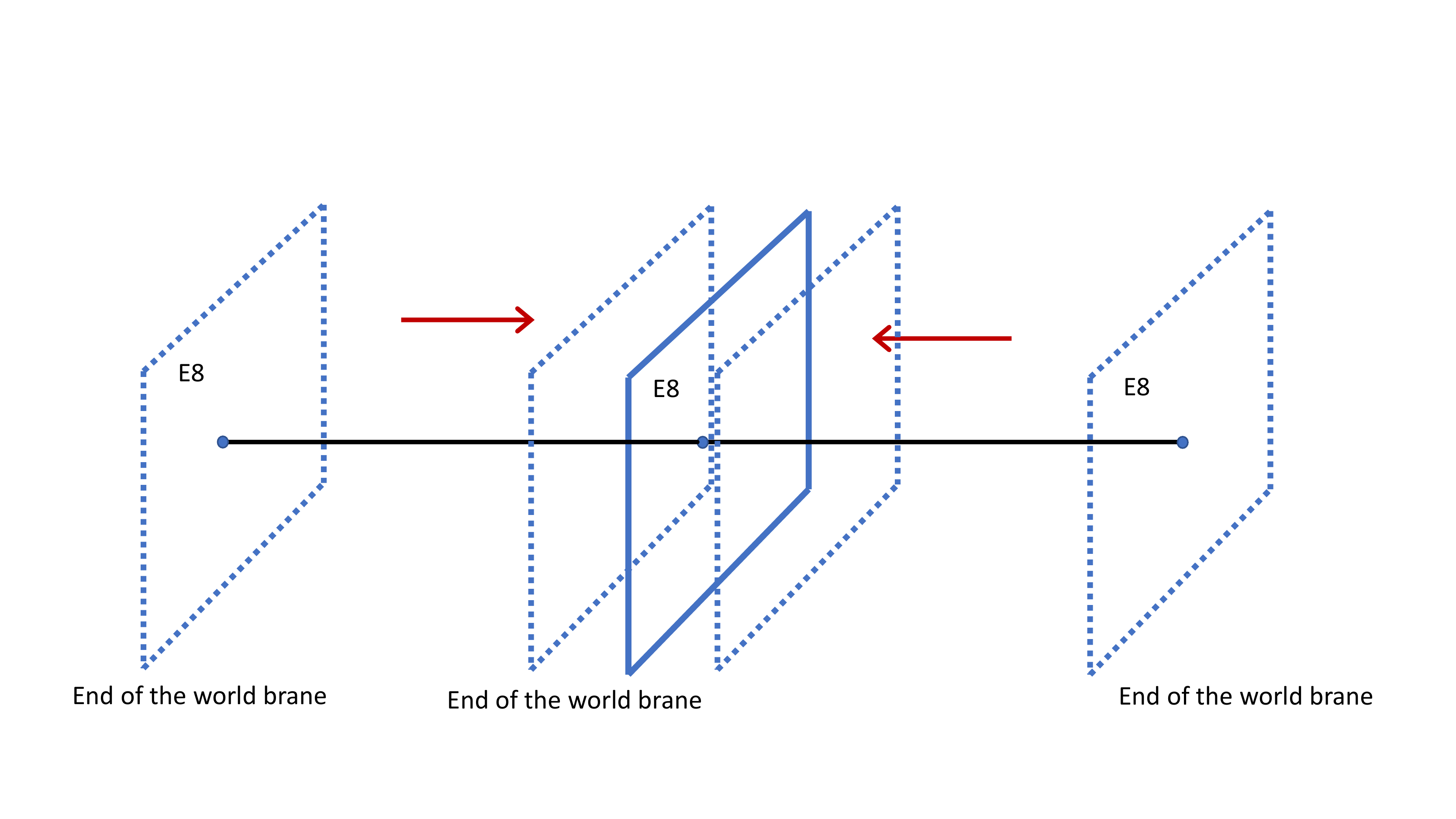} 
\caption{A variation of the original ekpyrotic universe scenario in which the end-of-the world branes approach each other, pass through their overlap point and at some point return, so completing the cycle.} \label{F:ekpyvar}
\end{center}
\end{figure}

In terms of a four-dimensional EFT,  this process can be understood in similar terms to the discussion of the
 Pre Big-Bang proposal. Neglecting the scalar potential and identifying  the separation between the branes with 
 the string dilaton (as it happens in the Horava-Witten scenario) we can use equations (\ref{veneziano})
for the case $d=4$. Out of the four possibilities provided by the choices of sign we can choose:
\be
a(t)\ = \ |t|^{1/2}, \qquad \phi\ = \ \phi_0\pm \sqrt{3}\log|t|.
\ee
The  scale factor $a(t)$ goes from contraction
at negative $t$  to expansion at positive $t$. This still leaves the choice of sign for the 
dilaton open. Since the string coupling is proportional to 
$e^{-\phi}$, the negative sign  choice that was taken in the pre big-bang scenario
corresponds to strong string coupling whereas the positive choice, 
chosen in the ekpyrotic scenario, implies weak coupling at $t=0$. Therefore 
Khoury and collaborators conjectured that the transition is smooth at the singular point and  afterwards the branes may start to separate again.

This leads us to a third version of this scenario that corresponds
 to the {\it cyclic universe} \cite{Steinhardt:2002ih}. In this case, the two branes keep separating and
 passing through each other an infinite number of 
times, as long as the interacting 
potential has a very particular form. For instance, for a potential like that of
 Fig. \ref{Fig:cyclic}, we may describe the universe's history by starting on the right
 hand side corresponding to the current time. The potential is taken to be slightly positive 
and with a slightly negative slope, reflecting the fact that the universe accelerates today as in quintessence.

 Since the slope of the potential is  negative, the scalar field
will start rolling towards smaller values representing the higher dimensional picture of the branes approaching each other.
At some point the field will cross the $V=0$ point and its energy density
 will mostly be kinetic. The potential rapidly becomes negative
 and then the energy density $\rho= V+\frac{1}{2}\dot\phi^2$ touches zero, implying 
that the universe starts contracting. Since the kinetic energy is also 
large, the 
field  passes through the minimum, towards the flat region at infinite
$\phi$ or zero string coupling, where the branes collide
and bounce back with enough energy to re-cross the steep minimum and
go to the right hand side of the 
potential, where it returns to a radiation dominated era,before repeating the
whole cycle again.
\begin{figure}[t]
\begin{center}
\includegraphics[width=100mm,height=60mm]{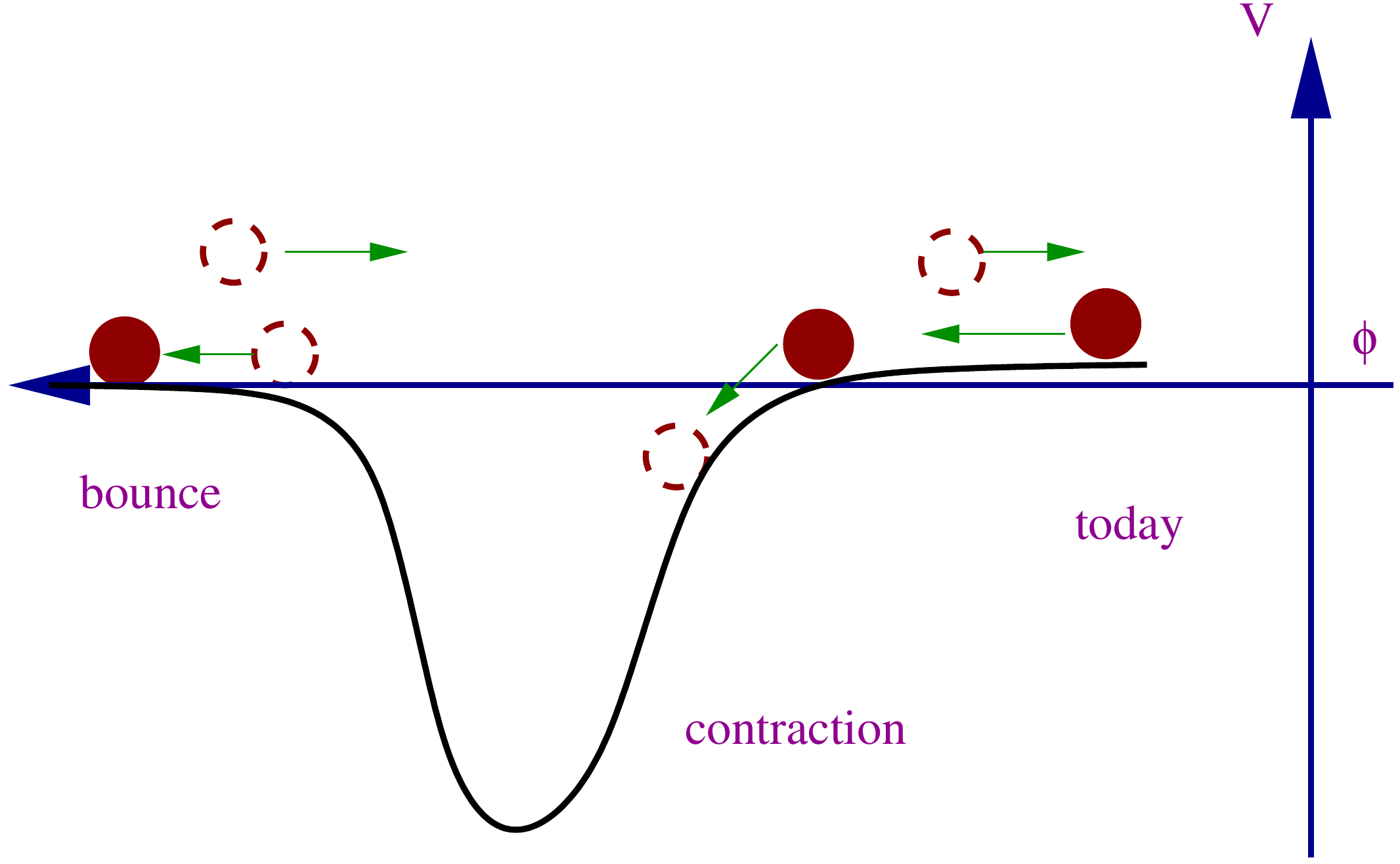} 
\caption{An illustration of the potential and
 trajectory of the field in the cyclic universe. Figure taken from  \cite{Quevedo:2002xw}. } \label{Fig:cyclic} 
\end{center}
\end{figure}

The different versions of this scenario claim to resolve the same questions addressed by inflation. 
For instance, the claim is that the horizon problem does not exist if there is a bounce, since there will be
 clear causal contact between different points. In the cyclic version,
the late period of mild inflation plays a similar role as the
original inflationary scenario by dissolving unwanted objects such as 
like magnetic monopoles, and emptying the universe for the next cycle, thereby also solving the
flatness problem.
 Furthermore, the spectrum of perturbations has been claimed to be 
consistent with observations. Although this issue has been debated, 
all parties seem to agree that the methods used so far are not entirely
conclusive one way or another.

There are some  interesting aspects to these proposals, 
especially regarding  the revival of the cyclic universe. A cyclic universe was originally proposed in the 1930's 
but it was immediately realised that the entropy increases on each cycle, requiring the length of
each cycle to increase. Extrapolating back in time, we reach 
an initial singularity rendering the model only semi-eternal,
similar to eternal inflation which also requires a beginning. So it is not really cyclic.

In the current version, the old entropy problem is claimed to be solved as follows. Although it is true that the total entropy 
 increases with each cycle, the entropy of matter is always the same 
at the end of each cycle. This is due to the present accelerated expansion
 which dilutes matter and so renders the universe essentially
 empty, actually one particle per Hubble radius, before restarting the cycle.
This scenario was found in the context of the ekpyrotic scenario,
but it is clearly independent of it and may have other  realisations.

Another interesting point of this scenario is that it connects the early
 universe and 
late universe in a coherent way. The current acceleration is used as a way 
 to prepare the universe to the next cycle.

Even though the scenarios were motivated in terms of string theory, the 
kind of scalar potentials that are needed for this scenario to work are relatively contrived and have not been 
derived from the underlying theory.
 This is certainly an important  question to be addressed before these models 
can be considered genuine string/M-theory models. In this sense, these scenarios are 
 presently at a similar stage as D-brane inflation was in 1998 where the scalar 
potential was only guessed, rather than explicitly calculated as in the 
brane/antibrane \cite{Burgess:2001fx, Kachru:2003sx}. 
Finding a potential with the 
proposed properties represents a clear open challenge for these models.
 
From a string theory perspective, there are several other problems with this scenario. 
In particular, the assumption that 
the moduli of the Calabi-Yau manifold are fixed and decoupled is not justified.
 Nonetheless, the main problem to deal with remains the big-bang singularity 
giving rise to the bounce. This is a strong assumption as description of the bounce relies on strong coupling and non-perturbative dynamics.
For further developments regarding the implementation of the ekpyrotic scenario see \cite{Buchbinder:2007ad}. For a comprehensive review of the subject see \cite{Lehners:2008vx}.

\medskip
In summary, what the three scenarios: of pre big-bang cosmology, string/brane gas cosmology and ekpyrotic/cyclic have in common is that they
contemplate a period of contraction, and represent examples of bouncing cosmologies.
For a nice recent review on such bouncing cosmologies, see \cite{Brandenberger:2016vhg}.

\subsection{The Rolling Tachyon}

As we have seen in section \ref{sec:infla} the open string tachyon plays an important role in brane anti-brane inflation. It provides the natural way to end inflation and is the source of production of lower dimensional branes like stringy cosmic strings.

From the formal perspective, there has been concrete progress in understanding from first principles the physics of the open string tachyon. In particular, using string field theory Sen managed to extract substantial information regarding the tachyon potential \cite{Sen:2002nu} (for a review see \cite{Sen:2004nf}). This is actually one of the only cases in which a scalar potential has been derived directly from string theory. It is therefore worth exploring the potential cosmological implications of the tachyon field, independent of brane inflation.
 
String calculations suggest that to all orders in derivative expansion these actions take a Born-Infeld form.
\be
\setlength\fboxsep{0.25cm}
\setlength\fboxrule{0.4pt}
\boxed{
{\cal L}\ = -\ V(T)\ \sqrt{1- g^{\mu\nu} \partial_\mu T \partial_\nu T}\ ,
}
\ee
where $V(T)$ can take different forms depending on the type of string theory, 
namely bosonic or supersymmetric. 

First, Sen studied the rolling of the tachyon to its asymptotic minimum 
$T\rightarrow \infty$ and concluded that, even though the vacuum should correspond to the closed string vacuum and the unstable D-brane system 
(such as brane/antibrane pairs or non BPS D-branes) has decayed, the energy density is still localised. Furthermore, 
he was able to prove that the resulting gas corresponds to a pressure-less gas. This is easy to see from the
effective action above for which the stress energy tensor
for a time-dependent tachyon implies
\be
\rho\ =\ \frac{V(t)}{\sqrt{1-\dot T^2}}\ , \qquad p\ =\ -V(T)\sqrt{1-\dot T^2}
\ .
\ee
 For constant energy density, the pressure behaves as $p=-V^2/\rho$ and at the
 minimum in $T\rightarrow \infty$ we know that $V\rightarrow 0$ and so 
$p\rightarrow 0$. The equation of state is $p= \omega\,\rho$ with $\omega=-(1-\dot T^2)$ and therefore
$-1\leq \omega\leq 0$.

For a time-dependent tachyon field, we should actually consider a time-dependent metric such as the one of  FLRW. 
In \cite{Gibbons:2002md} this was done, obtaining the 
Friedmann's equations for this Lagrangian coupled to 4D gravity:
\begin{subequations}
\begin{empheq}[box=\widefbox]{align}
H^2 & =  \frac{8\pi G}{3} \frac{V(T)}{\sqrt{1-\dot T^2}} - \frac{k}{a^2}\ , 
\nonumber \\
\frac{\ddot a}{a} & =   \frac{8\pi G}{3} \frac{V(T)}{\sqrt{1-\dot T^2}}
\left(1-\frac{3}{2} \dot T^2\right)\ .
\end{empheq}
\end{subequations}
Even without actually solving these equations, it can be easily seen that the
energy density decreases with time, while $T$ increases relaxing towards the
 asymptotic minimum of the potential. In the meantime the universe expands,
first accelerating ($|\dot T|<2/3$) and then decelerating ($|\dot T|>2/3$).
Depending on the value of the spatial curvature
 $k=0,1,-1$, the scale factor $a(t)$ goes to a constant for $k=0$, 
to a Milne universe $a(t)\rightarrow t $ for $k=-1$ or re-collapses, for 
$k=1$.

Another natural question is whether
this tachyonic potential can give rise to
 inflation by itself. However, this is challenging
The main reason is the absence of small parameters in the potential 
that can be tuned to give a sufficiently slow roll. See however \cite{Padmanabhan:2002cp,Frolov:2002rr,Fairbairn:2002yp,Cremades:2005ir}.

As well as open string tachyons, string spectra can also include tachyons in the closed string spectrum.
The situation with closed string tachyons is more complicated and less understood. One way to 
see this is that, while the open string tachyon triggers the disappearance of D-branes after their collision 
by settling to the minimum of the potential, closed string tachyons are intrinsically gravitational and so their condensation 
may represent the disappearance of spacetime itself. Some simplifying configurations have been studied in which this happens for localised regions of spacetime that somehow mimics the open string case but the general case
is not fully understood. This is also related to the fact that string field theory is better understood for open strings than for closed strings. For a detailed discussion
of some phenomenological aspects see e.g  \cite{Adams:2001sv,Choudhury:2002xu, Shiu:2002qe} and for comprehensive review on tachyon dynamics including cosmological implications see \cite{Sen:2004nf}.

\subsection{S-Branes}

 Closely connected to the rolling tachyon is the concept of an {\it S-brane} \cite{Gutperle:2002ai}. An S-brane is a topological defect for which 
all longitudinal dimensions are spacelike, and so it  exists
 only for an instant of time. There are several reasons to introduce these 
objects.  The simplest example  in field theory corresponds to 
a  potential for a real scalar field of the standard double-well form:
\be
V(\phi)\ = \lambda \left(\phi^2 - a^2\right)^2\ , 
\ee
with minima at $\phi_{\pm}=\pm a$. In 4D this has the standard domain wall solution $\phi(x)=a\tanh(\sqrt{2\lambda}ax)$  or 2-brane
topological defect interpolating between the regions where the field is in 
the
$\phi_+$ and $\phi_-$ vacua. 

For S-branes, we have a time-dependent configuration in which we start at the maximum 
of the potential  $\phi(x,t=0)=0$
but with nonzero velocity $\dot\phi(x,t=0)=v>0$. This will make the field  roll towards $\phi_+$, until it oscillates and eventually 
arrives at the minimum. A time reversal situation would have the field starting in $\phi_-$ and going to $\phi=0$. 
We can then have  the field evolving from $\phi_-$ at $t=-\infty$ to $\phi_+$ at $t=\infty$ looking like a kink in time and filling all spatial dimensions. This is an S2 brane. This  process requires some fine tuned exchange of energy in order for the field to climb the barrier.

Using the analogy with D$p$-branes, we expect that the S$p$-branes can
also be found as explicit solutions of Einstein's equations
 coupled to dilaton and antisymmetric tensor fields. 
 
We start with the Lagrangian for the metric, dilaton, antisymmetric tensor $F_{q+2}= 
dA_{q+1}$  (and setting $\kappa_p=1$),
\be
\setlength\fboxsep{0.25cm}
\setlength\fboxrule{0.4pt}
\boxed{
{\cal L} \ = \ \sqrt{-g}\left( R-\frac{1}{2} 
g^{\mu\nu}\partial_\mu\varphi\partial_\nu\varphi-\frac{1}{2(q+2)!}
 F_{q+2}^2\right).
\label{einstein}
}
\ee
The equations of motion have solutions similar to the ones found for black branes. 
In the same way  that $p$-brane solutions
are black hole-like, we expect that S-brane solutions correspond to
time-dependent backgrounds of the theory, and therefore may have a  cosmological interpretation. This is actually the case. 

The simplest example can be obtained for pure gravity. Starting with  the Schwarzschild solution in 4d corresponding to a black hole of mass $M$ we can perform the following analytic continuation:
 \be
 t\rightarrow ir,\qquad  r\rightarrow it, \qquad
 \theta\rightarrow i\theta, \qquad \phi\rightarrow i\phi
 \ee
  together 
with $M\rightarrow iP$. The metric becomes
\be
d\hat s_I^2\ = \ -\left[1-\frac{2P}{t}\right]^{-1}\ dt^2\ + 
\ \left[1-\frac{2P}{t}\right] \ dr^2\ + \ t^2\ \left(\sinh^2\theta\ \ 
d\phi^2\ + \ d\theta^2\right), 
\ee
whose surface of constant $r$ and $t$ is now the 
hyperbolic plane ${\cal H}_2$ rather than the two-sphere, consistent with the time-like nature of the S-brane. 

In addition to the symmetries of the hyperbolic space, the solution has
a spacelike Killing vector $\xi=\partial_r$ but is now time-dependent, 
again, as expected for a S-brane. The apparent singularity at $t=2P$ is
 again a horizon. For $t<2P$ the metric is:
\be
d\hat s_{II}^2\ = \ -\left[1-\frac{2P}{r}\right]\ dt^2\ 
+ \ \left[1-\frac{2P}{r}\right]^{-1} \ dr^2\ + \ r^2\ \left(\sinh^2\theta\ \
d\phi^2\ + \ d\theta^2\right), 
\ee
which is now static with a time-like singularity at $r=0$. The corresponding Penrose diagram is a $\pi/2$ rotation of the Schwarzschild black hole diagram as can be seen in figure (\ref{sbrane}).

\begin{figure}[t]
\begin{center}
\includegraphics[width=55mm,height=80mm]{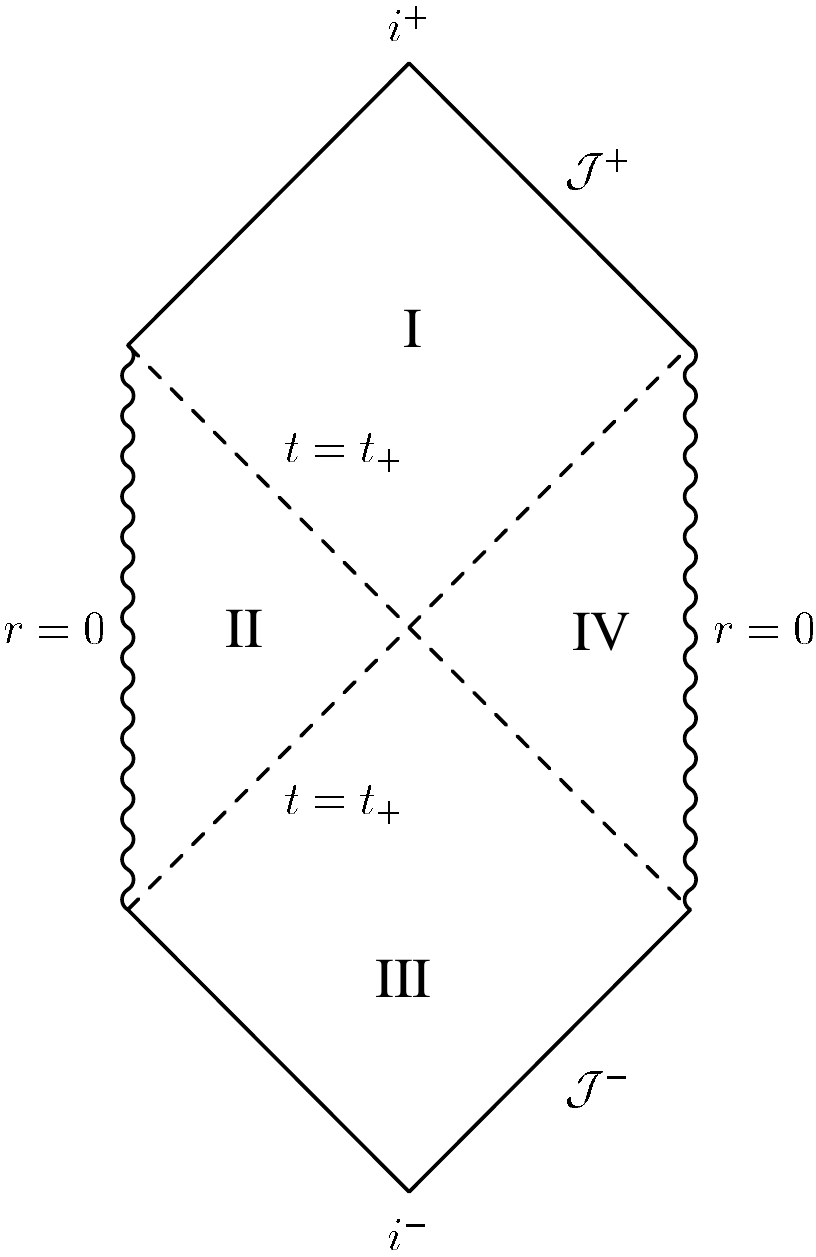} 
\caption{Penrose diagram for an S-brane. Note that this is a $\pi/2$ rotation of the Schwarzschild black hole Penrose diagram. Regions I and III are cosmological, representing expanding and contracting cosmologies, respectively, separated by smooth horizons which are identified with the S-brane. Regions II and IV are static and have time-like singularities, which can be identified with negative tension end-of-the-world brane-like objects similar to orientifolds. Their corresponding mass and charge can be computed explicitly.} 
\label{sbrane}\end{center}
\end{figure}

More general solutions  of (\ref{einstein}) will have both dilatonic  and  $F_{q+2}$ charges (see for instance \cite{Grojean:2001pv,Burgess:2002vu}). 
The static region provides us with a way to  identify 
this geometry correctly. It turns out that the singularities are the
 physical objects  to which mass, or tension and charge can be
 assigned unambiguously.
It was found that the two singularities correspond to negative tension 
objects, like end-of-the-world branes with opposite charge. 
Furthermore, the similarity with black
hole geometry 
indicates that there will be particle production 
 and we 
 can also compute a generalised Hawking temperature and entropy which could have 
interesting cosmological interpretations. Finally, just as in the case of the Pre Big-Bang, ekpyrotic/cyclic and brane gas scenarios, S-branes naturally have a period of contraction of the universe corresponding to region III of the Penrose diagram, followed by another period of expansion (region I). For further details on the cosmological interpretations of S-branes see for instance \cite{Grojean:2001pv,Burgess:2002vu,Cornalba:2002nv,Cornalba:2002fi,Burgess:2003tz,Kounnas:2011gz,Townsend:2003fx,Ohta:2003pu}.

\subsection{Swampland Conjectures}
\label{Sec:Swamp}

\begin{figure}[t]
\begin{center}
\includegraphics[width=140mm,height=90mm]{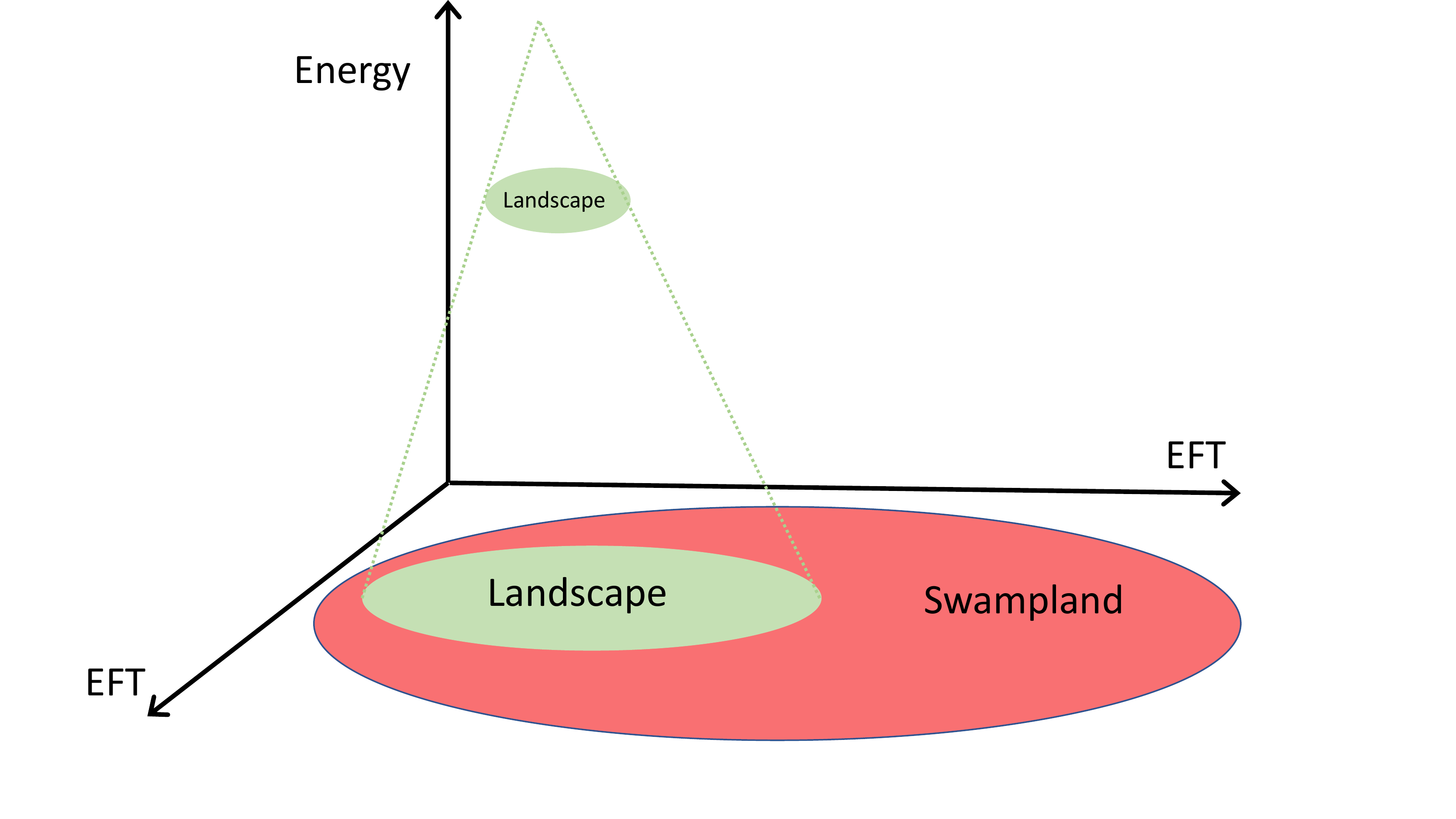} 
\caption{A cartoon representation of the swampland. At low energies there are many consistent effective field theories, but only a subset of them can be lifted to be UV complete inside a quantum gravity theory. These correspond to the landscape. The rest  are referred to as the swampland.} \label{swampland}
\end{center}
\end{figure}

The vast number of apparent string vacua has very interesting implications. As described in Chapter \ref{sec:DE}, it may be the only self-consistent way to explain the smallness of the dark energy and may provide a totally different approach to asking fundamental questions in physics, separating the `interesting questions' (those that do need an explanation from an underlying theory) from the `uninteresting questions' (those that may be explained by the presence of the multiverse). However, this may also lead to the belief that any theory at all may be derivable from string theory, resulting in a conclusion that it is impossible ever to test string theory, even in principle. This gives the idea of the swampland, illustrated in figure \ref{swampland}.

That said, since the early days of string theory we have known that this is not true: there are some, albeit only a few, general physics properties that can be extracted from string theory. Namely,
\begin{itemize}
\item The need for supersymmetry at the fundamental level (although the scale of its breaking is not known and it may take non-standard forms as in misaligned supersymmetry \cite{Dienes:1994np});
\item
 The existence of extra dimensions, and more concretely only 6 or 7 extra dimensions; 
 \item 
 The existence of moduli fields with specific properties, in particular gravitational-strength couplings, 
 which appear in very specific ways within the low-energy effective field theory; 
 \item The absence of an infinite number of continuous spin representations corresponding to massless particles. These are in principle allowed by the principles of quantum mechanics but have not been observed in nature, despite the lack of alternative explanations from basic principles;
 \item The general absence of global symmetries in the effective field theory.
\end{itemize}

These general results can be complemented with general properties of the landscape, for instance, if the landscape is dominated by Coleman-de Luccia vacuum transitions, a general claim exists that the curvature of the Universe has to be negative, implying an open universe. However, the power of such general results is limited and they still allow for the existence of the enormous landscape resulting in limited predictive power.

In the past few years a new approach towards addressing concrete questions from string theory has been developed and comes under the name of {\it swampland conjectures} \cite{Vafa:2005ui}. The idea is very simple: exploit our cumulative experience of string vacua to extract results that may actually be general. This program aims to state concrete conjectures whose validity may be tested through further exploration of string solutions, either to confirm or rule out the conjecture. The overall goal of the swampland conjectures is:

\begin{quotation}
\emph{Identify which effective field theories are consistent at low energies but cannot be consistent in a UV completion of the theory including gravity.}
\end{quotation}

This is a modern version of what used to be known as {\it vacuum cleaning} in the sense of having a systematic procedure to separate the proper vacua from those that cannot be UV completed. Such conjectures can be made much sharper by going beyond string theory and claiming that the corresponding conjectures will hold for {\it any} theory of quantum gravity, string theory or otherwise. In this sense, string theory is used only to identify the corresponding conjectures, while the swampland approach aims not only to select string vacua, but to identify general properties of any theory of gravity.

More generally, the study of UV constraints on IR physics is a  blooming field that has seen many new conceptual and technical developments recently.
While the swampland programme is one of these developments,  it is worth mentioning a powerful approach, which simply assumes  unitarity, locality, causality, and Lorentz invariance of the, otherwise unknown, UV completion to derive  constraints on the effective field theories, (see \cite{deRham:2022hpx} for a recent overview on these ideas). Combining these approaches may bring the swampland programme to a firmer footing, for example via positivity bounds or bootstrap arguments. 

Over the years, a range of swampland conjectures have been put forward. They range between those that are strongly motivated, but with limited phenomenological or cosmological impact, to conjectures that may have a major impact, but which lack a robust basis and may be considered extremely speculative. There are several excellent reviews on this field in which a detailed discussion of the conjectures has been explained and argued in much detail \cite{Brennan:2017rbf,Palti:2019pca,vanBeest:2021lhn, Grana:2021zvf} to which we refer the reader for further details and the majority of the original references. Here we content ourselves by briefly mentioning those conjectures that could be more relevant for cosmology:

\begin{enumerate}
\item{\it Absence of global symmetries}. A consistent theory of gravity with finite number of states cannot have exact global symmetries. General arguments in this direction have existed since the 1980s: both through arguments that -- contrary to local symmetries -- global symmetries are not protected by black holes after radiation, and also that within string theory, a conformal field theory that leads to a global symmetry also leads to a massless particle in the spectrum that corresponds to the gauge field of that symmetry and therefore the symmetry is not global but local in spacetime \cite{Banks:1988yz}.  More recently, further evidence has accumulated to support its validity. While the general impact of this result is strong, it provides only weak constraints on local symmetries with very weak couplings that to most practical particle physics purposes behave as global symmetries (see for instance \cite{Burgess:2008ri}).

\item {\it Weak gravity conjecture \cite{Arkani-Hamed:2006emk}}. The observational fact that we observe gravity to be the weakest force may not be a property only of our Universe but also of any possible universe described by a consistent theory of gravity. A concrete statement of this conjecture is that in any consistent theory of gravity, there should exist a particle of charge $Q$ and mass $M$ such that $Q^2>GM^2$ (for which the Coulomb force is stronger than the Newtonian force) although more precise and generalised formulations have been proposed \cite{Harlow:2022gzl}. This is the prime example of a  swampland conjecture  that has been argued and tested in sufficiently many ways that there exists a broad consensus that it is actually a correct statement. This has played a useful role in cosmology, especially when extended to forces mediated by scalar fields and  antisymmetric tensors, and in particular when they are dual to axion-like-particles. The weak gravity conjecture may also then be used to constrain the axion decay constant that plays a role in some models of inflation.

\item{\it Cobordism conjecture \cite{McNamara:2019rup}.} Two manifolds are called cobordant if their union is the boundary of another manifold of one extra dimension. This defines an equivalence relation. The corresponding equivalent classes may define a global (topological) charge. We may generalise the absence of global charges conjecture to this topological case and conjecture that  in  a consistent theory of gravity, all cobordism classes have to be trivial. If the corresponding manifolds are, for instance, the 6-dimensional compact spaces, the corresponding 4-dimensional EFTs would be separated by a domain wall (see figure (\ref{cobordism})). If the cobordism class is trivial then it must admit  an end-of-the-world configuration as in the Horava-Witten or bubble of nothing cases (see figure (\ref{cobordism2})). If this conjecture holds, it may have very important implications for cosmology due to the presence of the boundary-ending spacetime. For recent developments in this direction see for instance \cite{Angius:2022aeq,Angius:2022mgh}.

\begin{figure}[t]
\begin{center}
\includegraphics[width=140mm,height=90mm]{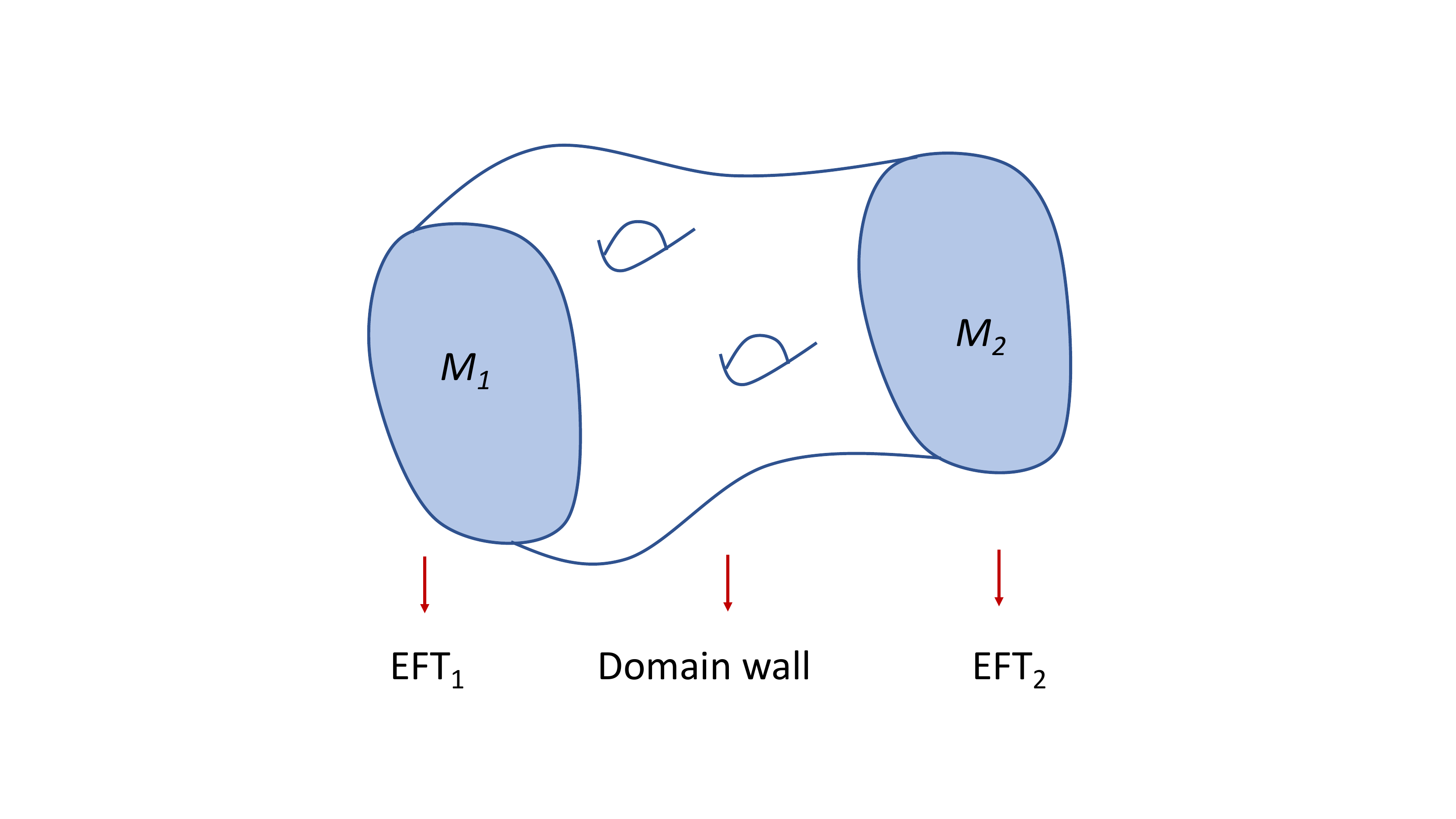} 
\vskip -15pt
\caption{A representation of cobordism among two manifolds for which their union is the boundary of another manifold of one extra dimension.} \label{cobordism}
\end{center}
\end{figure}

\item{\it Distance conjecture \cite{Ooguri:2006in}.} 
Consider an effective field theory coupled to gravity with a moduli space, ${\mathcal M}$, parameterised by {\em massless} scalar fields, $\phi_i$, and a metric $\gamma_{ij} (\phi_k)$ which determines the scalar's kinetic terms. 
In standard effective field theories, the consistency and trustworthiness of the dynamics of a scalar field potential $V(\phi)$ is determined by requiring that it does not excite modes with masses above the cutoff, $m \gtrsim \Lambda$.
 As long as this is satisfied, the range of possible values for $\phi$ is not bound by $\Lambda$. The swampland distance conjecture states that this no longer holds if the EFT is consistently uplifted to the UV. 

The swampland distance conjecture states that,  as some modulus approaches a point at infinite {\em geodesic distance} in moduli space, there is an infinite tower of states, which become exponentially massless with the geodesic distance $\Delta\phi$: $m\sim e^{-\Delta\phi}$. These states cannot be neglected from the EFT in this limit. The prime example of such behaviour is when $\phi$ represents the size of an extra dimension and the corresponding tower of states are either the Kaluza-Klein or the winding modes.

As well as infinite limits, the revised distance conjecture also states that for finite displacements, starting from a value $\phi_0$, at a point $\phi_0+\Delta \phi$ infinite towers of modes with mass of order $e^{-\Delta\phi}$ become lighter and lighter with the distance in field space $\Delta\phi$ and so can no longer be neglected from the EFT. Although the conjecture applies to massless scalar fields moving along geodesic trajectories, it could in principle have implications for the field range in single field inflationary models and/or the amount of non-geodesicity in multifield models. However, further work in this direction is needed to establish these possible constraints (see e.g. \cite{Kinney:2018nny, Kinney:2018kew, Palti:2019pca,vanBeest:2021lhn,Grana:2021zvf}).\footnote{ For example, in  \cite{Buratti:2018xjt} it has been shown  that  backgrounds with spacetime varying scalars can  lead to trans-Planckian motion without encountering exponentially falling towers of states. }

\begin{figure}[t]
\begin{center}
\includegraphics[width=140mm,height=90mm]{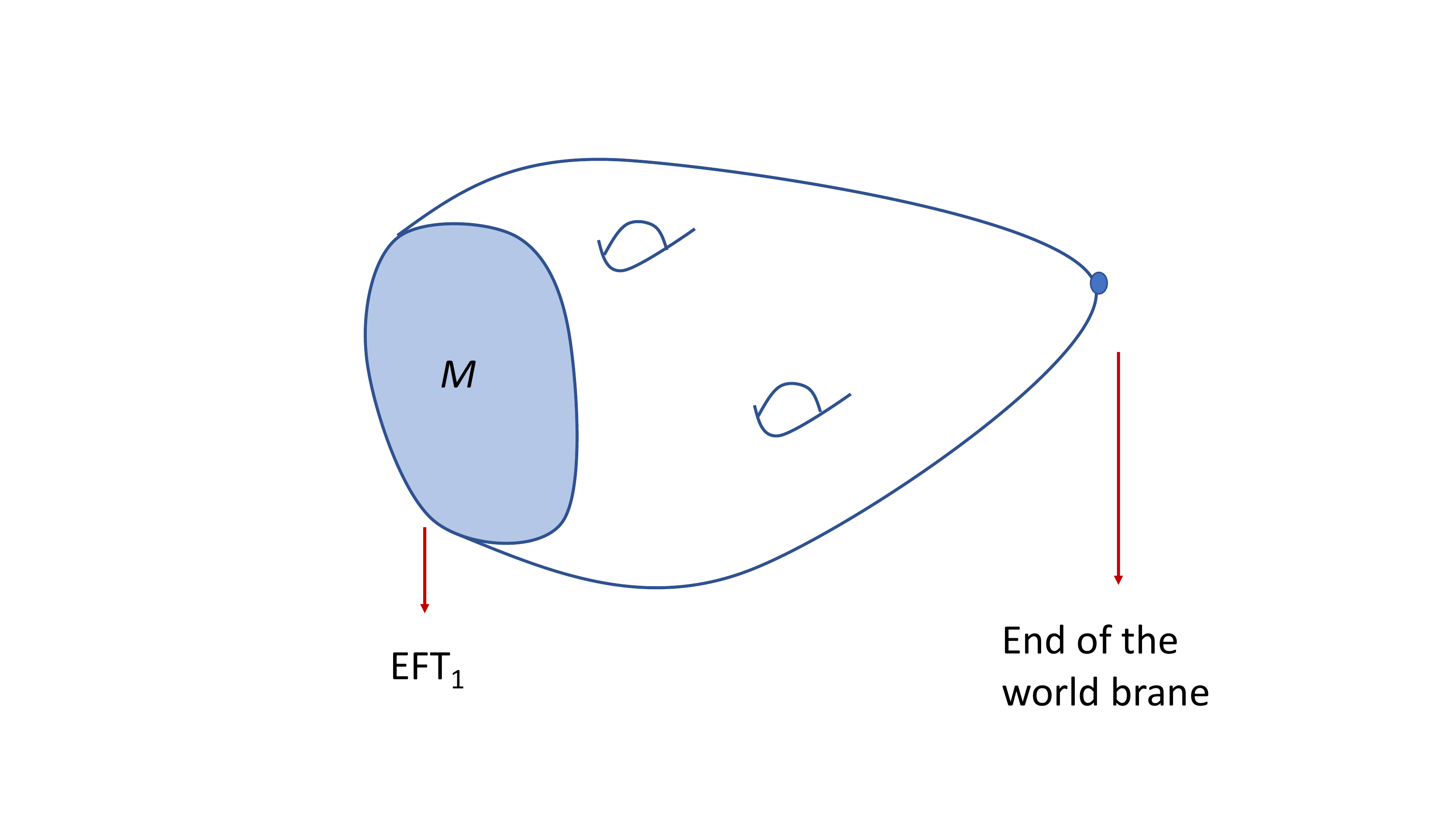} 
\vskip -15pt
\caption{Trivial cobordism with only one manifold in the boundary. The cobordism conjecture states that in this case there should be an end-of-the world configuration.} \label{cobordism2}
\end{center}
\end{figure}

\item{\it Conjectures on  AdS vacua (non-supersymmetric $\&$ supersymmetric):}  Swampland conjectures for non-supersymmetric AdS constructions were proposed in
\cite{Ooguri:2016pdq}. These conjectures are at two levels. The first is motivated by an extension of the weak gravity conjecture; the extension requires that
the equality between electric and gravitational forces is saturated  if and only if the underlying theory is supersymmetric and the states
 under consideration are BPS with respect to the supersymmetry. A consequence of this is that non-supersymmetric AdS solutions supported
 by flux\footnote{A d dimensional AdS solution is said to be supported by flux if a d-form flux field strength   space-fills the AdS space.}  are unstable as they can decay by a brane nucleation process which leads to flux depletion (in a process similar
 to that of \cite{Maldacena:1998uz}). The stronger form of the conjecture removes the requirement that the AdS solution is supported by
 flux and states that  there are no non-supersymmetric AdS solutions in a consistent quantum theory with low energy description in terms of Einstein gravity coupled to a finite number of matter fields. If correct, there will be important  implication not only for moduli stabilisation
 (the AdS vacuum in the LVS scenario is non-supersymmetric) but also for applications of the AdS/CFT correspondence to condensed matter physics, quantum information and hadron physics since the holographic models used in this context have no supersymmetry.  At present, the stronger form of the conjecture does not have much support (see \cite{Baykara:2022cwj} for good evidence in favour of non-supersymmetric AdS vacua in $O(16) \times O(16)$
 heterotic strings). The conjecture can be reformulated in the language of conformal field theories. Conformal field theories dual to Einstein gravity with a finite number of matter fields must satisfy the following (energy) gap condition: they  can have  only a small number of primary fields whose operator products generate all primary fields up to a  energy scale that can be  made parametrically large in the large N limit. The conjecture implies that this condition cannot be met in non-supersymmetric conformal field theories. For attempts to construct such non-supersymmetric conformal field theories meeting this condition see e.g. \cite{Giombi:2017mxl, Gurucharan:2014cva}. 
 
For the relationship of these conjectures of other swampland conjectures, see e.g. \cite{Bernardo:2021vfw}. More recently, it has be put forward that  some  supersymmetric AdS  vacua such as the KKLT AdS solution and pure supergravity AdS lie in the swampland \cite{Lust:2022lfc, Montero:2022ghl}.

\item{\it de Sitter conjecture \cite{Obied:2018sgi}.} It is well known that in string theory both AdS and Minkowski spaces are naturally obtained (mostly because they preserve supersymmetry), while de Sitter space is far more difficult to obtain: a fact which is very important for describing the current acceleration of the Universe through the $\Lambda CDM$ model in terms of the string landscape and early universe inflation. 

The concrete scenarios, such as KKLT and LVS presented in previous sections, provide realisations of de Sitter space from string theory. However, their validity relies on the EFT analysis of perturbative and non-perturbative corrections. As we discussed, the Dine-Seiberg problem that implies the runaway behaviour of the scalar potential for both the volume of the extra dimensions and the dilaton, suggests that in the regime where both perturbative expansions, $g_s$ and $\alpha'$, are under {\it full control}, we should be in the asymptotic runaway region, and so de Sitter vacua at any finite value of the volume or at any non-zero string coupling are in a regime where the approximations cannot be fully trustable. This has led to a bold conjecture stating that there are not actually any de Sitter vacua in a consistent theory of gravity and that the de Sitter vacua found in the literature are only artifacts of the fact that the approximations used are not under full control.\footnote{See also \cite{Cribiori:2020use} for an argument against trustable 4-dimensional dS solutions in $\mathcal{N}=2$ supergravity based on the magnetic weak gravity conjecture.} The original claim was:
\be
\setlength\fboxsep{0.25cm}
\setlength\fboxrule{0.4pt}
\boxed{
|\nabla V|\geq c \frac{V}{\Mp}\,,\qquad 0\leq c \sim {\mathcal{O}}(1)
}
\ee
This condition is clearly satisfied in the Dine-Seiberg runaway regime in which the approximations are under full control. However, the conjecture does not add further information in the weak coupling regions where vacua like KKLT and LVS are claimed to lie. Furthermore, the original conjecture was clearly violated by the Higgs potential (since at the maximum $|\nabla V|=0, V\geq 0$) \cite{Cicoli:2018kdo,Denef:2018etk} (see also \cite{Murayama:2018lie,Choi:2018rze,Hamaguchi:2018vtv}) and a refined version was proposed adding an extra condition on the Hessian of $V$ \cite{Garg:2018reu,Ooguri:2018wrx, Murayama:2018lie}. This conjecture is clearly speculative and less motivated than other swampland conjectures; however, it has motivated further work exploring in more detail  the validity and existence of de Sitter vacua in string theory, which is welcome given the importance of the claim and the challenge.

\item{\it Trans-Planckian censorship conjecture \cite{Bedroya:2019snp,Bedroya:2019tba}.}  It is well known that in inflationary models, microscopic modes redshift due to the expansion of the universe and may become macroscopic and observable at present. If these modes are below the Planck length, it seems to indicate a transfer of modes from the UV to the IR with the UV modes in a regime beyond the validity of the corresponding EFT. The Trans-Planckian censorship conjecture states that in a consistent theory of gravity this UV to IR transfer should not happen, though it does not explain why this should be the case. This puts, for instance, an upper bound on the number of e-folds of inflation. Even though this conjecture has interesting implications for early universe cosmology, the relevance of the constraints on the (time-dependent)  EFT has been questioned in e.g.~\cite{Kaloper:2018zgi,Dvali:2020cgt,Burgess:2020nec,Komissarov:2022gax, Lacombe:2023qfx}.

\end{enumerate}
In summary, the swampland approach brings an interesting perspective to be considered in general discussions of string cosmology. Confirming, discarding or refining these conjectures may lead to relevant progress in the field of string cosmology and, more generally, in principle, to progress in any cosmological implications of consistent theories of quantum gravity. For this much work will be needed.

\subsection{Bubbles of Nothing and the Wave Function of the Universe}

One of the deepest questions a theory of gravity should eventually address is why there is something rather than nothing. This appears a rather philosophical question, as how to define nothingness in a physical theory is unclear. Clearly it is not simply the vacuum state, as we know in 
quantum mechanics the vacuum is not empty. But over the years cosmologists have proposed concrete definitions of nothing (for example, the absence of space, time and matter). The first explicit example was Witten's bubble of nothing (BON) \cite{Witten:1981gj} illustrated in figure \ref{bon}. When considering simple compactifications of five-dimensional Kaluza-Klein theories, he found a transition from the flat five dimensional metric corresponding to a circular fifth dimension of radius $R$: $ds_5^2= ds_4^2+R^2 d\phi^2$, mediated by an instanton with Euclidean metric:
\be
ds^2=\frac{dr^2}{1-R^2/r^2}+r^2\left(d\theta^2+\cos^2\theta d\Omega_2^2 \right)+ \left(1-R^2/r^2\right) R^2 d\phi^2.
\ee
Similar to the Euclidean Schwarzschild metric, this is non-singular even at $r=R$. But this is a minimum value of the coordinate $r$. For large $r$ the metric is asymptotically flat. So this instanton mediates a transition from a flat spacetime with a circle of radius $R$ to a spacetime with maximum value of $r$ where the fifth dimension collapses which we may identify as a {\it bubble of nothing}. The interesting point is that after nucleation the further evolution of the bubble is through expansion at the speed of light, as can be seen by the Wick rotation $\theta\to it$. Since
$\cos\theta \to \cosh t$ the bubble radius increases exponentially with time eating up the full spacetime. In the 5d case, 
this transition depended on the non-existence of supersymmetry and was considered without taking into account moduli stabilisation of the extra dimension. Generalisations to 6d with moduli fixed by fluxes have been found with the similar dramatic outcome \cite{Blanco-Pillado:2010xww,Brown:2011gt}. Furthermore, it was recently argued that these bubbles of nothing are ubiquitous in string compactifications \cite{GarciaEtxebarria:2020xsr} and may represent eventual sources of instabilities (although being non-perturbative, the decay rate may be much suppressed).
\begin{figure}[t]
\begin{center}
\includegraphics[width=170mm,height=50mm]{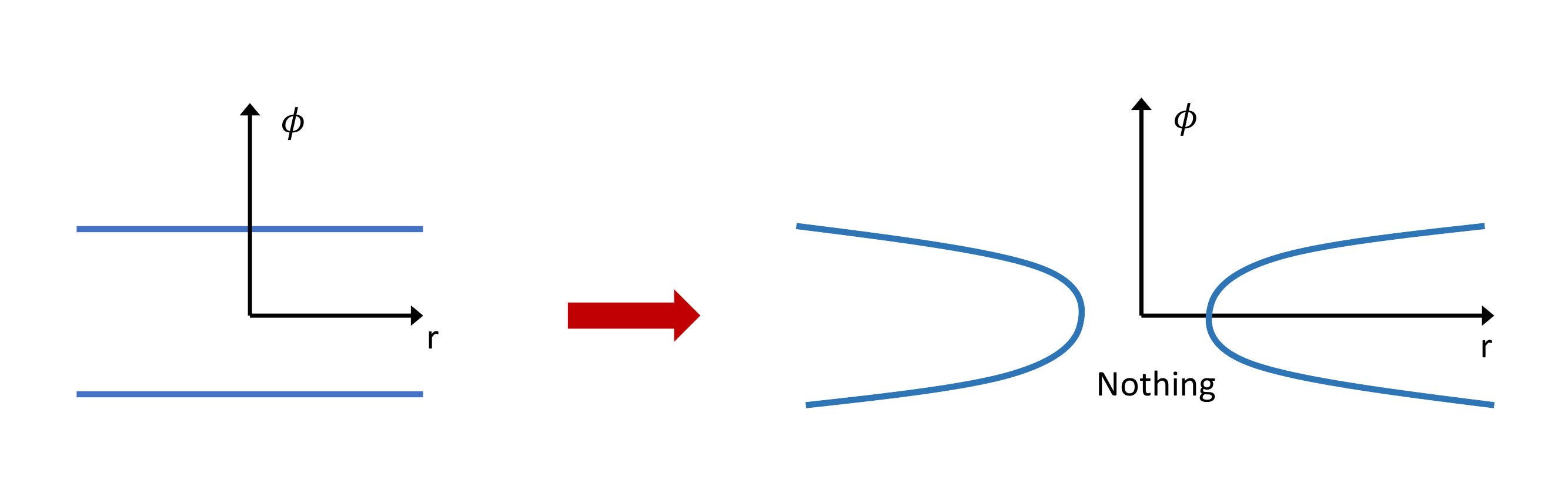} 
\caption{An illustration of the Bubble of Nothing scenario: a transition resulting to an expanding absence of space-time. On the left we have the flat 4d in the horizontal  and the extra dimension in the vertical with the two horizontal lines identified in a circle compactification. On the right the transition to the bubble of nothing. For large $r$ it is as in the left but for smaller values of $r$ at some point the extra dimension collapses and a bubble of nothing appears (a cross section is shown) and expands in time.} \label{bon}
\end{center}
\end{figure}

The second appearance of nothing was in the {\it creation out of nothing} scenario of Vilenkin \cite{Vilenkin:1982de,Vilenkin:1983xq,Vilenkin:1984wp} and the subsequent {\it wave function of the universe} of Hartle and Hawking \cite{Hartle:1983ai}. See also\cite{Linde:1983mx,Rubakov:1999qk,Hartle:2022jkt}. This  defines, within the domain of semiclassical gravity, a concrete proposal to describe the beginning of the universe from a state with no spacetime. So, in a sense it is the opposite of the bubble of nothing picture. 

From simple mini-superspace arguments the transition from nothing to a de Sitter space with cosmological constant given by $H^2>0$ is found to be of order 
\be
{\mathcal P}=|\Psi|^2\propto\, e^\frac{\eta\, \pi}{2G_4H^2},
\ee
with $\eta=+1$ for Hartle-Hawking boundary conditions (the no boundary proposal in which the nucleated universe is a superposition of expanding and contracting ones) and $\eta=-1$ for the Vilenkin or tunneling transition (for which the boundary conditions are such that they only give an expanding universe). Note that, up to normalisation factors, the Hartle-Hawking wave function favours smaller values of the cosmological constant, making it hard to justify inflation, whereas the Vilenkin normalisation prefers larger values of $H^2$. All these discussions are only within a simplified picture of mini-superspace and semiclassical gravity and, in principle, need to be further studied, especially once a proper quantum gravity theory is at hand. This has been a challenge for string theory and, despite several attempts, the study of such transitions is still in its infancy.

Note that this approach is similar in spirit to the vacuum transitions initiated by Coleman and de Luccia (CDL) which have been inherited in the discussions of the landscape. However, contrary to CDL transitions, here there is no bubble nucleated but instead a full spacetime. This can be done for de Sitter since it has finite volume but for AdS and Minkowski it may require introducing a cut-off in order to quantify the corresponding probability. Furthermore, contrary to the CDL case in which naturally the transition gives rise to an open universe, here, the de Sitter slicing that gives finite volume corresponds to a closed universe \cite{Hartle:2013oda,Cespedes:2020xpn}. This is important since there have been claims that the main general prediction of the string landscape is that it predicts open universes \cite{Freivogel:2005vv,Kleban:2012ph} and if eventually, in the maybe long future, open universes would be ruled out it could  be a strong argument against the landscape. But if closed universes can be created from nothing, settling the curvature of the universe may only differentiate between the two mechanisms. Clearly further studies in these directions are needed.\footnote{An important direction is
bubble stability, see e.g \cite{Johnson:2019tgc} for a recent analysis.} For string theoretical discussions of the wave function of the universe see for instance \cite{Ooguri:2005vr,Brustein:2005yn}.

\subsection{Holography and Cosmology}

The main theoretical development in string theory over the past 25 years has been holography. A gravitational theory in $d$ dimensions is equivalent to a non-gravitational theory in $(d-1)$-dimensions. One old motivation for this equivalence is the well known fact that the black hole entropy, being an extensive quantity, is proportional to the area of its horizon rather than to the internal volume of the black hole. 
The AdS/CFT correspondence now represents the best concrete definition of quantum gravity in anti-de Sitter space (AdS) since, at least in principle, the non-gravitational conformal field theory (CFT) on the boundary of AdS is understood, and any quantum aspects of the gravitational theory can be referred to concrete questions on the CFT side.

Although technically AdS/CFT remains a conjecture,
the evidence for the correspondence is by now overwhelming. In particular, this correspondence has provided a framework which 
allows the black hole loss of information puzzle stated by Hawking almost 50 years ago to be addressed. Holography 
 indicates that, as the CFT side is a standard quantum system, information cannot be lost. Therefore, if the equivalence is correct, 
information also cannot be lost within the gravitational system. Furthermore, using holographic arguments it has also recently been possible to calculate explicitly the entanglement entropy of black holes showing the Page curve behaviour that would be 
expected if the information is not lost\footnote{For studies of implication of entanglement (in particular Bell inequalities) in the cosmological context see \cite{Maldacena:2015bha, Choudhury:2016cso}.}.

It is therefore very appealing to try to extract cosmological implications of holography. Even though our universe is not of the AdS type, several approaches have been proposed, aiming to extend the success of AdS/CFT to cosmological questions. The general topic of holography is so vast 
that we will touch on it even more briefly than for previous subjects and refer the reader to the literature for more details. For recent reviews see for instance \cite{Anninos:2012qw, Flauger:2022hie}.

\begin{enumerate}
\item {\it AdS/CFT  and density perturbations in inflation.} 

One possibility to use AdS/CFT for inflation is to consider a simple potential with two minima with opposite signs of the cosmological constant. 
One minimum corresponds to a stable AdS vacuum and the other to a nearby metastable dS vacuum. The bulk geometry 
near the AdS vacuum is described by boundary CFT and so this can be used to extract information about bubbles of the dS phase from the CFT spectrum \cite{Freivogel:2005qh} (see however \cite{DeAlwis:2019rxg}) and, through analytic continuation, connect the correlators from AdS/CFT to a potential dS/CFT .

In a different direction, assuming that inflationary cosmology has a CFT dual there have been explicit efforts to quantify this 
relation by computing density perturbations using a QFT dual. In particular, known inflationary results regarding
 density perturbations are reproduced in the weak gravity regime. Exploring the domain of strongly coupled gravity by working with 
 the weakly coupled QFT offers an alternative phenomenology to inflation in which an almost scale invariant spectrum of 
 perturbations is also obtained. See e.g. \cite{Freivogel:2005qh, McFadden:2009fg, Bzowski:2012ih, Kundu:2014gxa} for concrete calculations in this
  interesting direction.

\item {\it dS/CFT and dS/dS proposals.} 

A dS/CFT correspondence
 was proposed in \cite{Strominger:2001pn}, following parallels with the
well established AdS/CFT correspondence,
 motivated by the close connection between dS and AdS spaces.
The argument is that a boundary at infinity of, say,  dS$_4$ corresponds to a Euclidean $R_3$
space for which the symmetry group of de Sitter space, $SO(4,1)$ acts as the conformal group
of the Euclidean $R_3$, suggesting that a conformal field theory on this boundary is dual to the full 4D gravity theory in de Sitter space.

 One of the interesting outcomes of
 this conjecture is that the renormalisation group parameter can be identified with
time, in much the same way it was identified with the extra spatial 
coordinate in the AdS/CFT case \cite{Strominger:2001gp}. One simple way to see this possibility is 
by writing the dS$_4$ metric in FRW coordinates ($k=0$) as
\be
ds^2\ = \ -dt^2\ + \ e^{Ht}\ d\vec{x}^{\,\,2},
\ee
with $\vec{x}$ the spatial coordinates 
 and $H$ the Hubble parameter. 
 
 The interesting observation is that this
 metric is invariant under
$t\rightarrow t+\lambda$, $\vec{x}\rightarrow e^{-\lambda H}\vec{x}$ which 
generates time evolution in the 4D bulk and scale transformations on the
 Euclidean boundary.
Late times (large values of $\lambda$) correspond to small distances (UV
 regime) whereas earlier 
times correspond to low energies and the IR regime.
Generic expressions for the scale factor $a(t)$ will not have this symmetry,
but if we assume that
$H(t)$ goes to a constant both in the infinite past and infinite future we can 
follow the time evolution
between two fixed points under the renormalisation group, which could
 eventually be identified
with early universe inflation and also current acceleration. 

The monotonic evolution in time fits well  with the 
expected c-theorem of field theories, holding in 2D, and the extension to the a-theorem in 4d.
The RG flow then corresponds to the direction from future to past. 

As a side note, the S-branes mentioned previously were
 an attempt to bring this correspondence closer to the AdS/CFT one, with the
 S-branes playing the role of the D-branes in the
boundary (the Euclidean $R_3$ in the example above).

A final independent and interesting alternative is the dS/dS correspondence in which a dS space is proposed to be dual to another dS space in one dimension less \cite{Alishahiha:2004md,Dong:2010pm}. This is partly based on the simple observation that the de Sitter dS$_d$ metric in $d$ dimensions can be seen as a warped compactification to dS$_{d-1}$:
\be
ds_d^2=d\omega^2+\sin^2\left(\frac{\omega}{R_{dS}}\right) ds_{d-1}^2.
\ee
Similarly to AdS/CFT, this implies an emergent spatial dimension and warped throats (two in this case since the warp factor vanishes at $\omega=0,\pi R_{dS}$), but contrary to AdS/CFT the dual is no longer a field theory without gravity but now a lower dimensional de Sitter space with a massless graviton.

Even though de Sitter space realisations in string theory are of interest, these potential dS/CFT or dS/dS dualities are not yet under firm grounds, unlike the AdS/CFT case, and much work needs to be done to turn these proposals into something useful for cosmology.

Further proposals for holography and cosmology have been put forward. See for instance \cite{Banks:2003ta}.

\item {\it Islands and cosmology}

A more recent development relates to the progress regarding the information loss paradox in black holes. The main new ingredient here is the concept of quantum extremal surfaces (QES) \cite{Ryu:2006bv,Hubeny:2007xt,Faulkner:2013ana,Engelhardt:2014gca} to generalise the expression for the von Neumann entropy. This prescription gives rise to a time evolution of the entropy such that it increases monotonically up to a maximum point where it starts decreasing, giving rise to what is called a Page curve as required for a unitary evolution  to address the information loss paradox. The key component of this calculation is an entanglement  {\it island} which is a region behind the black hole horizon that hosts the corresponding  degrees of freedom that avoid information loss \cite{Penington:2019npb,Almheiri:2019psf}.

The details of these results are beyond the scope of this review (for a review with all the relevant references see for instance \cite{Almheiri:2020cfm}). However, a notable point is that these results were achieved without the need for a full UV completion of semiclassical gravity. This gives more credibility to semiclassical approaches to quantum cosmology, such as the wave function of the Universe and vacuum transitions discussed above. Furthermore, concepts like quantum extremal surfaces and islands may also play an important role for cosmology. This has been recently explored in 
\cite{VanRaamsdonk:2020tlr,Hartman:2020khs,Chen:2020tes,VanRaamsdonk:2021qgv,Geng:2021wcq,Pasquarella:2022ibb, Bousso:2022gth} but it is fair to say that this direction is still in an infancy and there is much to be explored. 

\item{\it Emergence of spacetime}

One point that may be worth pointing out is the recent ideas regarding the emergence of time.  From the original AdS/CFT correspondence, it is clear that not only gravity but also at least one spatial dimension is emergent from the boundary CFT theory (in one less dimension and no gravity). More recent studies of the closed connection between gravitational theories and quantum entanglement have led to the proposals of getting spacetime from entanglement. The proposal of ER=EPR (Einstein-Rosen wormhole equivalent to the Einstein-Podolsky-Rosen entanglement) is a variation of this. Contrary to the emergence of spatial dimensions, the emergence of time has been more challenging to implement. But recent work has been done in this direction. See for instance \cite{Leutheusser:2021frk, Witten:2021unn, Chandrasekaran:2022eqq, Chandrasekaran:2022cip, Leutheusser:2022bgi,Cotler:2023xku}. These subjects fall beyond the scope of the present review and we refer the reader to the papers mentioned and references therein.

\end{enumerate}

Other approaches to string cosmology have been proposed with different levels of development. We would like to note the cosmology of matrix models, which, even though it has not been much explored, matrix models \cite{Banks:1996vh,Ishibashi:1996xs} are one of the few proposals (together with AdS/CFT)  for a non-perturbative formulation of string theory and may deserve further study. We refer to the recent review \cite{Brahma:2022ikl} for the progress and challenges of this approach. 

 \newpage

\section{Outlook}

  Observations in the last few decades  have completely transformed our view of the universe. They  have also raised puzzles on 
  all length scales: What drives acceleration of the present universe? Why are there super-horizon correlations in the CMB? What is dark matter?  The standard model of cosmology, together with inflation, provides a paradigm to accommodate these issues, but does not explain them from a microscopic theory and clearly needs to be put on a more solid theoretical footing. Some of
 these puzzles may have their roots in quantum gravity, and hence cosmology and quantum gravity need to be brought together. 
 
 String theory's remarkable mathematical
 structure has repeatedly shown that it contains all the ingredients needed for a quantum theory of gravity --  ultraviolet finiteness, 
 an understanding of black hole entropy  and explicit realisations of holography. However, string theory is not simply a theory of quantum gravity alone: it also comes with particles
 and interactions. This unison gives it the necessary elements to attack the present day questions in cosmology.  This review
is evidence of the tremendous progress that has been made over the years. At the same time, there remain many outstanding challenges; given the importance of these open questions, string cosmology requires further intense pursuit. 
 
 In terms of formal aspects,  understanding cosmological singularities remains the main open question. The level of progress made on other singular geometries (black holes), both in terms of matching microscopic counts of degrees of freedom with the entropy and also in terms of more quantitative understandings of how information is preserved in the process of Hawking radiation, makes us  optimistic that in the not too far future, a proper understanding of cosmological big-bang singularities may be achieved. We can hope for the same regarding the status of explicit realisations of de Sitter space within string theory. 
 
 It is often stated that string theory is decoupled from observational physics. Indeed, it is not possible to single out an in-out prediction accessible to current technology  that could rule out the theory. On one hand, there are strong indications that the theory is unique. On the other hand, there is an embarrassment of richness  once four-dimensional solutions are considered. Various ingredients (the possibly infinite diversity of compactification manifolds with different curvatures,  the huge number of quantised fluxes for each of these manifolds, the different brane configurations, the quantum corrections to leading order calculations, the possible non-critical solutions) all lead to a large multitude of vacua.
 And yet, despite many efforts, there is not a single string model that can be called fully realistic. Current experimental constraints are already enough to rule out most constructions, either through tests of fifth forces, Standard Model matter constraints, or the existence of relics or exotic particles.
 
 Extracting model independent properties has always been one of the main goals in the effort to confront string theory with data. Particle physics constraints are very important but are intrinsically model dependent as they depend on the
 nature of gauge symmetries, chiral matter content and couplings within a model. Cosmology offers greater chances for universality -- although some aspects are definitely model-dependent, as illustrated by the number of possible string candidates for inflation discussed in chapter 4. It is important to emphasise that each one of these represents a large class of models, as they refer to the nature of the corresponding inflaton candidate, and, in the best scientific tradition, they make concrete predictions that can be compared with experiments. Of the dozen or so scenarios listed, several of them are already in tension with data. This is a reminder that even if data cannot rule out string theory as such, the more standard and less ambitious programme of testing general classes of models is already under way. 
 
At this point we should emphasise that even though many string models of inflation have been constructed, none of them are particularly compelling. Much work therefore remains to be done regarding explicit realisations,  including moduli stabilisation and a potential for realistic matter. This task is closely related to the existence of de Sitter solutions in string theory, which also require careful accounts of moduli stabilisation. Furthermore, it is worth remarking that most alternative scenarios to string inflation include some period of contraction, although a contracting epoch does not exclude a subsequent period of inflation. 

One generic aspect of string theory is the existence of many possible low-energy solutions (often referred to as the string landscape). Associated to the string landscape is the controversial claim that this multiplicity of vacua, together with anthropic arguments, can be used to address the smallness of the cosmological constant, through the population of a vast discretuum of vacua.  Even leaving aside the controversial and unappealing aspects of anthropic arguments, before this approach can be claimed as successful it must also address important issues such as the population of the landscape and the measure problem. It has also been claimed that if our universe is the outcome of a vacuum transition from another universe as predicted by the landscape,  our universe should be that of an open universe.\footnote{This claim has been questioned recently \cite{Hawking:2017wrd,Cespedes:2020xpn}  but more work needs to be done before arriving to a definitive conclusion.} Even though current observations are consistent with a flat universe it may be possible that in not too far a future a non-zero curvature of the universe is determined. If this turns out to be closed rather than open, this would rule out this landscape scenario (if the open universe claim is correct).

One more concrete, but still generic, prediction from string theory is the existence of extra dimensions and the associated moduli fields. These may be heavy if compactifications have no residual supersymmetry, but in supersymmetric compactifications the moduli are typically light and their existence can imply substantial modifications to the epoch between inflation and BBN. We have discussed many implications of moduli, such as
 kination epochs, moduli domination, moduli reheating, dark radiation, and the possible existence of oscillons or oscillatons. We have also discussed possible ways to subject these ideas to potential experimental tests.

One appealing aspect of the swampland and bootstrap programmes is their model independent nature, as they aim to provide general 
constraints, not only on what is achievable within string theory but also on what can hold in any proposed theory of quantum gravity. As we discussed in the previous section, at the moment the associated constraints are limited.  The swampland conjectures on more solid footings 
(such as the absence of global symmetries or the weak gravity conjectures) are the ones with the least phenomenological or cosmological impact, while those with major observational consequences (like the de Sitter or trans-Planckian conjectures) are more speculative. However, even setting aside their implications for potential observations, conceptually these conjectures may help to shape and understand the underlying quantum theory of gravity, string theory or otherwise.  Progress in streamlining these conjectures and constructing
explicit string models in accordance with them (such as those of quintessence) would be most welcome.

For potential observations, just as in particle physics, any single experimental test of a string scenario can probably also be described within a standard effective field theory independent of string theory (as an example, moduli can always be described in quantum field theory simply as scalar particles with non-renormalisable interactions). The best we can expect at this stage is a correlation of different observations of physics beyond the Standard Model. For example, a string model has to explain not only the early universe but also BSM physics. Furthermore, string models of inflation do not end simply with the end of inflation but contain a richness of physics after inflation due to the moduli fields.   
  
 Finally, the detection of gravitational waves has opened a new era in cosmology. Just
 as in  the of case electromagnetic waves, we will eventually study the universe with gravitational waves across a wide range of frequencies.  Hopefully, these truly gravitational probes will give us the hints we need to make the relation between strings and cosmology a quantitative one.
 
 We finish with a quote of Steven Weinberg that is apt for string cosmology: 
 
 \begin{quote}{\it The test of a physical theory is not that everything in it should be observable and every prediction it makes should be testable, but rather that enough is observable and enough predictions are testable to give us confidence that the theory is right.}
 \end{quote} 
   
We hope that this review is found useful in order to identify the main achievements and challenges in this exciting field. Let us also be optimistic
-- and hope that this review is soon rendered out of date by new discoveries and new observations. 
 
 \section{Acknowledgements}
 We thank all of our colleagues  that have shaped our understanding of this field and especially our collaborators on the subjects related to this review, including Shehu Abdussalam, Steve Abel,  Bobby Acharya, Yashar Aghababaie, Gerardo Aldazabal, Rouzbeh Allahverdi, Stephen Angus, Stefan Antusch, Luis Aparicio, Fien Apers, Fabio Apruzzi, Vikas Aragam, Tassos Avgoustidis,
 Vijay Balasubramanian, Arka Banerjee, Luke Barclay, Bruno V.~Bento, Marcus Berg, Per Berglund, Sukannya Bhattacharya,  Johan Bl{\aa}b\"ack, Jose Blanco-Pillado, Ralph Blumenhagen, Andrea Borghese, Philippe Brax, Igor Broeckel,   Cliff Burgess, Nana Cabo-Bizet, Philip Candelas, Alberto Casas, Juan Cascales, Pablo Camara, Francesco Cefala, Sebastian Cespedes, Sarah Chadburn, Dibya Chakraborty, Athanasios Chatzistavrakidis, Roberta Chiovoloni, Debika Chowdhury, David Ciupke, Jim Cline, Katy Clough, Daniel Cremades, Niccol\'o Cribiori, Chiara Crino, Francesc Cunillera, Kumar Das, Subinoy Das, Francesca Day,
  Shanta de Alwis,  Beatriz de Carlos, Xenia de la Ossa, Jean-Pierre Derendinger, Mansi Dhuria, Cristina Diamantini, Victor Diaz, Giuseppe Dibitetto, Kostas Dimopoulos, Eleonora Di Valentino,  Danielle Dineen, Matthew Dolan, Sean D. Downes, Bhaskar Dutta, Koushik Dutta, Damien A.~Easson,  Encieh Erfani, Cristina Escoda, Hassan Firouzjahi, Anamaria Font, Gabriele Franciolini, Mayukh Gangopadhyay, Juan Garc\'a-Bellido, Inaki Garc\'ia Etxebarria, Maria Pilar Garc\'ia del Moral, Fridrik Gautason, Ghazal Geshnizjani, Joaquim Gomes, Nicolas Grandi, Christophe Grojean, Andrew Frey, Diptimoy Ghosh, Steve Giddings, Marta G\'omez-Reino, Mark Goodsell, Ruth Gregory, Veronica Guidetti, Rajesh Gupta, Ulrich Haisch, Ed Hardy, Ehsan Hatefi,  Arthur Hebecker, Johnny Holland, Tristan Hubsch, Janet Hung, Luis Iba\~nez, Joerg Jaeckel, Nicholas Jennings, Esteban Jimenez, Renata Kallosh, Denis Klevers, Tatsuo Kobayashi, Steve Kom, Karta Kooner, Tomi Koivisto, David Kraljic, Sven Krippendorf, Andrei Linde,  Matteo Licheri, Oscar Loaiza-Brito, Ratul Mahanta, Mahbub Majumdar, M.C. David Marsh, Damian Mayorga Pena, Christoph Mayrhofer, Anupam Mazumdar,  Martin Mosny, Sebastian Moster, David F.~Mota, Francesco Muia, Maria Mylova, Peter Nilles, Sirui Ning, Gustavo Niz, Amin Nizami, Detlef Nolte, Carlos N\'u\~nez, Eimear O'Callaghan, Yessenia Olgu\'in-Trejo, Stefano Orani, Ogan Ozsoy, Sonia Paban, Antonio Padilla, Eran Palti, Francisco Pedro, Nicola Pedron, Veronica Pasquarella, Andrew Powell, Cari Powell, Mariano Quiros, Norma Quiroz, Ra\'ul Rabadan, Sa\'ul Ramos-S\'anchez, Govindan Rajesh, Seif Randjbar-Daemi, Filippo Revello, Soo-Jong Rey, Andreas Ringwald, Diederik Roest, Robert Rosati, Esteban Roulet, Markus Rummel, Alberto Salvio, Marco Scalisi,  Andreas Schachner, Jonas Schmidt, Matthias Schmitz, Marco Serra, Kajal Singh, Kuver Sinha, Raffaele Savelli, Ravi Sharma, Pramod Shukla, Aninda Sinha, Yoske Sumitomo, Kerim Suruliz, Gianmassimo Tasinato, Flavio Tonioni, Carlo Trugenberger, Angel Uranga, Gian Paolo Vacca, Vivian Poulin,  Roberto Valandro, Leo Van Nierop, Gonzalo Villa, Alexander Westphal, Matthew Williams, Danielle Wills, Lukas Witkowski, Timm Wrase, Marco Zagermann, Ren-Jie Zhang. 
FQ wants to particularly thank Cliff Burgess for 40 years of very enjoyable collaborations, some of them reported here.
 AM is supported in part by the SERB, DST, Government of India by the grant MTR/2019/000267.
 The work of FQ has been partially supported by STFC consolidated grants ST/P000681/1, ST/T000694/1. JC has been partially supported by STFC consolidated grants ST/P000770/1 and ST/T000864/1.
IZ is partially funded by STFC grant ST/T000813/1 and thanks {\em Fondazione Cassa di Risparmio}, Bologna for financial support during her visit to Bologna, where part of this work was developed.\\

For the purpose of open access,
the authors have applied a Creative Commons Attribution (CC BY) licence to any Author Accepted Manuscript version arising. Data access statement: no new data were generated for this work.

 \newpage

\appendix

\addcontentsline{toc}{section}{References}
\bibliography{refsIZV2}

\end{document}